# Unified Theory of Bivacuum, Particles Duality, Time & Fields

## Bivacuum Mediated Interaction, as a Bridge between Normal and Paranormal

**Alex Kaivarainen**

**http://web.petrsu.ru/ ~alexk**
**H2o@karelia.ru**

**CONTENTS**

## SUMMARY

The concept of Bivacuum is introduced, as a dynamic matrix of the Universe, composed from non-mixing subquantum particles and antiparticles, with dimensions below Planck scale, representing vortical excitations with opposite direction of rotation, zero mass and charge. These basic virtual excitations form Bivacuum dipoles, representing the pairs: donut and anti-donut. Each pair of donuts is strongly correlated torus ($\mathbf{V}^+$) and antitorus ($\mathbf{V}^-$) of positive and negative energy, opposite charge and magnetic moments, separated by quantized energetic gap. Two kinds of such dipoles are named: Bivacuum fermions $\mathbf{BVF}^\uparrow$ and Bivacuum antifermions $\mathbf{BVF}^\downarrow$ with ($\mathbf{V}^+$) and ($\mathbf{V}^-$) both rotating clockwise or counterclockwise, correspondingly. The third kind of dipoles are Bivacuum bosons $\mathbf{BVB}^\pm$, representing intermediate structures between $\mathbf{BVF}^\uparrow$ and $\mathbf{BVF}^\downarrow$ with opposite rotation direction of torus and antitorus. The properties of tori ($\mathbf{V}^+$) and antitori ($\mathbf{V}^-$) in symmetric primordial Bivacuum compensate each other and make Bivacuum neutral with zero energy and charge density. The radiuses of torus and antitorus in symmetric Bivacuum dipoles are equal to each other. Their dimensions may have wide distribution values with most probable Compton radiuses of the regular electron, muon and tauon, correspondingly.

The emission ⇌ absorption of positive and negative Virtual Clouds ($\mathbf{VC}^+$ and $\mathbf{VC}^-$) by torus and antitorus of each Bivacuum dipole as a result of in-phase quantum transition between excited and ground states is accompanied by the in-phase oscillation of Bivacuum gap, separating tori and antitori of this dipoles. Such kind of coherent Bivacuum dipoles dynamics excite virtual pressure waves ($\mathbf{VPW}^\pm$), representing the density of sub-quantum particles oscillation.

The shift of equilibrium between population of Bivacuum fermions and antifermions, for example, in strong electric, magnetic or gravitational field, destroys part of virtual Cooper pairs:

$$[\mathbf{BVF}^\uparrow \bowtie \mathbf{BVF}^\downarrow]$$

If the same fields shift also a symmetry between tori and antitori of unpaired **BVF** or $\mathbf{BVB}^\pm$, this shift change the equality of positive and negative energies of Virtual Pressure Waves and make the difference between them nonzero:

$$\mathbf{\Delta VPW}^\pm = \mathbf{VPW}^+ - \mathbf{VPW}^-$$

This difference, induced by gravitational or magnetic field of the Earth, can be a source of free and pure energy of Bivacuum, which can be used in overunity devices.

Virtual spin waves $\mathbf{VirSW}^{\mathbf{S}=\pm1/2}$, with properties of massless collective Nambu-Goldstone modes, like a real spin waves in condensed matter, are resulted from the oscillation of spin orientation: $\mathbf{S} = +\mathbf{1/2} \rightleftharpoons \mathbf{S} = -\mathbf{1/2}$ of individual Bivacuum fermions and antifermions in Cooper pairs via "flip-flop" mechanism, accompanied by origination of intermediate states -

Bivacuum bosons $\mathbf{BVB}^{\pm}(\mathbf{V}^{+}\Updownarrow\mathbf{V}^{-})$:

$$\mathbf{BVF}^{\uparrow}(\mathbf{V}^{+}\upuparrows \mathbf{V}^{-}) \rightleftharpoons \mathbf{BVB}^{\pm}(\mathbf{V}^{+}\Updownarrow \mathbf{V}^{-}) \rightleftharpoons \mathbf{BVF}^{\downarrow}(\mathbf{V}^{+}\downdownarrows \mathbf{V}^{-})$$

The $\mathbf{VirSW}^{\mathbf{S=\pm 1/2}}$, like so-called torsion field, can serve as a carrier of the phase/spin (angular momentum) and information - qubits, but not the energy. The transmission of VirSW throw the volume of virtual BC domains, even with huge dimensions, can be instant or nonlocal.

The Nonlocality in the volume of real or virtual Bose condensate (**BC**) has been proved theoretically, using the Virial theorem. The Bivacuum, like liquid helium, contains two components: the superfluid with nonlocal properties and normal one, representing fraction of Bivacuum dipoles not involved in virtual **BC**.

In accordance to our model, the mass $\mathbf{m}^{+} = \pm\mathbf{m}_0/\sqrt{1-(\mathbf{v}/\mathbf{c})^2}$ of the actual torus has the conventional Lorenz dependence on the external tangential or translational velocity (**v**). On contrary, the mass $\mathbf{m}^{-}$ of complementary torus has the reciprocal relativistic dependence. As a consequence, the product of the actual and complementary relativistic masses is equal to the rest mass of muon or tauon squared: $\mathbf{m}^{+}\mathbf{m}^{-}= \mathbf{m}_0^2$ and the sum of positive and negative energies of asymmetric torus and antitorus is equal to doubled kinetic energy of corresponding dipole:

$$\mathbf{E}_{V^+} + \mathbf{E}_{V^-} = [\mathbf{m}^{+} - \mathbf{m}^{-}]\mathbf{c}^2 = \mathbf{m}^{+}\mathbf{v}^2$$

*The 1st stage* of elementary particles origination from Bivacuum dipoles is the formation of sub-elementary fermions or antifermions pairs with shape of truncated cones and opposite direction of the apexes. This is a result of Bivacuum fermions and antifermions symmetry shifts towards positive or negative energy, correspondingly, as a result their Cooper pairs $[\mathbf{BVF}^{\uparrow}\bowtie \mathbf{BVF}^{\downarrow}]$ rotation around common axis in absence of translational motion. Due to different relativistic dependencies of torus and antitorus of Bivacuum dipoles: ($\mathbf{V}^{+}$) and ($\mathbf{V}^{-}$) mass on tangential velocity (**v**) of rotation around common axe of rotating Cooper pairs $[\mathbf{BVF}^{\uparrow}\bowtie \mathbf{BVF}^{\downarrow}]$, formed by Bivacuum fermions and antifermions, their symmetry shift is accompanied by uncompensated rest mass and charge origination. It is important, that this asymmetry occur at Golden mean condition when $(\mathbf{v}/\mathbf{c})^2 = \mathbf{0.618}$ (http://arxiv.org/abs/physics/0207027). However, translational velocity on this stage of mass and charge origination is zero. The energy and charge of asymmetric sub-elementary fermions and antifermions in rotating Cooper pairs compensate each other.

*The 2nd stage* of elementary particles or antiparticles origination is a fusion of two triplets, like electron and positron or proton and antiproton from three above mentioned Cooper pairs of corresponding generation of sub-elementary fermions and antifermions. The unpaired sub-elementary fermion or antifermion determines the mass and charge of the whole triplet, as elementary fermion:

$$< [\mathbf{F}_{\uparrow}^{+} \bowtie \mathbf{F}_{\downarrow}^{-}]_{x,y} + \left(\mathbf{F}_{\updownarrow}^{\pm}\right)_z >^{e,p}$$

These triplets are stabilized by three factors: a) the resonance exchange interaction of Bivacuum virtual pressure waves ($\mathbf{VPW}^{\pm}$) with pulsing sub-elementary fermions of Compton frequency; b) the Coulomb attraction between sub-elementary fermions of the opposite charges; c) the gluons (pairs of cumulative virtual clouds in terms of our theory) exchange between sub-elementary fermions (quarks in the case of protons and neutrons).

The fusion of elementary fermions from sub-elementary ones can be accompanied by huge energy release, determined by the value of mass defect, determined by mass difference between sub-elementary fermions/antifermions and the triplets, formed by them as a result of fusion.

New scenario of Big Bang from primordial Bivacuum is based on chain reaction energy of elementary particles fusion The proposed model of elementary particles can be verified experimentally, using 3-beams collider of muons and antimuons. Theory predicts, that the colliding of 2 muons and 1 antimuon at Golden mean velocity: $\mathbf{v} = \sqrt{0.618}\,\mathbf{c}$ should be followed by the electrons origination and kinetic energy outburst, i.e. temperature jump.

It is shown, that the [**corpuscle**(**C**) $\rightleftharpoons$ **wave**(**W**)] duality of fermions is a result of modulation of quantum zero-point frequency of beats between the asymmetric 'actual' (torus) and 'complementary' (antitorus) states of rotating around common axis with tangential velocity $\mathbf{v} = \sqrt{0.618}\,\mathbf{c}$ sub-elementary fermions and antifermions of the triplets $< [\mathbf{F}_{\uparrow}^{+} \bowtie \mathbf{F}_{\downarrow}^{-}]_{x,y} + \left(\mathbf{F}_{\updownarrow}^{\pm}\right)_z >^{e,p}$ by de Broglie wave frequency of these fermions. The resulting energy of these two dynamic processes of quantum beats can be expressed as:

$$\mathbf{E}_{tot}^{res} = \mathbf{h}\nu_{\mathbf{C}\rightleftharpoons\mathbf{W}}^{\mathbf{res}} = (\hbar\omega_0)_{rot}^{in} + \mathbf{h}\nu_B^{ext} = (\mathbf{m}_0\omega_0^2\mathbf{L}_0^2)_{\mathbf{rot}}^{in} + \frac{\mathbf{h}^2}{\mathbf{m}_V^+\lambda_B^2}$$

The external - empirical frequency of de Broglie wave $\nu_B^{ext}$ is equal to the frequency of$[\mathbf{C} \rightleftharpoons \mathbf{W}]$ pulsations of the primary 'anchor' Bivacuum dipole, next to rotating sub-elementary fermion. The anchor dipole represents Bivacuum boson $\mathbf{BVB}^{\pm}(\mathbf{V}^{+} \Updownarrow \mathbf{V}^{-})$ with symmetry shift, provided by translational kinetic energy and momentum of triplets. The $[\mathbf{C}]$ phase of each sub-elementary fermion of triplets exists as a mass, electric and magnetic asymmetric dipoles. The total energy, mass, charge and spin of triplet, moving in space with velocity $(\mathbf{v})$ is determined by the unpaired sub-elementary fermion, since the paired ones in triplets compensate each other.

The boson like photon can be a result of fusion/annihilation of two triplets: [electron + positron], turning two asymmetric triplets to one or more pair of sub-elementary fermions. More common way of photons origination is due to acceleration of elementary charges - triplets, following by sufficient symmetry shift in pairs: $[\mathbf{BVF}^{\uparrow} + \mathbf{BVF}^{\uparrow}]$, interacting by head-to-tail principle in contrast to side-by-side in virtual Cooper pairs $[\mathbf{BVF}^{\uparrow} \bowtie \mathbf{BVF}^{\downarrow}]$. The pair of sub-elementary fermions forming photon with additive spins determines its integer value of spin$(\mathbf{S} = \pm \mathbf{1}\hbar)$. In the case of bosons, like photons, propagating in space with light velocity, the contribution of the rest mass to their energy is zero or very close to zero due absence of tangential velocity in pairs $[\mathbf{BVF}^{\uparrow} + \mathbf{BVF}^{\uparrow}]$, unlike in rotating fast Cooper pairs.

It is shown, that Principle of Least Action and 2nd and 3d laws of thermodynamics is a consequence of forced combinational resonance of elementary particles with basic virtual pressure waves $(\mathbf{VPW}^{\pm})$ of Bivacuum.

It is shown, that the *pace of time* $(\mathbf{dt}/\mathbf{t})$ for any closed *conservative system* is determined by the pace of its kinetic energy change $(-\mathbf{dT}/\mathbf{T}_k)_{x,y,z}$, *anisotropic* in general case (Kaivarainen, 2005; 2006):

$$\left[\frac{\mathbf{dt}}{\mathbf{t}} = \mathbf{d}\ln\mathbf{t} = -\frac{\mathbf{dT}_k}{\mathbf{T}_k} = -\mathbf{d}\ln\mathbf{T}_k\right]_{x,y,z}$$

Similar relation can be obtained from principle of uncertainty for free particle with kinetic energy $(\mathbf{T}_k)$ in coherent form: $\mathbf{T}_k\,\mathbf{t} = \hbar$.

By definition a *conservative system* is a system in which work done by a force is: 1. Independent of path; 2. Completely reversible.

It is important to note, that in closed conservative mechanical or quantum system the total energy is permanent:

$$\mathbf{E}_{tot} = \mathbf{V} + \mathbf{T}_k = const$$

$$or: \Delta\mathbf{E}_{tot} = 0 \quad and \quad \Delta\mathbf{V} = -\Delta\mathbf{T}_k$$

and the *time of action* is always the *external one*.

The time in each *closed conservative system, where* $\mathbf{E}_{tot} = const$, are characterized by its dimensionless *pace of time* and *time* itself:

$$\mathbf{t} = \left[-\frac{\vec{\mathbf{v}}}{\mathbf{d}\vec{\mathbf{v}}/\mathbf{dt}}\,\frac{1-(\mathbf{v}/\mathbf{c})^2}{2-(\mathbf{v}/\mathbf{c})^2}\right]_{x,y,z}$$

In contrast to formula for time from special relativity, our formula in addition to velocity contains also acceleration. The correctness of our formula is confirmed on number of examples.

A theory of Virtual Replica (VR) of macroscopic objects in Bivacuum and primary VR multiplication in space VRM(r) is proposed. The primary Virtual replica of the object can be subdivided on two kinds: the surface $\mathbf{VR}_{sur}$ and the volume $\mathbf{VR}_{vol}$. Their superposition contains the total information about any material object: $\mathbf{VR} = \mathbf{VR}_{sur} \bowtie \mathbf{VR}_{vol}$. The volume $\mathbf{VR}_{vol}$ reflects the internal spatial and dynamic structures in the volume of macroscopic object.

The surface $\mathbf{VR}_{sur}$, like the regular optical hologram, reflects a three-dimensional (3D) shape of the object. It represents a 3D interference pattern of modulated by the surface particles of the object Bivacuum virtual pressure waves $\mathbf{VPW}_m^{\pm}$ (the surface object waves) with basic reference virtual pressure waves of Bivacuum: $\mathbf{VPW}_0^{\pm}$.

The volume $\mathbf{VR}_{vol}$ is a result of 3D interference of modulated by the particles in the object's volume $\mathbf{VPW}_m^{\pm}$, named the volume object waves with all pervading reference virtual waves of Bivacuum $\mathbf{VPW}_0^{\pm}$. The regular optical holograms do not contain information about the internal properties of the object, like the volume Virtual Replica $\mathbf{VR}_{vol}$.
The interference of primary Virtual Replica, located in the volume of the object, depending on time, with Bivacuum reference waves - provides $\mathbf{VR}(\mathbf{t})$ multiplication in space and time $\mathbf{VRM}(\mathbf{r},\mathbf{t})$. To keep the energy, charge and spin conservation, it was assumed, that the mass/energy symmetry shifts of Bivacuum dipoles, involved in $\mathbf{VR}_{sur}$ and $\mathbf{VR}_{vol}$ formation and their multiplication (iteration), should compensate each other. This condition is satisfied if virtual standing waves of multiplicated Virtual Replicas $\mathbf{VRM}(\mathbf{r},\mathbf{t})$ are formed by the Cooper pairs of Bivacuum fermions and antifermions with opposite mass and charge symmetry shifts:

$$\mathbf{VRM}(\mathbf{r},\mathbf{t})_{sur,vol} = \sum N[\mathbf{BVF}^{\uparrow} \bowtie \mathbf{BVF}^{\downarrow}]_{sur,vol}$$

where **N** tends to infinity.
The absolute value of paired dipoles symmetry shift can vary from small to big, reflecting the properties of the object. The Virtual Replica spatial Multiplication $\mathbf{VRM}(\mathbf{r})$ can be named the Holoiteration (in Greece 'holo' means the 'whole' or 'total').
The Virtual Replica Multiplication: $\mathbf{VRM}(\mathbf{r})$ can be a result of linear superposition of primary **VR** of the object with corresponding amplitude of probability $\mathbf{c_n}$, like the orthogonal eigen values of wave function of elementary particle:

$$\mathbf{VRM}(\mathbf{r}) = \sum \mathbf{c_n} \mathbf{VR_n}$$

The condition of orthogonality means that only one of the infinitive numbers of virtual replicas can be displayed at the certain time, depending on the conditions of **VR** detection. The stability of the most probable secondary Virtual Replica as a hierarchical system of virtual standing waves could be responsible for so-called phantom or ghost effect of the object after its destruction or replacing. For individual elementary particles the notion of secondary virtual replica, as a result of primary **VR** multiplication, coincides with notion of secondary 'anchor sites', important for particle jump-way propagation in space (http://arxiv.org/abs/physics/0207027).

The dynamic Virtual Replica Multiplication $\mathbf{VRM}(\mathbf{r},\mathbf{t})$ is a process, filling all the volume around the object with secondary Virtual Replicas. Each selected region of this Holoiteration interference pattern of $\mathbf{VRM}(\mathbf{r},\mathbf{t})$ contains information about the external - shape/surface and the internal - volume properties of macroscopic object changing with time. The possibility of feedback action of Virtual Replica, generated by the object, on physical properties of this object or similar kinds of objects can be tested experimentally.

## Abbreviations and Definitions, Introduced in Unified theory*

- $(\mathbf{V}^+)$ and $(\mathbf{V}^-)$ are correlated actual torus and complementary antitorus (pair of 'donuts') of Bivacuum of the opposite energy, charge and magnetic moment, formed by collective excitations of non mixing subquantum particles and antiparticles of opposite angular momentums;
- $\mathbf{BVF}^{\uparrow} = [\mathbf{V}^+ \upuparrows \mathbf{V}^-]^i$ and $\mathbf{BVF}^{\downarrow} = [\mathbf{V}^+ \downdownarrows \mathbf{V}^-]^i$ are virtual dipoles of three opposite poles: actual (inertial) and complementary (inertialess) mass, positive and negative charge, positive and negative magnetic moments, separated by energetic gap, named Bivacuum fermions and Bivacuum antifermions. The opposite half integer spin $S = \pm\frac{1}{2}\hbar$ of $(\mathbf{BVF}^{\updownarrow})^i$, notated as ($\uparrow$ **and** $\downarrow$), depends on direction of clockwise or counterclockwise in-phase rotation of pairs of [torus $(\mathbf{V}^+) +$ antitorus $(\mathbf{V}^-)$], forming them. The index: $i = e, \mu, \tau$ define the energy and Compton radiuses of $(\mathbf{BVF}^{\updownarrow})^i$ of three electron generations;
- $\mathbf{BVB}^{\pm} = [\mathbf{V}^+ \uparrow\downarrow \mathbf{V}^-]^i$ are Bivacuum bosons, representing the intermediate transition state between Bivacuum fermions of opposite spins: $\mathbf{BVF}^{\uparrow} \rightleftharpoons \mathbf{BVB}^{\pm} \rightleftharpoons \mathbf{BVF}^{\downarrow}$;
- $|\mathbf{m}_V^+|\mathbf{c}^2$ and $|-\mathbf{m}_V^-|\mathbf{c}^2$ are the energies of torus and antitorus of Bivacuum dipoles: $[\mathbf{BVF}^{\updownarrow}]^i_{j,k}$ and $[\mathbf{BVB}^{\pm}]^i_{j,k}$;
- $\mathbf{VC}^+_{j,k} \sim (\mathbf{V}^+_j - \mathbf{V}^+_k)^i$ and $\mathbf{VC}^-_{j,k} \sim (\mathbf{V}^-_j - \mathbf{V}^-_k)^i$ are virtual clouds and anticlouds, composed from subquantum particles and antiparticles, correspondingly. Virtual clouds and anticlouds emission/absorption accompany the correlated transitions between different excitation energy states ($j$ and $k$) of tori $(\mathbf{V}^+_{j,k})^i$ and antitori $(\mathbf{V}^-_{j,k})^i$ of Bivacuum dipoles: $[\mathbf{BVF}^{\updownarrow}]^i_{j,k}$ and $[\mathbf{BVB}^{\pm}]^i_{j,k}$;

- *the notion of positive and negative space* is related to positive and negative Compton radiuses of conjugated torus and antitorus with opposite mass and charge, forming Bivacuum dipoles: $\pm\mathbf{L}_{\mu,\tau} = \frac{\hbar}{\pm m_{\mu,\tau}c}$
- *inter-space waves (ISW)* is the oscillation of gap between positive and negative mass/energy with opposite charges and magnetic moments, in-phase with emission⇌absorption of cumulative virtual clouds $\left[\mathbf{VC}_q^+ \bowtie \mathbf{VC}_q^-\right]^{\mu,\tau}$ by tori and antitori of Bivacuum dipoles.
- $\mathbf{VirP}^{\pm}$ is *virtual pressure* of subquantum particles, accompanied virtual clouds ($\mathbf{VC}_{j,k}^{\pm}$) emission and absorption in the process of torus and antitorus transitions between different *j* and *k* states;
- $\Delta\mathbf{VirP}_{j,k}^{\pm} = |\mathbf{VirP}^{+} - \mathbf{VirP}^{-}|_{j,k..} \sim ||\mathbf{m}_V^+| - |\mathbf{m}_V^-||\mathbf{c}^2 \geq 0$ is excessive virtual pressure, being the consequence of Bivacuum dipoles asymmetry. It determines the *kinetic energy* of Bivacuum dipoles, dependent on tangential or translational velocity;
- $\sum \mathbf{VirP}_{j,k}^{\pm} = (\mathbf{VirP}^{+} + \mathbf{VirP}^{-})_{j,k} \sim |\mathbf{m}_V^+ + \mathbf{m}_V^-|\mathbf{c}^2 > 0$ is a total virtual pressure. It determines the potential energy of Bivacuum;
- $\mathbf{VPW}_{q=1,2..}^{+}$ and $\mathbf{VPW}_{q=1,2..}^{-}$ are the *positive and negative virtual pressure waves,* representing $\mathbf{VirP}_{j,k}^{\pm}$. The polarizations of virtual pressure waves, excited by Bivacuum fermions and antifermions are opposite. In symmetric primordial Bivacuum the energy of these oscillations compensate each other;
- $\mathbf{F}_{\updownarrow}^{+}$ and $\mathbf{F}_{\updownarrow}^{-}$ are sub-elementary *fermions and antifermions* of the opposite charge (+/-) and energy. They emerge due to stable symmetry shift of the *mass and charge* between the *actual* ($\mathbf{V}^{+}$) and *complementary* ($\mathbf{V}^{-}$) torus of $\mathbf{BVF}^{\updownarrow}$ dipoles, providing the rest mass and charge origination: $[\mathbf{m}_V^+ - \mathbf{m}_V^-]^{\phi} = \pm\mathbf{m}_0$ and $[\mathbf{e}_V^+ - \mathbf{e}_V^-]^{\phi} = \pm\mathbf{e}_0$ to the left or right, correspondingly. Their stabilization and fusion to triplets, represented by electrons and protons, is accompanied by big energy release, determined by mass defect, occur when the velocity of rotation of Cooper pairs $[\mathbf{BVF}^{\uparrow} \bowtie \mathbf{BVF}^{\downarrow}]$ around the common axis corresponds to Golden mean: $(\mathbf{v}/\mathbf{c})^2 = 0.618$;
- *Hidden Harmony* condition means the equality of the internal and external group and phase velocities of Bivacuum fermions and Bivacuum bosons: $\mathbf{v}_{gr}^{in} = \mathbf{v}_{gr}^{ext}$; $\mathbf{v}_{ph}^{in} = \mathbf{v}_{ph}^{ext} \equiv \mathbf{v}$. It is proved that this condition is a physical background of Golden mean realization in natural systems: $\boldsymbol{\phi} = (\mathbf{v}^2/\mathbf{c}^2)^{ext,in} = 0.6180339887$;
- $\langle[\mathbf{F}_{\downarrow}^{+} \bowtie \mathbf{F}_{\uparrow}^{-}] + \mathbf{F}_{\updownarrow}^{\pm}\rangle^{e^-,p^+}$ are the coherent triplets of fused sub-elementary fermions and antifermions of $\mu$ and $\tau$ generations, representing the electron/positron or proton/antiproton. In the latter case a sub-elementary fermions and antifermions corresponds to *u* and *d* quarks;
- $\mathbf{CVC}^{+}$ and $\mathbf{CVC}^{-}$ are the *cumulative virtual clouds* of subquantum particles and antiparticles, standing for [W] phase of sub-elementary fermions and antifermions, correspondingly. Quantum beats [**C** ⇌ **W**] between asymmetric torus and antitorus of sub-elementary fermions are accompanied by [emission ⇌ absorption] of $\mathbf{CVC}^{\pm}$. The stability of triplets is determined by resonant exchange interaction between sub-elementary fermions and antifermions with $\mathbf{CVC}^{\pm}$ in the process of their [**Corpuscle** ⇌ **Wave**] pulsations. The virtual pairs $[\mathbf{CVC}^{+} \bowtie \mathbf{CVC}^{-}]_{e,p,n}$ display the gluons (bosons) properties, stabilizing the electrons, protons and neutrons;
- **VirBC** means *virtual Bose condensation* of Cooper - like pairs $[\mathbf{BVF}^{\uparrow} \bowtie \mathbf{BVF}^{\downarrow}]$ and/or $[\mathbf{BVB}^{\pm}]$ with external translational momentum close to zero: $\mathbf{p} \simeq \mathbf{0}$ and corresponding de Broglie wave length close to infinity: $\boldsymbol{\lambda}_B = (\mathbf{h}/\mathbf{p}) \simeq \infty$, providing the nonlocal properties of huge Bivacuum domains;
- $\mathbf{VirSW}^{\pm 1/2}$ are the *Virtual spin waves*, excited as a consequence of angular momentums of cumulative virtual clouds ($\mathbf{CVC}^{\pm}$) of sub-elementary particles in triplets $\langle[\mathbf{F}_{\uparrow}^{-} \bowtie \mathbf{F}_{\downarrow}^{+}] + \mathbf{F}_{\updownarrow}^{\pm}\rangle$ due to angular momentum conservation law. The $\mathbf{VirSW}^{\pm 1/2}$ are highly anisotropic, depending on orientation of triplets in space and their rotational/librational dynamics, being the physical background of torsion field;
- $\mathbf{VirG}_{\mathbf{SME}}^{i}$ is the nonlocal virtual spin-momentum-energy guide (quasi-1D virtual microtubule), formed primarily by standing $\mathbf{VirSW}_S^{S=+1/2} \overset{\mathbf{BVB}^{\pm}}{\underset{\mathbf{BVF}^{\uparrow}\bowtie\mathbf{BVF}^{\downarrow}}{<\!=\!\rightleftharpoons\!=\!>}} \mathbf{VirSW}_R^{S=-1/2}$ of opposite spins and induced self-assembly of Bivacuum bosons $(\mathbf{BVB}^{\pm})^{i}$ or Cooper pairs of $[\mathbf{BVF}^{\uparrow} \bowtie \mathbf{BVF}^{\downarrow}]^{i}$, representing quasi one-dimensional Bose condensate. The bundles of virtual guides $[\mathbf{N}(\mathbf{t},\mathbf{r}) \times \sum \mathbf{VirG}_{SME}(\mathbf{S} <\!=\!=\!> \mathbf{R})]_{x,y,z}^{i}$ connect the remote coherent triplets $\langle[\mathbf{F}_{\uparrow}^{-} \bowtie \mathbf{F}_{\downarrow}^{+}] + \mathbf{F}_{\updownarrow}^{\pm}\rangle^{e,p}$, representing elementary particles, like protons and electrons in free state or in composition of atoms or their coherent groups, providing remote nonlocal interaction - microscopic and macroscopic ones;

- (**mBC**) means *mesoscopic molecular Bose condensate* in the volume of condensed matter with dimensions, determined by the length of 3D standing de Broglie waves of molecules, related to their librations and translations;

- **VR** means three-dimensional (3D) *Virtual Replica* of elementary, particles, atoms, molecules and macroscopic objects, including living organisms. The primary **VR** of macroscopic object is a consequence of complex system of excitations of Bivacuum dipoles. It represents a superposition of Bivacuum virtual standing waves $\mathbf{VPW}_m^{\pm}$ and $\mathbf{VirSW}_m^{\pm 1/2}$, modulated by $[\mathbf{C} \rightleftharpoons \mathbf{W}]$ pulsation of elementary particles and translational and librational de Broglie waves of molecules of macroscopic object. The **VR** can be subdivided on the surface and volume virtual replicas;

- $\mathbf{VRM}^i(\mathbf{r},\mathbf{t})$ means the *primary* **VR** multiplication/iteration in space and time. The infinitive multiplication of primary $\mathbf{VR}^i$ in space in form of 3D packets of virtual standing waves is a result of interference of all pervading external coherent basic *reference waves* - Bivacuum Virtual Pressure Waves $(\mathbf{VPW}_{q=1}^{\pm})^i$ and Virtual Spin Waves $(\mathbf{VirSW}_{q=1}^{\pm 1/2})^i$ with similar kinds of modulated standing waves, like that, forming the primary **VR**. The latter has a properties of the *object waves* in terms of optical holography. Consequently, the **VRM** can be named ***Holoiteration*** by analogy with hologram (in Greece *'holo'* means the 'whole' or 'total'). The spatial $\mathbf{VRM(r)}$ may stand for *remote vision* of psychic. The ability of enough complex system of $\mathbf{VRM(t)}$ to self-organization in nonequilibrium conditions, make it possible multiplication of primary **VR** not only in space but as well, in time in both time direction - positive (evolution) and negative (devolution). The feedback reaction between most probable/stable $\mathbf{VRM(t,r)}$ and nerve system of psychic, including visual centers of brain, can be responsible for *clairvoyance;*

- **The entanglement − channels** are virtual beams, representing multiple correlated bundles of virtual guides $[\mathbf{N}(\mathbf{t},\mathbf{r}) \times \sum \mathbf{VirG}_{SME}\,(\mathbf{S} <\!\!==\!\!> \mathbf{R})]_{x,y,z}^{i}$, connecting coherent elementary particles of the nerve cells of [S]- *psychic* and [R] - *target* in superimposed $\mathbf{VRM}(\mathbf{r},\mathbf{t})_S \bowtie \mathbf{VRM}(\mathbf{r},\mathbf{t})_R$. This combination of Bivacuum mediated interactions (BMI), providing the transmission of not only information, but as well the momentum and energy, can be responsible for *telekinesis and remote healing*;

- **BMI** is a new fundamental *Bivacuum Mediated Interaction,* additional to electromagnetic, gravitational, weak and strong ones. It is a result of superposition of Virtual replicas of Sender [**S**] and Receiver [**R**] in nonequilibrium state, provided by $\mathbf{VRM(r,t)}$ and formation of virtual guides bundles $[\mathbf{N}(\mathbf{t},\mathbf{r}) \times \sum \mathbf{VirG}_{SME}\,(\mathbf{S} <\!\!==\!\!> \mathbf{R})]_{x,y,z}^{i}$ between coherent atoms of [**S**] and [**R**]. Just **BMI** is responsible for remote ultraweak nonlocal interaction between entangled systems and, so-called, paranormal phenomena, which turns to 'normal' in the framework of Unified theory.

*********************************************************************

**The abbreviations are not in alphabetic, but in logical order to make this glossary more useful for perception of new notions, introduced in Unified theory.*

# Introduction

The Dirac's approach to vacuum admits the existence of positive and negative energy (Dirac, 1947). In Dirac's vacuum its realm of negative energy is saturated with infinitive number of electrons. However, it was assumed that these electrons, following Pauli principle, have not any gravitational or viscosity effects. The positrons in his model represent the 'holes', originated in realm of negative energy, as a result of the electrons jumps to realm of positive energy over the energetic gap between two realms: $\Delta = \mathbf{2m_0c^2}$.

Currently it becomes clear, that the Dirac type model of vacuum is not general enough to explain all known experimental data of elementary particles, for example, the bosons emergency.

The model of Bivacuum, described in this book and previous works of this author (Kaivarainen, 1995; 2000; 2004; 2005, 2006, 2008), is more flexible and able to explain the origination of all kind of elementary particles without Higgs bosons. One of the consequences of our model is absence of Dirac's monopole.

Our approach use the same starting point of equal probability of positive and negative energy. However, in contrast to Dirac model, the positive and negative energies are confined in each of Bivacuum elements, named Bivacuum dipoles.

Few modern models of vacuum have common features with our one. The brief description of these models and comparison with Bivacuum concept is presented below.

Aspden (2003) introduced in his ether theory the basic unit, named Quon, as a pair of virtual muons of opposite charges, i.e. [muon + antimuon]. This idea has some common with our model of Bivacuum dipoles. Each Bivacuum dipole represents collective excitations of sub-quantum particles and antiparticles, composing vortical pair: *torus* + *antitorus* of opposite energy/mass, charge and magnetic moments with three Compton radiuses, corresponding to two lepton generation: muon and tauon (Kaivarainen, 2004-2006).

Our notions of strongly correlated torus ($\mathbf{V}^+$) and antitorus ($\mathbf{V}^-$) of Bivacuum dipoles have also some similarity with '*phytons*', introduced by Akimov and Shipov for explanation of torsion field action. After Akimov (1995): "In non polarized condition, physical vacuum contains in each of its elements a *phyton*, which is a kind of circle, rotating in *opposite* directions, corresponding to right and left spin. The spins of phytons are compensated, as far the sum of their angular momentums is zero. This is a reason, why the vacuum does not manifest nonzero angular momentum. But if the spinning object appears in space, then the *phytons*, with axes of rotation, coinciding with that of the object, will keep the same rotation, and phytons which' rotational axes were originally in the opposite direction, will be inverted partly under the influence of the spinning object. As a result, so-called torsion field is exited.

However, this model do not consider the symmetry shift between circles, forming the pairs, and failed to explain the mass and charge origination in contrast to our approach.

Three subclasses of Bivacuum dipoles where introduced: Bivacuum bosons $(\mathbf{BVB}^{\pm})_{S=0}$ with torus and antitorus, rotating in opposite direction, Bivacuum fermions $\mathbf{BVF}^{\uparrow}$ and antifermions $\mathbf{BVF}^{\downarrow}$ with [torus + antitorus] both rotating clockwise or counterclockwise, correspondingly. The latter Bivacuum dipoles of opposite spins are tending to formation of virtual Cooper pairs $[\mathbf{BVF}^{\uparrow} \bowtie \mathbf{BVF}^{\downarrow}]_{S=0,\pm1}$. The ability of Bivacuum dipoles to form virtual Bose condensate from $(\mathbf{BVB}^{\pm})_{S=0}$ and $[\mathbf{BVF}^{\uparrow} \bowtie \mathbf{BVF}^{\downarrow}]_{S=0,\pm1}$ is demonstrated in our theory. The bundles of these self-assemblies of Bivacuum excitations, like vortical structures in superfluid $^4\mathbf{He}$ and $^3\mathbf{He}$, allow consider Bivacuum, as a two-component liquid with superfluid and normal properties.

The superfluid model of vacuum, composed from pairs of fermions of opposite spins and charge where discussed earlier by Sinha et. al., (1976; 1976a; 1978) and later by Boldyreva and Sotina (1999).

In accordance with Planck aether hypothesis of Winterberg (2002), the vacuum is a superfluid, made up of positive and negative Planck mass particles. The Planck mass plasma model makes the following assumptions:

1. The ultimate building blocks are positive and negative Planck mass particles. The interaction obeys the laws of Newtonian mechanics, which can be violated during the collision between a positive and a negative Planck mass particle. These violation means that during the collision between a positive and a negative Planck mass particle, the momentum, not the energy, fluctuates.

2. Planck mass particles of the same sign - repel and those of opposite sign - attract each other, with the magnitude and range of the force equal to the Planck force $\mathbf{M}_p\mathbf{c}^2/\mathbf{R}_P = \mathbf{c}^4/\mathbf{G}$ and the Planck length $\mathbf{R}_P = \mathbf{h}/(\mathbf{M}_p\mathbf{c})$.

3. Space - vacuum is filled with an equal number of positive and negative Planck mass particles whereby each Planck length volume is in the average occupied by one Planck mass particle. The collision of positive and negative Planck mass particles is a source of *zitterbewegung* in Winterberg model of vacuum.

In its ground state, the Planck aether is a two component positive-negative mass superfluid with a phonon - roton energy spectrum for each component. Assuming that the phonon - roton spectrum measured in superfluid helium is universal, this would mean that in the Planck aether this spectrum has the same shape.

Rotons can be viewed as a small vortex rings with the ring radius of the same order as the vortex core radius. A fluid with cavitons is in a state of negative pressure, and the same is true for a fluid with vortex rings. In vortices the centrifugal force creates a vacuum in the vortex core, making a vortex ring to behave like a caviton.

In Winterberg model, in contrast to ours, the positive and negative masses are not considered as a unified mass dipoles with possibility of polarization and symmetry shift. The mechanism of origination of mass, charge, magnetic moment and spin of elementary particles, also where not considered.

Nonetheless of some common features with models of Aspden, Akimov - Shipov's ' and Winterberg, the developed by this author concept of Bivacuum and it elements: Bivacuum bosons ($\mathbf{BVB}^{\pm}$) and fermions ($\mathbf{BVF}^{\updownarrow}$), is more advanced. Our approach explains the origination of *mass and charge* of sub-elementary fermions, as a result of torus $\mathbf{V}^+$ and antitorus $\mathbf{V}^-$ of Bivacuum dipoles symmetry shift, the mechanism of *corpuscle* $\rightleftharpoons$ *wave* pulsation as a result of quantum beats between asymmetric $\mathbf{V}^+$ and $\mathbf{V}^-$ and fusion of elementary particles as a triplets of sub-elementary fermions and antifermions at Golden mean conditions. *In contrast with Standard Model our approach do not need the Higgs bosons for explanation of mass origination.*

The electric, magnetic and gravitational fields are shown to be a result of elastic *recoil* $\rightleftharpoons$ *antirecoil* effects in Bivacuum matrix, accompanied the *corpuscle* $\rightleftharpoons$ *wave* pulsation of elementary particles. So called *zitterbewegung* is a result of elementary particles vibrations, accompanied their $\mathbf{C} \rightleftharpoons \mathbf{W}$ pulsation.

In the framework of our approach all fundamental physical phenomena are hierarchically interrelated and unified. This new approach to UNIFICATION in physics includes also a new fundamental *Bivacuum mediated interaction*, filling a logical gap between normal and paranormal phenomena by extending the limits of existing paradigm.

David Bohm was the first one, who made an attempt to explain the wholeness of the Universe, without loosing the causality principle. Experimental discovery of Aharonov-Bohm effect (1950), pointing that the electron is able to "feel" the presence of a magnetic field even in a regions where the probability of field is zero, was stimulating. For explanation of nonlocality Bohm introduced in 1952 the notion of *quantum potential*, which pervaded all of space. But unlike gravitational and electromagnetic fields, its influence did not decrease with distance. All the particles are interrelated by very sensitive to any perturbations quantum potential. This means that signal transmission between particles may occur instantaneously. The idea of *quantum potential or active information* is close to notion of *pilot wave*, proposed by de Broglie at the Solvay Congress in 1927. In fact, Bohm develops the de Broglie idea of pilot wave, applying it for many-body system.

In 1957 Bohm published a book: Causality and Chance in Modern Physics. Later he comes to conclusion, that Universe has a properties of giant, flowing hologram. Taking into account its dynamic nature, he prefer to use term: *holomovement*. In his book: Wholeness and the Implicate Order (1980) Bohm develops an idea that our *explicated unfolded reality is a product of enfolded (implicated) or hidden order of existence. He consider the manifestation of all forms in the universe, as a result of enfolding and unfolding exchange between two orders, determined by super quantum potential.* This idea corresponds well to our dynamic model of *Corpuscle* $\rightleftharpoons$ *Wave* pulsation, accompanied by reversible emission and absorption of Cumulative virtual clouds $\mathbf{CVC}^{\pm}$ and excitation of Virtual Pressure Waves ($VPW^{\pm}$), propagating in space.

In book, written by D. Bohm and B. Hiley (1993): ”THE UNDIVIDED UNIVERSE. An ontological interpretation of quantum theory” the electron is considered, as a particle with well- defined position and momentum which are, however, under influence of special wave (quantum potential). Elementary particle, in accordance with these authors, is a *sequence of incoming and outgoing waves*, which are very close to each other. However, particle itself does not have a wave nature. Interference pattern in double slit experiment after Bohm is a result of periodically ”bunched” character of quantum potential.

After Bohm, the manifestation of corpuscle - wave duality of particle is dependent on the way, which observer interacts with a system. He claims that both of this properties are *always* enfolded in particle. *It is a basic difference with our model, assuming that the wave and corpuscle phase are realized alternatively with high frequency during two different semiperiods of sub-elementary particles, forming particles of quantum beats between sublevels of positive (actual) and negative (complementary) energy. This frequency is amplitude and phase modulated by experimentally revealed de Broglie wave of particles, determined by their external translational velocity and momentum.*

The important point of Bohmian philosophy, coinciding with our concepts of Bivacuum, elementary particles and Virtual Replica, is that everything in the Universe is a part of dynamic continuum. Neurophysiologist Karl Pribram does made the next step in the same direction as Bohm: *”The brain is a hologram enfolded in a holographic Universe”.*

The good popular description of Bohm and Pribram ideas are presented in books: ”The Bell’s theorem and the curious quest for quantum reality” (1990) by David Peat and ”The Holographic Universe” (1992) by Michael Talbot. Such original concepts are interesting and stimulating, indeed, but should be considered as a first attempts to transform intuitive perception of duality and quantum wholeness into clear geometrical and mathematical models.

Some common features with our model of duality has a Unitary Quantum Theory (UQT), proposed by Sapogin (1982). In the UQT any elementary particle is not a point and source of field like in the ordinary quantum mechanics, but represents a wave packet of a certain unified field (Sapogin and Boichenko, 1991). The dispersion equation of such a nonlinear field turned out to be such, that the wave packet (particle) during its movement periodically appears and disappears, and the envelope of this process coincides with the de Broglie wave. The periodic disappearance (spreading in the Universe) and repeated appearance of numerous particles - represent vacuum fluctuations. The corresponding transversal self-focusing of the wave packet is possible only in conditions if the refraction index of space/vacuum is dependent of particle velocity. The square of wave packet describes the oscillating charged particle mass and energy (Sapogin, et.al., 2002), following the conventional Newton equations. The essential negative consequence of UQT is the absence of the energy and the momentum conservation laws for single particles.

In 1950 John Wheeler and Charles Misner published Geometrodynamics, a new description of space-time properties, based on topology. Topology is more general than Euclidean geometry and deeper than non-Euclidean, used by Einstein in his General theory of relativity. Topology does not deal with distances, angles and shapes. Drawn on a sheet of stretching rubber, a circle, triangle and square are indistinguishable. A ball, pyramid and a cube also can be transformed into the other. However, objects with holes in them can never be transformed by stretching and deforming into objects without holes. For example black hole can be described in terms of topology. It means that massive rotating body behave as a space-time hole. Wheeler supposed that *elementary particles and antiparticles, their spins, positive and negative charges can be presented as interconnected black and white holes.* Positron + electron pair correspond to such model. The energy, directed to one of the hole, goes throw the connecting particles and antiparticles tube -"handle" and reappears at the other. The connecting tube exist in another space-time than holes itself. Such a tube is undetectable in normal space and the process of energy transmission looks as instantaneous. In conventional space-time two ends of tube, termed 'wormholes' can be a vast distant apart. This model gives one of possible explanations of quantum nonlocality.

The same is true for introduced in our theory nonlocal Virtual spin-momentum-energy guides (**VirG**$_{SME}$). The **VirG**$_{SME}$, formed by virtual Cooper pairs of Bivacuum fermions, may connect a Cooper pairs of real particles: electrons, protons and neutrons with opposite spins and counterphase Corpuscle ⇌ Wave pulsation.

Sidharth (1998, 1999) considered *elementary particle as a relativistic vortex of Compton radius, from which he recovered its mass and quantized spin* ($s = \frac{1}{2}\hbar$). He pictured a particle as a fluid vortex steadily circulating with light velocity along a 2D ring or spherical 3D shell with radius

$$L = \frac{\hbar}{2mc} \qquad 1$$

Inside such vortex the notions of negative energy, superluminal velocities and nonlocality are acceptable without contradiction with conventional theory.

Bohm's hydrodynamic formulation and substitution of wave function with

$$\psi = \mathrm{R}e^{iS} \qquad 2$$

where $R$ and $S$ are real function of space $\vec{r}$ and time $t$, transforms the Schrödinger equation to

$$\frac{\partial\rho}{\partial t} + \vec{\nabla}(\rho\vec{\mathbf{v}}) = 0 \quad or: \qquad 3$$

$$\hbar\frac{\partial S}{\partial t} + \frac{\hbar^2}{2m}(\vec{\nabla}S)^2 + V = \frac{\hbar^2}{2m}(\nabla^2 R/R) \equiv Q \qquad 4$$

where: $\rho = R^2$; $\vec{\mathbf{v}} = \frac{\hbar^2}{2m}\vec{\nabla}S$ *and* $Q = \frac{\hbar^2}{2m}(\nabla^2 R/R)$

Sidharth comes to conclusion that the energy of nonlocal quantum potential ($Q$) is determined by inertial mass ($m$) of particle:

$$Q \equiv -\frac{\hbar^2}{2m}(\nabla^2 R/R) = mc^2 \qquad 5$$

He treated also a charged Dirac fermions, as a Kerr-Newman black holes. Within the region of Compton vortex the superluminal velocity and negative energy are possible after Sidharth. If measurements are averaged over time $t \sim mc^2/\hbar$ and over space $L \sim \hbar/mc$, the imaginary part of particle's position disappears and we are back in usual Physics (Sidharth, 1998).

Barut and Bracken (1981) considered *zitterbewegung* - rapidly oscillating imaginary part of particle position, leading from Dirac theory (1947), as a harmonic oscillator in the Compton wavelength region of particle.

The Einstein (1971, 1982) and Shrödinger (1930) also spoke about oscillation of the electron with frequency: $\mathbf{\nu} = \mathbf{m_0 c}^2/h$ and the amplitude: $\boldsymbol{\zeta}_{\max} = \hbar/(\mathbf{2mc})$. It was demonstrated by Shrödinger, that position of free electron can be presented as: $\mathbf{x} = \overline{\mathbf{x}} + \boldsymbol{\zeta}$, where $\overline{\mathbf{x}}$ characterize the average position of the free electron, and $\zeta$ its instant position, related to its oscillations.

Hestness (1990) proposed, that *zitterbewegung* arises from self-interaction, resulting from wave-particle duality.

This ideas are close to our explanation of elementary particles zero-point oscillations and *zitterbewegung*, as a recoil ⇌ antirecoil vibrations, accompanied particles corpuscle ⇋ wave pulsations. Corresponding oscillations of each particle's kinetic energy, in accordance to our theory of time (Kaivarainen, 2005, 2006), is related with oscillations of *internal* time for this system. We came here to concept of *space-time-energy discreet trinity*, generated by corpuscle – wave duality.

Serious attack on problem of quantum nonlocality was performed by Roger Penrose (1989) with his twister theory of space-time. After Penrose, quantum phenomena can generate space-time. The twisters, proposed by him, are lines of infinite extent, resembling twisting light rays. Interception or conjunction of twisters lead to origination of particles. In such a way the local and nonlocal properties and particle-wave duality are interrelated in twisters geometry.

The analysis of main paradoxes in quantum mechanics was presented by Asher Peres (1992) and Charles Bennett et. al., (1993).

In our Unified model, the *local* properties of sub-elementary particles are resulted from their tori and antitori symmetry shift, accompanied by uncompensated mass and charge origination.

The *nonlocal* interaction of two or more particles of the same kinds (photons, electrons, protons, neutrons), providing the entanglement, can be mediated by virtual guides ($\mathbf{VirG}_{SME}$) of spin, momentum and energy, connecting these particles with close frequency of counterphase [**C** ⇌ **W**] pulsation (Kaivarainen, 2006 a,b).

The approach developed by Daniel Dubois (1999) using computational derivation of quantum relativistic systems with forward-backward space-time shifts, led to some results, similar to ours (Kaivarainen, 1995, 2001, 2003, 2004, 2006). For example, the group and phase masses, introduced by Dubois, related to internal group and phase velocities, has analogy with actual and complementary masses, introduced in our Unified theory (UT). In both approaches, the product of these masses is equal to the particle's rest mass squared. The notion of discrete time interval, used in Dubois approach corresponds to period of [$C \rightleftharpoons W$] pulsation of sub-elementary particles, determined by their rest mass. The positive internal time interval, in accordance to our model, corresponds to forward $C \to W$ transition, accompanied by the actual mass and kinetic energy decreasing and the negative one to the backward $W \to C$ transition, i.e. increasing of particle mass and kinetic energy.

Puthoff (2001) developed the idea of 'vacuum engineering', using hypothesis of polarizable vacuum (PV). The electric permittivity ($\boldsymbol{\varepsilon}_0$) and magnetic permeability ($\boldsymbol{\mu}_0$) is interrelated in 'primordial' symmetric vacuum, as: $\boldsymbol{\varepsilon}_0\boldsymbol{\mu}_0 = \mathbf{1/c^2}$. It is shown that changing of vacuum refraction index: $\mathbf{n} = \mathbf{c/v} = \boldsymbol{\varepsilon}^{1/2}$ , for example in gravitational or electric fields, is accompanied by variation of lot of space-time parameters.

Fock (1964) and Puthoff (2001), explained the bending of light beam, induced by gravitation near massive bodies also by vacuum refraction change, i.e. in another way, than General theory of relativity. However, the mechanism of vacuum polarization and corresponding refraction index changes in electric and gravitational fields remains obscure. Our Unified theory of Bivacuum, matter and field propose such mechanism.

The transformation of neutron to proton and electron, in accordance to Electro - Weak (EW) theory, developed by Glashov (1961), Weinberg (1967) and Salam (1968), is mediated by negative *massless* $W^-$ boson. The reverse reaction in EW theory: proton $\to$ neutron is mediated by positive *massless* $W^+$ boson. Scattering of the electron on neutrino, not accompanied by charge transferring, is mediated by third *massless* neutral boson $Z^0$.

In (EW) theory the Higgs field was introduced for explanation of spontaneous symmetry violation of intermediate vector bosons: charged $W^\pm$ and neutral $Z^0$ with spin 1, accompanied by origination of big mass of these particles. The EW theory needs also the quantum of Higgs field, named Higgs bosons with big mass, zero charge and integer spin.

The fusion of Higgs bosons with $W^\pm$ and $Z^0$ particles is accompanied by increasing of their mass up to 90 mass of protons. The experimental discovery of heavy $W^\pm$ and $Z^0$ particles in 1983 after their separation, accompanied getting the system a big external energy, was considered as a conformation of EW theory.

The spontaneous symmetry violation of vacuum, in accordance to Goldstone theorem, turns two virtual particles with imaginary masses ($i\mathbf{m}$) to one real particle with mass: $\mathbf{M}_1 = \sqrt{2}\,\mathbf{m}$ and one real particle with zero mass: $\mathbf{M}_2 = \mathbf{0}$. However, the Higgs field and Higgs bosons are still not found.

"We have eliminated most of hunting area", confirms Neil Calder from CERN recently. *This author propose another explanation of mass and charge origination.*

In conventional approach, described above, two parameters of $\mathbf{W}^\pm$ particles: charge and mass are considered, as independent. In our new approach they are interrelated in certain way.

Thomson, Heaviside and Searl supposed that mass is an electrical phenomena. In theory of Haisch, Rueda and Puthoff (1994), Rueda and Haish (1998) it was proposed, that the inertia is a reaction force, originating in a course of dynamic interaction between the electromagnetic zero-point field (ZPF) of vacuum and charge of elementary particles. However, it's not clear in this approach, how the charge itself originates.

Our Unified theory is an attempt to unify mass and charge with magnetic moment, spin and symmetry shift of sub-elementary fermions, induced by external translational-rotational motion. This theory unifies the origination of elementary particles, their rest mass and charge, electromagnetism and gravitation with particles corpuscle-wave duality, standing also for their zero-point oscillations. In accordance to formalism of our theory, the rest mass and charge of elementary fermions origination are both the result of Bivacuum fermions (BVF) symmetry shift, corresponding to Golden mean conditions, i.e. equality of the ratio of external tangential velocity of Cooper pairs of Bivacuum fermions and antifermions rotation, or of translational velocity of Bivacuum bosons to light velocity squared to: $(\mathbf{v/c})^2 = 0.618 = \phi$. At this condition the asymmetric Bivacuum dipoles turns to sub-elementary fermions.

The electric, magnetic and gravitational fields are the result of huge number of Bivacuum dipoles symmetry shift oscillation, excited by *recoil⇌ antirecoil* dynamics, accompanied the corpuscle ⇌ wave pulsation of sub-elementary fermions and antifermions, forming the elementary particles and rotation of elementary fermions (triplets) as a whole, depending on direction of triplets propagation.

In our approach, the resistance of particle to acceleration (i.e. inertia force), proportional to its mass (second Newton's law) is a consequence of resistance of frequency of particle's **C** ⇌ **W** pulsation to change, keeping the equilibrium (tuned state) with frequency of surrounding Bivacuum dipoles symmetry - energy oscillation. We named this resistance to equilibrium shift between dynamics of particles and dynamics of Bivacuum - *"The generalized principle of Le Chatelier's"*.

In contrast to nonlocal Mach's principle, our theory of particle - Bivacuum interaction explains the existence of inertial mass of even single particle in the empty Universe.

*The main goals and achievements of our work can be formulated as follows:*

1. Development of superfluid Bivacuum model, as the dynamic matrix of dipoles, formed by pairs of virtual tori and antitori of the opposite energy/mass, charge and magnetic moments, compensating each other. The explanation of fusion of the fermions, like electrons, protons and neutrons from asymmetric Bivacuum sub-elementary fermions/antifermions, representing muons/antimuons ($\mu$) and tauons/antitauons. ($\tau$). The *external* properties of such elementary particles are still described by the existing formalism of quantum mechanics and Maxwell equations, in contrast to new formalism for internal properties of sub-elementary and elementary particles, proposed in our Unified theory;

2. Finding the equations, unifying the internal (hidden) and external parameters of sub-elementary particles. Derivation of the conditions of the rest mass and charge of sub-elementary fermions and antifermions (muons and tauons) origination. Proving that it happens at certain symmetry shift between torus and antitorus of corresponding generation of Bivacuum fermions, determined by Golden mean conditions. Finding out the fusion conditions of sub - elementary fermions and antifermions to the triplets - elementary fermions. Understanding the mechanisms of triplets stabilization;

3. Explanation of the absence of Dirac monopole in Nature;

4. Development of the dynamic model of wave-corpuscle duality of sub-elementary particles/antiparticles, forming elementary particles and antiparticles;

7. The mechanism of elementary particles jump-way propagation in space;

8. Unification of the Principle of least action, the time, the 2nd and 3d laws of thermodynamics with Principle of least action and influence of Bivacuum virtual pressure waves (**VPW**$^{\pm}$), on the dynamics of elementary particles and trajectory of their propagation in space;

5. The mechanism of elementary particles jump-way propagation in space;

6. Unification of the Principle of least action, the time, the 2nd and 3d laws of thermodynamics with Principle of least action and influence of Bivacuum virtual pressure waves (**VPW**$^{\pm}$), on the dynamics of elementary particles and trajectory of their propagation in space;

7. New theory of time for conservative systems, when total energy of system is permanent and the absolute values of variations of kinetic and potential energies are equal to each other. Derivation of formula for time, containing not only velocity, but also acceleration;

8. The interrelation of zero-point oscillations and recoil⇌antirecoil effects, accompanied the [**Corpuscle** ⇌ **Wave**] pulsation of fermions, with electric, magnetic and gravitational fields excitation. The electromagnetic field, as a consequence of photons, as a pair of sub-elementary fermions with additive spins [**C** $\rightleftharpoons$ **W**] pulsation;

9. Theory of Virtual Replica (**VR**) of any material object and its multiplication in space and time: $\mathbf{VRM(r,t)}$, as a consequence of superposition of the *reference* Bivacuum virtual pressure waves ($\mathbf{VPW}^{\pm}_{q=1}$) with the *object* virtual waves ($\mathbf{VPW}^{\pm}_{\mathbf{m}}$) representing $\mathbf{VPW}^{\pm}_{q=1}$, modulated by de Broglie waves of particles (nucleons), forming this object;

10. Working out the new mechanism of Bivacuum mediated nonlocal interaction between remote coherent mesoscopic and macroscopic systems as a result of superposition of their $\mathbf{VRM(r,t)}$ and via Virtual guides of spin, momentum and energy ($\mathbf{VirG}_{S,M,E}$), assembling to virtual bundles, connecting remote protons and neutrons with the same frequency of [**Corpuscle** $\rightleftharpoons$ **Wave**] pulsation;

11. Explanation of 'paranormal' results of Kozyrev's, Shnoll and Tiller experiments and the mechanism of overunity devices action and other "paranormal" phenomena, extending the limits of mainstream paradigm.

## 1. The Bivacuum Concept

Our Bivacuum concept, like Paul Dirac theory of vacuum, admit the equal probability of positive and negative energy. The Unified theory (UT) represents the efforts of this author to create the Hierarchical picture of the World, starting from postulated specific Bivacuum superfluid matrix properties, providing the elementary particles origination and their **Corpuscle** $\rightleftharpoons$ **Wave** pulsation, exiting the fields.

The Bivacuum is introduced, as a dynamic matrix of the Universe, composed from non mixing subquantum particles and antiparticles. The subquantum particles and antiparticles are considered, as the minimum stable vortical structures of Bivacuum with dimensions less than Planck length: $1.61605 \times 10^{-35}$ m, when the quantization and quantum mechanics do not work anymore.

In accordance to our conjecture, the subquantum particles and antiparticles have zero mass and charge and opposite directions of rotation - clockwise and counterclockwise. Their spontaneous collective paired vortical excitations (self-organization) follows by Bivacuum dipoles origination in form of strongly correlated pairs:
$\left[ tori\ (\mathbf{V}^+) + antitori\ (\mathbf{V}^-) \right]$, separated by energetic gap.

The 2-dimensional tori and antitori in primordial Bivacuum are characterized by the opposite mass, charge and magnetic moments, compensating each other and making Bivacuum neutral with zero energy and charge density. The 3-d dimension of the space is provided by the axis of energetic gap between $(\mathbf{V}^+)$ and $(\mathbf{V}^-)$, normal to plane of tori and antitori.

The most probable radiuses of tori and antitori of symmetric primordial Bivacuum dipoles are equal to each other and determined by Compton radiuses of *muons* ($\mu$) *and tauons* ($\tau$).

However, the distribution of radiuses of Bivacuum dipoles interrelated with separation between tori and antitori can be very wide without violation of symmetry and energy conservation. In special case the radiuses of correlated tori and antitori of pairs can be much bigger, corresponding, for example, the Compton radius of regular electron or positron.

Three kinds of Bivacuum dipoles are named Bivacuum fermions, antifermions and Bivacuum bosons.

The infinitive number of Bivacuum fermions and antifermions: $\mathbf{BVF}^{\uparrow} \equiv [\mathbf{V}^+ \upuparrows \mathbf{V}^-]^{\mu,\tau}$, $\mathbf{BVF}^{\downarrow} \equiv [\mathbf{V}^+ \downdownarrows \mathbf{V}^-]^{\mu,\tau}$ and Bivacuum bosons: $\mathbf{BVB}^{\pm} \equiv [\mathbf{V}^+ \uparrow\downarrow \mathbf{V}^-]^{\mu,\tau}$, as intermediate state between $\mathbf{BVF}^{\uparrow}$ and $\mathbf{BVF}^{\downarrow}$ fill all 'empty' space and free space in the atoms and molecules.

The virtual Cooper pairs of Bivacuum fermions and antifermions:

$$\left\{\mathbf{BVF}^{\uparrow} \equiv [\mathbf{V}^{+} \upuparrows \mathbf{V}^{-}]^{\mu,\tau} \bowtie \mathbf{BVF}^{\downarrow} \equiv [\mathbf{V}^{+} \downdownarrows \mathbf{V}^{-}]^{\mu,\tau}\right\} \qquad 1.1$$

may form huge superfluid domains of virtual Bose condensate with dimensions, determined by translational de Broglie wave length of virtual pairs, which tends to infinitive in primordial Bivacuum of the ideal symmetry and minimum translational dynamics.

The Bivacuum boson $\mathbf{BVB}^{\pm}$ has two polarization, depending on direction of rotation of its torus and antitorus as respect to direction of translational propagation in space or external field gradient:

$$\mathbf{BVB}^{+} = [\mathbf{V}^{+} \uparrow\downarrow \mathbf{V}^{-}]^{\mu,\tau} \qquad 1.2$$

$$\mathbf{BVB}^{-} = [\mathbf{V}^{+} \downarrow\uparrow \mathbf{V}^{-}]^{\mu,\tau} \qquad 1.2a$$

The $\mathbf{BVB}^{\pm}$ is the short living intermediate state in the process of interconversions between Bivacuum fermions and antifermions in virtual Cooper pairs:

$$[\mathbf{V}^{+} \upuparrows \mathbf{V}^{-}]^{\mu,\tau} \rightleftharpoons \mathbf{BVB}^{\pm} \rightleftharpoons [\mathbf{V}^{+} \downdownarrows \mathbf{V}^{-}]^{\mu,\tau} \qquad 1.3$$

At certain conditions $\mathbf{BVB}^{+}$ and $\mathbf{BVB}^{-}$ can be assembled in 'virtual trains' propagating in opposite directions and carrying the momentum and energy (see Fig.1).

The *positive and negative energies of torus and antitorus* ($\pm\mathbf{E}_{\mathbf{V}^{\pm}}$) of two lepton generations ($i = \mu,\tau$), interrelated with their radiuses ($\mathbf{L}^{n}_{\mathbf{V}^{\pm}}$), are quantized as quantum harmonic oscillators of opposite energies:

$$[\mathbf{E}^{n}_{\mathbf{V}^{\pm}} = \pm\mathbf{m}_0\mathbf{c}^2(\frac{1}{2}+\mathbf{n}) = \pm\hbar\omega_0(\frac{1}{2}+\mathbf{n})]^{i} \qquad \mathbf{n} = 0,1,2,3\ldots \qquad 1.4$$

$$or: \left[\mathbf{E}^{n}_{\mathbf{V}^{\pm}} = \frac{\pm\hbar\mathbf{c}}{\mathbf{L}^{n}_{\mathbf{V}^{\pm}}}\right]^{i} \qquad where: \left[\mathbf{L}^{n}_{\mathbf{V}^{\pm}} = \frac{\pm\hbar}{\pm\mathbf{m}_0\mathbf{c}(\frac{1}{2}+\mathbf{n})} = \frac{\mathbf{L}_0}{\frac{1}{2}+\mathbf{n}}\right]^{i} \qquad 1.4a$$

where: $[\mathbf{L}_0 = \hbar/\mathbf{m}_0\mathbf{c}]^{e,\mu,\tau}$ can be a Compton radii of the electron of corresponding lepton generation, i.e. regular electron, muon and tauon ($i = e,\mu,\tau$) and $\mathbf{L}^{e}_0 >> \mathbf{L}^{\mu}_0 > \mathbf{L}^{\tau}_0$. The Bivacuum fermions $(\mathbf{BVF}^{\updownarrow})^{\mu,\tau}$ with smaller Compton radiuses can be located inside the bigger ones $(\mathbf{BVF}^{\updownarrow})^{e}$.

The absolute values of increments of torus and antitorus energies ($\Delta\mathbf{E}^{i}_{\mathbf{V}^{\pm}}$), interrelated with increments of their radii ($\Delta\mathbf{L}^{i}_{\mathbf{V}^{\pm}}$) in primordial Bivacuum (i.e. in the absence of matter and field influence), resulting from in-phase symmetric fluctuations are equal:

$$\Delta\mathbf{E}^{i}_{\mathbf{V}^{\pm}} = -\frac{\hbar c}{\left(\mathbf{L}^{i}_{\mathbf{V}^{\pm}}\right)^2}\Delta\mathbf{L}^{i}_{\mathbf{V}^{\pm}} = -\mathbf{E}^{i}_{\mathbf{V}^{\pm}}\frac{\Delta\mathbf{L}^{i}_{\mathbf{V}^{\pm}}}{\mathbf{L}^{i}_{\mathbf{V}^{\pm}}} \qquad or: \qquad 1.5$$

$$-\Delta\mathbf{L}^{i}_{\mathbf{V}^{\pm}} = \frac{\pi\left(\mathbf{L}^{i}_{\mathbf{V}^{\pm}}\right)^2}{\pi\hbar c}\Delta\mathbf{E}^{i}_{\mathbf{V}^{\pm}} = \frac{\mathbf{S}^{i}_{\mathbf{BVF}^{\pm}}}{2h\mathbf{c}}\Delta\mathbf{E}^{i}_{\mathbf{V}^{\pm}} = \mathbf{L}^{i}_{\mathbf{V}^{\pm}}\frac{\Delta\mathbf{E}^{i}_{\mathbf{V}^{\pm}}}{\mathbf{E}^{i}_{\mathbf{V}^{\pm}}} \qquad 1.5a$$

where: $\mathbf{S}^{i}_{\mathbf{BVF}^{\pm}} = \pi\left(\mathbf{L}^{i}_{\mathbf{V}^{\pm}}\right)^2$ is a square of the cross-section of torus and antitorus, forming Bivacuum fermions ($\mathbf{BVF}^{\updownarrow}$) and Bivacuum bosons ($\mathbf{BVB}^{\pm}$).

The virtual *mass*, *charge* and *magnetic moments* of torus and antitorus of $\mathbf{BVF}^{\updownarrow}$ and $\mathbf{BVB}^{\pm}$ are opposite and in symmetric *primordial* Bivacuum compensate each other in their basic ($\mathbf{n} = \mathbf{0}$) and excited ($\mathbf{n} = \mathbf{1},\mathbf{2},\mathbf{3}\ldots$) states.

The Bivacuum 'atoms': $\mathbf{BVF}^{\updownarrow} = [\mathbf{V}^{+} \Updownarrow \mathbf{V}^{-}]^{i}$ and $\mathbf{BVB}^{\pm} = [\mathbf{V}^{+}\uparrow\downarrow \mathbf{V}^{-}]^{i}$ represent dipoles of three different poles - the mass $(\mathbf{m}_V^+ = |\mathbf{m}_V^-| = \mathbf{m}_0)^i$, electric ($e_+$ and $e_-$) and magnetic ($\boldsymbol{\mu}_+$ and $\boldsymbol{\mu}_-$) dipoles.

The torus and antitorus $(\mathbf{V}^+ \Updownarrow \mathbf{V}^-)^i$ of Bivacuum fermions of opposite spins $\mathbf{BVF}^\uparrow$ and $\mathbf{BVF}^\downarrow$ are both rotating in the same direction: clockwise or counterclockwise. This determines the positive and negative spins ($\mathbf{S} = \pm\mathbf{1/2}\hbar$) of Bivacuum fermions. Their opposite spins may compensate each other, forming virtual Cooper pairs: $[\mathbf{BVF}^\uparrow \bowtie \mathbf{BVF}^\downarrow]$ with neutral boson properties.

The *energy gap* between the torus and antitorus of symmetric $(\mathbf{BVF}^\updownarrow)^i$ or $(\mathbf{BVB}^\pm)^i$ is:

$$[\mathbf{A}_{BVF} = \mathbf{E}_{\mathbf{V}^+} - (-\mathbf{E}_{\mathbf{V}^-}) = \hbar\boldsymbol{\omega}_0(1+2\mathbf{n})]^i = \mathbf{m}_0^i\mathbf{c}^2(1+2\mathbf{n}) = \frac{h\mathbf{c}}{[\mathbf{d}_{\mathbf{V}^+\Updownarrow\mathbf{V}^-}]_n^i} \qquad 1.6$$

where the characteristic separation (distance) between torus $(\mathbf{V}^+)^i$ and antitorus $(\mathbf{V}^-)^i$ of Bivacuum dipoles *(gap dimension)* is a quantized parameter:

$$[\mathbf{d}_{\mathbf{V}^+\Updownarrow\mathbf{V}^-}]_n^i = \frac{h}{\mathbf{m}_0^i\mathbf{c}(1+2\mathbf{n})} \qquad 1.7$$

From (1.6) and (1.2a) we can see, that at $\mathbf{n} \to \mathbf{0}$, the energy gap $\mathbf{A}_{BVF}^i$ is decreasing till $\hbar\boldsymbol{\omega}_0 = \mathbf{m}_0^i\mathbf{c}^2$ and the spatial separation between torus and antitorus $[\mathbf{d}_{\mathbf{V}^+\Updownarrow\mathbf{V}^-}]_n^i$ is increasing.

On the contrary, the infinitive symmetric excitation of torus and antitorus is followed by tending the spatial gap between them to zero: $[\mathbf{d}_{\mathbf{V}^+\Updownarrow\mathbf{V}^-}]_n^i \to 0$ at $\mathbf{n} \to \infty$. This means that the quantization of space and energy of Bivacuum dipoles are interrelated and discreet.

The ratio of separation between torus and antitorus to their to radius of dipoles in primordial (symmetric) Bivacuum is a permanent value, independent on the amplitude of gap oscillation:

$$\frac{[\mathbf{d}_{\mathbf{V}^+\Updownarrow\mathbf{V}^-}]_n}{\mathbf{L}_{\mathbf{V}^\pm}^n} = \pi \qquad 1.8$$

The gap and radius oscillations are accompanied by the emission and absorption of virtual clouds ($\mathbf{VC}^\pm$), providing exchange interaction between counterphase oscillating Bivacuum fermions or excitation of Virtual Pressure Waves ($\mathbf{VPW}^\pm$).

The frequency of gap oscillation is equal to frequency of Virtual Clouds exchange in Cooper pairs or frequency of $\mathbf{VPW}^\pm$ if Bivacuum dipoles are independent - unpaired.

## 2. The Dynamics of Bivacuum dipoles, exiting Virtual Pressure ($\mathbf{VPW}^\pm$) and Virtual Spin Waves ($\mathbf{VirSW}_q$)

The emission and absorption of Virtual clouds $(\mathbf{VC}_{j,k}^+)^i$ and anti-clouds $(\mathbf{VC}_{j,k}^-)^i$ by Bivacuum dipoles are the result of correlated transitions between different excitation states ($n = j,k$) of tori $(\mathbf{V}_{j,k}^+)^i$ and antitori $(\mathbf{V}_{j,k}^-)^i$, forming Bivacuum dipoles: $\left[\mathbf{BVF}^\updownarrow\right]^{\mu,\tau}$ and $\left[\mathbf{BVB}^\pm\right]^{\mu,\tau}$ :

$$(\mathbf{VC}_q^+)^{\mu,\tau} \equiv [\mathbf{V}_j^+ - \mathbf{V}_k^+]^{\mu,\tau} \quad - \textit{ virtual cloud} \qquad 2.1$$

$$(\mathbf{VC}_q^-)^{\mu,\tau} \equiv [\mathbf{V}_j^- - \mathbf{V}_k^-]^{\mu,\tau} \quad - \textit{ virtual anticloud} \qquad 2.1a$$

where: $j > k$ are the integer quantum numbers of torus and antitorus excitation states; $q = j - k$.

The virtual clouds: $(\mathbf{VC}_q^+)^{\mu,\tau}$ and $(\mathbf{VC}_q^-)^{\mu,\tau}$ exist in form of collective excitations - 'drops' of *subquantum* particles and antiparticles of opposite energies, correspondingly.

The process of emission of Virtual Cloud of positive and negative energy by Bivacuum fermion in each Cooper pair is accompanied by simultaneous absorption of this cloud by the antifermion and vice verse. Because of such local exchange interaction, stabilizing the pairs, the external virtual pressure waves can not be excited. The emission ⇌ absorption of positive and negative Virtual clouds by each Bivacuum dipole is accompanied by the in-phase oscillation of Bivacuum gap, separating tori and antitori of this dipoles. So, the Bivacuum fermions and antifermions, forming virtual Cooper pairs $[\mathbf{BVF}^{\uparrow} \bowtie \mathbf{BVF}^{\downarrow}]_q$, are pulsing between excited and basic states in counterphase, like the gap between their tori and antitori. However, these pulsation do not change the spin state of Bivacuum fermions.

The interaction between symmetric Bivacuum fermion: $\mathbf{BVF}^{\uparrow} = [\mathbf{V}^+ \uparrow\uparrow \mathbf{V}^-]$ and Bivacuum antifermion: $\mathbf{BVF}^{\downarrow} = [\mathbf{V}^+ \downarrow\downarrow \mathbf{V}^-]$, by means of Virtual Clouds ($VC^+$ and $VC^-$) exchange, stabilizing Cooper pairs in symmetric primordial Bivacuum without spin inversion is illustrated below:

$$(\mathbf{E}_{\mathbf{V}^+\uparrow\uparrow\mathbf{V}^-} + \Delta\mathbf{E}_{V^\pm})^A \underset{-\mathbf{VC}^\pm}{\overset{+\mathbf{VC}^\pm}{\rightleftharpoons}} (\mathbf{E}_{\mathbf{V}^+\downarrow\downarrow\mathbf{V}^-} - \Delta\mathbf{E}_{V^\pm})^B \qquad 2.2$$

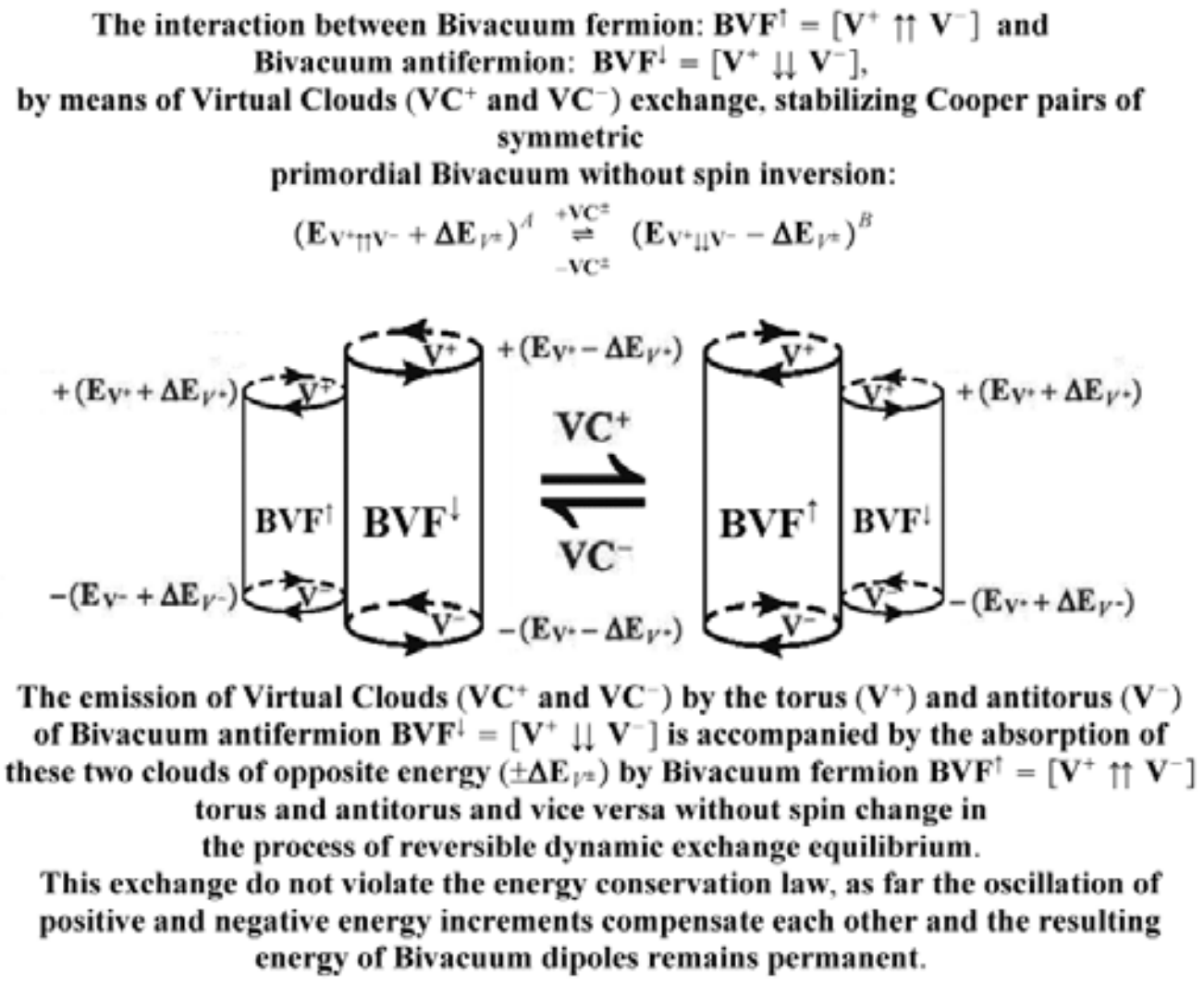

**Figure 1**. The emission of Virtual Clouds ($VC^+$ and $VC^-$) by the torus ($\mathbf{V}^+$) and antitorus ($\mathbf{V}^-$) in primordial symmetric Bivacuum antifermion $\mathbf{BVF}^{\downarrow} = [\mathbf{V}^+ \downarrow\downarrow \mathbf{V}^-]$ is accompanied by the absorption of these two clouds of opposite energy ($\pm\Delta\mathbf{E}_{V^\pm}$) by Bivacuum fermion $\mathbf{BVF}^{\uparrow} = [\mathbf{V}^+ \uparrow\uparrow \mathbf{V}^-]$ torus and antitorus and vice versa without spin change in the process of reversible dynamic exchange equilibrium. This exchange do not violate the energy conservation law, as far the oscillation of positive and negative energy increments compensate each other and the resulting energy of Bivacuum dipoles remains permanent.

The virtual multilayer membrane, as a main structure of Bivacuum coherent domains, is composed by Virtual bilayers of Cooper pairs of Bivacuum fermions and antifermions in state of virtual Bose condensation. The transition between two states of virtual domain of Bose condensate with opposite polarization of **N** Virtual Cooper pairs via the intermediate Bivacuum bosons can be presented like:

$$\begin{gathered} \mathbf{N}\left\{\mathbf{BVF}^{\uparrow} \equiv [\mathbf{V}^+ \uparrow\uparrow \mathbf{V}^-]^{\mu,\tau} \bowtie \mathbf{BVF}^{\downarrow} \equiv [\mathbf{V}^+ \downarrow\downarrow \mathbf{V}^-]^{\mu,\tau}\right\} \\ \Updownarrow \;\; 2\mathbf{N} \times \mathbf{BVB}^{\pm} \\ \mathbf{N}\left\{\mathbf{BVF}^{\downarrow} \equiv [\mathbf{V}^+ \downarrow\downarrow \mathbf{V}^-]^{\mu,\tau} \bowtie \mathbf{BVF}^{\uparrow} \equiv [\mathbf{V}^+ \uparrow\uparrow \mathbf{V}^-]^{\mu,\tau}\right\} \end{gathered} \qquad 2.3$$

In state of virtual Bose condensation of Cooper pairs, all transition between ground and excited states of all Bivacuum dipoles should be coherent. In each virtual Cooper pair the emission of Virtual Clouds $(\mathbf{VC}_q^+)^{\mu,\tau}$ and $(\mathbf{VC}_q^-)^{\mu,\tau}$ by first $\mathbf{BVF}^\uparrow$, as a result of its transition from excited to ground state, is accompanied by its absorption by second $\mathbf{BVF}^\downarrow$. This means that virtual Cooper pairs can not be a source of external Virtual Pressure Waves ($\mathbf{VPW}^\pm$). In domains of virtual Bose condensation, where Bivacuum dipoles can not be distinguished, the $\mathbf{BVB}^\pm$ may be the mediator of spin state not only between the $\mathbf{BVF}^\uparrow$ and $\mathbf{BVF}^\downarrow$ of the same Cooper pair but also between the remote pairs in the volume of this coherent virtual domain.

In symmetrical primordial Bivacuum the in-phase oscillation of energy of torus ($\mathbf{V}^+$) and antitorus ($\mathbf{V}^-$) of each Bivacuum dipole compensate each other. However, when the $[\mathbf{R} \times \mathbf{BVF}^\uparrow \rightleftharpoons \mathbf{L} \times \mathbf{BVF}^\downarrow]_q$ equilibrium symmetry is shifted to the left or right ($\mathbf{R} \neq \mathbf{L}$), for example by external magnetic, electric or gravitational field, the part of virtual Cooper pair will be disassembled to single Bivacuum fermions or antifermions. This should be accompanied by the origination of *uncompensated* virtual pressure waves of positive or negative energy: $\mathbf{VPW}^+$ or $\mathbf{VPW}^-$, propagating in space with light velocity. Just these waves participate in Virtual Replicas (**VR**) of material objects formation, which will be described later.

The energies of positive and negative $\mathbf{VPW}_q^+$ and $\mathbf{VPW}_q^-$, emitted ⇌ absorbed by Bivacuum dipoles, as a result of their torus ($V^+$) and antitorus ($V^-$) transitions between **j** and **k** quantum states can be presented as:

$$\mathbf{E}^i_{\mathbf{VPW}_q^+} = \hbar\omega_0^i(\mathbf{j}-\mathbf{k})_{V^+} = \mathbf{m}_0^i\mathbf{c}^2(\mathbf{j}-\mathbf{k}) \qquad 2.4$$

$$\mathbf{E}^i_{\mathbf{VPW}_q^-} = -\hbar\omega_0^i(\mathbf{j}-\mathbf{k})_{\mathbf{V}^-} = -\mathbf{m}_0^i\mathbf{c}^2(\mathbf{j}-\mathbf{k}) \qquad 2.4a$$

The quantized fundamental Compton frequency of $\mathbf{VPW}_q^\pm$ is:

$$\mathbf{q}\omega_0^i = \mathbf{q}\mathbf{m}_0^i\mathbf{c}^2/\hbar \qquad 2.5$$

where: $\mathbf{q} = \mathbf{j} - \mathbf{k} = \mathbf{1}, \mathbf{2}, \mathbf{3}..$ is the quantization number of $\mathbf{VPW}_{j,k}^\pm$ energy;

In symmetric primordial Bivacuum the total compensation of positive and negative Virtual Pressure Waves takes a place:

$$\mathbf{q}\mathbf{E}^i_{\mathbf{VPW}_{j,k}^+} = \left|-\mathbf{q}\mathbf{E}^i_{\mathbf{VPW}_{j,k}^-}\right| = \mathbf{q}\hbar\omega_0^i \qquad 2.6$$

This means that the coherent excitation of $\mathbf{VPW}_{j,k}^+$ and $\mathbf{VPW}_{j,k}^-$ do not violate the energy conservation law.

The density oscillation of $\mathbf{VC}_{j,k}^+$ and $\mathbf{VC}_{j,k}^-$, composed from subquantum particles, represent *positive and negative virtual pressure waves* ($\mathbf{VPW}_{j,k}^+$ and $\mathbf{VPW}_{j,k}^-$).

If the applied fields also shift the equality of energies of tori and antitori of the unpaired Bivacuum fermions or antifermions, this is followed by origination of inequality between energies of positive and negative Virtual Pressure waves:

$$\Delta\mathbf{VPW}^\pm = |\mathbf{VPW}^+ - \mathbf{VPW}^-|^{\mu,\tau} \qquad 2.7$$

Corresponding excessive energy can be used, as a source of pure 'free' energy and designing the overunity devices, based on resonant mechanism of consuming this energy.

The examples of Virtual microtubules (microfilaments), formed by Bivacuum Cooper pairs, assembled and disassembled presented below:

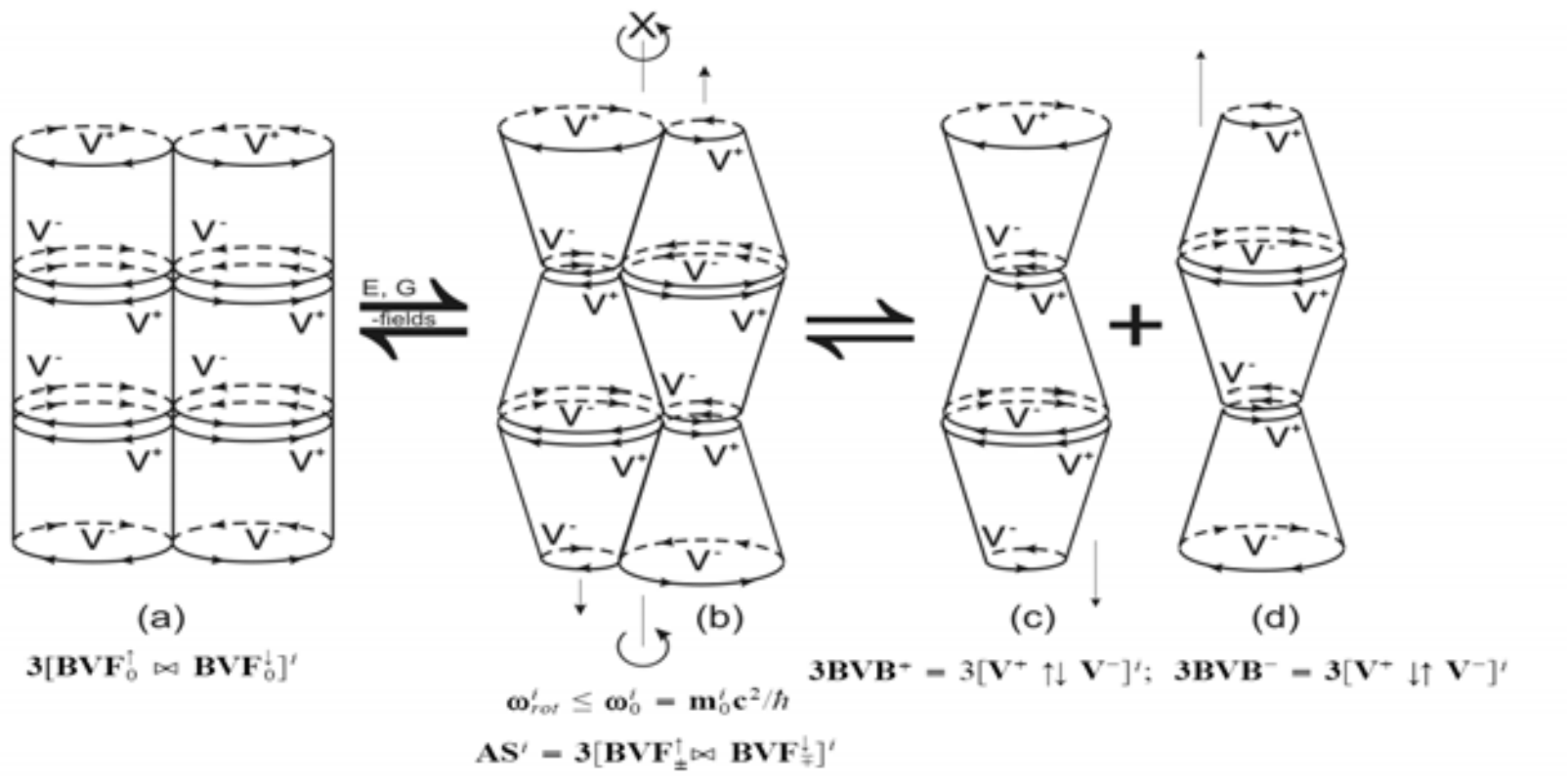

**Figure 2**. (**a**): three Cooper pairs of symmetric Bivacuum fermions in primordial Bivacuum: $\mathbf{3[BVF_0^{\uparrow} \bowtie BVF_0^{\downarrow}]^i}$ forming a double coherent Virtual microtubules (**VirMT**$^i$). This symmetric structures do not rotate around the main common axis and their external tangential and translational velocity is zero: $\mathbf{v} = \mathbf{0}$, like resulting mass and charge. Only the internal rotation of torus ($\mathbf{V}^+$) and antitorus ($\mathbf{V}^-$) of Bivacuum dipoles takes a place;

(**b**): the same structure as (**a**) in strong electric, magnetic or gravitational fields. The symmetric Bivacuum fermions and antifermions pairs turns to asymmetric three Cooper pairs of Bivacuum fermions, rotating around the common axis (X). The rotation and translation of these charged virtual filaments should be accompanied by excitation of the magnetic field.

(**c**) + (**d**) are the result of dissociation of double virtual microtubules $\mathbf{VirMT}^i_{[\mathbf{BVF}_\pm^\uparrow \bowtie \mathbf{BVF}_\mp^\downarrow]}$, presented at (**b**):

$$\mathbf{3[BVF_\pm^\uparrow \bowtie BVF_\mp^\downarrow]^{\mu,\tau}} \rightleftharpoons \mathbf{3(BVB^\pm)^{\mu,\tau}} + \mathbf{3(BVB^\mp)^{\mu,\tau}}$$

in strong electric or magnetic field, to single virtual microtubules $\mathbf{VirMT}^i_{\mathbf{BVB}}$ of Bivacuum bosons of opposite polarization. At this dissociation, the rotational kinetic energy of double $\mathbf{VirMT}^{\mu,\tau}_{[\mathbf{BVF}_\pm^\uparrow \bowtie \mathbf{BVF}_\mp^\downarrow]}$ turns to translational kinetic energy of two single **VirMT**.

If these two single virtual microtubules are formed by Bivacuum bosons with symmetry shift, corresponding to Golden mean, they propagate in opposite direction with velocity $\mathbf{v} = \sqrt{\phi}\,\mathbf{c}$, consuming the energy of resonant exchange interaction of asymmetric dipoles with the excessive virtual pressure waves: $\Delta\mathbf{VPW}^\pm = |\mathbf{VPW}^+ - \mathbf{VPW}^-|^{\mu,\tau}$ of Bivacuum, excited and absorbed by huge amount of elementary particles. The selective trapping of the energy of such individual trains can be used as a source for free energy of Bivacuum.

## 2.1. Virtual multi - bilayers (membranes) formed by virtual Cooper pairs

The virtual multi-layer membranes (**VirMem**) can be formally presented as a result of association infinitive number of Bivacuum fermions Cooper pairs of opposite polarization:

$$\mathbf{VirMem} = \sum^{\infty} \left[ \frac{BVF^\uparrow \bowtie BVF^\downarrow}{BVF^\downarrow \bowtie BVF^\uparrow} \right]^{\mu,\tau}_{x,y} \qquad 2.8$$

where Bivacuum fermions and antifermions, rotating in the opposite direction are interacting with each other side-by-side. Each of such Cooper pair $[BVF^\uparrow \bowtie BVF^\downarrow]$ has the counterpart $[BVF^\downarrow \bowtie BVF^\uparrow]$. These pairs are interacting with each other by 'head-to-tail' principle, forming virtual microtubules (Fig.2 a) in symmetric unperturbed by fields Bivacuum. Such systems may compose the virtual Bose condensation.

The oscillation of such two Cooper pairs between the symmetrically excited and the ground states are counterphase. This means that the virtual *gap oscillation* between tori and antitori of corresponding Bivacuum dipoles are also counterphase. This process may be accompanied by the exchange of pairs of Virtual Clouds $\left[\mathbf{VC}_q^+ \bowtie \mathbf{VC}_q^-\right]^{\mu,\tau}$ between virtual Cooper pairs.

The new notion of *positive and negative space* is related to positive and negative Compton radiuses of conjugated torus and antitorus with opposite mass and charge, forming Bivacuum dipoles: $\pm\mathbf{L}_{\mu,\tau} = \frac{\hbar}{\pm m_{\mu,\tau}c}$. The oscillation of gap between positive and negative mass/energy, opposite charges and magnetic moments can be named as *inter-space waves (ISW)*, in-phase with $\left[\mathbf{VC}_q^+ \bowtie \mathbf{VC}_q^-\right]^{\mu,\tau}$ emission⇌absorption by Bivacuum dipoles.

The virtual multilayer membranes in Bivacuum can be considered also, as the assembly of huge number of virtual microtubules of Bivacuum side-by-side.

Each of the layer of virtual membranes, can pulse between the excited and ground state in counterphase with the next one, interacting with each other via dynamic exchange by pairs of Virtual Clouds $\left[\mathbf{VC}_q^+ \bowtie \mathbf{VC}_q^-\right]^{\mu,\tau}$ or local virtual pressure waves $[\mathbf{VPW}^+ \bowtie \mathbf{VPW}^-]^{\mu,\tau}$. This process occur without violation of the energy conservation law because of positive and negative energy oscillation compensation. Such pulsations are accompanied by nonlocal Bivacuum gap oscillation over the space of virtual Bose condensation (BC) of Bivacuum dipoles.

## 2.2 Bivacuum Virtual Spin Waves Excitation

**The virtual spin waves $\mathbf{VirSW}_{j,k}^{\pm 1/2}$**, with properties of massless collective Nambu-Goldstone modes, like a real spin waves, represent the oscillation of angular momentum equilibrium of individual Bivacuum fermions or in composition of Cooper pairs with opposite spins via "flip-flop" mechanism, accompanied by origination of intermediate states - Bivacuum bosons $\mathbf{BVB}^{\pm} \equiv [\mathbf{V}^+ \uparrow\downarrow \mathbf{V}^-]^{\mu,\tau}$:

$$\mathbf{VirSW}_{j,k}^{\pm 1/2} \Longrightarrow \qquad 2.9$$

$$\mathbf{BVF}^{\uparrow} \equiv [\mathbf{V}^+ \upuparrows \mathbf{V}^-]^{\mu,\tau} \rightleftharpoons \mathbf{BVB}^{\pm} \equiv [\mathbf{V}^+ \uparrow\downarrow \mathbf{V}^-]^{\mu,\tau} \rightleftharpoons \mathbf{BVF}^{\downarrow} \equiv [\mathbf{V}^+ \downdownarrows \mathbf{V}^-]^{\mu,\tau} \qquad 2.9a$$

The $\mathbf{VirSW}_{j,k}^{\pm 1/2}$, like so-called torsion field, can serve as a carrier of the phase/spin (angular momentum) and information - qubits, but not the energy. The collective spin inversion in coherent Bivacuum domains of coherent bivacuum dipoles in state of virtual Bose condensate, is accompanied by collective and instant Virtual spin wave excitation. This nonlocal collective process in domains may induce the opposite conversion in the neighboring virtual domain. The corresponding chain reaction can be responsible for one of the mechanism of macroscopic entanglement or nonlocality.

The Virtual Spin Waves ($\mathrm{VirSW}^{\pm 1/2}$) propagation in space, as a result of spin states exchange between asymmetric Bivacuum fermion ($\mathrm{BVF}^{\uparrow}$) and antifermion ($\mathrm{BVF}^{\downarrow}$) forming virtual Cooper pairs is illustrated below.

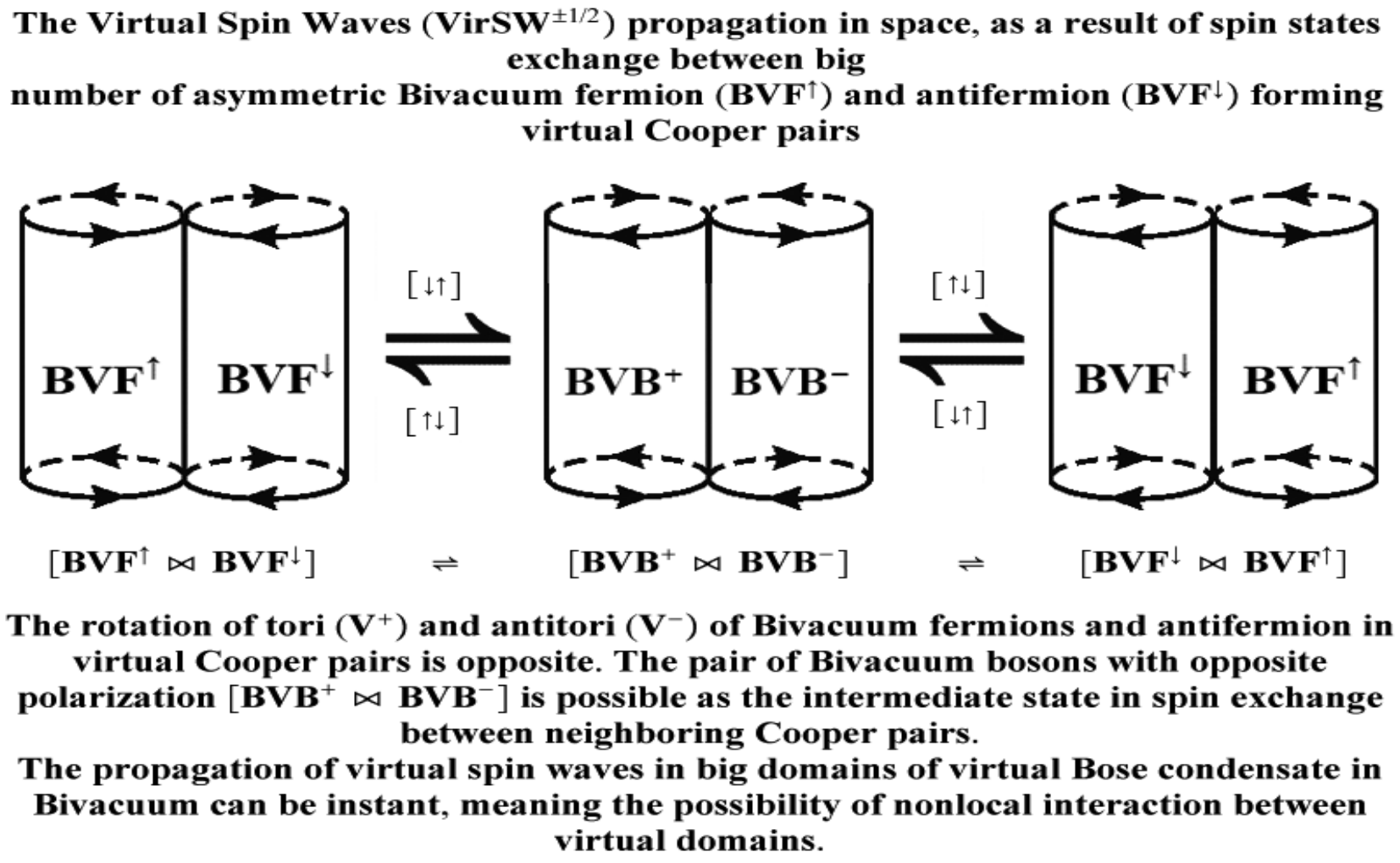

**Figure 3**. The rotation of tori ($V^+$) and antitori ($V^-$) in virtual Cooper pairs of the opposite spins is also opposite. The propagation of virtual spin waves in big domains of virtual Bose condensate in Bivacuum is not accompanied by the energy/momentum transmission and can be instant. This means the possibility of nonlocal interaction between neighboring virtual domains of Bivacuum.

## 3. Virtual Bose condensation (VirBC), as a condition of Bivacuum superfluid and nonlocal properties

It follows from our model of Bivacuum, that the infinite number of Cooper pairs of Bivacuum fermions $[\mathbf{BVF}^{\uparrow}\bowtie\mathbf{BVF}^{\downarrow}]^i_{S=0}$ and their intermediate states - Bivacuum bosons $(\mathbf{BVB}^{\pm})^i$, as elements of Bivacuum, have zero or very small (in presence of fields and matter) translational momentum: $\mathbf{p}^i_{\mathbf{BVF}^{\uparrow}\bowtie\mathbf{BVF}^{\downarrow}} = \mathbf{p}^i_{\mathbf{BVB}} \to 0$ and corresponding de Broglie wave length tending to infinity: $\lambda^i_{\mathbf{VirBC}} = \mathbf{h}/\mathbf{p}^i_{\mathbf{BVF}^{\uparrow}\bowtie\mathbf{BVF}^{\downarrow},\mathbf{BVB}} \to \infty$.

This condition leads to origination of 3D system of virtual double virtual microtubules from Cooper pairs of Bivacuum fermions $[\mathbf{BVF}^{\uparrow}\bowtie\mathbf{BVF}^{\downarrow}]_{S=0}$, and single virtual microtubules, formed by Bivacuum bosons $(\mathbf{BVB}^{\pm})_{S=0}$, closed or open, connecting remote coherent elementary particles.

The longitudinal momentum of Bivacuum dipoles forming virtual microfilaments and their bundles/beams can be close to zero and corresponding de Broglie wave length:

$$\lambda = \frac{h}{|\mathbf{m}_V^+ - \mathbf{m}_V^-|c} \qquad 3.1$$

$$\lambda \to \infty \quad \text{at } \mathbf{m}_V^+ \to \mathbf{m}_V^-$$

exceeding the distance between neighboring dipoles many times.

The 3D system of these double and single microtubules (see Fig. 50) represents Bose condensate with superfluid properties.

The Bivacuum, like liquid helium, can be considered as a liquid, containing two components: the described superfluid and normal, representing fraction of Bivacuum dipoles not involved in domains of virtual Bose condensation and virtual guides (VirG). The radii of VirG are determined by the Compton radii of the electrons, muons and tauons, interconnecting similar remote particles with opposite spins.

Their length is limited by decoherence effects, related to Bivacuum symmetry shift. In highly symmetric Bivacuum the length of **Virtual Guides** with nonlocal properties, connecting remote coherent elementary particles, may have the order of stars and galactic separation.

In some cases virtual microfilaments/microtubules (**VirMT**) may form a closed rotating rings with perimeter, determined by resulting standing de Broglie wave length of Bivacuum dipoles forming the rings. The life-time of such closed structures can be big, if they have a properties of standing and non-dissipating systems of virtual de Broglie waves of Bivacuum dipoles.

### *3.1 Virtual Bose condensation (VirBC), as a base of Bivacuum superfluid properties and nonlocality*

**Nonlocality**, as the independence of potential energy on the distance from energy source in 3D net filaments of virtual (and real) Bose condensate, follows from application of the Virial theorem to systems of Cooper pairs of Bivacuum fermions $[\mathbf{BVF}^{\uparrow}\bowtie\ \mathbf{BVF}^{\downarrow}]_{S=0}$ and Bivacuum bosons ($\mathbf{BVB}^{\pm}$) (Kaivarainen, 1995; 2004-2006).

The Virial theorem in general form is correct not only for classical, but also for quantum systems. It relates the averaged kinetic $\overline{\mathbf{T}}_k(\vec{\mathbf{v}}) = \sum_i \overline{\mathbf{m}_i \mathbf{v}_i^2/2}$ and potential $\overline{\mathbf{V}}(\mathbf{r})$ energies of particles, composing these systems:

$$2\overline{\mathbf{T}}_k(\vec{\mathbf{v}}) = \sum_i \overline{\mathbf{m}_i \mathbf{v}_i^2} = \sum_i \vec{\mathbf{r}}_i \partial \overline{\mathbf{V}}/\partial \vec{\mathbf{r}}_i \qquad 3.2$$

If the potential energy $\overline{\mathbf{V}}(\mathbf{r})$ is a homogeneous $\mathbf{n}$ – *order* function like:

$$\overline{\mathbf{V}}(\mathbf{r}) \sim \mathbf{r}^{\mathbf{n}}, \quad \text{then } \mathbf{n} = \frac{2\overline{\mathbf{T}}_k}{\overline{\mathbf{V}}(\mathbf{r})} \qquad 3.3$$

For example, for a harmonic oscillator, when $\overline{\mathbf{T}}_k = \overline{\mathbf{V}}$, we have $\mathbf{n} = \mathbf{2}$. For Coulomb interaction: $\mathbf{n} = -\mathbf{1}$ and $\bar{\mathbf{T}} = -\bar{\mathbf{V}}/\mathbf{2}$.

The important consequence of the Virial theorem is that, if the average kinetic energy and momentum ($\overline{\mathbf{p}}$) of particles in a certain volume of a Bose condensate (BC) tends to zero:

$$\overline{\mathbf{T}}_k = \overline{\mathbf{p}}^2/\mathbf{2m} \to \mathbf{0} \qquad 3.4$$

the interaction between particles in the volume of BC, characterized by the radius: $\mathbf{L}_{BC} = (\hbar/\overline{\mathbf{p}}) \to 0$, becomes nonlocal, as independent on distance between them:

$$\overline{\mathbf{V}}(\mathbf{r}) \sim \mathbf{r}^{\mathbf{n}} = \mathbf{1} = \mathbf{const} \quad \mathbf{at} \quad \mathbf{n} = 2\overline{\mathbf{T}}_k/\overline{\mathbf{V}}(\mathbf{r}) = \mathbf{0} \qquad 3.5$$

Consequently, it is shown, that nonlocality, as independence of potential on the distance from potential source, is the inherent property of macroscopic Bose condensate. The individual particles (real, virtual or subquantum) in a state of Bose condensation are spatially indistinguishable due to the uncertainty principle. The Bivacuum dipoles $[\mathbf{BVF}^{\uparrow}\bowtie\ \mathbf{BVF}^{\downarrow}]_{S=0}$ and $(\mathbf{BVB}^{\pm})_{S=0}$ due to quasi one-dimensional Bose condensation are tending to self-assembly by 'side-to-side' or 'head-to-tail' principle. They compose very long virtual microtubules - Virtual Guides with wormhole properties. In special cases they form a closed structures - rotating rings with radius, dependent on velocity of rotation. The 3D net of these two kind of Virtual Guides (double $\mathbf{VirG}^{\mathbf{BVF}^{\uparrow}\bowtie\mathbf{BVF}^{\downarrow}}$ and mono $\mathbf{VirG}^{\mathbf{BVB}^{\pm}}$) bundles represent the nonlocal and superfluid fraction of Bivacuum..

## 4. Virtual Particles and Antiparticles

Generally accepted difference of virtual particles from the actual ones, is that the former, in contrast to latter, does not follow the laws of relativistic mechanics:

$$\mathbf{m} = \frac{\mathbf{m}_0}{[1 - (\mathbf{v}/\mathbf{c})^2]^{1/2}} \qquad 4.1$$

For actual free particle with rest mass ($\mathbf{m}_0$) and relativistic mass ($\mathbf{m}$), the formula, following from (4.1) is:

$$\mathbf{E}^2 - \vec{p}^{\,2}\mathbf{c}^2 = \mathbf{m}_0^2\mathbf{c}^4 \qquad 4.2$$

where $\mathbf{E}^2 = (\mathbf{mc}^2)^2$ is the total energy squared and $\vec{\mathbf{p}} = \mathbf{m}\,\vec{\mathbf{v}}$ is the momentum of particle.

In accordance to our model of Bivacuum, virtual particles represent asymmetric Bivacuum dipoles $(\mathbf{BVF})^{as}$ and $(\mathbf{BVB})^{as}$ of three electron's generation ($i = e, \mu, \tau$) in unstable state far from Golden mean conditions (see sections 5.3 and 6.1).

The virtual particles, like the real sub-elementary particles, may exist in two phase: Corpuscular [C]- phase, representing correlated pairs of asymmetric torus ($\mathbf{V}^+$) and antitorus ($\mathbf{V}^-$) of two different energy states and Wave [W]- phase, resulting from quantum beats between these states (see chapter 8). Corresponding transitions are accompanied by emission ⇌ absorption of Cumulative Virtual Clouds ($\mathbf{CVC}^+$ or $\mathbf{CVC}^-$), formed by subquantum particles and antiparticles.

Virtual particles differs from real sub-elementary ones by their low stability and inability for fusion to triplets, as far their symmetry shift is not big enough to follow the Golden Mean condition (see Chapter 7).

For virtual particles the equality (4.2) in [C]-phase is invalid in contrast to real ones.

For [W]-phase of particles, represented by Cumulative Virtual Clouds ($\mathbf{CVC}^\pm$) and excited by them subquantum particles and antiparticles density oscillation in Bivacuum - virtual pressure waves ($\mathbf{VPW}_q^+ \bowtie \mathbf{VPW}_q^-$), the relativistic mechanics and equality (4.2) are not valid also. *Consequently, the causality principle also do not works in a system of* $\mathbf{VPW}_q^\pm$. This is true for both: virtual and real particles.

The [electron - proton] interaction is generally considered, as a result of virtual photons exchange (cumulative virtual clouds $\mathbf{CVC}^\pm$ exchange in terms of our theory), when the electron and proton total energies do not change. Only the directions of their momentums are changed. In this case the energy of virtual photon in the equation (4.2) $E = 0$ and:

$$\mathbf{E}^2 - \vec{p}^{\,2}\mathbf{c}^2 = -\vec{p}^{\,2}\mathbf{c}^2 < \mathbf{0} \qquad 4.3$$

The measure of virtuality (**Vir**) after Dirac is a measure of deviation from 4.2:

$$(\mathbf{Vir}) \sim \left|\ \mathbf{m}_0^2\mathbf{c}^4 - (\mathbf{E}^2 - \vec{p}^{\,2}\mathbf{c}^2)\ \right| \geq 0 \qquad 4.4$$

In contrast to actual particles, the virtual ones have a more limited radius of action, determined by dimension of $\mathrm{CVC}^\pm$. The more is the virtuality (**Vir**), the lesser is the action radius. Each of emitted virtual quantum (cumulative virtual cloud) must be absorbed by the same particle or another in a course of their [**C** ⇌ **W**] pulsation.

A lot of process in quantum electrodynamics can be illustrated by Feynman diagrams (Feynman, 1985). In these diagrams, *actual* particles are described as infinitive rays (lines) and virtual particles as stretches connecting these lines (Fig. 4).

Each peak (or angle) in Feynman diagrams means emission or absorption of quanta or particles. The energy of each process (electromagnetic, weak, strong) is described using correspondent fine structure constants.

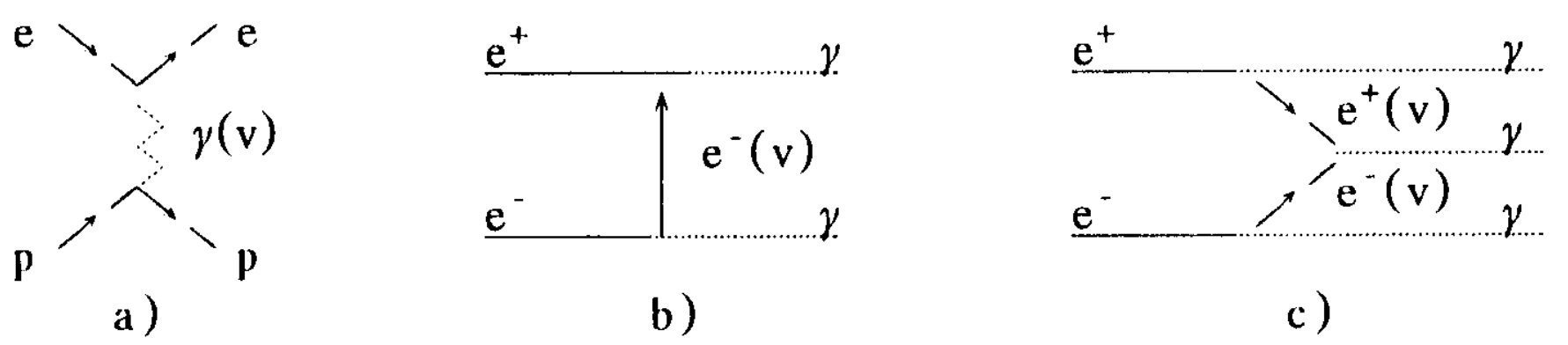

**Fig. 4**. Feynman diagrams describing electron-proton scattering (interaction), mediated by virtual photons - (cumulative virtual clouds in our terminology): **a**) - annihilation of electron and positron by means of virtual electron $e^-(v)$ and virtual positron $e^+(v)$ with origination of *two* and *three* actual photons ($\gamma$) : diagrams **b**) and **c**) correspondingly.

The background of our Unified Theory is presented by three postulates, which has been verified and confirmed in the course of this theory development.

## 5. Three postulates and related conservation rules for Bivacuum fermions $(\mathbf{BVF}^{\updownarrow})_{as}$ and Bivacuum bosons $(\mathbf{BVB}^{\pm})_{as}$

**Postulate I**. The absolute values of the *internal* rotational kinetic energies of torus and antitorus are permanent equal to each other and to the half of the rest mass energy of the electrons of corresponding lepton generation, independently on the *external* rotational and translational group velocity (**v**), turning the symmetric Bivacuum fermions ($\mathbf{BVF}^{\updownarrow}$) to asymmetric ones:

$$[\mathbf{I}]: \quad \left(\frac{1}{2}\mathbf{m}_V^+(\mathbf{v}_{gr}^{in})^2 = \frac{1}{2}|-\mathbf{m}_V^-|(\mathbf{v}_{ph}^{in})^2 = \frac{1}{2}\mathbf{m}_0\mathbf{c}^2 = \mathbf{const}\right)_{in}^{\mu,\tau} \qquad 5.1$$

where the positive $\mathbf{m}_V^+$ and negative $-\mathbf{m}_V^- = i^2\mathbf{m}_V^-$ are the 'actual' - inertial and 'complementary' - inertialess masses of torus ($\mathbf{V}^+$) and antitorus ($\mathbf{V}^-$); the $\mathbf{v}_{gr}^{in}$ and $\mathbf{v}_{ph}^{in}$ are the *internal* angular group and phase velocities of subquantum particles and antiparticles, forming torus and antitorus, correspondingly. In symmetric conditions of *primordial* Bivacuum and its virtual dipoles, when the influence of matter and fields is absent: $\mathbf{v}_{gr}^{in} = \mathbf{v}_{ph}^{in} = \mathbf{c}$.

It follows from my theory of time, described later, that the *pace of time* ($\mathbf{dt/t}$) for any closed *conservative system* is determined by the pace of its kinetic energy change $(-\mathbf{dT/T}_k)_{x,y,z}$, *anisotropic* in general case (Kaivarainen, 2006, 2007):

$$\left[\frac{\mathbf{dt}}{\mathbf{t}} = \mathbf{d}\ln\mathbf{t} = -\frac{\mathbf{dT}_k}{\mathbf{T}_k} = -\mathbf{d}\ln\mathbf{T}_k\right]_{x,y,z} \qquad 5.2$$

*Consequently, the Postulate I: $T_k^{in} = const$ - means zero pace of time and the infinitive life-time of torus and antitorus of Bivacuum dipoles $BVF^{\updownarrow}$ and $BVB^{\pm}$, independently of their symmetry shift.*

The *actual* (inertial) mass has the regular relativistic dependence on the external rotational and translational velocity $\mathbf{v} = \mathbf{v}^{ext}$ of Bivacuum dipoles:

$$\pm\mathbf{m}_V^+ = \frac{\mathbf{m}_0}{\pm\sqrt{1-(\mathbf{v/c})^2}} = \mathbf{m} \quad \text{(inertial mass)} \qquad 5.3$$

while the *complementary* (inertialess) mass ($\mp\mathbf{m}_V^-$) of antitorus $\mathbf{V}^-$ with sign, opposite to that of the actual one ($\pm\mathbf{m}_V^+$), has the reciprocal relativistic dependence on external velocity:

$$\mp \mathbf{m}_V^- = \mp\mathbf{m}_0\sqrt{1-(\mathbf{v/c})^2} \quad \text{(inertialess mass)} \qquad 5.4$$

It is important result, that in the case of nonzero external velocity (tangential or translational) of Bivacuum dipole ($\mathbf{v} > \mathbf{0}$) its total energy,equal to sum of of its positive and negative energies of actual torus and complementary antitorus is equal to doubled kinetic energy of dipole, anisotropic in general case:

$$\left[\begin{array}{c} \mathbf{E}_{BVF,BVB} = (\mathbf{m}_V^+ - \mathbf{m}_V^-)\mathbf{c}^2 = \mathbf{m}_V^+\mathbf{v}^2 = \mathbf{m}_V^+\mathbf{L}^2\boldsymbol{\omega}^2 \\ = 2\mathbf{T}_k = \frac{\mathbf{m}_0\mathbf{v}^2}{\sqrt{1-(\mathbf{v/c})^2}} \end{array}\right]_{x,y,z}^{\mu,\tau} \qquad 5.5$$

The ratio of absolute values of energies/mass of antitorus and torus (5.3) to (5.4) is:

$$\frac{|-\mathbf{m}_V^-|}{\mathbf{m}_V^+} = \frac{\mathbf{m}_0^2}{(\mathbf{m}_V^+)^2} = 1 - \left(\frac{\mathbf{v}}{\mathbf{c}}\right)^2 \qquad 5.6$$

**Postulate II**. The absolute *internal* magnetic moments of torus ($\mathbf{V}^+$) and antitorus ($\mathbf{V}^-$) of asymmetric Bivacuum fermions $\mathbf{BVF}_{as}^{\uparrow} = [\mathbf{V}^+\uparrow\uparrow\ \mathbf{V}^-]$ and antifermions: $\mathbf{BVF}_{as}^{\downarrow} = [\mathbf{V}^+\downarrow\downarrow\ \mathbf{V}^-]$ are permanent, equal to each other and to Bohr magneton ($\boldsymbol{\mu}_B$):

$$[\mathbf{II}]: \quad \left(\begin{array}{c} |\pm\boldsymbol{\mu}_+| \equiv \frac{1}{2}|\mathbf{e}_+|\frac{|\pm\hbar|}{|\mathbf{m}_V^+|(\mathbf{v}_{gr}^{in})_{rot}} = |\pm\boldsymbol{\mu}_-| \equiv \frac{1}{2}|-\mathbf{e}_-|\frac{|\pm\hbar|}{|-\mathbf{m}_V^-|(\mathbf{v}_{ph}^{in})_{rot}} = \\ = \boldsymbol{\mu}_B \equiv \frac{1}{2}|\mathbf{e}_0|\frac{\hbar}{\mathbf{m}_0\mathbf{c}} = \mathbf{const} \end{array}\right)^{\mu,\tau} \qquad 5.7$$

The magnetic moments of torus and antitorus are independent on their internal $(\mathbf{v}_{gr,ph}^{in})_{rot}$ and external velocity ($\mathbf{v} > \mathbf{0}$) and mass and charge symmetry shifts.

The actual and complementary masses $\mathbf{m}_V^+$ and $|-\mathbf{m}_V^-|$, internal angular velocities ($\mathbf{v}_{gr}^{in}$ and $\mathbf{v}_{ph}^{in}$) and electric charges $|\mathbf{e}_+|$ and $|\mathbf{e}_-|$ of $\mathbf{V}^+$ and $\mathbf{V}^-$ are dependent on the external and internal velocities, however, in such a way, that their changes compensate each other and magnetic moments of torus and antitorus remain invariant:

$$|\pm\boldsymbol{\mu}_+| = |\pm\boldsymbol{\mu}_-| = \boldsymbol{\mu}_B = \mathbf{const} \qquad 5.7a$$

This postulate reflects the condition of the invariance of magnetic moments $|\pm\boldsymbol{\mu}_\pm|$ and spin values ($\mathbf{S} = \pm\frac{1}{2}\hbar$) of torus and antitorus of Bivacuum dipoles with respect to their internal and external velocity, i.e. the absence of these parameters symmetry shifts.

One may see also that **Postulate II** means in fact that the resulting spins of Bivacuum fermion or antifermion, equal to that of torus and antitorus, correspondingly, are permanent and equal to:

$$\mathbf{S} = \pm\frac{1}{2}\hbar = \mathbf{const} \qquad 5.8$$

This follows from known relation between the spin and Bohr magneton ($\boldsymbol{\mu}_B$):

$$|\pm\boldsymbol{\mu}_+| = |\pm\boldsymbol{\mu}_-| = \boldsymbol{\mu}_B = \pm\frac{1}{2}\hbar\frac{\mathbf{e}_0}{\mathbf{m}_0\mathbf{c}} = \mathbf{S}\frac{\mathbf{e}_0}{\mathbf{m}_0\mathbf{c}} \qquad 5.9$$

where: $\mathbf{e}_0/\mathbf{m}_0\mathbf{c}$ is gyromagnetic ratio of the electron.

*We may conclude that in fact the Postulate II reflects the permanent half-integral value of spin* $\pm\frac{1}{2}\hbar$ *of Bivacuum fermions, independent on their internal and external velocity.*

**Postulate III**. *The equality of Coulomb attraction force between torus and antitorus of primordial Bivacuum dipoles*: $[\mathbf{V}^+ \Updownarrow \mathbf{V}^-]^{\mu,\tau}$ of $\mu$ and $\tau$ kinds (muons and tauons), providing uniform electric energy distribution in Bivacuum:

$$[\mathbf{III}] : \ \mathbf{F}_0^i = \left( \frac{\mathbf{e}_0^2}{[\mathbf{d}^2_{\mathbf{V}^+\Updownarrow\mathbf{V}^-}]_n} \right)^{\mu} = \left( \frac{\mathbf{e}_0^2}{[\mathbf{d}^2_{\mathbf{V}^+\Updownarrow\mathbf{V}^-}]_n} \right)^{\tau} \qquad 5.10$$

where: $[\mathbf{d}_{\mathbf{V}^+\Updownarrow\mathbf{V}^-}]_n^{\mu,\tau} = \frac{\hbar}{\mathbf{m}_0^{\mu,\tau}\mathbf{c}(1+2\mathbf{n})}$ is the separation between torus and antitorus of Bivacuum dipoles at the same state of excitation $(n)$ and $\left(\mathbf{e}_0^2 = |\mathbf{e}_+|\,|\mathbf{e}_-|\right)^{\mu,\tau}$.

The important consequences of *Postulate III* are the following relations, unifying the rest mass and charges of the tori and antitori of Bivacuum dipoles - the precursors of sub-elementary fermions: muons and tauons with mass and charge of the regular electron $(\mathbf{m}_0$ and $\mathbf{e}_0)^e$:

$$(\mathbf{e}_0\mathbf{m}_0)^e = (\mathbf{e}_0\mathbf{m}_0)^{\mu} = (\mathbf{e}_0\mathbf{m}_0)^{\tau} \qquad 5.11$$

$$or : \ (\mathbf{e}_0\mathbf{m}_0)^{e,\mu,\tau} = \sqrt{|\mathbf{e}_+\mathbf{e}_-||\mathbf{m}_V^+\mathbf{m}_V^-|}^{\,e,\mu,\tau} = const \qquad 5.11a$$

The other forms of dependence of the charges of tori and antitori $(\mathbf{e}_0^{\mu,\tau})$ of Bivacuum fermions on their mass $(\mathbf{m}_0^{\mu,\tau})$ are:

$$\mathbf{e}_0^{\mu} = \mathbf{e}_0^e(\mathbf{m}_0^e/\mathbf{m}_0^{\mu}) = \frac{\mathbf{e}_0^e}{206,7} \qquad 5.12$$

$$\mathbf{e}_0^{\tau} = \mathbf{e}_0^e(\mathbf{m}_0^e/\mathbf{m}_0^{\tau}) = \frac{\mathbf{e}_0^e}{3487,28} \qquad 5.12a$$

where $\mathbf{e}_0^e$ is the charge of the regular electron.

It follows from Postulate III and eqs.(5.12 and 5.12a), that the tori and antitori of symmetric $[\mathbf{V}^+ \Updownarrow \mathbf{V}^-]^{\mu,\tau}$ with bigger mass: $\mathbf{m}_0^{\mu} = 206,7\,\mathbf{m}_0^e$; $\mathbf{m}_0^{\tau} = 3487,28\,\mathbf{m}_0^e$ and smaller separation (gap) have correspondingly smaller charges, providing the uniform charge density distribution in Bivacuum.

Just these conditions provide *the same charge* of muon and tauon, equal to that of regular electron, notwithstanding of their different mass.

In other words, Postulate III explains why sub-elementary fermions: muons and tauons, fusing the electron/positron and proton/antiproton have same by the absolute value of electric charges, nonetheless of their big mass difference.

### *5.1. Three compensation principles, following from postulates I and II*

[**A**]. The *Mass Compensation Principle* follows from different relativistic dependencies of the actual and complementary mass of Bivacuum dipoles, reciprocal to each other (eqs 14 and 14a):

$$|\pm\mathbf{m}_V^+|\,|\mp\mathbf{m}_V^-| = \mathbf{m}_0^2 \qquad 5.13$$

The important consequence of our approach is that the product of the absolute values of the actual and complementary energies of Bivacuum dipoles is a permanent value, independent on the external velocity of asymmetric Bivacuum fermions, forming fields and matter:

$$\mathbf{E}_0^2 = \mathbf{m}_V^+\mathbf{c}^2 \times \mathbf{m}_V^-\mathbf{c}^2 = \mathbf{m}_0^2\mathbf{c}^4 = const \qquad 5.13a$$

The famous Einstein relation: $\mathbf{E} = \mathbf{mc^2}$ is equal to *actual* energy of tori of Bivacuum dipoles: $\mathbf{E}_V^{\dagger} = \mathbf{m}_V^{\dagger}\mathbf{c}^2 = \mathbf{E}$. However, the *complementary* energy and mass of antitori of Bivacuum dipoles are hidden and not revealed by this relation, being the part of more general formula 5.13a.

We may see, that the Einstein's energy/mass relation is a part of our *Mass Compensation Principle,* determined by the invariant rest energy/mass of elementary particle, squared ($\mathbf{E}_0^2$ ).

[**B**]. The *internal Group and Phase Velocities Compensation Principle* for internal dynamics of torus ($\mathbf{v}_{gr}^{in}$) and antitorus ($\mathbf{v}_{ph}^{in}$) of Bivacuum dipoles follows from the **Postulate** (**I**) in form:

$$\mathbf{v}_{gr}^{in}\,\mathbf{v}_{ph}^{in} = \mathbf{c}^2 \qquad 5.14$$

Similar relation is well known already for external group and phase velocities of relativistic de Broglie waves:

$$\mathbf{v}_{gr}^{ext}\,\mathbf{v}_{ph}^{ext} = \mathbf{c}^2 \qquad 5.15$$

[**C**]. The *Charge Compensation Principle* has the same simple shape, as previous compensation principles, is a consequence of **Postulate II**:

$$|\mathbf{e}_+|\,|\mathbf{e}_-| = \mathbf{e}_0^2 \qquad 5.16$$

*5.2. The solution of Dirac Monopole problem, following from Unified theory*

The Dirac theory, searching for elementary magnetic charges ($g^-$ and $g^+$), symmetric to electric ones ($e^-$ and $e^+$), named *monopoles*, leads to the following relation between the magnetic monopole and electric charge of the same signs:

$$\mathbf{ge} = \frac{n}{2}\hbar\mathbf{c} \quad \text{or:} \quad \mathbf{g} = \frac{n}{2}\frac{\hbar\mathbf{c}}{\mathbf{e}} = \frac{n}{2}\frac{\mathbf{e}}{\alpha} \qquad 5.17$$

$n = 1,2,3$ is the integer number

where $\alpha = \mathbf{e}^2/\hbar\mathbf{c} \simeq \mathbf{1/137}$ is the electromagnetic fine structure constant.

It follows from this definition, that minimal magnetic charge (*at* $\mathbf{n} = \mathbf{1}$) is as big as $g \cong 67.7e$. The mass of monopole should be huge $\sim 10^{16}\,GeV$. All numerous and very expensive attempts to reveal the Monopoles experimentally has failed.

*From our the Postulate II it follows that the uncompensated magnetic charge (monopole) simply is absent.*

In contrast to electric and mass dipoles symmetry shifts of Bivacuum dipoles, providing electric charge and mass origination of sub-elementary fermions, depending on their velocity (tangential or translational), the symmetry shift between the internal actual $|\mu_+|$ and complementary $|\mu_-|$ *magnetic charges* of Bivacuum dipoles torus and antitorus is absent and velocity independent.

This consequence follows from the *Postulate II* of our theory (eq.5.7):

$$\Delta|\mu_\pm| = (|\mu_+| - |\mu_-|) = \mathbf{0} \qquad 5.18$$

The magnetic charges of torus ($\mathbf{V}^+$) and antitorus ($\mathbf{V}^-$) are permanent values, equal to the Bohr magneton:

$$|\mu_+| = |\mu_-| = \mu_B \equiv \frac{1}{2}|\mathbf{e}_0|\frac{\hbar}{\mathbf{m}_0\mathbf{c}} = \mathbf{const} \qquad 5.19$$

In accordance to *Postulate II*, the magnetic charges of torus and antitorus are independent on the external velocity ($\mathbf{v}$) of symmetric and asymmetric Bivacuum dipoles and sub-elementary particles. The equality of the actual (torus) and complementary (antitorus) magnetic moments of Bivacuum dipoles - independent on their external velocity and the absence of symmetry shift, creating uncompensated moment, explains the *absence of magnetic monopoles in Nature.*

*5.3. The rest mass origination without Higgs bosons*

Due to different relativistic dependencies of masses ($\mathbf{m}_V^+$ and $\mathbf{m}_V^-$) of torus and antitorus ($\mathbf{V}^+$) and ($\mathbf{V}^-$) of Bivacuum fermions ($\mathbf{BVF}^{\uparrow}$) and antifermions ($\mathbf{BVF}^{\downarrow}$), rotating around common axis, on the external tangential velocity ($\mathbf{v}$), the symmetry shift of Bivacuum dipoles is accompanied by uncompensated mass and charge origination.

The process of symmetry shift is followed by turning of pair of symmetric Bivacuum fermion and antifermion to Cooper pairs of fermions: muons ($\mu$) and tauons ($\tau$) with antimuons and antitauons. of opposite spin, charge and energy. However, the resulting energy, mass and charge in such pairs is equal to zero.

It is easy to find out from our formulas for actual and complementary masses: ($\mathbf{m}_V^+$) and ($\mathbf{m}_V^-$), that the rest masses $\mathbf{m}_0^{\mu,\tau}$ of asymmetric Bivacuum dipoles originate at their certain symmetry shift, determined by Golden Mean condition $(\mathbf{v}/\mathbf{c})^2 = \phi = 0.618$.

The sum of the positive (actual) and negative (complementary) energies of torus and antitorus of Bivacuum fermion or antifermion at Golden mean condition can be presented as:

$$\mathbf{E}^{\phi}_{\mathbf{BVF}} = \mathbf{E}^{\phi}_{\mathbf{V}^+} + \mathbf{E}^{\phi}_{\mathbf{V}^-} = [\mathbf{m}_V^+ - \mathbf{m}_V^-]^{\phi}\mathbf{c}^2 = \pm\mathbf{m}_0^{\mu,\tau}\mathbf{c}^2 \qquad 5.20$$

$$or: \left[\frac{\mathbf{m}_0}{\pm\sqrt{1-\phi}} - \mathbf{m}_0\left(\pm\sqrt{1-\phi}\right)\right]\mathbf{c}^2 = \mathbf{m}_0\frac{\phi}{\pm\sqrt{1-\phi}}\mathbf{c}^2 = \pm\mathbf{m}_0\mathbf{c}^2 \qquad 5.21$$

From known quadratic Golden mean equation: $\phi^2 + \phi = 1$ it follows that: $\frac{\phi}{\pm\sqrt{1-\phi}} = \pm 1$.

The described process of mass origination is beyond Standard Model, which use the hypothesis of Higgs bosons (see Introduction). We have to stress, that nonetheless of big efforts of laboratories over the world, involved in high energy physics, no experimental indications to existing of Higgs bosons where revealed.

## 6. The relation between the internal parameters of Bivacuum dipoles and their external velocity

The important relativistic formula, unifying a lot of internal and external parameters of asymmetric Bivacuum fermions ($\mathbf{BVF}^{\updownarrow}_{as}$) was derived from the Postulates I and II:

$$\left(\frac{\mathbf{c}}{\mathbf{v}_{gr}^{in}}\right)^2 = \frac{1}{[1-(\mathbf{v}^2/\mathbf{c}^2)^{ext}]^{1/2}} \qquad 6.1$$

$$= \left(\frac{\mathbf{m}_V^+}{\mathbf{m}_V^-}\right)^{1/2} = \frac{\mathbf{v}_{ph}^{in}}{\mathbf{v}_{gr}^{in}} = \frac{\mathbf{L}_V^-}{\mathbf{L}_V^+} = \frac{\mathbf{L}_0}{(\mathbf{L}_V^+)^2} = \frac{|\mathbf{e}_+|}{|\mathbf{e}_-|} = \frac{\mathbf{e}_+^2}{\mathbf{e}_0^2} \qquad 6.1a$$

where the radiuses of torus ($\mathbf{L}_V^+$) and antitorus ($\mathbf{L}_V^-$), as a basis of truncated cone, as a shape of asymmetric Bivacuum fermions, have the following relativistic dependencies on their external rotational or translational group velocity ($\mathbf{v} \equiv \mathbf{v}_{gr}$):

$$\left[\mathbf{L}_V^+ = \mathbf{L}_0[1 - (\mathbf{v}^2/\mathbf{c}^2)^{ext}]^{1/4}\right]^i \qquad 6.2$$

$$\left[\mathbf{L}_V^- = \frac{\mathbf{L}_0}{[1 - (\mathbf{v}^2/\mathbf{c}^2)^{ext}]^{1/4}}\right]^i \qquad 6.2a$$

$$\text{where: } \left[\mathbf{L}_0 = (\mathbf{L}_V^+\mathbf{L}_V^-)^{1/2} = \hbar/\mathbf{m}_0\mathbf{c}\right]^i \; - \; \textit{Compton radius} \qquad 6.2b$$

*The absolute external velocity of Bivacuum dipoles squared (*$\mathbf{v}^2$*) as respect to primordial Bivacuum (absolute reference frame)*, can be expressed, using 26 and 26a, as a criteria of asymmetry of these dipoles torus and antitorus, accompanied their external motion:

$$\left[\mathbf{v}^2 = \mathbf{c}^2\left(1 - \frac{\mathbf{m}_V^-}{\mathbf{m}_V^+}\right) = \mathbf{c}^2\left(1 - \frac{\mathbf{e}_-^2}{\mathbf{e}_+^2}\right) = \mathbf{c}^2\left(1 - \frac{\mathbf{S}_+}{\mathbf{S}_-}\right)\right]_{x,y,z} \qquad 6.3$$

where: $\mathbf{S}_+ = \pi(\mathbf{L}_V^+)^2$ and $\mathbf{S}_- = \pi(\mathbf{L}_V^-)^2$ are the squares of cross-sections of torus and antitorus of Bivacuum dipoles as the truncated cones.

*The existence of absolute velocity in our Unified theory (anisotropic in general case) and the Universal reference frame of Primordial Bivacuum, pertinent for Ether concept, is an important difference with Special relativity theory.*

The relativistic dependences of the actual charge $\mathbf{e}_+$ and actual mass ($\mathbf{m}_V^+$) on external *absolute* velocity of Bivacuum dipole, following from (6.1) and (5.3) are:

$$\mathbf{e}_+ = \frac{\mathbf{e}_0}{[1 - (\mathbf{v}^2/\mathbf{c}^2)]^{1/4}} \qquad 6.4$$

$$\mathbf{m}_V^+ = \frac{\mathbf{m}_0}{\sqrt{1 - (\mathbf{v}/\mathbf{c})^2}} \qquad 6.4a$$

The ratio of the actual charge to the actual inertial mass from (4.5 and 4.5a) has also the relativistic dependence:

$$\frac{\mathbf{e}_+}{\mathbf{m}_V^+} = \frac{\mathbf{e}_0}{\mathbf{m}_0}[1 - (\mathbf{v}^2/\mathbf{c}^2)]^{1/4} \qquad 6.4b$$

The decreasing of this ratio with velocity increasing is weaker, than it follows from the generally accepted statement, that charge has no relativistic dependence in contrast to the actual mass $\mathbf{m}_V^+$. The direct experimental investigation of relativistic dependence of this ratio on the external velocity ($\mathbf{v}$) may confirm the validity of our formula (6.4b) following from our approach.

From eqs. (5.5) and (6.1) we find for mass and charge symmetry shift:

$$\Delta\mathbf{m}_\pm = \mathbf{m}_V^+ - \mathbf{m}_V^- = \mathbf{m}_V^+\left(\frac{\mathbf{v}}{\mathbf{c}}\right)^2 \qquad 6.5$$

$$\Delta\mathbf{e}_\pm = \mathbf{e}_+ - \mathbf{e}_- = \frac{\mathbf{e}_+^2}{\mathbf{e}_+ + \mathbf{e}_-}\left(\frac{\mathbf{v}}{\mathbf{c}}\right)^2 \qquad 6.5a$$

These mass and charge symmetry shifts determines the relativistic dependence of the *effective* mass and charge of the fermions. In direct experiments only the actual mass ($\mathbf{m}_V^+$) and charge ($\mathbf{e}_\pm$) can be registered. It means that the complementary mass ($\mathbf{m}_V^-$) and charge are *hidden* for observation.

The ratio of charge shift to mass shifts (the *uncompensated* charge and mass ratio) is:

$$\frac{\Delta \mathbf{e}_{\pm}}{\Delta \mathbf{m}_{\pm}} = \frac{\mathbf{e}_{+}^{2}}{\mathbf{m}_{V}^{+}\,(\mathbf{e}_{+} + \mathbf{e}_{-})} \qquad 6.6$$

The mass symmetry shift can be expressed via the squared charges symmetry shift also in the following manner:

$$\Delta \mathbf{m}_{\pm} = \mathbf{m}_{V}^{+} - \mathbf{m}_{V}^{-} = \mathbf{m}_{V}^{+}\frac{\mathbf{e}_{+}^{2} - \mathbf{e}_{-}^{2}}{\mathbf{e}_{+}^{2}} \qquad 6.7$$

or using (6.5) this formula turns to:

$$\frac{\mathbf{e}_{+}^{2} - \mathbf{e}_{-}^{2}}{\mathbf{e}_{+}^{2}} = \frac{\mathbf{v}^{2}}{\mathbf{c}^{2}} \qquad 6.8$$

The overall shape of asymmetric $\left(\mathbf{BVF}_{as}^{\updownarrow} = [\mathbf{V}^{+} \Updownarrow \mathbf{V}^{-}]\right)^{i}$ is a *truncated cone* (Fig.2) with plane, parallel to the base with radiuses of torus ($L^{+}$) and antitorus ($L^{-}$), defined by eq. (4.3).

*6.1 The Hidden Harmony of Bivacuum dipoles,*
*as a background of Golden Mean and the rest mass and charge origination*

One of the multiple forms of general expression (6.1 and 6.1a) is formula, unifying the internal and external group and phase velocities of asymmetric Bivacuum dipoles:

$$\left(\frac{\mathbf{v}_{gr}^{in}}{\mathbf{c}}\right)^{4} = 1 - \left(\frac{\mathbf{v}_{gr}^{ext}}{\mathbf{c}}\right)^{2} \qquad 6.9$$

where: $(\mathbf{v}_{gr}^{ext}) \equiv \mathbf{v}$ is the external translational-rotational group velocity of Bivacuum dipole.

*The conditions of "Hidden Harmony"* were introduced in our approach as:

a) the equality of the internal and external group velocities and

b) the equality of the internal and external phase velocities of asymmetric Bivacuum dipoles:

$$(\mathbf{v}_{gr}^{in})_{\mathbf{V}^{+}}^{rot} = (\mathbf{v}_{gr}^{ext})^{tr} \equiv \mathbf{v} \qquad 6.10$$

$$(\mathbf{v}_{ph}^{in})_{\mathbf{V}^{-}}^{rot} = (\mathbf{v}_{ph}^{ext})^{tr} \qquad 6.10a$$

and introducing the notation:

$$\left(\frac{\mathbf{v}_{gr}^{in}}{\mathbf{c}}\right)^{2} = \left(\frac{\mathbf{v}}{\mathbf{c}}\right)^{2} \equiv \phi \qquad 6.11$$

formula (6.9) turns to a simple quadratic *Golden mean equation:*

$$\boldsymbol{\phi}^{2} + \boldsymbol{\phi} - \mathbf{1} = \mathbf{0} \qquad 6.12$$

$$\text{which has a few modes :} \quad \phi = \frac{1}{\phi} - 1 \quad or: \quad \frac{\phi}{(1-\phi)^{1/2}} = 1 \qquad 6.12a$$

$$or: \quad \frac{1}{(1-\phi)^{1/2}} = \frac{1}{\phi} \qquad 6.12b$$

The solution of the Golden mean equation is equal to Golden mean: $(\mathbf{v}/\mathbf{c})^{2} = \phi = 0.618$. *It is remarkable, that the Golden Mean, which plays so important role on different Hierarchic* *levels of matter organization: from elementary particles to galactic and even in our perception of beauty (i.e. our mentality), has so deep physical roots as Hidden Harmony conditions* (6.10 and 6.10a).

Our theory is the first one, elucidating these roots (Kaivarainen, 1995; 2000; 2005). This important fact points, that we are on the right track searching the mechanism of mass and charge origination from Bivacuum dipoles and elementary particles origination.

We can see, that the unified formula (6.1 and 6.1a) at Golden Mean conditions: $(\mathbf{v}/\mathbf{c})^2 = \phi = 0.618$ and using the Golden Mean equation in shape: $\frac{1}{(1-\phi)^{1/2}} = \frac{1}{\phi}$ turns to:

$$\left[\left(\frac{\mathbf{m}_V^+}{\mathbf{m}_V^-}\right)^{1/2} = \frac{\mathbf{m}_V^+}{\mathbf{m}_0} = \frac{\mathbf{v}_{ph}^{in}}{\mathbf{v}_{gr}^{in}} = \frac{\mathbf{L}^-}{\mathbf{L}^+} = \frac{|\mathbf{e}_+|}{|\mathbf{e}_-|} = \left(\frac{\mathbf{e}_+}{\mathbf{e}_0}\right)^2\right]^{\phi} = \frac{1}{\phi} \qquad 6.13$$

where the actual ($e_+$) and complementary ($e_-$) charges and corresponding mass at Golden Mean conditions are:

$$\mathbf{e}_+^{\phi} = \mathbf{e}_0/\phi^{1/2}; \qquad \mathbf{e}_-^{\phi} = \mathbf{e}_0\phi^{1/2} \qquad 6.14$$

$$(\mathbf{m}_V^+)_{\mu,\tau}^{\phi} = \mathbf{m}_0^{\mu,\tau}/\phi; \qquad (\mathbf{m}_V^-)_{\mu,\tau}^{\phi} = \mathbf{m}_0^{\mu,\tau}\phi \qquad 6.14a$$

using (6.14a) it is easy to see, that the difference between the actual and complementary mass of tori and antitori of asymmetric Bivacuum fermions of two generations $[\mathbf{V}^+ \Updownarrow \mathbf{V}^-]^{\mu,\tau}$ at Golden Mean conditions is equal to the rest mass of corresponding sub-elementary fermions: muon and tauon:

$$\left[\ |\Delta\mathbf{m}_V|^{\phi} = \mathbf{m}_V^+ - \mathbf{m}_V^- = \mathbf{m}_0(1/\phi - \phi) = \mathbf{m}_0\right]^{\mu,\tau} \qquad 6.15$$

One of the form of Golden Mean equation $\phi^2 + \phi - \mathbf{1} = \mathbf{0}$ is: $(1/\phi - \phi) = \mathbf{1}$

This is an important result, confirming our approach, that the Bivacuum fermions symmetry shift, responsible for origination of the rest mass of unstable muons and tauons before their fusion to the electrons and protons, correspondingly, is determined by the universal for all hierarchical levels of Nature - Golden mean condition, based on Hidden Harmony (eqs.6.5 and 6.5a).

*The same is true for the charge origination.* The Golden Mean symmetry shift between actual and complementary charges of Bivacuum dipoles of two generations $[\mathbf{V}^+ \Updownarrow \mathbf{V}^-]^{\mu,\tau}$ stands for elementary charge of sub-elementary fermions or antifermions. From (6.14) we get:

$$\left[\phi^{3/2}\mathbf{e}_0 = |\Delta\mathbf{e}_\pm|^{\phi} = |\mathbf{e}_+ - \mathbf{e}_-|^{\phi} \equiv |\mathbf{e}|^{\phi}\right]^{e,\mu,\tau} \qquad 6.16$$

$$\text{where:}\quad [(|\mathbf{e}_+||\mathbf{e}_-|) = \mathbf{e}_0^2]^{\mu,\tau} \qquad 6.16a$$

where the charges of tori and antitori in symmetric Bivacuum fermions $[\mathbf{V}^+ \Updownarrow \mathbf{V}^-]^{\mu,\tau}$ are related to their mass like it follows from *Postulate III* of this theory eqs (5.12 and 5.12a).

These conditions (5.12 and 5.12a) provide the same charge symmetry shift of $[\mathbf{V}^+ \Updownarrow \mathbf{V}^-]^{\mu,\tau}$ of muon and tauon (sub-elementary fermions), equal to that of regular electron.

*This is a reason, why sub-elementary fermions: muons and tauons, creating the electrons and protons, correspondingly, have the same absolute electric charges, nonetheless of their big mass difference:* $\mathbf{m}_0^{\mu} = 206,7\ \mathbf{m}_0^{e}$ and $\mathbf{m}_0^{\tau} = 3487,28\ \mathbf{m}_0^{e}$.

*6.2. The dynamic and geometrical conditions for sub-elementary fermions (muons and tauons) origination*

Let us analyze the dynamics and geometry of Bivacuum fermions conversion to muons and tauons of rotating virtual Cooper pairs at Golden Mean (GM) conditions:

$$[\mathbf{BVF}^{\uparrow} \bowtie \mathbf{BVF}^{\downarrow}]^{\mu,\tau}_{GM} = [\mathbf{V}^{+} \upuparrows \mathbf{V}^{-}]^{\mu,\tau}_{GM} \bowtie [\mathbf{V}^{+} \downdownarrows \mathbf{V}^{-}]^{\mu,\tau}_{GM} \qquad 6.17$$

before their fusion to triplets of the electron or proton $< [\mathbf{F}^{+}_{\uparrow} \bowtie \mathbf{F}^{-}_{\downarrow}]_{x,y} + \mathbf{F}^{\pm}_{\updownarrow} >^{e,p}_{z}$ .

This fusion converts the unstable muons and tauons to sub-elementary fermions in composition of stable triplets (electrons and protons), with smaller mass (see next chapter).

The dynamic and spatial properties of Bivacuum fermions and antifermions of two generation, turning them to muons or tauons, are the result of participation in two rotational process simultaneously:

*1) rotation of asymmetric* $\mathbf{F}^{\pm}_{\updownarrow} >^{\mu,\tau}$ around its own axis with spatial image of *truncated cone* with resulting radius:

$$\mathbf{L}^{\phi}_{\mathbf{BVF}_{as}} = \frac{\hbar}{|\mathbf{m}^{+}_{V} + \mathbf{m}^{-}_{V}|^{\phi}\mathbf{c}} = \frac{\hbar}{\mathbf{m}_0(1/\phi + \phi)\mathbf{c}} = \frac{\hbar}{2.236\,\mathbf{m}_0\mathbf{c}} = \frac{\mathbf{L}_0}{2.236} \qquad 6.18$$

where the total mass of Bivacuum fermion at Golden Mean condition is

$$|\mathbf{m}^{+}_{V} + \mathbf{m}^{-}_{V}|^{\phi} = \mathbf{m}_0(1/\phi + \phi) = 2.236\,\mathbf{m}_0 \qquad 6.19$$

$\mathbf{L}_0 = \hbar/\mathbf{m}_0\mathbf{c}$ is the Compton radius of the rest mass of muon or tauon.

It was shown above (6.14a), that at Golden Mean conditions:
$(\mathbf{m}^{+}_{V})^{\phi} = \mathbf{m}_0/\phi$ and $(\mathbf{m}^{-}_{V})^{\phi} = \mathbf{m}_0\,\phi$.

*2) rolling of truncated cones* of each of two Bivacuum fermions and antifermions around the axis, common for both of them, inside a larger vorticity with radius of Cooper pair $[\mathbf{BVF}^{\uparrow} \bowtie \mathbf{BVF}^{\downarrow}]^{\mu,\tau}_{GM}$, equal to *Compton radius of* muons or tauons:

$$\left[\mathbf{L}^{\phi}_{\mathbf{BVF}^{\uparrow}_{as} \bowtie \mathbf{BVF}_{as}^{\downarrow}} = \frac{\hbar}{|\mathbf{m}^{+}_{V} - \mathbf{m}^{-}_{V}|^{\phi}\mathbf{c}} = \frac{\hbar}{\mathbf{m}_0\mathbf{c}} = \mathbf{L}_0\right]^{\mu,\tau} \qquad 6.20$$

with Golden mean angular frequency:

$$\left[(\omega^{i}_{\mathrm{v},\tilde{\mathrm{v}}})^{\phi}_{rot} = \frac{\mathbf{c}}{\mathbf{L}_0} = \omega_0 = \frac{\mathbf{m}_0\mathbf{c}^2}{\hbar}\right]^{\mu,\tau} \qquad 6.21$$

The ratio of radius $\mathbf{L}^{\phi}_{\mathbf{BVF}^{\uparrow}_{as} \bowtie \mathbf{BVF}_{as}^{\downarrow}}$ of Cooper pair to the radius $\mathbf{L}^{\phi}_{\mathbf{BVF}_{as}}$ of single Bivacuum fermion is equal to the ratio of potential ($\mathbf{V}$) to kinetic ($\mathbf{T}_k$) energy of relativistic de Broglie wave, as it follows from known formula for relativistic de Broglie wave $\left(\frac{\mathbf{V}}{\mathbf{T}_k} = 2\frac{\mathbf{v}_{ph}}{\mathbf{v}_{gr}} - 1\right)$ at GM conditions:

$$\frac{\mathbf{L}^{\phi}_{\mathbf{BVF}^{\uparrow}_{as} \bowtie \mathbf{BVF}_{as}^{\downarrow}}}{\mathbf{L}^{\phi}_{\mathbf{BVF}_{as}}} = \frac{|\mathbf{m}^{+}_{V} + \mathbf{m}^{-}_{V}|^{\phi}}{|\mathbf{m}^{+}_{V} - \mathbf{m}^{-}_{V}|^{\phi}} = \left(\frac{\mathbf{V}}{\mathbf{T}_k}\right)^{\phi} = 2\left(\frac{\mathbf{v}_{ph}}{\mathbf{v}_{gr}}\right)^{\phi} - 1 = 2.236 \qquad 6.22$$

where: $\left(\frac{\mathbf{v}_{ph}}{\mathbf{v}_{gr}}\right)^{\phi} = \left(\frac{\mathbf{c}^2}{\mathbf{v}^2}\right)^{\phi} = \frac{1}{\phi} = 1.618$

From (6.22) we may see, that the internal *rotational* potential and kinetic energy of asymmetric Bivacuum fermions are equal, correspondingly, to the sum and difference of the absolute values of energies of torus and antitorus:

$$\mathbf{V} = \frac{1}{2}|\mathbf{m}_V^+ + \mathbf{m}_V^-|\mathbf{c}^2 \qquad and \qquad \mathbf{T}_k^+ = \frac{1}{2}|\mathbf{m}_V^+ - \mathbf{m}_V^-|\mathbf{c}^2 = \frac{1}{2}\mathbf{m}_V^+\mathbf{v}^2 \qquad 6.23$$

This result is a good evidence in proof for our expression for the *actual* energy of torus ($\mathbf{V}^+$) sub-elementary particle, as a sum of its internal potential and kinetic energies:

$$\mathbf{E}_V^+ = \mathbf{m}_V^+\mathbf{c}^2 = \mathbf{V} + \mathbf{T}_\mathbf{k}^+ = \frac{1}{2}(\mathbf{m}_V^+ + \mathbf{m}_V^-)\mathbf{c}^2 + \frac{1}{2}(\mathbf{m}_V^+ - \mathbf{m}_V^-)\mathbf{c}^2 \qquad 6.24$$

The complementary negative energy of antitorus ($\mathbf{V}^-$) can be expressed in similar way as:

$$\mathbf{E}_V^- = -\mathbf{m}_V^-\mathbf{c}^2 = \mathbf{V} + \mathbf{T}_\mathbf{k}^- = -\Big[\frac{1}{2}(\mathbf{m}_V^+ + \mathbf{m}_V^-)\mathbf{c}^2 + \frac{1}{2}(\mathbf{m}_V^- - \mathbf{m}_V^+)\mathbf{c}^2\Big] \qquad 6.25$$

The total energy of asymmetric Bivacuum dipoles, including muons or tauons, is equal to sum of $\mathbf{E}_V^+$ and $\mathbf{E}_V^-$ :

$$\mathbf{E}_{tot} = \mathbf{E}_V^+ + \mathbf{E}_V^- = |\mathbf{m}_V^+ - \mathbf{m}_V^-|\mathbf{c}^2 = \mathbf{m}_V^+\mathbf{v}^2 \qquad 6.26$$

In the absence of external motion of Bivacuum dipoles, when $\mathbf{v} = \mathbf{0}$, their resulting total energy is zero: $\mathbf{E}_{tot} = 0$ and symmetry shift of tori and antitori is absent: $\mathbf{m}_V^+\mathbf{c}^2 = -\mathbf{m}_V^-\mathbf{c}^2$.

## 7. Fusion of triplets of elementary fermions from muons and tauons

The next *stage* of elementary fermions (electrons, protons and neutrons) formation is a fusion of triplets $< [\mathbf{F}_\uparrow^+ \bowtie \mathbf{F}_\downarrow^-] + \mathbf{F}_\updownarrow^\pm >^{e,p,n}$ from the above described Cooper pairs of pairs of muons, tauons and their antiparticles of opposite spin and charge $[\mathbf{BVF}^+ \bowtie \mathbf{BVF}^-]_{\mu,\tau}^{\phi}$ with unpaired muon or tauon.

The corresponding reaction of fusion, following by origination of triplets of elementary fermions and antifermions: electron+positron, protons+antiproton, neutron+antineutron, can be presented as a reaction of two Cooper pairs with one Cooper pair of Bivacuum fermions with properties of muons and tauons:

$$2[\mathbf{BVF}^+ \bowtie \mathbf{BVF}^-]_{\mu,\tau}^{\phi} + [\mathbf{BVF}^+ \bowtie \mathbf{BVF}^-]_{\mu,\tau}^{\phi} \rightleftharpoons \qquad 7.1$$

$$\langle[\mathbf{F}_\downarrow^+ \bowtie \mathbf{F}_\uparrow^-] + \mathbf{F}_\downarrow^+\rangle^{e^-,p^+,n} + \langle[\mathbf{F}_\downarrow^+ \bowtie \mathbf{F}_\uparrow^-] + \mathbf{F}_\uparrow^-\rangle^{e^+,p^-,n} \qquad 7.1a$$

The fusion of elementary fermions and antifermions from sub-elementary ones can be accompanied by huge energy release, determined by the value of mass defect. Corresponding kinetic energy (high temperature) push the elementary particles and antiparticles away from each other, preventing their annihilation just after origination. So, the equilibrium 7.1 - 7.1a is strongly shifted to the right at Golden Mean conditions.

*Just the unpaired sub-elementary fermions:* $\mathbf{F}_\downarrow^+\rangle$ *or* $\mathbf{F}_\uparrow^-\rangle$ *determines the mass, spin and charge of the whole triplet, as elementary particle, as far as the properties of paired sub-elementary fermions compensate each other.*

Both of stages of elementary particles origination:

the *1st - symmetry shift* of rotating Bivacuum fermions of $\mu$ and $\tau$ generation, turning them to muons and tauons and the *2nd - fusion* of the muons and tauons and their antiparticles to the triplets of the electrons, protons and neutrons, occur at Golden mean (GM) condition.

This condition is defined as a ratio of tangential velocity of Cooper pairs rotation around common axis to light velocity, equal to GM: $(\mathbf{v}/\mathbf{c})^2 = 0.618 = \phi$.

*The triplets (elementary particles)* $< [\mathbf{F}_{\uparrow}^{+} \bowtie \mathbf{F}_{\downarrow}^{-}]_{x,y} + \mathbf{F}_{\updownarrow}^{\pm} >^{e,p}$ *are stabilized by three factors:*

a) the resonance exchange interaction with Bivacuum virtual pressure waves $(\mathbf{VPW}_{q=1}^{\pm})^{e,p,}$ and sub-elementary fermions pulsing with Compton angular frequency: $\boldsymbol{\omega}_{q=1}^{e,p} = \mathbf{m}_{q=1}^{e,p}\mathbf{c}^2/\hbar$;

b) the Coulomb and magnetic attraction between sub-elementary fermions of the opposite electric and magnetic charges;

c) the gluons (cumulative virtual clouds $\mathbf{CVC}^{+}$ and $\mathbf{CVC}^{-}$ in our terminology) exchange between unpaired sub-elementary fermion of the electron or proton $\mathbf{F}_{\updownarrow}^{\pm} >$ and one of the paired of the same charge, but opposite spin and opposite phase of $\mathbf{C} \to \mathbf{W}$ and $\mathbf{W} \to \mathbf{C}$ pulsation.

So, the electrons and positrons also have the triplet structure, as the protons and neutrons, in our approach. The proposed model of the electrons, as the triplets can be verified experimentally, using 3-beam collider of muons and antimuons. *The simultaneous scattering of 2 muons and 1 antimuon at the same space point with Golden mean velocity:* $\mathbf{v} = \sqrt{\phi}\,\mathbf{c} = \mathbf{0.786c}$ *should be followed by the electrons origination and energy outburst. The colliding of two antimuons and one muon should be accompanied by the positron fusion and micro-explosion.*

Similar experiment should be managed with tauon - antitauon collider, following by the protons and neutrons origination.

The model of the triplet of truncated cones (asymmetric sub-elementary fermions) of the electron is presented at Figure below.

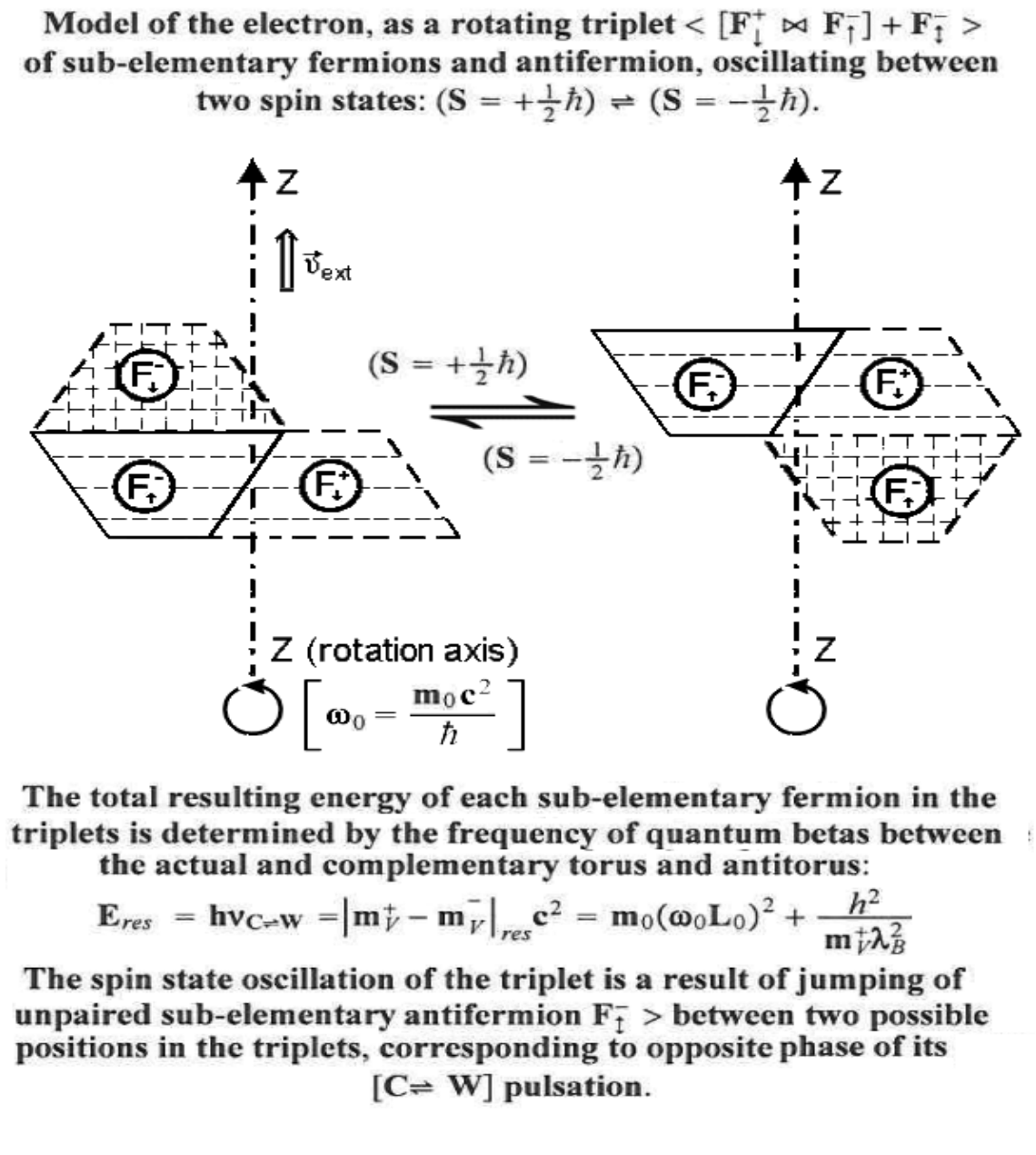

**Figure 5**. Model of the electron or positron, as a triplets $< [\mathbf{F}_{\uparrow}^{+} \bowtie \mathbf{F}_{\downarrow}^{-}]_{x,y} + \left(\mathbf{F}_{\updownarrow}^{\pm}\right)_z >^{e,p}$ oscillating between two spin state. These elementary particles are the result of fusion of the unpaired Bivacuum antifermion (muon) or antifermion (antimuon) with Cooper pair $[\mathbf{BVF}_{\pm}^{\uparrow} \bowtie \mathbf{BVF}_{\mp}^{\downarrow}]_{GM}^{\mu,\tau}$ at Golden mean condition.

Similar model, as for electron, is valid for proton and neutron. For explanation of neutral charge of neutron we suppose, that the neutron is a complex of proton with virtual electron with zero spin and negative charge. This complex is stable in stable atoms but is unstable in free neutrons

The tangential velocity of rotating unpaired sub-elementary fermion $[\mathbf{F}_{\downarrow}^{-}]$ in triplet around the same axis, as each of paired, is also the same $(\mathbf{v}/\mathbf{c})^2 = \phi = 0.618$. This provides the origination of similar rest mass $(\mathbf{m}_0)$ and charge $|\mathbf{e}^{\pm}|$, as has each of the paired sub-elementary fermions $[\mathbf{F}_{\uparrow}^{+} \bowtie \mathbf{F}_{\downarrow}^{-}]^{e,p}$ after fusion to triplet $< [\mathbf{F}_{\uparrow}^{+} \bowtie \mathbf{F}_{\downarrow}^{-}]_{x,y} + \mathbf{F}_{\updownarrow}^{\pm} >^{e,p}$ . The properties of paired $[\mathbf{F}_{\uparrow}^{+}$ and $\mathbf{F}_{\downarrow}^{-}]^{e,p}$ totally compensate each other and the mass, charge and spin of elementary particle (triplet) is determined only by the unpaired sub-elementary fermion $\mathbf{F}_{\updownarrow}^{\pm} >^{e,p}$ .

*7.1 The energy of fusion of the elementary fermions, like electrons and protons and their antiparticles from muons and tauons*

It follows from our approach, that the regular electrons and positrons are the result of fusion of triplets from odd number of *muons* and *antimuons* in relation 2:1 and 1:2, correspondingly. The protons and antiprotons are resulted from fusion of the same number *tauons* and *antitauons.* in the same proportion. A single muons and tauons, as asymmetric Bivacuum fermions, are existing also, however, they have very short life-time. The experimental values of life-times of unstable muons and tauons with properties of Bivacuum fermions at Golden Mean (GM) conditions $\left(\mathbf{BVF}_{as}^{\updownarrow}\right)_{GM}^{\mu,\tau}$, are very small: $2.19 \times 10^{-6}s$ and $3.4 \times 10^{-13}s$, respectively. *The stability of monomeric muons and tauons, strongly increases, as a result of their fusion to triplets of the electrons, protons and neutrons, since this process is accompanied by huge energy release, determined by their mass decreasing (mass defect).*

Different superpositions of three sub-elementary fermions (former tauons) after fusion to triplets, like different combinations of three interlacing *Borromean rings* (symbol, popular in Medieval Italy) and their different dynamics, can be responsible for different properties of the protons and neutrons.

The mass of tauon and antitauons. is: $\mathbf{m}_{\tau^{\pm}} = 1782(3)$ MeV. For the other hand, the mass of proton and neutron are: $\mathbf{m_p} = 938,280(3)$ MeV and $\mathbf{m_n} = 939,573\ (3)$ MeV, correspondingly. They are about two times less, than the mass of tauon or antitauon, equal, in accordance to our model, to mass of one of the quark or antiquark after fusion to triplets. This mass/energy difference between the tauon and one of quark, i.e. unpaired one, is close to the energy of neutral massless *gluons* (exchange bosons), stabilizing the triplets of protons and neutrons.

In the case of neutrons this difference is a bit less, providing, however, much shorter life-time of isolated neutrons (918 sec.), than that of protons (>$10^{31}$ years). We suppose that this huge difference in the life-span is determined by different dynamic structure of these two triplets, providing the positive charge of proton and neutrality of the neutron.

One of possible explanation of neutrality, is the conjecture, that neutron is a metastable complex of proton and virtual electron, which turns to real electron as a result of $\beta$ –decay:

$$\mathbf{n} \rightarrow \mathbf{p} + \mathbf{e}^{-} + \overline{\nu}_e \qquad 7.2$$

where $\overline{\nu}_e$ is the electron antineutrino.

The neutrino, in our approach can be a pair of Bivacuum bosons, rotating around common axes clockwise or counterclockwise, depending on direction of propagation. The energy of such neutrino is determined by velocity of rotation.

In contrast to standard model in our approach the charge of quark (sub-elementary fermion in term of our approach) is supposed to be the integer number $(\mathbf{e}^{\pm})$, not the fractional one.

The additional mass defect of the paired tauons should be twice the same as one of the unpaired. The total big difference between the mass of 3 independent sub-elementary particles (tauons) and the mass of triplets (protons and neutrons), determines the sum of the gluons energy and the excessive kinetic energy - thermal energy release, as a result of these elementary particles fusion.

The mass of the regular electron is: $\mathbf{m}_{e^\pm} = 0,511003(1)$ MeV and the mass of $\mu -$ electron (muon) is: $\mathbf{m}_{\mu^\pm} = 105,6595(2)\,MeV$. The relative difference in these masses, about 200, is much higher that for protons and tauons. This provides very strong stability of the electron as a triplet. This is a reason, why it is generally accepted, that the electron is a real elementary particle.

Like in the case of protons, the fusion of the electrons and positrons from muons and antimuons should be accompanied by the release of huge amount of kinetic thermal energy. Part of the fusion energy is used for *electronic – gluons* origination, similar to hadronic gluons in protons and neutrons, is responsible for strong exchange interaction between quarks of triplets.

The electronic gluons (*e – gluons*) are responsible for the yet unknown electronic strong interaction.

*7.2 Correlation between our model of hadrons and conventional quark model of protons and neutrons in Standard Model*

We suppose that the *proton* ($Z = +1;\ S = \pm 1/2$) *is constructed* by the same principle as the electron (Fig.5). It is a result of fusion of pair of sub-elementary fermion and antifermion of $\boldsymbol{\tau}$ –generation $[\mathbf{F}_\updownarrow^- \bowtie \mathbf{F}_\updownarrow^+]^p_{S=0}$ and one unpaired sub-elementary fermion $\left(\mathbf{F}_\updownarrow^+\right)^\tau_{S=\pm 1/2}$, accompanied by huge energy release, corresponding to mass defect: $\Delta\mathbf{E} \sim (m^\tau - m^p)\mathbf{c}^2$. These three components of proton correspond to three quarks: $\left(\mathbf{F}_\updownarrow^+\right)^p_{S=\pm 1/2} \sim \mathbf{q}^+$ and antiquarks $\left(\mathbf{F}_\updownarrow^-\right)^p_{S=\pm 1/2} \sim \mathbf{q}^-$.

The difference with standard quark model is that we do not need to use the notion of fractional charge in our model of proton with spin $S = \pm 1/2$:

$$\mathbf{p} \equiv\ < [\mathbf{F}_\uparrow^- \bowtie \mathbf{F}_\downarrow^+]^{x,y}_{S=0} + (\mathbf{F}_\uparrow^+)^z_{S=\pm 1/2} >^\tau \qquad 7.3$$

$$or:\ \mathbf{p} \sim \left\langle [\mathbf{q}^- \bowtie \mathbf{q}^+]^{x,y}_{S=0} + (\mathbf{q}^+)^z_{S=\pm 1/2} \right\rangle \qquad 7.3a$$

The charges, spins and mass/energy of sub-elementary particles and antiparticles in pairs $[\mathbf{F}_\uparrow^- \bowtie \mathbf{F}_\downarrow^+]^\tau_{x,y}$ compensate each other. The resulting properties of protons ($\mathbf{p}$) are determined by unpaired/uncompensated sub-elementary particle $\mathbf{F}_\uparrow^+ >^\tau$ of heavy $\tau$ –electrons generation, including the recoil effects, responsible for charge effect of proton. The mutual recoil and charge compensation effects of two paired sub-elementary fermions is possible, if they are oriented opposite to each other and are pulsing in 2-dimensional plane (x,y) in the the same phase.

The absence of charge in the *neutron* ($Z = 0;\ S = \pm 1/2$) can be explained in two ways:

1) as a result of complex of proton with oppositely charged sub-elementary fermion of $e$ - generation:

$$\mathbf{n} \equiv\ < [\mathbf{F}_\uparrow^- \bowtie \mathbf{F}_\downarrow^+]^\tau_{S=0} + [(\mathbf{F}_\uparrow^+)^\tau \bowtie (\mathbf{F}_\downarrow^-)^e]_{S=\pm 1/2} > \qquad 7.4$$

$$or:\ \mathbf{n} \sim\ [\mathbf{q}^+ \bowtie \mathbf{q}^-]^{\boldsymbol{\tau}}_{S=0} + \left(\mathbf{q}_\uparrow^0\right)^{\boldsymbol{\tau e}}_{S=\pm 1/2} \qquad 7.4a$$

$$or:\ \mathbf{n} \sim\ [\boldsymbol{\tau}^+ \bowtie \boldsymbol{\tau}^-]^{\boldsymbol{\tau}}_{S=0} + ([\boldsymbol{\tau}_\uparrow^+]^{\boldsymbol{\tau}} \bowtie [\mathbf{F}_\updownarrow^-]^e) \qquad 7.4b$$

where the neutral quark $\left(\mathbf{q}_\uparrow^0\right)^{\boldsymbol{\tau e}}_{S=\pm 1/2}$ is introduced, as a metastable complex of positive sub-elementary $\tau$ –fermion $\left(\mathbf{F}_{\updownarrow}^{+}\right)^{\tau}$ with negative electron's sub-elementary fermion $\mathbf{F}_{\uparrow}^{-} >^{e}$ sub-elementary fermion of opposite charge $[\mathbf{F}_{\updownarrow}^{-}]^{e}$:

$$\left(\mathbf{q}_{\uparrow}^{0}\right)_{S=\pm 1/2}^{\tau\mathbf{e}} = \left(\left[\mathbf{q}_{\uparrow}^{+}\right] \bowtie [\mathbf{F}_{\updownarrow}^{-}]^{e}\right) \qquad 7.5$$

This means that the positive charge of unpaired heavy sub-elementary particle $(\mathbf{F}_{\uparrow}^{+})^{\tau}$ in neutron (**n**) is compensated by the charge of the light sub-elementary fermion $(\mathbf{F}_{\downarrow}^{-})^{e}$. In contrast to charge, the spin of unpaired $(\mathbf{F}_{\uparrow}^{+})^{\tau}$ is not compensated (totally) by spin of $(\mathbf{F}_{\downarrow}^{-})^{e}$ in neutrons, because of strong mass and angular momentum difference in conditions of the $(\mathbf{F}_{\downarrow}^{-})^{e}$ confinement.

The 2nd possible explanation of zero charge of the neutron is such relative 3D configuration of sub-elementary fermions, which provides the recoilless $\mathbf{C} \rightleftharpoons \mathbf{W}$ pulsation of all three sub-elementary fermions, like in Mössbauer effect (see section 8.10).

$$\langle[(\mathbf{F}_{\uparrow}^{+})^{x} \bowtie (\mathbf{F}_{\downarrow}^{-})^{y}]_{W} \bowtie (\mathbf{F}_{\uparrow}^{-})_{C}^{z}\rangle_{n} \rightleftharpoons \langle[(\mathbf{F}_{\uparrow}^{+})^{x} \bowtie (\mathbf{F}_{\downarrow}^{-})^{y}]_{C} \bowtie (\mathbf{F}_{\uparrow}^{-})_{W}^{z}\rangle_{n} \qquad 7.5a$$

In this configuration all three sub-elementary fermions in [C] phase are oriented normal to each other and the recoil and charge effects, accompanied $\mathbf{C} \rightleftharpoons \mathbf{W}$ pulsation of all three sub-elementary fermion totally compensate each other.

Different superpositions of three sub-elementary fermions, like different combinations of three interlacing *Borromean rings* (symbol, popular in Medieval Italy) can be responsible for different properties of the electrons, protons and neutrons.

The mass of $\tau$- electron, equal to that of $\tau$-positron is: $\mathbf{m}_{\tau^{\pm}} = 1782(3)$ MeV, the mass of the regular electron is: $\mathbf{m}_{e^{\pm}} = 0{,}511003(1)$ MeV and the mass of $\mu$ – electron is: $\mathbf{m}_{\mu^{\pm}} = 105{,}6595(2)\,MeV.$

For the other hand, the mass of proton and neutron are correspondingly: $\mathbf{m_p} = 938{,}280(3)$ MeV and $\mathbf{m_n} = 939{,}573\ (3)$ MeV. They are about two times less, than the mass of $\tau$- electron, equal, in accordance to our model, to mass of its unpaired sub-elementary fermion $(\mathbf{F}_{\uparrow}^{+})^{\tau}$. This difference characterize the energy of neutral massless *gluons* (exchange bosons), stabilizing the triplets of protons and neutrons. In the case of neutrons this difference is a bit less (taking into account the mass of $[\mathbf{F}_{\updownarrow}^{-}]^{e}$), providing, however, much shorter life-time of isolated neutrons (918 sec.), than that of protons ($>10^{31}$ years).

In accordance to our *hadrons* models, each of three quarks (sub-elementary fermions of $\tau$ – generation) in protons and neutrons can exist in 3 states (*red*, *green* and *blue*), but not simultaneously:

1. The *red* state of quark/antiquark means that it is in corpuscular [C] phase;
2. The *green* state of quark/antiquark means that it is in wave [W] phase;
3. The *blue* state means that quark/antiquark $(\mathbf{F}_{\updownarrow}^{\pm})^{\tau}$ is in the transition [C]⇔[W] state.

The 8 different combinations of the above defined states of 3 quarks of protons and neutrons correspond to *8 gluons colors*, stabilizing the these hadrons. The triplets of quarks are stabilized by the emission ⇌ absorption of cumulative virtual clouds ($\mathbf{CVC}^{\pm}$) in the process of quarks $[\mathbf{C} \rightleftharpoons \mathbf{W}]$ pulsation.

The known experimental values of life-times of **μ** and **τ** electrons, corresponding in our model, to monomeric asymmetric sub-elementary fermions $\left(\mathbf{BVF}_{as}^{\updownarrow}\right)^{\mu,\tau}$, are equal only to $2{,}19 \times 10^{-6}s$ and $3{,}4 \times 10^{-13}s$, respectively. We assume here, that stability of monomeric sub-elementary particles/antiparticles of **e**, **μ** and **τ** generations, strongly increases, as a result of their fusion in triplets, possible at Golden mean conditions.

The well known example of weak interaction, like $\beta$ – *decay* of the neutron to proton, electron and **e** –antineutrino:

$$\mathbf{n} \rightarrow \mathbf{p} + \mathbf{e}^- + \overline{\boldsymbol{\nu}}_e \qquad 7.6$$

$$or: \left\langle [\mathbf{q}^+ \bowtie \overline{\mathbf{q}}^-] + \left(\mathbf{q}_\uparrow^0\right)^{\boldsymbol{\tau}\mathbf{e}}_{S=\pm 1/2} \right\rangle \rightarrow ([\mathbf{q}^+ \bowtie \overline{\mathbf{q}}^-] + \mathbf{q}^+) + \mathbf{e}^- + \overline{\boldsymbol{\nu}}_e \qquad 7.6a$$

is in accordance with our model of elementary particles and theory of neutrino (section 8.4).

The sub-elementary fermion of $\tau$ – generation in composition of proton or neutron can be considered, as a quark and the sub-elementary antifermion, as antiquark:

$$\left(\mathbf{F}_\updownarrow^+\right)^\tau \sim \mathbf{q}^+ \quad and \quad \left(\mathbf{F}_\updownarrow^-\right)^\tau \sim \overline{\mathbf{q}}^- \qquad 7.7$$

In the process of $\beta$ –decay of neutron the unpaired negative sub-elementary fermion $[\mathbf{F}_\updownarrow^-]^e$ in (5.11) forms a complex - triplet (electron) with complementary virtual Cooper pair $[\mathbf{F}_\uparrow^- \bowtie \mathbf{F}_\downarrow^+]^e_{S=0}$ from the vicinal to neutron polarized Bivacuum:

$$[\mathbf{F}_\updownarrow^-]^e + [\mathbf{F}_\uparrow^- \bowtie \mathbf{F}_\downarrow^+]^e_{S=0} \rightarrow \mathbf{e}^- \qquad 7.8$$

If we accept the explanation of zero charge of neutron, as a result of total compensation of recoil dynamics in the process of correlated $\mathbf{C} \rightleftharpoons \mathbf{W}$ pulsation of three of its sub-elementary fermions, then $\beta$ –decay can be considered as conversion of such specific configuration of neutron (5.11a) to another configuration, pertinent for proton:

$$\langle [\mathbf{F}_\uparrow^+ \bowtie \mathbf{F}_\downarrow^-] \bowtie (\mathbf{F}_\uparrow^-) \rangle_n \rightarrow \mathbf{p} + \mathbf{e}^- + \overline{\boldsymbol{\nu}}_e \qquad 7.8a$$

where the configuration of proton is described by (7.3). This conversion is accompanied by excitation of vicinal virtual electron($\widetilde{\mathbf{e}}^-$) and its transition to the real pair [electron + antineutrino] $\mathbf{e}^- + \overline{\boldsymbol{\nu}}_e$.

The energy of 8 gluons, corresponding to different superposition of $[\mathbf{CVC}^+ \bowtie \mathbf{CVC}^-]_{S=0,1}$, emitted and absorbed with in-phase $[\mathbf{C} \rightleftharpoons \mathbf{W}]$ pulsation of pair [quark + antiquark] in triplets (5.9 - 5.9b):

$$[\mathbf{F}_\uparrow^+ \bowtie \mathbf{F}_\downarrow^-]^\tau_{S=0,1} = [\mathbf{q}^+ + \widetilde{\mathbf{q}}^-]_{S=0,1} \qquad 7.9$$

is about 50% of energy/mass of quarks and antiquarks ($\boldsymbol{\tau}$ sub-elementary fermions and antifermions).

These 8 gluons, responsible for strong interaction, can be presented as a following combinations of transitions states of $\tau$ – sub-elementary fermions (quarks $q_2$ and $q_3$) and antifermion (antiquark $\widetilde{q}_1$), corresponding to two spin states of proton ($S = \pm 1/2\hbar$) of unpaired quark.

For its spin state: $S_{q_3} = +1/2\hbar$ we have following 4 transition combinations of triplets transition states, corresponding to four types of gluons:

$$1) \quad \left\langle \left( [C \rightarrow W]^{S=1/2}_{\widetilde{q}_1} \bowtie [C \rightarrow W]^{S=-1/2}_{q_2} \right) + [C \rightarrow W]^{S=1/2}_{q_3} \right\rangle \qquad 7.10$$

$$2) \quad \left\langle \left( [W \rightarrow C]^{S=1/2}_{\widetilde{q}_1} \bowtie [W \rightarrow C]^{S=-1/2}_{q_2} \right) + [C \rightarrow W]^{S=1/2}_{q_3} \right\rangle \qquad 7.10a$$

$$3) \quad \left\langle \left( [C \rightarrow W]^{S=1/2}_{\widetilde{q}_1} \bowtie [C \rightarrow W]^{S=-1/2}_{q_2} \right) + [W \rightarrow C]^{S=1/2}_{q_3} \right\rangle \qquad 7.10b$$

$$4) \quad \left\langle \left( [W \rightarrow C]^{S=1/2}_{\widetilde{q}_1} \bowtie [W \rightarrow C]^{S=-1/2}_{q_2} \right) + [W \rightarrow C]^{S=1/2}_{q_3} \right\rangle \qquad 7.10c$$

and for the opposite spin state of unpaired quark: $S_{q_3} = -1/2\hbar$ we have also 4 transition states combinations, representing another four types of gluons:

$$5)\ \left\langle \left( [C \to W]^{S=1/2}_{\tilde{q}_1} \bowtie [C \to W]^{S=-1/2}_{q_2} \right) + [C \to W]^{S=-1/2}_{q_3} \right\rangle \quad 7.11$$

$$6)\ \left\langle \left( [W \to C]^{S=1/2}_{\tilde{q}_1} \bowtie [W \to C]^{S=-1/2}_{q_2} \right) + [C \to W]^{S=-1/2}_{q_3} \right\rangle \quad 7.11a$$

$$7)\ \left\langle \left( [C \to W]^{S=1/2}_{\tilde{q}_1} \bowtie [C \to W]^{S=-1/2}_{q_2} \right) + [W \to C]^{S=-1/2}_{q_3} \right\rangle \quad 7.11b$$

$$8)\ \left\langle \left( [W \to C]^{S=1/2}_{\tilde{q}_1} \bowtie [W \to C]^{S=-1/2}_{q_2} \right) + [W \to C]^{S=-1/2}_{q_3} \right\rangle \quad 7.11c$$

One of our versions of elementary particle fusion have some similarity with thermonuclear *fusion* and can be as follows. The rest mass of *isolated* sub-elementary fermion/antifermion *before* fusion of the electron or proton, is equal to the rest mass of unstable muon or tauon, correspondingly. The 200 times decrease of muons mass is a result of mass defect, equal to the binding energy of triplets: electrons or positrons. It is provided by origination of electronic *e-gluons* and release of the huge amount of excessive kinetic (thermal) energy, for example in form of high energy photons or *e-neutrino* beams.

In protons, as a result of fusion of three $\tau$ –electrons/positrons, the contribution of hadron *h-gluon* energy to mass defect is only about 50% of their mass. However, the absolute hadron fusion energy yield is higher, than that of the electrons and positrons.

*Our hypothesis of stable electron/positron and hadron fusion from short-living $\mu$ and $\tau$ - electrons, as a precursor of electronic and hadronic quarks, correspondingly, can be verified using special collider.*

In accordance to our Unified Theory, there are two different mechanisms of stabilization of the electron and proton structures in form of triplets of sub-elementary fermions/antifermions of the reduced $\mu$ and $\tau$ generations, correspondingly, preventing them from exploding under the action of self-charge:

1. Each of sub-elementary fermion/antifermion, representing asymmetric pair of torus ($\mathbf{V}^+$) and antitorus ($\mathbf{V}^-$), as a charge, magnetic and mass dipole, is stabilized by the Coulomb, magnetic and gravitational attraction between torus and antitorus;

2. The stability of triplet, as a whole, is provided by the exchange of Cumulative Virtual Clouds ($CVC^+$ and $CVC^-$) between three sub-elementary fermions/antifermions in the process of their [$\mathbf{C} \rightleftharpoons \mathbf{W}$] pulsation. In the case of proton and neutron, the 8 transition states corresponds to 8 *h-gluons* of hadrons, responsible for strong interaction. In the case of the electron or positron, the stabilization of triplets is realized by 8 lighter *e-gluons*. The process of internal energy exchange of pairs $[\mathbf{F}^-_\uparrow \bowtie \mathbf{F}^+_\downarrow]^{e,p}_{S=0,1}$ with unpaired sub-elementary fermion in triplets of hadrons is accompanied also by the energy exchange with external Bivacuum medium. It is resulted in modulation of positive and negative virtual pressure waves [$\mathbf{VPW}^+ \bowtie \mathbf{VPW}^-$] of Bivacuum, generating the Virtual Replica Multiplication of nucleons (see chapter 13). The feedback reaction between Bivacuum and elementary particles is also existing.

*7.3 Possible structure of mesons, $W^\pm$ and $Z^0$ bosons of electroweak interaction*

By definition of Standard Model, the *mesons* are a family of subatomic particles (about 140) that participate in strong interactions and have masses intermediate between leptons and baryons. When the mass of such particles, formed by quarks like baryons, exceeds the mass of baryons (proton, neutron, lambda and omega), they named *bosonic hadrons.* It is generally assumed, that they are composed of a quark and an antiquark. They are bosons with spin, equal to zero or 1 and possible charge: 0, +1 and -1.

In our approach (see 5.15) the pairs of sub-elementary fermions of $\tau$ or $\mu$ generations $[\mathbf{F}^-_\uparrow + \mathbf{F}^+_\downarrow]^{\tau,\mu}_{S=0,1} = [\mathbf{q}^- + \mathbf{q}^+]^{\tau,\mu}_{S=0,1}$ (see 5.6 - 5.9a), have a properties of *mesons,* as a neutral [quark + antiquark] pair with bosonic integer spin. However these sub-elementary fermions are not symmetric necessarily, like $[\mathbf{F}_{\uparrow}^{-} \bowtie \mathbf{F}_{\downarrow}^{+}]_{S=0,1}^{\tau,\mu}$ of triplets. The coherent cluster of such pairs - from one to four pairs: $(\mathbf{n}[\mathbf{q}^{+} + \mathbf{q}^{-}])_{S=0,1,2,3,4}$ can provide the experimentally revealed integer spins of mesons - from zero to four.

We assume also that some of experimentally revealed charged mesons, like *pions* $(\pi^{+})$, standing for nucleons interaction, may represent the intermediate bosonic state of spin exchange process between sub-elementary fermion and antifermion of muon generation $(\mathbf{BVB})_{S=0}^{z=+1}$:

$$[\mathbf{F}_{\uparrow}^{-} + \mathbf{F}_{\downarrow}^{+}]_{S=0,1}^{\mu} \rightleftharpoons \left[\mathbf{F}_{\uparrow}^{-} \rightleftharpoons (\mathbf{BVB})_{\mathbf{S=0}}^{\mathbf{z=\pm1}} \rightleftharpoons \mathbf{F}_{\downarrow}^{+}\right]^{\mu} \qquad 7.12$$

If *pion* with mass (0.140 GeV/c$^2$), is the intermediate state between muon and antimuon, indeed, this explains the decay of pion and antipion on muon (antimuon) and muonic neutrino (antineutrino):

$$(\mathbf{BVB})_{\mathbf{S=0}}^{\mathbf{z=\pm1}} \to \boldsymbol{\mu}^{\pm} + \boldsymbol{\nu}_{\mu}(\overline{\boldsymbol{\nu}}_{\mu}) \qquad 7.13$$

The negatively charged *kaon* $(K^{-})$ and antikaon $(\overline{K}^{+})$with mass (0.494 GeV/c$^2$) about 5 times bigger than that of muon (0.106 GeV/c$^2$), can represent a small cluster of the odd number of Bivacuum bosons of $\mu$ – generation, like:

$$[2(\mathbf{BVB}^{+} \bowtie \mathbf{BVB}^{-}) + \mathbf{BVB}^{\pm}]^{z=\pm1} \qquad 7.14$$

The neutral heavy B-zero meson ($\mathbf{B}^0$) with mass (5.279 GeV/c$^2$) and eta-c meson (2.980 GeV/c$^2$) can be a clusters of *even* number of Bivacuum bosons of $\tau$ – generation of opposite symmetry shift, compensating the opposite charges of each other in pairs.

The interrelation between muon and the electron follows from two decay reactions of muon and antimuon:

$$\boldsymbol{\mu}^{-} \to \mathbf{e}^{-} + \overline{\boldsymbol{\nu}}_{e} + \boldsymbol{\nu}_{\mu} \qquad 7.15$$

$$\boldsymbol{\mu}^{+} \to \mathbf{e}^{+} + \boldsymbol{\nu}_{e} + \overline{\boldsymbol{\nu}}_{\mu} \qquad 7.15a$$

In terms of our Unified theory (UT), the neutrino and antineutrino are stable carriers of the excessive Bivacuum dipoles mass/energy symmetry shifts - positive ($\boldsymbol{\nu}_{e,\mu}$) or negative ($\overline{\boldsymbol{\nu}}_{e,\mu}$) see section 8.4.

The existence of heavy charged ($\mathbf{W}^{+}$, $\mathbf{W}^{-}$) and neutral ($\mathbf{Z}^{0}$) force carriers bosons with integer spin **0**, **1**, **2**... and mass: $(80.4;\ 80.4\ \textit{and}\ 91.187)\ GeV/c^2$, correspondingly, introduced in electroweak theory is confirmed experimentally. These bosons complex structure differs strongly from that of photons. This author suggest, that the charged bosons $\mathbf{W}^{+}$, $\mathbf{W}^{-}$ are the 'rings' constructed from the *odd* number of asymmetric Bivacuum bosons of $\tau$ – generation of opposite symmetry shift and charge and the neutral bosons ($\mathbf{Z}^{0}$) from the *even* number of paired Bivacuum bosons $(\mathbf{BVB}^{+} \bowtie \mathbf{BVB}^{-})_{as}^{\tau}$, compensating the charges of each other. These heavy bosons belongs to class of very unstable particles, named *resonances*, as far their decay/disassembly is related to strong interaction. Their life times $\boldsymbol{\tau} = \hbar/\boldsymbol{\Gamma}$ interrelated with *width of resonance* ($\boldsymbol{\Gamma}$) are very short $\sim 2\times10^{-25}$ s.

The rotating around common axes $\mathbf{BVB}^{+}$ and $\mathbf{BVB}^{-}$ forming virtual microtubules has a positive and negative charge and mass symmetry shift, corresponding to Golden mean condition $(\mathbf{v}^2/\mathbf{c}^2) = \phi = 0.618$. These dipoles interact *side-by-side* in the same pairs and by *head-to-tail* principle when forming doubled microtubules from adjacent pairs:

$$\mathbf{n} \times (\mathbf{BVB}^{+} \bowtie \mathbf{BVB}^{-})_{S=0,1,..}^{\tau} = \mathbf{n} \times \left[(\mathbf{V}^{+}\uparrow\downarrow \mathbf{V}^{-})^{i} \bowtie (\mathbf{V}^{+}\downarrow\uparrow \mathbf{V}^{-i})\right]_{S=0,1,..}^{\tau} \qquad 7.16$$

We suppose, that these pairs polymerize in ring structures, different from that of photon and providing the uncompensated mass of such rotating virtual rings, equal to mass of $W^{\pm}$ and $Z^0$ bosons. The positive and negative charges in each pair $(\mathbf{BVB}^+ \bowtie \mathbf{BVB}^-)^{\tau}_{S=0,1,..}$ compensate each other and the resulting charge of the 'ring' is equal to charge ($\mathbf{e}^{\pm}$) of one excessive unpaired $(\mathbf{BVB}^+)^{\tau}_{S=0,1,..}$ or $(\mathbf{BVB}^-)^{\tau}_{S=0,1,..}$.

It is possible to evaluate the velocity of bosonic 'ring' rotation, taking its mass, equal to: $\mathbf{M}_{W^{\pm}} = 80.4\ \mathrm{GeV/c^2}$ and the ring radius, equal to Compton radius of neutron: $\mathbf{L_n} = \hbar/\mathbf{m}_n c$, the region of electroweak interaction action. Then using the obtained earlier formula (3.14) for de Broglie radius of Bivacuum dipoles circulation, we get the condition for bosonic 'ring' ($\mathbf{L}^{W^{\pm}}_{\mathbf{Vir}}$):

$$\mathbf{L}^{W^{\pm}}_{\mathbf{Vir}} = \frac{\hbar \mathbf{c}}{\mathbf{M}_{W^{\pm}} \mathbf{v}^2} = \frac{\hbar}{\mathbf{m}_n \mathbf{c}} = \mathbf{L_n} \qquad 7.17$$

where the mass of neutron $\mathbf{m}_n = 0.940\ \mathrm{GeV/c^2}$.

From this formula we may get the velocity of 'ring' rotation:

$$\mathbf{v} = \mathbf{c} \times \left( \frac{\mathbf{m}_n}{\mathbf{M}_{W^{\pm}}} \right)^{1/2} = \mathbf{c} \times \mathbf{0.1081} \qquad 7.18$$

The corresponding velocity for $\mathbf{Z}^0$ boson is very close to that. We may see, that rotation of these ring - shape bosons is nonrelativistic. However, it becomes equal to light velocity, at the assumption, that radius of heavy bosons is determined by their Compton radius.

Otherwise, the heavy bosons and other *resonances* can be considered as the intermediate - gluonic state, when the asymmetric Bivacuum boson and antiboson with zero charge, but opposite polarization, exchange their cumulative virtual clouds, being simultaneously in the wave [W] phase. In this case the equality (7.16) turns to:

$$\mathbf{n} \times (\mathbf{BVB}^+ \bowtie \mathbf{BVB}^-)^{\tau}_{S=0,1,..} \overset{\mathbf{C \to W}}{\Rightarrow} \mathbf{n} \times (\mathbf{CVC}^+ \bowtie \mathbf{CVC}^-)^{\tau}_{S=0,1,..} \qquad 7.19$$

The proposed approach to analysis of hadrons and mesons intrinsic features can be developed further to explain the general roots of all know elementary particles, taking into account their duality of sub-elementary fermions of all three generation and combination of their different states. It looks that it is possible to do without strong contradictions with Standard model. However our theory explains the origination of mass of elementary particles without Higgs field and corresponding bosons, not detected experimentally.

*7.5 New Scenario of the Big Bang*

Our approach leads to new scenario of Big Bang, taking into account the experimental data of acceleration of the Universe expansion. It can be anticipated, that after hundred of billions years of such expansion, dying of stars and black holes evaporation, the mass and energy density of the Universe will tend to zero and slightly asymmetric Bivacuum tends to primordial, symmetric one.

The big collective symmetry fluctuations of vast number of Bivacuum dipoles toward the positive and negative energy simultaneously in domains of virtual Bose condensation without violation of energy conservation may happen. The spontaneous fusion of elementary fermions - the triplets from asymmetric Bivacuum sub-elementary particles: tauons, muons and their antiparticles, can be considered as a kind of coherent chain reaction, accompanied by huge energy release.

*Just this energy of the avalanche chain reaction of elementary fermions fusion from sub-elementary ones, provided by of enormous mass defect, could be a driving energy of Big Bang* (Kaivarainen, 2006: http://arxiv.org/ftp/physics/papers/0207/0207027.pdf). The

still-remaining minor asymmetry of Bivacuum before Big Fluctuation (the trace of the dyed Universe) may be responsible for small difference between probability of matter and antimatter origination, providing the relict radiation after annihilation.

*So we may conclude, that the huge outburst of energy and the Big Bang, following the expansion of the Universe, is a result of correlated and opposite Bivacuum dipoles symmetry fluctuation, following by elementary particles fusion from sub-elementary ones.*

## 8. The dynamic mechanism of Corpuscle - Wave duality of elementary particles

The dynamic model of counterphase Corpuscle ⇌ Wave pulsation of paired and unpaired sub-elementary fermions in composition of triplets (elementary fermions, like electrons and protons), following from our approach, is presented below.

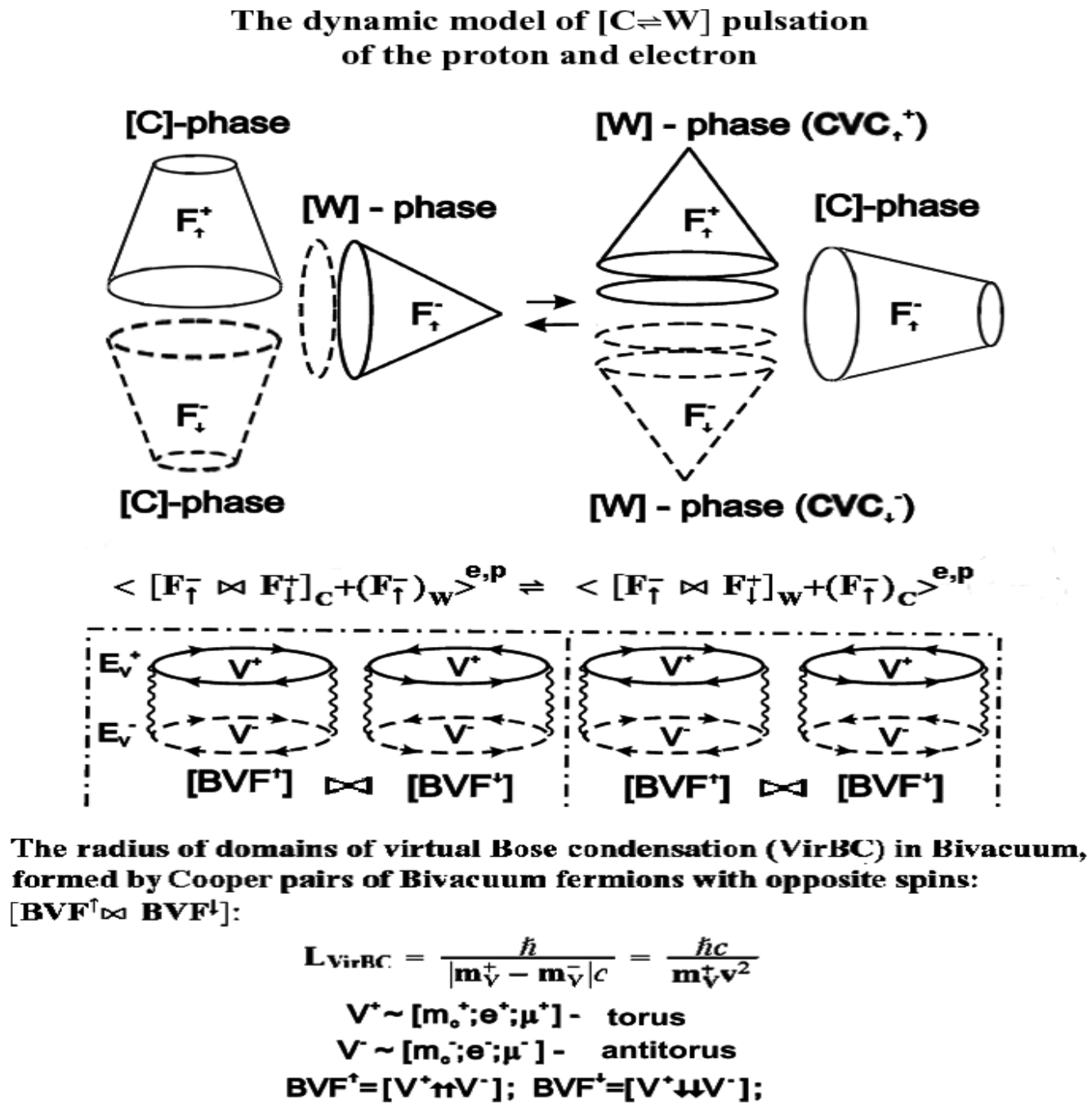

**Figure. 6.** Dynamic model of [**C** ⇌ **W**] pulsation of triplets of sub-elementary fermions/antifermions of the electrons or protons $< [\mathbf{F}_\uparrow^+ \bowtie \mathbf{F}_\downarrow^-] + \mathbf{F}_\updownarrow^\pm >^{e,p}$ . The pulsation of the paired sub-elementary fermions $[\mathbf{F}_\uparrow^- \bowtie \mathbf{F}_\downarrow^+]^{e,p}$ is counterphase to pulsation of unpaired sub-elementary fermion/antifermion $\mathbf{F}_\updownarrow^\pm >^{e,p}$ . The unpaired $\mathbf{F}_\updownarrow^\pm >^{e,p}$ determines the properties of triplet. The pair $[\mathbf{F}_\uparrow^- \bowtie \mathbf{F}_\downarrow^+]^{e,p}$ is responsible for interaction of triplet with Bivacuum virtual pressure waves.

The rotation of triplets $< [\mathbf{F}_\uparrow^+ \bowtie \mathbf{F}_\downarrow^-] + \mathbf{F}_\updownarrow^\pm >^{e,p}$ around main axis with Golden mean tangential velocity ($\mathbf{v}_{rot} = \sqrt{\phi}\,\mathbf{c}$), provides the rest mass and charge origination of triplet.

The propagation of triplets in space with translational velocity ($\mathbf{v}_{tr}^{ext} \equiv \mathbf{v}$) determines its external kinetic energy, momentum and the empirical de Broglie wave length. The absolute values of energies, spins and charges of all three sub-elementary fermions of triplets are interdependent and equal to each other. However the paired sub-elementary fermions $[\mathbf{F}_\uparrow^+ \bowtie \mathbf{F}_\downarrow^-]$ compensate the properties of each other totally. They are responsible for interaction of triplets with Bivacuum virtual pressure waves.

The total energy of each sub-elementary fermion or antifermion in the triplet: $\mathbf{E}_{tot}^{res} = |\mathbf{m}_V^+ - \mathbf{m}_V^-|^{res}\mathbf{c}^2$ can be presented as a sum of two contributions:

a) the internal rotational rest mass energy: $\mathbf{m}_0\mathbf{c}^2 = |\mathbf{m}_V^+ - \mathbf{m}_V^-|_{rot}^{\phi}\mathbf{c}^2 = (\mathbf{m}_0\boldsymbol{\omega}_0^2\mathbf{L}_0^2)_{\mathbf{rot}}^{in}$ and

b) the external doubled translational kinetic energy of the triplet:

$\mathbf{m}_V^+\mathbf{v}_{tr}^2 = \frac{\mathbf{h}^2}{\mathbf{m}_V^+\boldsymbol{\lambda}_B^2} = |\mathbf{m}_V^+ - \mathbf{m}_V^-|_{tr}^{ext}\mathbf{c}^2$:

$$\mathbf{E}_{tot}^{res} = \mathbf{E}_{rot}^{in} + \mathbf{E}_{tr}^{ext} \qquad 8.1$$

$$or: \quad \mathbf{E}_{tot}^{res} = |\mathbf{m}_V^+ - \mathbf{m}_V^-|^{res}\mathbf{c}^2 = \mathbf{m}_V^+\mathbf{v}_{res}^2 = \mathbf{m}_0\mathbf{c}^2 + \mathbf{m}_V^+\mathbf{v}_{ext}^2 \qquad 8.1a$$

$$or: \quad \mathbf{E}_{tot}^{res} = \mathbf{h}\boldsymbol{\nu}_{\mathbf{C}\rightleftharpoons\mathbf{W}}^{\mathbf{res}} = (\mathbf{m}_0\boldsymbol{\omega}_0^2\mathbf{L}_0^2)_{\mathbf{rot}}^{in} + \frac{\mathbf{h}^2}{\mathbf{m}_V^+\boldsymbol{\lambda}_B^2} \qquad 8.1b$$

where: $\mathbf{L}_0 = \hbar/\mathbf{m}_0\mathbf{c}$ is the Compton radius of sub-elementary fermion; $\boldsymbol{\lambda}_B = h/\mathbf{m}_V^+\mathbf{v}_{tr}^{ext} = \mathbf{v}/\boldsymbol{\nu}_B$ is the empirical de Broglie wave length of the triplet, dependent on its external translational velocity and momentum.

The right parts of $(8.1 - 8.1b)$ contain also small contributions of electric, magnetic and gravitational energies, which will be considered in chapter 17. They are not influence the dynamic mechanism of duality, proposed in this book.

The resulting velocity squared ($\mathbf{v}_{res}^2$) of sub-elementary fermions in composition of elementary fermion is related to external translational velocity squared $\mathbf{v}_{ext}^2$ as:

$$\mathbf{v}_{res}^2 = \phi\mathbf{c}^2 + \mathbf{v}_{ext}^2 \qquad 8.2$$

The resulting energy of sub-elementary fermions in triplets $\mathbf{E}_{tot}^{res}$ can be expressed via frequency of quantum beats:

$\boldsymbol{\nu}_{\mathbf{C}\rightleftharpoons\mathbf{W}}^{\mathbf{res}} = |\mathbf{m}_V^+ - \mathbf{m}_V^-|^{res}\mathbf{c}^2/h$ between their actual (torus) and complementary (antitorus) states, as a sum of angular frequency of rotation of triplet $(\boldsymbol{\omega}_0)_{rot}^{in}$, providing the rest mass origination, and empirical de Broglie wave frequency ($\boldsymbol{\nu}_B$):

$$\mathbf{h}\boldsymbol{\nu}_{\mathbf{C}\rightleftharpoons\mathbf{W}}^{res} = (\hbar\boldsymbol{\omega}_0)_{rot}^{in} + \mathbf{h}\boldsymbol{\nu}_B^{ext} \qquad 8.3$$

The internal rotational energy formula of elementary fermions, responsible for the rest mass and charge origination has few shapes:

$$\mathbf{E}_{rot}^{in} = (\hbar\boldsymbol{\omega}_0)_{rot}^{in} = (\mathbf{m}_0\boldsymbol{\omega}_0^2\mathbf{L}_0^2)_{\mathbf{rot}}^{in} = \mathbf{m}_0\mathbf{c}^2 \qquad 8.4$$

The internal Golden mean angular frequency of rotation of sub-elementary fermions in triplets is equal to Compton frequency of beats between their torus and antitorus:

$$\left[ (\boldsymbol{\omega}_0)_{rot}^{in} = \frac{|\mathbf{m}_V^+ - \mathbf{m}_V^-|^{\phi}\mathbf{c}^2}{\hbar} = \frac{(\mathbf{m}_0\mathbf{c}^2)_{rot}^{in}}{\hbar} \right]^{e,p} \qquad 8.5$$

The external translational energy of de Broglie wave can be introduced via its frequency ($\boldsymbol{\nu}_B^{ext}$), and contribution to resulting truncated cone asymmetry: $(\mathbf{m}_V^+ - \mathbf{m}_V^-)^{anc}\mathbf{c}^2 = 2\mathbf{T}_k$, determined by the external translational kinetic energy of the triplet:

$$\mathbf{E}_{tr}^{ext} = \mathbf{T}_k + \mathbf{V} = \mathbf{h}\boldsymbol{\nu}_B^{ext} = |\mathbf{m}_V^+ - \mathbf{m}_V^-|_{tr}^{ext}\mathbf{c}^2 = \mathbf{m}_V^+\mathbf{v}^2 \qquad 8.6$$

At the harmonic oscillator or standing wave condition the averaged kinetic and potential energies are equal: $\overline{\mathbf{T}_k} = \overline{\mathbf{V}} = \frac{1}{2}\mathbf{m}_V^+\mathbf{v}^2$.

It follows from our approach, that the fundamental phenomenon of *corpuscle – wave* duality is a result of the modulation of the internal $[C \rightleftharpoons W]^{in}$ pulsation frequency ($\boldsymbol{\omega}_0^{in}$) of sub-elementary fermions by the frequency of the external empirical de Broglie wave of triplet: $\nu_B^{ext}$ :

$$\nu_B^{ext} = \frac{\mathbf{m}_V^+ \mathbf{v}^2}{h} = \frac{h}{\mathbf{m}_V^+ \boldsymbol{\lambda}_B^2} \qquad 8.7$$

Beside the frequency $\nu_B^{ext}$ and length $\boldsymbol{\lambda}_B = h/\mathbf{m}_V^+\mathbf{v}$ the de Broglie wave should be characterized by the amplitude ($\mathbf{A}_B$).

We introduce amplitude $\mathbf{A}_B$, equal to that of Cumulative virtual cloud ($CVC^\pm$), from eq. 8.6 as:

$$\mathbf{A}_B = \frac{h}{|\mathbf{m}_V^+ - \mathbf{m}_V^-|_{tr}^{ext}\mathbf{c}} = \frac{h}{\mathbf{m}_V^+\mathbf{v}}\frac{\mathbf{c}}{\mathbf{v}} = \boldsymbol{\lambda}_B \frac{\mathbf{c}}{\mathbf{v}} \qquad 8.7a$$

So, the amplitude of de Broglie wave in form of cumulative virtual cloud ($\mathbf{CVC}^\pm$) is equal to its length only at $\mathbf{v} = \mathbf{c}$, i.e. for the photon.

For the fermions: $\mathbf{v} < \mathbf{c}$ and $\mathbf{A}_B > \boldsymbol{\lambda}_B$. Consequently, the shape of $\mathbf{CVC}^\pm$ can be approximated by the ellipsoid of rotation with axis $\mathbf{A}_B$ and $\boldsymbol{\lambda}_B$. The amplitude $\mathbf{A}_B$ characterize separation between positive and negative charges of $\mathbf{CVC}^\pm$, as a virtual dipole.

The expression (8.7a) coincides with dimension of the 'inerton cloud' introduced by Krasnoholovets (2008).

We may see that our formula for total energy of sub-elementary particle (8.1a and 8.1b) differs strongly from the conventional formula for energy of relativistic particle, following from Lorenz formula:

$$\mathbf{m}_V^+ = \mathbf{m} = \frac{\mathbf{m}_0}{\sqrt{1 - (\mathbf{v}/\mathbf{c})^2}} \qquad 8.8$$

$$\mathbf{E} = \mathbf{mc}^2 = \mathbf{m}_V^+\mathbf{c}^2 = \pm\sqrt{\mathbf{m}_0^2\mathbf{c}^4 + \mathbf{m}^2\mathbf{v}^2\mathbf{c}^2} \qquad 8.9$$

The latter expression is valid for the energy of actual (inertial) torus of sub-elementary particle only.

### *8.1 The stages of Corpuscle → Wave transition of sub-elementary fermions in composition of triplets (elementary fermions)*

*The 1st stage* is a reversible dissociation of [C] phase of sub-elementary fermion $(\mathbf{F}_\updownarrow^\pm)_\mathbf{C}$ to:

- the Cumulative virtual cloud $\mathbf{CVC}_0^{S=\pm 1/2}$ of subquantum particles with energy, spin and charge, determined by the rest mass, rest charge and half - integer spin $S = \pm 1/2$ of sub-elementary fermions with asymmetry, determined by Golden mean, and to:

- the remnant '*anchor*' Bivacuum boson ($\mathbf{BVB}_{an}^\pm$) which [C] phase can be considered, as the additional truncated cone with asymmetry determined by tangential velocity and translational energy of the triplet:

$$(\mathbf{I}): \left[ (\mathbf{F}_\updownarrow^\pm)_\mathbf{C} \overset{\textbf{Recoil/Antirecoil}}{<\!=\!=\!=\!=\!=\!=\!=\!=\!=\!=\!>} \left(\mathbf{CVC}_0^{S=\pm 1/2} + \mathbf{BVB}_{an}^\pm\right)_\mathbf{W} \right]^{e,p} \qquad 8.10$$

*The 2nd stage* of [$\mathbf{C} \to \mathbf{W}$] transition is a *reversible* dissociation of the anchor Bivacuum boson $[\mathbf{BVB}_{an}^\pm]_\mathbf{C}^{e,p} = [\mathbf{V}^+ \Updownarrow \mathbf{V}^-]_{an}^{e,p}$ to symmetric Bivacuum boson $\mathbf{BVB}^\pm$ and to corresponding anchor cumulative virtual cloud $(\mathbf{CVC}_{S=0}^\pm)_{\mathbf{BVF}_{an}^\updownarrow}$, with linear dimension of the empirical de Broglie wave of elementary fermion (triplet), carrying the energy (virtual mass), charge and spin of $[\mathbf{BVB}_{an}^\pm]_\mathbf{C}^{e,p}$

$$(\mathbf{II}) : [\mathbf{BVB}_{an}^{\pm}]_{\mathbf{C}}^{e,p} \overset{\textbf{Recoil/Antirecoil}}{<==========>} [\mathbf{BVB}^{\pm} + (\mathbf{CVC}_{an}^{S=0})_{\mathbf{BVB}_{an}^{\pm}}]_{W}^{e,p} \quad 8.11$$

This second stage of reaction of transition of [C] phase to [W] phase determines the empirical parameters of de Broglie wave of elementary particle. The extra-asymmetry of the truncated cone, is provided by the external translational velocity of the triplets - elementary particles.

The empirical external de Broglie frequency is equal to frequency of the anchor Bivacuum fermion beats ($\boldsymbol{\nu}_{ext} = \mathbf{m}_V^{+}\mathbf{v}_{tr}^2/\mathbf{h} = \mathbf{h}/\mathbf{m}_V^{+}\boldsymbol{\lambda}_B$). Relativistic contribution to properties of particle is determined by its translational velocity ($\mathbf{v}_{tr}$), empirical de Broglie wave length $\lambda_B = \mathbf{h}/\mathbf{m}_V^{+}\mathbf{v}_{tr}$ and amplitude (8.7a): $\mathbf{A}_B = \boldsymbol{\lambda}_B(\mathbf{c}/\mathbf{v})$.

The shape of $(\mathbf{CVC}^{\pm})_{\mathbf{BVF}_{an}^{\updownarrow}}$ can be approximated by the ellipsoid with main axis $\mathbf{A}_B$, normal to $\boldsymbol{\lambda}_B$ and direction of particle propagation.

In regular nonrelativistic conditions, when the velocity of particle: $\mathbf{v} << \mathbf{c}$, the frequency of the 1st stage is much higher than that of 2nd stage. However, sometime both quantum transitions are in-phase, like at Fig.7d. The spatial jump to another corpuscular phase Fig.7a, occur just in this general Wave state, when both cumulative virtual clouds are superimposed:

$$\mathbf{CVC}_0^{S=\pm 1/2} \bowtie \mathbf{CVC}_{an}^{S=0}$$
$$or: \quad [\mathbf{W}_0] \bowtie [\mathbf{W}_{an}]$$

Four intermediate configurations of sub-elementary fermion in the process of its Corpuscle - Wave pulsation are presented below. The image (a) of totally Corpuscular phase $[C_0] \bowtie [C_{an}]$ is the superposition of two truncated cones of asymmetric Bivacuum dipoles. The higher truncated cone is responsible for the rest mass ($m_0$) and charge ($e_0$) origination, determined by rotation of pairs of sub-elementary fermions around common axes. The lower one is the 'anchor' Bivacuum bosons $[C_{an}]$ with asymmetry, determined by translational kinetic energy of particle.

The image (d) of totally Wave phase $[\mathbf{W}_0] \bowtie [\mathbf{W}_{an}]$ of sub-elementary fermion is a superposition of two cylinders (symmetric Bivacuum bosons $BVB^{\pm}$) + two cones (cumulative virtual cloud $CVC^{\pm}$). For $[W_0]$ phase the cone is on the top of cylinder and for $[W_{an}]$ on the bottom of down cylinder.

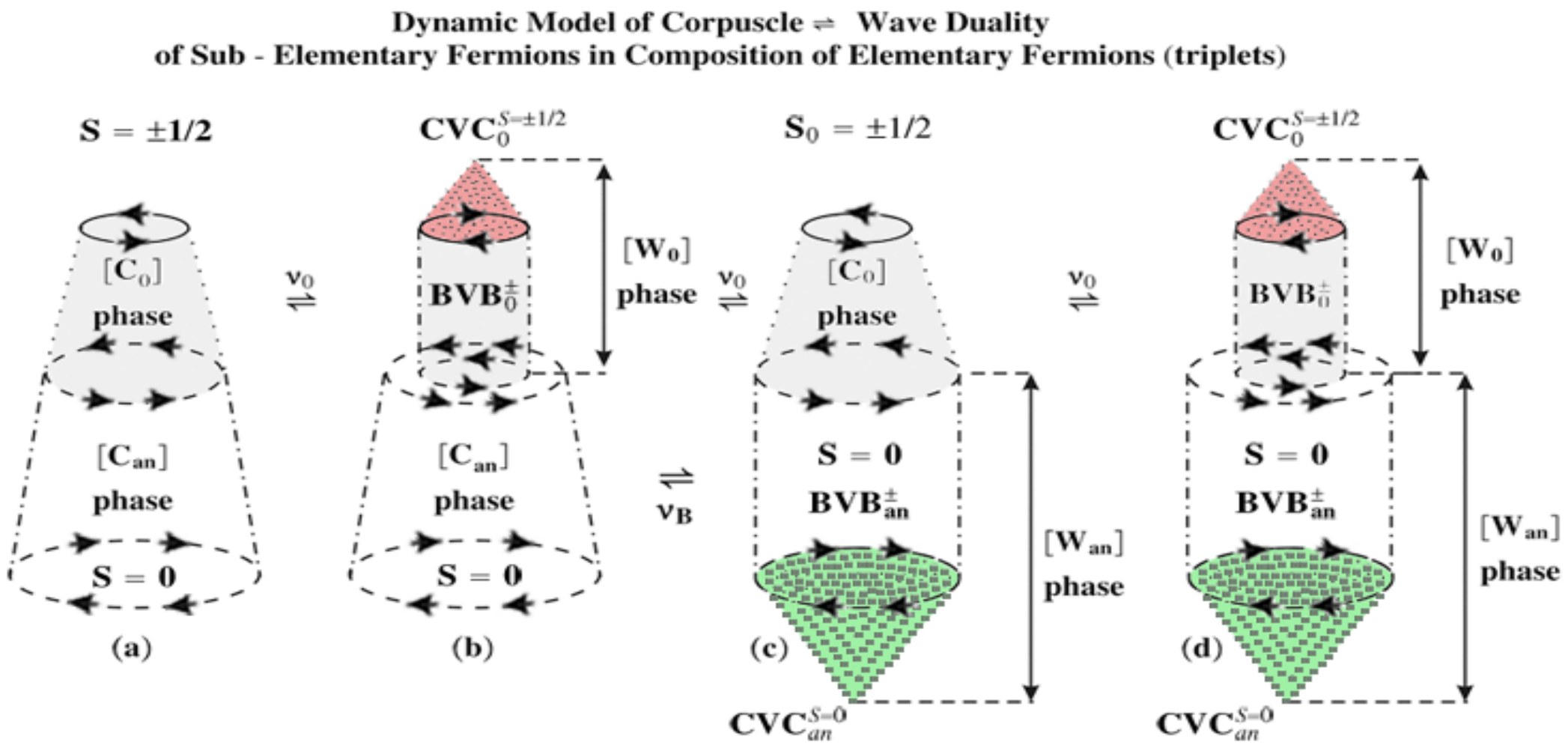

**Figure 7**. Possible stages of Corpuscle - Wave pulsation: (a) - the configuration of sub-elementary fermion in pure Corpuscular phase, presented as the superposition of rotational contribution (upper), responsible for the rest mass and energy $[\mathbf{C}_0]$ (upper truncated cone) and translational contribution $[\mathbf{C}_{an}]$ - the 'anchor' asymmetric Bivacuum boson ($BVB^{\pm}_{an}$) (down truncated cone); (b) - the intermediate stage, when the rest mass contribution transforms to the wave phase: $[\mathbf{C}_0] \to [\mathbf{W}_0]$, however the anchor contribution is still in corpuscular $[C_{an}]$ phase; (c) - possible intermediate stage, when the rotational contribution is in $[\mathbf{C}_0]$ phase and translational contribution is in the wave $[\mathbf{W}_{an}]$ phase; (d) - final stage of $\mathbf{C} \to \mathbf{W}$ transition, when both: rotational and translational contributions to sub-elementary fermion are in the wave phase $[\mathbf{W}_0] \bowtie [\mathbf{W}_{an}]$.

The superposition of the wave phases $[W_0]$ and $[W_{an}]$ have the images of two cylinders - symmetric Bivacuum bosons $BVB^{\pm}_0$ and $BVB^{\pm}_{an}$ with corresponding cones on the top and bottom . The cones from dots (sub-quantum particles) mean cumulative virtual clouds $CVC_0^{S=\pm 1/2}$, emitted in the process of $[\mathbf{C}_0] \to [W_0]$ transition and and $CVC_{an}^{S=0}$, emitted in the process of $[\mathbf{C}_{an}] \to [W_0]$ transition. The formed virtual cloud $CVC_0^{S=\pm 1/2}$ is a carrier of fermions spin $S = \pm 1/2$ and the latter $CVC_{an}^{S=0}$ - is a carrier of bosonic zero spin $S = 0$.

## 9. Dynamic model of the photon

The photon can be a result of fusion/annihilation of pair of triplets: electron + positron:

$$< [\mathbf{F}_{\uparrow}^{+} \bowtie \mathbf{F}_{\downarrow}^{-}] + \mathbf{F}_{\uparrow}^{-} >^{e,p} \; + \; < [\mathbf{F}_{\uparrow}^{+} \bowtie \mathbf{F}_{\downarrow}^{-}] + \mathbf{F}_{\downarrow}^{+} >^{e,p} \; \rightleftharpoons \qquad 9.1$$

$$<[\mathbf{BVF}^{\uparrow} \bowtie \mathbf{BVF}^{\downarrow}]_{S=0} + (\mathbf{F}_{\updownarrow}^{-} + \mathbf{F}_{\updownarrow}^{+})_{S=\pm 1} + [\mathbf{BVF}^{\uparrow} \bowtie \mathbf{BVF}^{\downarrow}]_{S=0}> \rightleftharpoons \qquad 9.1a$$

$$[\mathbf{F}_{\mathbf{W}}^{+} \Leftrightarrow \mathbf{F}_{\mathbf{C}}^{-}]_{S=\pm 1} \qquad \textit{photon} \qquad 9.1b$$

The main consequence of the electron + positron fusion is origination of pair of asymmetric sub-elementary fermions and antifermion of the same spin and opposite charges, oscillating counterphase between Corpuscle and Wave state:

$$[\mathbf{F}_{\mathbf{W}}^{+} \Leftrightarrow \mathbf{F}_{\mathbf{C}}^{-}]_{S=\pm 1} \rightleftharpoons [\mathbf{F}_{\mathbf{C}}^{+} \Leftrightarrow \mathbf{F}_{\mathbf{W}}^{-}]_{S=\pm 1} \qquad 9.2$$

This process is accompanied also by conversion of two pairs of sub-elementary particles antiparticles $2[\mathbf{F}_{\uparrow}^{+} \bowtie \mathbf{F}_{\downarrow}^{-}]$ to Cooper pairs of Bivacuum fermions $2[\mathbf{BVF}^{\uparrow} \bowtie \mathbf{BVF}^{\downarrow}]_{S=0}$, as a regular element of Bivacuum, not participating directly in the photon structure.

Two tori and two antitori of pair of the photon $(\mathbf{F}_{\updownarrow}^{-} \Leftrightarrow \mathbf{F}_{\updownarrow}^{+})_{S=\pm 1}$, conjugated by "tail-to-tail" or "head-to-tail" principle, like at Fig.8, rotate in the same direction (clockwise or counterclockwise). This provide the additivity of their half-integer spins and the integer value of spin of photon: $S = [(\pm\frac{1}{2}\hbar) + (\pm\frac{1}{2}\hbar)] = \pm 1\hbar$. The average (resulting) charge of the photon as a symmetric dipole is zero.

Our general formula for energy of sub-elementary fermions, as a parts of elementary fermions, like electron, takes into account both - the rotational and translational contributions:

$$\mathbf{E}_{tot} = (\mathbf{m}_V^{+} - \mathbf{m}_V^{-})\mathbf{c}^2 = \mathbf{V}_{tot} + \mathbf{T}_k = (\mathbf{m}_0\boldsymbol{\omega}_0^2\mathbf{L}_0^2)_{\mathbf{rot}}^{in} + \left(\frac{\mathbf{h}^2}{\mathbf{m}_V^{+}\boldsymbol{\lambda}_B^2}\right) \qquad 9.3$$

where total potential and kinetic energy of sub-elementary fermion are:

$$\mathbf{V}_{tot} = (\mathbf{m}_0\boldsymbol{\omega}_0\mathbf{L}_0)^{in} + \frac{1}{2}\mathbf{m}_V^+\boldsymbol{\lambda}_B^2\mathbf{v}_B^2 = (\mathbf{m}_0\boldsymbol{\omega}_0\mathbf{L}_0)^{in} + \frac{1}{2}(\mathbf{m}_V^+ - \mathbf{m}_V^-)_{tr}\mathbf{c}^2 \qquad 9.4$$

$$\mathbf{T}_k = \frac{1}{2}\mathbf{m}_V^+\boldsymbol{\lambda}_B^2\mathbf{v}_B^2 = \frac{\mathbf{h}^2}{2\mathbf{m}_V^+\boldsymbol{\lambda}_B^2} = \frac{1}{2}\mathbf{m}_V^+\mathbf{v}_B^2 = \frac{1}{2}(\mathbf{m}_V^+ - \mathbf{m}_V^-)_{tr}\mathbf{c}^2 \qquad 9.4a$$

In the case of bosons, like photon, the rest mass contribution in energy is zero: $(\mathbf{m}_0\boldsymbol{\omega}_0^2\mathbf{L}_0^2)^{in}_{\mathbf{rot}} = 0$ as a consequence of the absence of pairs rotation around common axis (the tangential velocity, angular frequency $\boldsymbol{\omega}_0$ and the rest mass $\mathbf{m}_0$ are zero). Such rotation is possible only for Cooper pairs of asymmetric Bivacuum fermion and antifermion, conjugated side-by-side, but not 'tail-to-tail', like in photon (see Fig.8).

At such conditions, the external, translational only, velocity of the photon $[\mathbf{F}_{\mathbf{W}}^+ \Leftrightarrow \mathbf{F}_{\mathbf{C}}^-]_{\mathbf{S}=\pm 1}$ is equal to the light velocity $\mathbf{v} = \mathbf{c}$. From 9.4 and 9.4a we can see that for the photon the kinetic and potential energy are equal to each other, like in the case of harmonic oscillator or standing wave:

$$\mathbf{V}_{tot} = \mathbf{T}_k = \frac{1}{2}\mathbf{m}_V^+\mathbf{v}_B^2 = \frac{1}{2}(\mathbf{m}_V^+ - \mathbf{m}_V^-)_{tr}\mathbf{c}^2 \qquad 9.5$$

These conditions turn the energy of the photon to couple of simple shapes, corresponding to Corpuscular and Wave phases of photon:

$$(\mathbf{E}_{ph})_{\mathbf{C}} = \mathbf{V}_{tot} + \mathbf{T}_k = (\mathbf{m}_V^+ - |\mathbf{m}_V^-|)_{tr}\mathbf{c}^2 = \mathbf{m}_{ph}\mathbf{c}^2 \qquad 9.6$$

$$(\mathbf{E}_{ph})_{\mathbf{W}} = \mathbf{h}\boldsymbol{\nu}_{ph} = \frac{hc}{\boldsymbol{\lambda}_{ph}} \qquad 9.6a$$

The effective mass/energy of photon is $\mathbf{m}_{ph} = (\mathbf{m}_V^+ - |\mathbf{m}_V^-|)_{tr} = \mathbf{h}\boldsymbol{\nu}_{ph}/\mathbf{c}^2 = \frac{\mathbf{h}}{\mathbf{c}\boldsymbol{\lambda}_{ph}}$, where: $\boldsymbol{\lambda}_{ph} = \mathbf{c}/\boldsymbol{\nu}_{ph}$ is the photon wave length; $\boldsymbol{\nu}_{ph}$ is the frequency of photon, equal to frequency of beats between tori and antitori of sub-elementary fermions, forming photon (Figure 8).

Since the phase ($\mathbf{v}_{ph}$) and group ($\mathbf{v}$) velocities of the photon are equal to light velocity ($\mathbf{v} = \mathbf{c}$), the amplitude of the photon cumulative virtual cloud ($\mathbf{CVC}^{\pm}$) is equal to its wave length:

$$\mathbf{A}_{ph} = \lambda_{ph}\frac{\mathbf{c}}{\mathbf{v}} = \lambda_{ph} \qquad 9.7$$

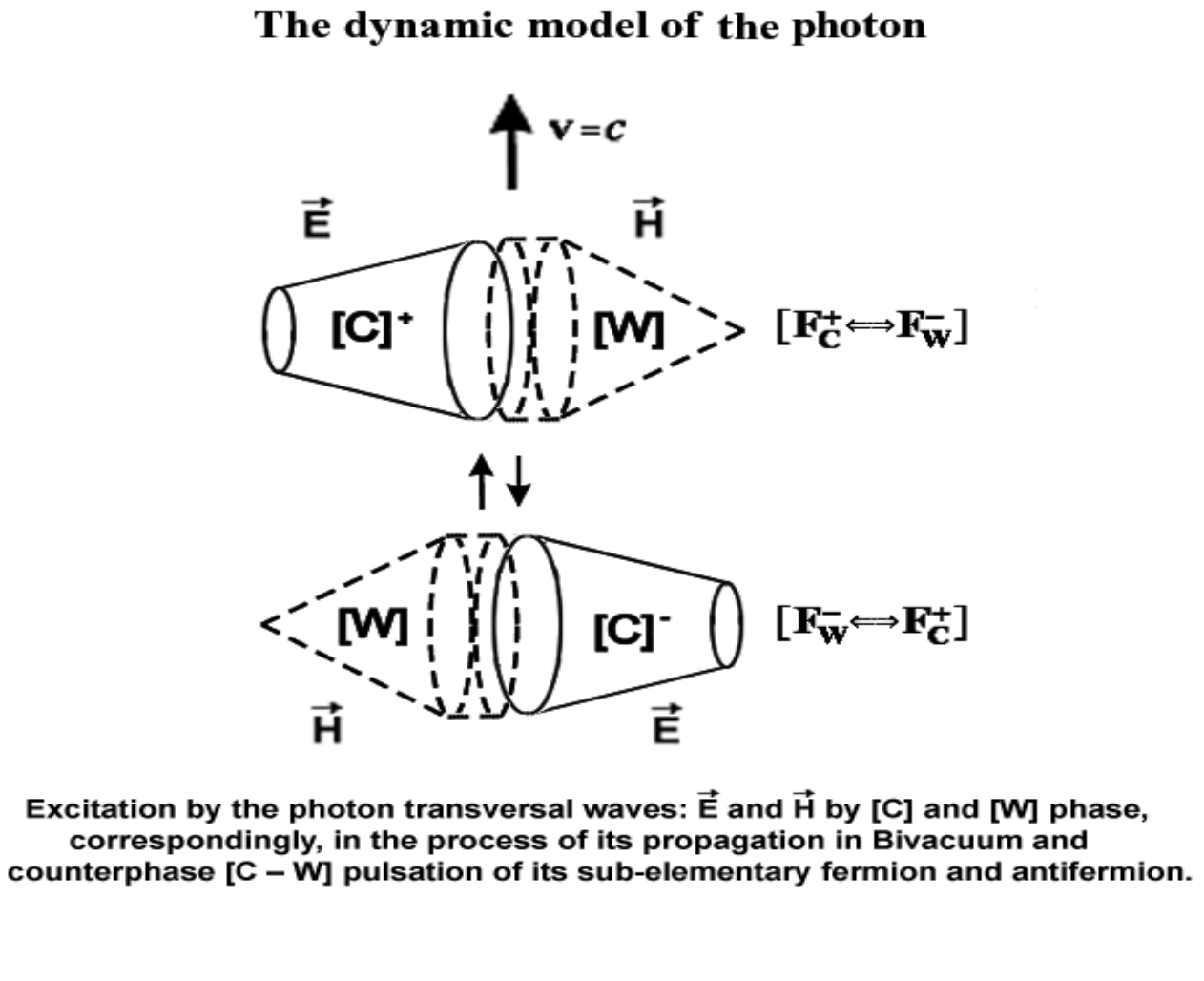

**Figure 8**. Dynamic model of the photon Corpuscle ⇌ Wave pulsation in the process of propagation in space, exciting transversal electric and magnetic Bivacuum perturbations. Four tori and antitori of conjugated sub-elementary fermion and antifermion of the photon rotate clockwise ($\mathbf{S} = +\hbar$) or counter clockwise ($\mathbf{S} = -\hbar$) direction, normal to direction of photon propagation in space.

There are two possibilities of $\mathbf{C} \rightleftharpoons \mathbf{W}$ pulsation of the photon, accompanied by electric and magnetic properties of photon realization:

1. Both sub-elementary fermions and antifermions are pulsing in-phase;
2. They are pulsing counterphase, exchanging the energy of cumulative virtual clouds $\mathbf{CVC}^+$ and $\mathbf{CVC}^-$ in the process of their propagation in space, like presented at Fig.8 above.

*In 1st case,* the wave [W] phase $(\mathbf{E}_{ph})_\mathbf{W}$ is a result of simultaneous quantum transitions - beats between four asymmetric tori and antitori of pair of Bivacuum fermions in Corpuscular phase $\{\mathbf{F}_\updownarrow^- + \mathbf{F}_\updownarrow^+\}_{S=\pm 1}$, accompanied by emission of unified cumulative virtual cloud with energy $2|\varepsilon_{CVC^+} - \varepsilon_{CVC^-}|$, where $|\varepsilon_{CVC^+} - \varepsilon_{CVC^-}|$ is the energy of one sub-elementary fermion.

In [C] phase the photon may display its electric properties, because the electric dipole properties of $\{\mathbf{F}_\updownarrow^- + \mathbf{F}_\updownarrow^+\}_{S=\pm 1}$ induce corresponding symmetry shift in surrounding Bivacuum dipoles pairs.

In [W] phase of photon, the rotating Cumulative virtual cloud $\mathbf{CVC}^\pm$ of sub-elementary fermions in [W] phase with uncompensated energy $2|\varepsilon_{CVC^+} - \varepsilon_{CVC^-}|$ and uncompensated charge, shift the equilibrium between surrounding Bivacuum fermions and antifermions $[\mathbf{BVF}^\uparrow \rightleftharpoons \mathbf{BVB}^\pm \rightleftharpoons \mathbf{BVF}^\downarrow]$ to the left or right, generating in such a way the magnetic field.

*In 2nd case,* the above processes occur alternatively, when the positive and negative asymmetric Bivacuum dipoles oscillate between [C] and [W] states counterphase, generating the electric and magnetic properties simultaneously. The stability of the photon structure and the permanent total energy are the consequence of dynamic exchange interaction of $\mathbf{CVC}^+$ and $\mathbf{CVC}^-$ between asymmetric Bivacuum fermion and antifermion in the process of the photon propagation in space.

The 2nd model means that the corpuscular and wave properties of photon display themselves simultaneously and half of photon wave length behave, as the wave and other half as the corpuscle (Fig.8).This dynamic model of photon is more probable, that the 1st one.

The most common way of photon origination is not the [electron + positron] annihilation, but the acceleration of charged fermions - triplets, accompanied by their relativistic mass/energy increasing and generation if multiplicated secondary 'anchor sites' in Bivacuum.

The velocity jump (acceleration) of the elementary charges, following by the photons origination, occur at:

- the non-uniform acceleration of the electrons in the undulators or
- acceleration of the electrons in the process of quantum transitions in the atoms and molecules from the excited state to the ground one.

### *9.1 Possible nature of neutrino and antineutrino*

Following our approach to elementary particles structure, the neutrino $(\nu)^i$ and antineutrino $(\overline{\nu})^i$ of three lepton generation ($i = \mathbf{e}, \mu, \tau$) can be considered, as a stable neutral pair of asymmetric charged Bivacuum fermion $\mathbf{BVF}_-^{S=1/2} \equiv [\mathbf{V}^+ \uparrow\uparrow \mathbf{V}^-]$ or antifermion$\mathbf{BVF}_+^{S=-1/2} \equiv [\mathbf{V}^+ \downarrow\downarrow \mathbf{V}^-]$ and asymmetric Bivacuum antiboson or boson, correspondingly. The energy and charge of $(\mathbf{BVB}^\pm)^i \equiv [\mathbf{V}^+ \uparrow\downarrow \mathbf{V}^-]$, compensate that of $\mathbf{BVF}_{\pm}^{S=\pm 1/2}$ in each neutrino or antineutrino. Both asymmetric Bivacuum dipoles pulse counterphase between [C] and [W] phase, providing stability of neutrino due to exchange interaction:

$$(\nu)^i \sim \left[ \left(\mathbf{BVF}_{\mp}^{\uparrow}\right)_W \Leftrightarrow \left(\overline{\mathbf{BVB}}^{\pm}\right)_C \right]^i \rightleftharpoons \left[ \left(\mathbf{BVF}_{\mp}^{\uparrow}\right)_C \Leftrightarrow \left(\overline{\mathbf{BVB}}^{\pm}\right)_W \right]^i \quad 9.8$$

$$(\bar{\nu})^i \sim \left[ \left(\overline{\mathbf{BVF}}_{\mp}^{\uparrow}\right)_W \Leftrightarrow (\mathbf{BVB}^{\pm})_C \right]^i \rightleftharpoons \left[ \left(\overline{\mathbf{BVF}}_{\mp}^{\uparrow}\right)_C \Leftrightarrow (\mathbf{BVB}^{\pm})_W \right]^i \quad 9.8a$$

These two Bivacuum dipoles are conjugated with each other by "tail-to-tail" principle, like in the case of the photon (Fig.8), providing the rest mass and charge very close to zero in the absence of rotation.

The frequency of beats between asymmetric and symmetric states of pairs $[\mathbf{BVF}_{\mp}^{S=\pm 1/2} \Leftrightarrow \mathbf{BVB}^{\pm}]^i$, equal to neutrino frequency, is determined by slight difference in the energy of sub-elementary fermion $(\mathbf{BVF}_{\mp}^{S=\pm 1/2})^i$ and sub-elementary antiboson $(\mathbf{BVB}^{\pm})^i$ in pairs:

$$\omega_{\nu,\tilde{\nu}}^i = \left| \omega_{C\rightleftharpoons W}^{\mathbf{BVF}_{\mp}^{S=\pm 1/2}} - \omega_{C\rightleftharpoons W}^{\mathbf{BVB}^{\pm}} \right|^i = \frac{\left| \mathbf{E}_{\mathbf{BVF}_{\mp}^{S=\pm 1/2}}^i - \mathbf{E}_{\mathbf{BVB}^{\pm}}^i \right|}{\hbar} \quad 9.9$$

The mass/energy of each of three generation of neutrino can be estimated as (Kaivarainen 2006b):

$$\mathbf{m}_{\nu,\tilde{\nu}}^i = \frac{\hbar\omega_{\nu,\tilde{\nu}}^i}{\mathbf{c}^2} = \beta^i(\mathbf{m}_V^{\mp})_{e,\mu,\tau}^{\phi} = \frac{1}{\phi}\frac{(\mathbf{m}_0^i)^3}{\mathbf{M}_{Pl}^2} = 1.618\frac{(\mathbf{m}_0^i)^3}{\mathbf{M}_{Pl}^2} \quad 9.10$$

where the dimensionless *gravitational fine structure constant* $\beta^i$ is different for each lepton generation, introduced in chapter 15:

$$\beta^i = \left(\frac{\mathbf{m}_0^i}{\mathbf{M}_{Pl}}\right)^2 \quad 9.11$$

$(\mathbf{m}_V^{\mp})_{e,\mu,\tau}^{\phi} = (\mathbf{m}_0^{e,\mu,\tau}/\phi)$ are the actual mass of the electrons or positrons of three generation at Golden mean conditions, participating in weak interaction, following by corresponding neutrino and antineutrino emission; $\mathbf{M}_{Pl}^2 = \hbar\mathbf{c}/\mathbf{G}$ is a Planck mass.

Corresponding mass evaluation fit the currently existing ones in form of inequalities, i.e. mass of the electron neutrino is less than $1 \times 10^{-8}$ Ge/c$^2$, mass of muon neutrino is less than $0.0002 Ge/c^2$ and mass of the tau neutrino - less, than $0.02 Ge/c^2$. These vales correspond well to description of neutrino properties at: http://en.wikipedia.org/wiki/neutrino.

*Neutrino oscillation* between different lepton flavor (electron, muon, or tau) follows from experimental data. For example, the solution of the *solar neutrino problem*, as a major discrepancy between measurements of the neutrinos flowing through the Earth and theoretical models of the solar interior needs the neutrino oscillation. The probability of measuring a particular flavor for a neutrino varies periodically as it propagates. In accordance to our model of neutrino these interconversions can be a result of simultaneous reversible excitation of pair $(\nu)^i \sim [\mathbf{BVF}_{\mp}^{\uparrow} \Leftrightarrow \overline{\mathbf{BVB}}^{\pm}]^e$ from it ground state with minimum energy of torus and antitorus to their certain excited states, corresponding to muon and tau neutrinos $[\mathbf{BVF}_{\mp}^{\uparrow} \Leftrightarrow \overline{\mathbf{BVB}}^{\pm}]^{\mu,\tau}$. Consequently, the neutrino oscillation between different generations can be a result of *absorbtion* or *emission* by one type of neutrino the high frequency pair of standing Bivacuum virtual pressure waves (8.18a) of corresponding generation $[\mathbf{VPW}^{+} \bowtie \mathbf{VPW}^{-}]_{q>>1}^{\mu,\tau}$. These neutrino oscillations:

$$(\nu)^e \rightleftharpoons (\nu)^\mu \rightleftharpoons (\nu)^\tau \quad 9.12$$

do not violate the energy conservation due to compensation of positive and negative Bivacuum energies.

**10. The jump-way mechanism of the elementary particles propagation in space**

The propagation of elementary particles, like triplets-fermions (electrons, protons, etc.):

$$< [\mathbf{F}_\uparrow^+ \bowtie \mathbf{F}_\downarrow^-]_W + (\mathbf{F}_\updownarrow^\pm)_C >^{e,p} \overset{jump}{\to} < [\mathbf{F}_\uparrow^+ \bowtie \mathbf{F}_\downarrow^-]_C + (\mathbf{F}_\updownarrow^\pm)_W >^{e,p} \quad 10.1$$

or doublets of Bivacuum fermions - bosons (photons):

$$[\mathbf{F}_\mathbf{W}^+ \Leftrightarrow \mathbf{F}_\mathbf{C}^-]_{S=\pm1} \overset{jump}{\to} [\mathbf{F}_\mathbf{C}^+ \Leftrightarrow \mathbf{F}_\mathbf{W}^-]_{S=\pm1} \quad 10.2$$

can be considered as a jumping process between multiple *secondary anchor sites*. The most probable length and frequency of jumps in Wave [W] phase of particles are determined by the empirical de Broglie wave length and frequency of propagating particles. The conversion of [W] phase of paired sub-elementary particles to Corpuscular [C] phase: $[\mathbf{F}_\uparrow^+ \bowtie \mathbf{F}_\downarrow^-]_W \to [\mathbf{F}_\uparrow^+ \bowtie \mathbf{F}_\downarrow^-]_C$ occurs on *secondary anchor sites.*

The energy and charge conservation law demands the resulting energy of all secondary anchor sites should be zero. It is possible, if we assume that all *secondary anchor sites* **SAS**(**F**) or **SAS**(**B**) are composed from one or more pairs of asymmetric Bivacuum bosons $\mathbf{BVB}^\pm$ or Bivacuum fermions $\mathbf{BVF}_\pm^{\uparrow\downarrow}$ with energy, spin and charge compensating each other:

$$\mathbf{SAS(B)} = \sum^N [\mathbf{BVB}^\pm \bowtie \mathbf{BVB}^\pm]_n \quad 10.3$$

$$\mathbf{SAS(F)} = \sum^N [\mathbf{BVF}_\pm^\uparrow \bowtie \mathbf{BVF}_\mp^\downarrow]_n \quad 10.3a$$

**The stage I** *of elementary fermion, like the electron or proton,* propagation - the excitation of *multiple secondary anchor sites* in Bivacuum by Virtual Pressure Waves ($\mathbf{VPW}_m^\pm$), modulated by paired sub-elementary particle in **W** –phase, when the the unpaired sub-elementary fermion of the triplet is in Corpuscular [**C**] phase.

The absolute values of asymmetry and energy of all sub-elementary fermions (paired and unpaired) are the same in triplets. These values are dependent on external kinetic energy, momentum and velocity of the triplet.

The asymmetry of paired sub-elementary fermions of elementary particles in [C] phase and the asymmetry and energy of Virtual Pressure Waves $(\mathbf{VPW}^+ \bowtie \mathbf{VPW}^-)_{e,p}$, radiated by their [W] -phase are directly interrelated.

So, the asymmetry of pairs of Bivacuum dipoles: $[\mathbf{BVB}^\pm \bowtie \mathbf{BVB}^\pm]_n$ or $[\mathbf{BVF}_\pm^\uparrow \bowtie \mathbf{BVF}_\mp^\downarrow]_n$ forming **SAS** and their spatial multiplication, can be induced by the basic Bivacuum virtual pressure waves $\mathbf{VPW}_0^\pm$, modulated by de Broglie wave amplitude $\mathbf{A_B} = \boldsymbol{\lambda}_B(\mathbf{c}/\mathbf{v})$ of asymmetric primary anchor site $[\mathbf{F}_\uparrow^+ \bowtie \mathbf{F}_\downarrow^-]_{anc}^{e,p}$ of propagating elementary fermion (fig. 7).

The most probable separation between the *excited secondary anchor sites* is determined by the empirical de Broglie wave length of the particle ($\boldsymbol{\lambda}_B = h/\mathbf{p}$), equal to Cumulative Virtual Clouds $\mathbf{CVC}^\pm$ length, emitted by unpaired sub-elementary fermion.

**The stage II** of the fermion propagation represents its jumps to one of most probable secondary anchor site, turning this site to primary one. At this stage of jump, the unpaired/uncompensated sub-elementary fermions are in the external Wave [W] phase and paired in [C] phase. Immediately after such '*anchoring' of particle* the unpaired/uncompensated sub-elementary fermions turn to [C] phase and paired to [W] phase. This is last stage of one cycle of 'jump' and the next cycle starts. This jumping mechanism provides the particles propagation with average velocity of particle, equal to it group velocity.

The most probable direction of jumping coincides with particle's momentum vector in its [C] phase. However the new location of particle, as only one of many possible multiplicated *secondary anchor sites*, is not rigidly predetermined and the 'jumps' can be considered as the stochastic process. The described mechanism of elementary particles propagation in space can be named "the kangaroo effect".

*The propagation of photon* in Bivacuum also represent a 'jumping' from one secondary anchor site to another **SAS**, separated from each other by the photon wave length and with frequency, equal to frequency of its Corpuscle $\rightleftharpoons$ Wave pulsation. The modulation of basic symmetric $\mathbf{VPW}_0^{\pm}$ of Bivacuum by asymmetric Virtual Pressure Waves $(\mathbf{VPW}^{+} \bowtie \mathbf{VPW}^{-})_{ph}^{as}$ excited by one of sub-elementary fermions of the photon being in [W] phase $[\mathbf{F}_{\mathbf{W}}^{+} \Leftrightarrow \mathbf{F}_{\mathbf{C}}^{-}]_{S=\pm 1}$. The absorption of modulated virtual pressure waves $\mathbf{VPW}_m^{\pm}$ by *secondary anchor site*: $[\mathbf{BVB}^{\pm} \bowtie \mathbf{BVB}^{\pm}]_n$ is followed by their Bivacuum dipoles in pairs opposite symmetry shift and turning these Bivacuum pairs to the photon $[\mathbf{F}_{\mathbf{W}}^{+} \Leftrightarrow \mathbf{F}_{\mathbf{C}}^{-}]_{S=\pm 1}$.

*10.1 The possibility of appearance of particle in Corpuscular phase in any point of space*

The principle of superposition in quantum mechanics has the same formal expression as the waves superposition in classical mechanics:

$$\Phi(\mathbf{r},t) = c_1\Phi(\mathbf{r},t)_1 + c_2\Phi(\mathbf{r},t)_2 + \ldots c_n\Phi(\mathbf{r},t)_n \qquad 10.4$$

where: $c_n$ are arbitrary complex numbers; $\Phi(\mathbf{r},t)_n$ is wave function, describing different and alternative/orthogonal $(n)$ states of quantum system. In accordance to our approach, these quantum states correspond to multiple *secondary anchor sites* of moving in space particle. It will be shown later in this work, that the notion of multiple anchor site of elementary particles coincide with notion of their Virtual Replica multiplication in space.

However, in contrast to waves superposition in classical systems, any state of quantum system is not a result of 'mixing' of other states, but always the alternative or orthogonal. This means that only one state of many spatially separate is allowed for quantum system at given time moment. It is so-called collapsing of the wave function, describing this system. Just this collapsing accompanies the jumping of elementary particle between multiple secondary anchor sites on the way of its propagation. As was mentioned already, the most probable length of jump is equal to de Broglie wave length of elementary particle.

## 11. New interpretation of Shrödinger equation. General shape of wave function, describing both the external and internal dynamics of elementary particle

The stationary Shrödinger equation can be easily derived from universal for homogeneous medium wave equation:

$$\nabla^2\Phi(r,t) - \frac{1}{\mathbf{v}^2}\frac{\partial\Phi(r,t)}{\partial t^2} = 0 \qquad 11.1$$

where $\Phi(r,t)$ is the wave amplitude (scalar), depending distance from source (r) and time (t) in the process of its propagation with permanent velocity $(\mathbf{v})$. One of possible form of time and space dependent wave function:

$$\Phi(r,t) = \mathbf{C}\exp\left[i\left(\frac{\mathbf{x}}{\mathbf{L}_B} - \boldsymbol{\omega}_\mathbf{B}\mathbf{t}\right)\right] = \mathbf{C}\exp\left(i\frac{\mathbf{x}}{\mathbf{L}_B}\right)\exp(-i\boldsymbol{\omega}_\mathbf{B}\mathbf{t}) \qquad 11.2$$

In the case of harmonic dependence of the wave amplitude on time with angle frequency $\omega$, it can be presented as:

$$\Phi(r,t) = \Phi(r)\exp(-i\omega t) \qquad 11.3$$

Putting 11.3 to 11.1 we get the following equation:

$$\nabla^2\Phi^{m,e}(r) + \mathbf{k}^2\Phi^{m,e}(r) = 0 \qquad 11.4$$

where $\mathbf{k}$ is a wave number ($\mathbf{k} = \omega/\mathbf{v} = \mathbf{2\pi}/(\mathbf{vT}) = \mathbf{2\pi}/\boldsymbol{\lambda} = \mathbf{1}/\mathbf{L}$).

The conversion of (11.4) to form describing corpuscle-wave duality can be done using de Broglie relations:

$$\mathbf{k} = \mathbf{p}/\hbar = \mathbf{2\pi}/\mathbf{L}_B; \quad \mathbf{L}_B = \hbar/\mathbf{p} \qquad 11.5$$

$$\mathbf{k}^2 = \mathbf{p}^2/\hbar^2 = (\mathbf{2\pi}/\mathbf{L}_B)^2 = 1/\boldsymbol{\lambda}_B^2 \qquad 11.5a$$

in stationary conditions, when the total energy of de Broglie wave, equal to sum of its external kinetic ($\mathbf{T}_k$) and potential ($\mathbf{V}$) energies, is time-independent, like in standing waves, for example:

$$\mathbf{E} = \mathbf{T}_k + \mathbf{V} = \frac{\mathbf{p}^2}{2\mathbf{m}} + \mathbf{V} = \mathbf{const} \qquad 11.6$$

$$or: \ \mathbf{p}^2 = 2\mathbf{m}(\mathbf{E} - \mathbf{V}) \qquad 11.6a$$

The de Broglie wave number squared from 11.5a and 11.6a is

$$\mathbf{k}^2 = (2\mathbf{m}/\hbar)(\mathbf{E} - \mathbf{V}) \qquad 11.7$$

Combining 11.7 with 11.4, we get the *stationary* Shrödinger equation:

$$\nabla^2\Phi(r) + (2\mathbf{m}/\hbar)(\mathbf{E} - \mathbf{V})\Phi(r) = 0 \qquad 11.8$$

It has solutions for continuous wave function, existing as *eigenfunctions* only at certain discreet *eigenvalues* of energy ($\mathbf{E}_n$). It was shown by Shrödinger, that spectra of these energies of the electron in potential electric field ($\mathbf{V}$) describes correctly the absorption spectra of hydrogen atoms.

The time-dependent form of Shrödinger equation includes the time and space dependent wave function, like (11.2):

$$\Phi(\mathbf{r},t) = \Phi(\mathbf{r})\exp(-i\mathbf{E}\mathbf{t}/\hbar) = \mathbf{C}\exp\left(i\frac{\mathbf{x}}{\mathbf{L}_B}\right)\exp(-i\boldsymbol{\omega}_\mathbf{B}\mathbf{t}) \qquad 11.9$$

The corresponding equation can be presented as:

$$-\frac{\hbar}{i}\frac{\partial\Phi(\mathbf{r},t)}{\partial t} = \left(-\frac{\hbar}{2\mathbf{m}}\nabla^2 + \mathbf{V}\right)\Phi(\mathbf{r},t) \qquad 11.10$$

The inertial mass in 11.10 in accordance to our Unified theory, is equal to the actual mass of unpaired/uncompensated sub-elementary fermion of elementary particle: $\mathbf{m} = \mathbf{m}_V^+$.

The properties of stationary wave function $\Phi(\mathbf{r})$ and time-dependent $\Phi(\mathbf{r},t)$ should be the same, i.e. they are *continuous, single-valued and finitesmall*. The product of wave function with its *complex conjugate* function, characterize the density of probability of particle location in this point of space at certain time moment:

$$\Phi(\mathbf{r},t)\ \Phi^*(\mathbf{r},t) = |\Phi(\mathbf{r},t)|^2 \qquad 11.11$$

In solutions of Shrödinger equation the certain eigenvalues of energy ($\mathbf{E}_n$) corresponds to eigenfunctions ($\Phi_n$), describing *anchor sites (primary and secondary)* of elementary particles in their corpuscular [C] phase (see 10.4).

It follows from our theory of wave-corpuscle duality, that de Broglie wave length ($\boldsymbol{\lambda}_B = 2\pi\mathbf{L}_B$), its amplitude $\mathbf{A}_B = \boldsymbol{\lambda}_B(\mathbf{c}/\mathbf{v})$ and frequency ($\boldsymbol{\omega}_\mathbf{B}$), as a crucial parameters of wave function (11.9), are determined by properties of the *anchor Bivacuum fermions* of uncompensated sub-elementary fermions of the electron or proton or bosons, like photon.

The *external* de Broglie wave frequency ($\boldsymbol{\omega}_B^{ext}$) and wave number ($\mathbf{k}_B$) of particle can be expressed via *internal* ($\boldsymbol{\omega}_0^{in}$), *total* ($\boldsymbol{\omega}_{\mathbf{C}\rightleftharpoons\mathbf{W}}$) frequencies and corresponding energies as:

$$\boldsymbol{\omega}_B^{ext} = \frac{1}{\hbar}[(\mathbf{m}_V^+ - \mathbf{m}_V^-)_{anc}^{ext}\mathbf{c}^2]_{tr} = \boldsymbol{\omega}_{\mathbf{C}\rightleftharpoons\mathbf{W}}^{res} - \boldsymbol{\omega}_0 \qquad 11.12$$

$$or: \quad \mathbf{k}_B = \frac{\mathbf{1}}{\mathbf{L}_B} = \frac{\mathbf{c}}{\hbar}\,[\mathbf{m}_V^+(\mathbf{m}_V^+ - \mathbf{m}_V^-)]^{1/2} \qquad 11.13$$

where: $\mathbf{m}_V^+ = \mathbf{m}_0/\mathbf{R}$; $\quad \mathbf{m}_V^- = \mathbf{R}\mathbf{m}_0$; $\mathbf{R} = \sqrt{1-(\mathbf{v}/\mathbf{c})^2}$ is relativistic factor decreasing with external translational group velocity ($\mathbf{v}$) increasing.

At $\mathbf{v} \to \mathbf{c}$, the, the rest mass contribution tends to zero, $\mathbf{m}_V^+ \gg \mathbf{m}_V^-$ and $\boldsymbol{\omega}_B^{ext} \to \boldsymbol{\omega}_{\mathbf{C}\rightleftharpoons\mathbf{W}}$ and $\mathbf{k}_B \to (\mathbf{m}_V^+\mathbf{c}/\hbar)$.

The mass and charge symmetry shifts of asymmetric Bivacuum fermions and antifermions are interrelated (see Chapter 6):

$$\Delta\mathbf{m}_V^{\pm} = (\mathbf{m}_V^+ - \mathbf{m}_V^-) = \mathbf{m}_V^+\left(\frac{\mathbf{v}}{\mathbf{c}}\right)^2 \qquad 11.14$$

$$\Delta\mathbf{e}_{\pm} = (\mathbf{e}_+ - \mathbf{e}_-) = \frac{\Delta\mathbf{m}_V^{\pm}\,\mathbf{e}_+^2}{\mathbf{m}_V^+(\mathbf{e}_+ + \mathbf{e}_-)} = \left(\frac{\mathbf{v}}{\mathbf{c}}\right)^2\frac{\mathbf{e}_+^2}{\mathbf{e}_+ + \mathbf{e}_-} \qquad 11.15$$

where the *actual* charge ($\mathbf{e}_+$), in accordance to eqs. 6.1 and 6.1a, has the following relativistic dependence on the external velocity of Bivacuum dipoles:

$$\mathbf{e}_+ = \frac{\mathbf{e}_0}{[1-\mathbf{v}^2/\mathbf{c}^2]^{1/4}} \qquad 11.16$$

The complementary charge ($\mathbf{e}_- = \mathbf{e}^2/\mathbf{e}_+$) can be calculated from the charge compensation principle (eq. 5.25): $|\mathbf{e}_+\mathbf{e}_-| = \mathbf{e}_0^2$.

Using the relations above, we may present the dimensionless coefficient of wave function (C) in (11.9), as a maximum symmetry shift of the *anchor* Bivacuum fermion, reduced to the rest mass ($\mathbf{m}_0$) and rest charge ($\mathbf{e}_0$):

$$\mathbf{C_m} = \Delta\mathbf{m}_V^{\pm}/\sqrt{2}\,\mathbf{m}_0 = \frac{|\mathbf{m}_V^+ - \mathbf{m}_V^-|}{\sqrt{2}\,\mathbf{m}_0} = \frac{\mathbf{m}_V^+}{\sqrt{2}\,\mathbf{m}_0}\left(\frac{\mathbf{v}}{\mathbf{c}}\right)^2 \qquad 11.17$$

$$\mathbf{C_e} = \Delta\mathbf{e}_{\pm}/\sqrt{2}\,\mathbf{e}_0 = \frac{|\mathbf{e}_+ - \mathbf{e}_-|}{\sqrt{2}\,\mathbf{e}_0} = \left(\frac{\mathbf{v}}{\mathbf{c}}\right)^2\frac{\mathbf{e}_+^2}{\sqrt{2}\,\mathbf{e}_0(\mathbf{e}_+ + \mathbf{e}_-)} \qquad 11.18a$$

We assume here, that as far the complementary mass and charge are undetectable directly and we may consider them as imaginary ones: $i\mathbf{m}_V^-$ and $i\mathbf{e}_-$. Consequently, using 11.12; 11.13 and 11.17, we may present the wave function (11.9) and its complex conjugate in terms of Bivacuum dipoles symmetry shifts. This is important for understanding the mechanism of particle internal dynamics and its propagation in space:

$$\Phi(\mathbf{r},t) = \mathbf{C}\exp\left(i\frac{\mathbf{x}}{\mathbf{L}_B}\right)\exp(-i\boldsymbol{\omega}_\mathbf{B}\mathbf{t}); \qquad \Phi^*(\mathbf{r},t) = \mathbf{C}^*\exp\left(-i\frac{\mathbf{x}}{\mathbf{L}_B}\right)\exp(i\boldsymbol{\omega}_\mathbf{B}\mathbf{t}) \qquad 11.19$$

$$\Phi(\mathbf{r},t) = \frac{\mathbf{m}_V^+ - i\mathbf{m}_V^-}{\sqrt{2}\,\mathbf{m}_0}\exp\left[i\frac{\mathbf{x}}{\hbar}\mathbf{c}\,[\overset{+}{\mathbf{m}}_V(\mathbf{m}_V^+ - i\mathbf{m}_V^-)]^{1/2}\right]\exp\left\{-i\frac{1}{\hbar}[(\mathbf{m}_V^+ - i\mathbf{m}_V^-)\mathbf{c}^2]_{tr}\mathbf{t}\right\} = \qquad 11.20$$

$$\Phi(\mathbf{r},t) = \frac{\mathbf{m}_V^+ - i\mathbf{R}\mathbf{m}_0}{\sqrt{2}\,\mathbf{m}_0}\exp\left[i\frac{\mathbf{x}}{\hbar}\mathbf{c}\,[\overset{+}{\mathbf{m}}_V(\mathbf{m}_V^+ - i\mathbf{R}\mathbf{m}_0)]^{1/2}\right]\times \qquad 11.20a$$

$$\times\exp\left\{-i\frac{1}{\hbar}[(\mathbf{m}_V^+ - i\mathbf{R}\mathbf{m}_0)\mathbf{c}^2]_{tr}\mathbf{t}\right\} \qquad 11.20b$$

$$\Phi^*(\mathbf{r},t) = \frac{\mathbf{m}_V^+ + i\mathbf{m}_V^-}{\sqrt{2}\,\mathbf{m}_0}\exp\left[i\frac{\mathbf{x}}{\hbar}\mathbf{c}\,[\overset{+}{\mathbf{m}}_V(\mathbf{m}_V^+ + i\mathbf{m}_V^-)]^{1/2}\right]\exp\left\{-i\frac{1}{\hbar}[(\mathbf{m}_V^+ + i\mathbf{m}_V^-)\mathbf{c}^2]_{tr}\mathbf{t}\right\} = \qquad 11.21$$

$$\Phi^*(\mathbf{r},t) = \frac{\mathbf{m}_V^+ + i\mathbf{R}\mathbf{m}_0}{\sqrt{2}\,\mathbf{m}_0}\exp\left[i\frac{\mathbf{x}}{\hbar}\mathbf{c}\,[\overset{+}{\mathbf{m}}_V(\mathbf{m}_V^+ + i\mathbf{R}\mathbf{m}_0)]^{1/2}\right]\exp\left\{-i\frac{1}{\hbar}[(\mathbf{m}_V^+ + i\mathbf{R}\mathbf{m}_0)\mathbf{c}^2]_{tr}\mathbf{t}\right\} \qquad 11.21a$$

From 11.20a and 11.21a it follows, that at $\mathbf{v} = \mathbf{c}$ and $\mathbf{R} = \mathbf{0}$ these wave functions turn to that, describing *photons* with effective mass $\mathbf{m}_V^+ = \hbar\boldsymbol{\omega}/\mathbf{c}^2$; and frequency $\boldsymbol{\omega} = \frac{1}{\hbar}[\mathbf{m}_V^+\mathbf{c}^2]_{tr}$.

$$[\Phi(\mathbf{r},t) = \Phi^*(\mathbf{r},t)]_{ph} = \frac{\mathbf{m}_V^+}{\sqrt{2}\,\mathbf{m}_0}\exp\left[i\frac{\mathbf{x}}{\hbar}\mathbf{m}_V^+\mathbf{c}\right]\exp\left\{-i\frac{1}{\hbar}[\mathbf{m}_V^+\mathbf{c}^2]_{tr}\mathbf{t}\right\} \qquad 11.22$$

where: $\mathbf{m}_V^+\mathbf{c}^2 = \mathbf{h}\boldsymbol{\nu}_{ph}$ is the photon energy.

The product of the conventional forms of complex conjugate wave functions (11.19) gives the space and time independent pre-exponential coefficient squared: $|\Phi(\mathbf{r},t)|^2 = \mathbf{C}^*\mathbf{C} = const.$

From product of 11.20a and 11.21a, we get the *new general formula for density of probability* of location of particle in Corpuscular [C] phase, dependent on space and time $|\Phi(\mathbf{r},t)|^2$:

$$|\Phi(\mathbf{r},t)|^2 = \Phi(\mathbf{r},t)\Phi^*(\mathbf{r},t) = \qquad 11.23$$

$$= \frac{(\mathbf{m}_V^+)^2 + (\mathbf{m}_V^-)^2}{2\mathbf{m}_0^2}\exp\left[i\frac{\sqrt{2}\,\mathbf{x}}{\mathbf{L}_C}\right]\exp\{-i\,2\omega_{C\rightleftharpoons W}\mathbf{t}\}$$

where the resulting frequency of $\mathbf{C} \rightleftharpoons \mathbf{W}$ pulsation of uncompensated sub-elementary fermions is $\boldsymbol{\omega}_{C\rightleftharpoons W} = \mathbf{m}_V^+\mathbf{c}^2/\hbar$ and $\mathbf{L}_C = \hbar/\mathbf{m}_V^+\mathbf{c}$ is the characteristic dimension of elementary particle in [C] phase.

The resulting energy of this state is characterized by the length of hypotenuse of triangle with adjacent cathetus squared:

$$\mathbf{E}^{Res}_{\mathbf{V}^+\Updownarrow\mathbf{V}^-} = \mathbf{m}^{\pm}_{\mathbf{V}^+\Updownarrow\mathbf{V}^-}\mathbf{c}^2 = \sqrt{(\mathbf{m}_V^+)^2 + (\mathbf{m}_V^-)^2}\,\mathbf{c}^2 \qquad 11.24$$

It is important to point out, that in state of rest, when the external translational velocity of elementary particle is zero ($\mathbf{v} = \mathbf{0}$), the real and complementary mass are equal to the rest mass: $\mathbf{m}_V^+ = \mathbf{m}_V^- = \mathbf{m}_0$, the external de Broglie wave length tends to infinity ($\lambda_B = 2\pi L_B = \infty$) and its frequency to zero ($\omega_B = 0$), the wave function, described by *conventional* expression (11.2) becomes equal to coefficient $\mathbf{C}$. This coefficient itself, as a square root of pre-exponential factor $\mathbf{C} = \sqrt{\frac{(\mathbf{m}_V^+)^2+(\mathbf{m}_V^-)^2}{2\mathbf{m}_0^2}}$ at these conditions is equal to $\mathbf{C} = \mathbf{1}$. The corresponding density of probability describing only the external properties of particle $\mathbf{C}^2 = \mathbf{1}$ is a permanent value, independent on space and time.

However, the general expression of density of probability (11.23) of particle location in selected point of space-time, when its external translational velocity is equal to zero ($\mathbf{v}^{ext} = \mathbf{0}$), following *from our theory*, turns to:

$$|\Phi(\mathbf{r},t)|^2 = \exp\left(i\sqrt{2}\,\frac{\mathbf{x}}{\mathbf{L}_0}\right)\exp(-i2\boldsymbol{\omega}_0\mathbf{t}) \qquad 11.25$$

where the Compton wave length and frequency of particle are equal, correspondingly, to:

$$\mathbf{L}_0 = \frac{\mathbf{c}}{\omega_0} = \frac{\hbar}{\mathbf{m}_0\mathbf{c}} \quad \text{and} \quad \boldsymbol{\omega}_0 = \frac{\mathbf{m}_0\mathbf{c}^2}{\hbar} \qquad 11.25a$$

We can see, that the general expression of density probability of particle in [C] phase location (7.45), in contrast to conventional, the permanent one, is *oscillating* due to internal $[\mathbf{C} \rightleftharpoons \mathbf{W}]_{in}$ pulsation of sub-elementary fermions, rotating around common axes, as presented in Fig.1 and Fig.3. At fixed coordinate (**x**), the probability of particle in [C] phase location is dependent on time, i.e. phase of pulsation. At fixed time (**t**) this probability is dependent on coordinate of particle in [C] phase.

Our notion of the 'anchor' sites is in accordance with interpretation of wave function as a cohomological measure of quantum vorticity by Kiehn (1989, 1998). *An exact complex mapping of the wave function has been found by him.*

The wave function after separation into real and imaginary parts, transforms the two-dimensional Schrödinger equation for a charged particle interacting with an electromagnetic field into two differential systems.

The first partial differential system is exactly the evolutionary equation for the vorticity of a compressible, viscous two dimensional Navie-Stokes fluid.

The second system is related to the Beltrami equation defining a minimal surface in terms of the kinetic and potential energy.

The absolute square of the wave function is exactly the vorticity distribution in a fluid.

This distribution corresponds to distribution of secondary anchor sites in our model of particle propagation. This interpretation of the wave function offers an alternative to the Copenhagen dogma.

## 12. The Principle of least action. Second and Third laws of Thermodynamics as a consequence of Bivacuum interaction with elementary particles

### *12.1 Principle of least action*

Let us analyze the formula of *action* in Maupertuis-Lagrange form:

$$\mathbf{S}_{ext} = \int_{t_0}^{t_1} 2\mathbf{T}_k^{ext}\,\mathbf{dt} \qquad 12.1$$

The *principle of Least action*, responsible for choosing one of number of possible particles trajectories from one configuration to another has a form:

$$\Delta\mathbf{S}_{ext} = 0 \qquad 12.2$$

This means, that the optimal trajectory of each particle corresponds to *minimum variations of its external kinetic energy and time.*

The time interval: $\mathbf{t} = \mathbf{t}_1 - \mathbf{t}_2 = \mathbf{n}\mathbf{t}_B$ we take as a quantized period of the de Broglie wave of particle ($\mathbf{t}_B = 1/\boldsymbol{\nu}_B$):

$$\mathbf{t} = \mathbf{t}_1 - \mathbf{t}_2 = \mathbf{n}\mathbf{t}_B = \mathbf{n}/\mathbf{v}_B \qquad 12.3$$
$$\mathbf{n} = \mathbf{1}, \mathbf{2}, \mathbf{3} \ldots$$

Using eqs.(12.1), we get for the action in Maupertuis-Lagrange form

$$\mathbf{S}_{ext} = 2\mathbf{T}_k^{ext} \times \mathbf{t} = \mathbf{m}_\mathbf{V}^+ \mathbf{v}^2 \times \mathbf{t} = \mathbf{t}|\mathbf{m}_\mathbf{V}^+ - \mathbf{m}_\mathbf{V}^-|\mathbf{c}^2 \qquad 12.4$$

$\mathbf{m}_\mathbf{V}^+ \mathbf{v}^2 = 2\mathbf{T}_\mathbf{k}^\mathbf{ext}$ is the doubled kinetic energy of particle. It can be expressed via difference between resulting energy of particle:

$$\mathbf{E}^{res} = \hbar\omega_{\mathbf{C}\rightleftharpoons\mathbf{W}}^{\mathbf{res}} = |\mathbf{m}_V^+ - \mathbf{m}_V^-|\mathbf{c}^2 = \mathbf{m}_0\mathbf{c}^2 + \mathbf{m}_\mathbf{V}^+\mathbf{v}^2 \qquad 12.5$$

and the rest mass energy and via difference of corresponding frequencies:

$$2\mathbf{T}_k^{ext} = \mathbf{m}_\mathbf{V}^+\mathbf{v}^2 = \mathbf{E}^{res} - \mathbf{m}_0\mathbf{c}^2 = \hbar[\omega_{\mathbf{C}\rightleftharpoons\mathbf{W}}^{\mathbf{res}} - \omega_0]^{e,p,n} = \frac{\mathbf{S}}{\mathbf{t}} \qquad 12.6$$

the forced resonance of Bivacuum Virtual Pressure Waves $(\mathbf{VPW}_q^\pm)^{e,p,n}$ at condition $\mathbf{v} << \mathbf{c}$with basic frequency: $\omega_0^{e,p,n}$ interacting with paired sub-elementary fermions of triplets $[\mathbf{F}_\downarrow^- \bowtie \mathbf{F}_\uparrow^+]$ in the process of $[\mathbf{C} \rightleftharpoons \mathbf{W}]$ pulsation should decrease the kinetic energy of elementary particles:

$$< [\mathbf{F}_\downarrow^- \bowtie \mathbf{F}_\uparrow^+]_W + (\mathbf{F}_\updownarrow^\pm)_C > \overset{\omega_0^{e,p,n}}{\rightleftharpoons} < [\mathbf{F}_\downarrow^- \bowtie \mathbf{F}_\uparrow^+]_C + (\mathbf{F}_\updownarrow^\pm)_W > \qquad 12.7$$

We can see from 12.6, that minimization of action $\mathbf{S} \to 0$ and $\Delta\mathbf{S} \to \mathbf{0}$ during permanent time range ($\mathbf{t} = const$), means minimization of kinetic energy of the protons and neutrons, composing nuclear of the atoms of the objects, means minimization of kinetic energy of particles $\mathbf{T}_k \to 0$. This process of tending of resulting frequency of of $\mathbf{C} \rightleftharpoons \mathbf{W}$ pulsation to the rest mass frequency $[\omega_{\mathbf{C}\rightleftharpoons\mathbf{W}}^{\mathbf{res}} \to \omega_0]^{e,p,n}$ can be stimulated by virtual pressure waves of positive and negative energy: $[\mathbf{VPW}^+ \bowtie \mathbf{VPW}^-]^{p,n}$ with basic frequency: $\omega_0^{p,n} = \mathbf{m}_0^{p,n}\mathbf{c}^2/\hbar$, corresponding to the rest mass energy of the protons and neutrons.

*We may conclude, that tending of* $\Delta\omega = [\omega_{C\rightleftharpoons W} - \omega_0]^{e,p,n}$ *to zero due to influence of basic* $VPW_{q=1}^\pm$ *at* $q = 1$ *on triplets dynamics (forced resonance), minimizing translational velocity and kinetic energy of elementary particles, provides realization of principle of Least action.*

### *12.2 Bivacuum - mediated interaction as a background of 2nd and 3d laws of thermodynamics*

At the velocity of triplets, like protons and neutrons ($\mathbf{v} > \mathbf{0}$), the interaction of their pulsing paired sub-elementary fermions with basic ($\mathbf{q} = \mathbf{1}$) virtual pressure waves of Bivacuum $[\mathbf{VPW}^+ \bowtie \mathbf{VPW}^-]^{p,n}$ due to forced resonance should slow down atoms thermal velocity $\mathbf{v} \to \mathbf{0}$.

*The second law of thermodynamics*, formulated as a spontaneous irreversible transferring of the heat energy from the warm body to the cooler body or surrounding medium, also means decreasing of kinetic energy of particles, composing this body.

Consequently, the 2nd law of thermodynamics, as well as Principle of Least Action, can be a consequence of forced resonance of $[\mathbf{VPW}^+ \bowtie \mathbf{VPW}^-]^{p,n}$ with $\mathbf{C} \rightleftharpoons \mathbf{W}$ pulsation of paired sub-elementary fermions and antifermions, slowing down particles thermal *translational* dynamics at pull-in-range synchronization conditions and minimization of resulting frequency and energy of $\mathbf{C} \rightleftharpoons \mathbf{W}$ pulsation $\omega_{\mathbf{C}\rightleftharpoons\mathbf{W}}^{\mathbf{res}}$

$$\hbar(\omega^{res}_{C \rightleftharpoons W} \overset{v \to 0}{\to} \omega_0)) \qquad 12.11$$

*The third law of thermodynamics* states, that the entropy of equilibrium system is tending to zero at the absolute temperature close to zero. Again, this may be a consequence of forced combinational resonance between basic $[\mathbf{VPW}^+ \bowtie \mathbf{VPW}^-]^{p,n}$ and paired sub-elementary fermions $[\mathbf{C} \rightleftharpoons \mathbf{W}]$ pulsation (12.7), when *translational* velocity of particles $\mathbf{v} \to 0$ and $\omega_{C \rightleftharpoons W} \to \omega_0$. At these conditions in accordance with Hierarchic theory of condensed matter (Kaivarainen, 1995; 2001; 2001a) the de Broglie wave length of atoms is tending to infinity and to state of macroscopic Bose condensation of ultimate coherence and order, i.e. minimum entropy.

This result of our Unified theory could explain the energy conservation, notwithstanding of the Universe cooling. *Decreasing* of thermal kinetic energy of particles in the process of cooling is compensated by increasing of potential energy of particles interaction, accompanied the *increasing* of particles de Broglie wave length and their Bose condensation.

## 13 New approach to problem of Time, as a "Time of Action"

Using formula for *action* in Maupertuis-Lagrange form:

$$\mathbf{S} = 2\mathbf{T}_k^{ext} \times \mathbf{t} = \mathbf{m}_V^+\mathbf{v}^2 \times \mathbf{t} = \mathbf{min} \qquad 13.1$$

it is easy to show, that the *pace of time* ($\mathbf{dt/t}$) for any closed *conservative system* is determined by the pace of its kinetic energy change $(-\mathbf{dT}/\mathbf{T}_k)_{x,y,z}$, *anisotropic* in general case (Kaivarainen, 2004; 2005):

$$\left[\frac{\mathbf{dt}}{\mathbf{t}} = \mathbf{d}\ln\mathbf{t} = -\frac{\mathbf{dT}_k}{\mathbf{T}_k} = -\mathbf{d}\ln\mathbf{T}_k\right]_{x,y,z} \qquad 13.2$$

Similar relation can be obtained from principle of uncertainty for free particle with kinetic energy ($\mathbf{T}_k$) in coherent form: $\mathbf{T}_k\,\mathbf{t} = \hbar$.

From formula (12.12) it is easy to derive a formula for *"Time of Action"* for conservative mechanical systems.

It is important to note, that in closed conservative mechanical or quantum system the total energy is permanent:

$$\mathbf{E}_{tot} = \mathbf{V} + \mathbf{T}_k = const \qquad 13.3$$
$$or: \Delta\mathbf{E}_{tot} = 0 \quad and \quad \Delta\mathbf{V} = -\Delta\mathbf{T}_k$$

and the *time of action* is always the *external one*.

By definition a *conservative system* is a system in which work done by a force is:

1. Independent of path;
2. Completely reversible.

Using relations (13.2) and relativistic expression for kinetic energy of *system or mechanical object*:

$$\mathbf{T}_k = \mathbf{m}_V^+\mathbf{v}^2/2 = \frac{\mathbf{m}_0\mathbf{v}^2}{2\sqrt{1-(\mathbf{v/c})^2}} \qquad 13.4$$

the *pace of time* and *time of action* for closed system can be presented via *acceleration and velocity* of one or more parts, composing this system (Kaivarainen, 2004, 2005):

$$\left[\frac{\mathbf{dt}}{\mathbf{t}} = \mathbf{d}\ln\mathbf{t} = -\frac{\mathbf{d}\vec{\mathbf{v}}}{\vec{\mathbf{v}}}\frac{2-(\mathbf{v}/\mathbf{c})^2}{1-(\mathbf{v}/\mathbf{c})^2}\right]_{x,y,z} \quad 13.5$$

We proceed from the fact, that the true *inertial frames* in our accelerating, rotating and gravitating Universe and in all of its lower hierarchical levels and subsystems - *are nonexisting.*

The dynamics and accelerations in each *closed conservative system, where* $\mathbf{E}_{tot} = const$, are characterized by its dimensionless *pace of time* (13.5) and *time* itself:

$$\mathbf{t} = \left[-\frac{\vec{\mathbf{v}}}{\vec{\mathbf{a}}}\frac{1-(\mathbf{v}/\mathbf{c})^2}{2-(\mathbf{v}/\mathbf{c})^2}\right]_{x,y,z} \quad 13.6$$

where the acceleration in different kinds of motion can be expressed in different forms:

$$\vec{\mathbf{a}} = \mathbf{d}\vec{\mathbf{v}}/\mathbf{dt} = \frac{\mathbf{v}^2}{\mathbf{r}} = \omega^2\mathbf{r} \quad 13.7$$

$$or: \quad \vec{\mathbf{a}} = \mathbf{G}\frac{\mathbf{M}}{r^2} \text{ - free fall acceleration} \quad 13.7a$$

The external reference frame for selected conservative system can be only the another inertialess system/frame, including the former one as a part and with other relativistic factor: $\mathbf{R}^2 = \mathbf{1} - (\mathbf{v}/\mathbf{c})^2$ . In such approach the *internal time* ($\mathbf{t}^{in}$) of smaller system can be analyzed as a part of external time of conservative system ($\mathbf{t}^{ext}$) :

$$\mathbf{t}^{in} = \frac{\mathbf{t}^{ext}}{\sqrt{1-(\mathbf{v}^{ext}/\mathbf{c})^2}} = -\frac{\vec{\mathbf{v}}\sqrt{1-(\mathbf{v}/\mathbf{c})^2}}{\vec{\mathbf{a}}\,[2-(\mathbf{v}/\mathbf{c})^2]} \quad 13.8$$

The shape of this formula in conditions, when the external time (13.6) is invariant:

$$\mathbf{t}^{ext} = \mathbf{t} = const$$

is close to to conventional formula of special relativity (13.11) for time or clock, moving with velocity ($\mathbf{v} \lesssim \mathbf{c}$) relatively to the clock in rest ($\mathbf{v} \ll \mathbf{c}$).

From our formula for time we can see, that the time for selected object (microscopic or macroscopic) of conservative system is positive at velocity: $0 < \mathbf{v} < \mathbf{c}$, if its acceleration is negative ($\mathbf{d}\vec{\mathbf{v}}/\mathbf{dt} < \mathbf{0}$). On contrary, time is negative, if acceleration is positive ($\mathbf{d}\vec{\mathbf{v}}/\mathbf{dt} > \mathbf{0}$). For example, if temperature of conservative system and its kinetic energy are decreasing, the time and its pace are positive.

Thermal oscillations of atoms and molecules in condensed matter, like pendulums oscillation, are accompanied by alternation the sign of acceleration and, consequently, sign of time ($\pm\mathbf{t}^{ext}$ and $\pm\mathbf{t}^{in}$).

The **Corpuscle → Wave** transition of elementary particle, as it follows from Unified theory, is accompanied by decreasing of mass and kinetic energy of unpaired sub-elementary fermion and converting the kinetic energy of [C] phase to potential energy of $\mathbf{CVC}^{\pm}$ of [W] phase. Consequently, this semiperiod of pulsation is characterized by positive time $(\mathbf{t} > \mathbf{0})_{C\to W}$. On contrary, the reverse [**W** → **C**] transition corresponds to negative time $(\mathbf{t} < \mathbf{0})_{W\to C}$.

In the absence of particles acceleration ($\mathbf{a} = \mathbf{d}\vec{\mathbf{v}}/\mathbf{dt} = \mathbf{0}$; $\mathbf{dT}_k/\mathbf{T}_k = 0$ and $\mathbf{c} > \mathbf{v} > \mathbf{0}$; ), the time of action (**t**) is infinitive and its pace (**dt/t**) is zero:

$$\mathbf{t} \to \infty \qquad \textbf{and} \qquad \left(\frac{\mathbf{dt}}{\mathbf{t}}\right) \to 0 \qquad 13.9$$

$$at \quad \left(\vec{\mathbf{a}} = \mathbf{d}\vec{\mathbf{v}}/\mathbf{dt}\right) \to \mathbf{0} \quad \textbf{and} \quad \mathbf{v} = \mathbf{const}$$

The infinitive life-time of the system means its absolute stability. The postulated by this author principle of conservation of internal kinetic energy of torus ($\mathbf{V}^+$) and antitorus ($\mathbf{V}^-$) of symmetric and asymmetric Bivacuum fermions/antifermions: $\left(\mathbf{BVF}_{as}^{\updownarrow}\right)^{\phi} \equiv \mathbf{F}_{\updownarrow}^{\pm}$ (eq.2.1), independently on their external velocity, in fact reflects the condition of infinitive life-time of Bivacuum dipoles in symmetric and asymmetric states. The latter means a stability of sub-elementary fermions and elementary particles, formed by them.

The permanent collective motion of the electrons in superconductors and atoms of $^4\mathbf{He}$ in superfluid liquids with constant velocity ($\mathbf{v} = \mathbf{const}$) and $\left(\mathbf{d}\vec{\mathbf{v}}/\mathbf{dt}\right) = \mathbf{0}$ in the absence of collisions and accelerations are good examples, confirming validity of our formula (12.14), as far in these conditions $\mathbf{t} \to \infty$.

When the external translational velocity and external accelerations of Bivacuum dipoles (**BVF** and $\mathbf{BVB}^{\pm}$) are zero: $\mathbf{v} = 0$ and $\mathbf{d}\vec{\mathbf{v}}/\mathbf{dt} = \mathbf{0}$, like *in primordial Bivacuum*, the notion of time is uncertain: $\mathbf{t} = \mathbf{0}/\mathbf{0}$.

Interesting, that similar uncertainty in time (12.14) corresponds to opposite limit condition, pertinent for photon or neutrino in primordial Bivacuum, when $\mathbf{v} = \mathbf{c} = \mathbf{const}$ and $\mathbf{d}\vec{\mathbf{v}}/\mathbf{dt} = \mathbf{0}$. Just in such conditions when causality principle do not work the anomalous time effects are possible.

*In our approach, the velocity of light is the absolute value, determined by physical properties of Bivacuum, like sound velocity in any medium is determined by elastic properties of medium. The primordial Bivacuum superfluid matrix represents the Universal Reference Frame (URF) in contrast to conventional Relative Reference Frame (RRF). Consequently the Bivacuum has the Ether properties and Bivacuum dipoles - the properties of* ***ethons*** *- elements of the Ether.*

The positive acceleration of the Universe expansion ($\mathbf{d}\vec{\mathbf{v}}/\mathbf{dt} > \mathbf{0}$) at $\mathbf{c} > \mathbf{v} > 0$, in accordance to (12.13 and 12.14), means negative pace of external time and time itself for this highest Hierarchical level of Bivacuum organization. For the other hand, the process of cooling of each regular star system, like our Solar system, following gradual cooling of star, means slowing down the internal kinetic energy of thermal motion of atoms and molecules in such system, i.e. negative acceleration ($\mathbf{d}\vec{\mathbf{v}}/\mathbf{dt} < \mathbf{0}$) at $\mathbf{c} > \mathbf{v} > 0$. It corresponds to positive internal time and its pace in star systems. These opposite sign and the 'arrow' direction of *time of action* on different hierarchical levels of Universe organization, possibly is a consequence of tending of the Universe to keep its total energy permanent, following energy conservation law.

In accordance with Einstein relativistic theory (Landau and Lifshitz, 1988), the time of clock in the rest state ($\mathbf{t}^{ext}$), which can be considered, as the *external inertial frame* is interrelated with time ($\mathbf{t}^{in}$) in other inertial frame, moving relatively to former with velocity ($\mathbf{v}$) as:

$$\mathbf{t}^{ext} = (t_2' - t_1')^{ext} = \mathbf{t}^{in}\sqrt{1 - (\mathbf{v}/\mathbf{c})^2} \qquad 13.10$$

$$\mathbf{t}^{in} = \frac{\mathbf{t}^{ext}}{\sqrt{1 - (\mathbf{v}/\mathbf{c})^2}} \qquad 13.11$$

where: $\mathbf{t}^{ext} \equiv (t_2' - t_1')^{ext}$ is the characteristic time of clock in the external reference rest frame; $\mathbf{t}^{in} \equiv (t_2 - t_1)^{in}$ is the *internal proper time* of clock, moving with velocity: $\mathbf{v} \lesssim \mathbf{c}$, relatively to clock in the rest frame.

It is easy to see, that in relativistic conditions, when $\mathbf{v}^{in} \rightarrow \mathbf{c}$, the *proper time* of moving system/clock is tending to infinity: ($T \sim \mathbf{t}^{in} \rightarrow \infty$). This means that the moving clock is slower, than similar clock in state of rest relatively to moving one. In contrast to our formula for time, this equation do not contain acceleration.

**If we consider the imaginary system**, containing only two clock in empty space, moving as respect to each other with permanent velocity, and use the 1st postulate of Special Relativity, i.e. similar laws of physics in any inertial system, we should get the similar time delay in both clocks, even if they move with different velocities in our Universal Reference Frame (URF) - Bivacuum. In other words, both clocks should display the *same time delay,* independently of difference of their velocities ratio to the light velocity $(\mathbf{v}/\mathbf{c})^2$. This result of special relativity is a consequence of assumption of the absence of Ether and absolute velocity. It sounds like a nonsense and has no experimental confirmation. It follows from our Unified theory, that the interpretation, given by Einstein to Michelson-Morley experiments, as the evidence of the Ether absence, was wrong in contrast to explanation, provided by the authors of this experiment themselves.

*Our formulas (13.5) and 13.6), describing the properties of time for conservative systems, are more advanced, than Einstein's (13.11), as far they are not limited by inertialess frames and contain not only the relativistic factor, but also the velocity itself and acceleration. It will be demonstrated below, that our* ***time of action*** *concept better describe the dynamic processes on microscopic - quantum and macroscopic - cosmic scales.*

Different closed conservative systems of particles/objects, rotating around common center on stable orbits with radius ($r$), like in Cooper pairs of sub-elementary fermions, atoms, planetary systems, galactics, etc. are characterized by *centripetal* ($\mathbf{a}_{cp}$) and *centrifugal* ($\mathbf{a}_{cen}$) acceleration, equal by absolute value:

$$\mathbf{a}_{cp} = -\frac{\mathbf{d}\vec{\mathbf{v}}}{\mathbf{dt}} = \frac{\vec{\mathbf{v}}^2}{\vec{\mathbf{r}}} = \omega^2\vec{\mathbf{r}} = -\mathbf{a}_{cen} \qquad 13.12$$

where the *tangential* velocity of rotation is related to the radius $\vec{\mathbf{r}}$ and angular frequency of orbital rotation ($\omega$) as:

$$\left[\vec{\mathbf{v}} = 2\pi\vec{\mathbf{r}} \times \nu = \omega\vec{\mathbf{r}}\right] \qquad 13.13$$

Consequently, we get for the ratio of tangential velocity of particle/object to its centripetal acceleration:

$$-\frac{\vec{\mathbf{v}}}{\mathbf{d}\vec{\mathbf{v}}/\mathbf{dt}} = \frac{1}{\omega} = \frac{\vec{\mathbf{r}}}{\vec{\mathbf{v}}} \qquad 13.14$$

Putting (13.14 and 13.13) into (13.6), we get the dependence of *time of action* for Corpuscular phase of elementary particle, characterizing period of rotation of structure, like Fig.6 (electron) or Fig.8 (photon) around internal main axes with radius of rotation ($\mathbf{r}$) and angular frequency ($\omega = \vec{\mathbf{v}}/\vec{\mathbf{r}}$):

$$\mathbf{t} = \left[\frac{\vec{\mathbf{r}}}{\vec{\mathbf{v}}}\frac{1-(\mathbf{v}/\mathbf{c})^2}{2-(\mathbf{v}/\mathbf{c})^2}\right]_{\mathbf{W}} = \left[\frac{1}{\omega}\frac{1-(\vec{\mathbf{r}}\omega/\mathbf{L}_0\omega_0)^2}{2-(\vec{\mathbf{r}}\omega/\mathbf{L}_0\omega_0)^2}\right]_{\mathbf{C}} \qquad 13.15$$

For sub-elementary fermion in [C] phase, when the *translational* energy of elementary particle, pertinent for [W] phase, turns to *rotational* one, we have, using (13.12):

$$(\mathbf{v}/\mathbf{c})^2 = (\vec{\mathbf{r}}\omega/\mathbf{L_0}\omega_0)^2 \qquad 13.16$$

$$where: \mathbf{L_0} = \hbar/\mathbf{m}_0\mathbf{c} \text{ and } \omega_0 = \mathbf{m}_0\mathbf{c}^2/\hbar$$

From (13.15) we can see, that for *nonrelativistic* conditions of orbital rotation of system/object, when its tangential velocity $\mathbf{v} << \mathbf{c}$ and permanent angular frequency: $\omega = \mathbf{v}/\mathbf{r} = \mathbf{const}$, we get the relation between characteristic time of this system and period of orbital rotation ($T$):

$$\mathbf{t}^{ext}_{\mathbf{v}<<\mathbf{c}} \simeq \left|\frac{1}{2\omega}\right| = \frac{1}{4\pi}T \qquad 13.17$$

For relativistic conditions of the same system, when $\mathbf{v} \simeq \mathbf{c}$ at angular velocity $(\omega = \mathbf{v}/\mathbf{r}) = \mathbf{const}$, we see from (13.15), that characteristic time and period of orbiting elementary particle or macroscopic object is tending to zero, as far $\left[1-(\mathbf{v}/\mathbf{c})^2\right] \overset{\mathbf{v}\to\mathbf{c}}{\to} 0$ and $\left[2-(\mathbf{v}/\mathbf{c})^2\right] \overset{\mathbf{v}\to\mathbf{c}}{\to} 1$:

$$\mathbf{t}_{\mathbf{v}\lesssim\mathbf{c}} \to 0 \text{ and the period } (T = \mathbf{1/\nu}) \to 0 \text{ at } \mathbf{v}\to\mathbf{c} \qquad 13.18$$

$$\text{and } \mathbf{r}\to\mathbf{r}_{max} \text{ as far } \left(\omega = \frac{\mathbf{v}}{\vec{\mathbf{r}}}\right) = \mathbf{const} \qquad 13.18a$$

For the case, under consideration, the increasing of radius of orbit ($\mathbf{r}$) proportional to increasing of tangential velocity of orbiting particle/object at permanent angular frequency is a consequence of condition (13.18a).

Formula (13.14) shows, that at very low acceleration: $\mathbf{a} = \mathbf{d}\vec{\mathbf{v}}/\mathbf{dt} \to \mathbf{0}$, the ratio [$\mathbf{v}/\mathbf{a}$] should dominate on ratio:

$$\frac{1-(\mathbf{v}/\mathbf{c})^2}{2-(\mathbf{v}/\mathbf{c})^2} << \left[-\frac{\mathbf{v}}{\mathbf{a}}\right] \qquad 13.19$$

Consequently, at condition (13.19) the time of action should increase with velocity of rotating or pulsing object. The same qualitative result follows from special relativity (13.11). *Consequently, at these condition the time delay in moving system, following from special relativity, is in accordance with our theory of time.*

The formula for time (13.15), determined by internal rotational degrees of freedom of stationary systems, like sub-elementary fermions in elementary particles, the electron orbiting in atom of hydrogen or any planet, rotating around the star, can be transformed to:

$$\mathbf{t} = \frac{1}{\omega}\frac{\mathbf{m}_V^+\mathbf{c}^2\left[1-(\mathbf{v}/\mathbf{c})^2\right]}{\mathbf{m}_V^+(2\mathbf{c}^2-\mathbf{v}^2)} = \frac{1-(\mathbf{v}/\mathbf{c})^2}{\omega}\frac{\mathbf{E}_{tot}}{2\mathbf{V}} \qquad 13.20$$

where: $\mathbf{E}_{tot} = \mathbf{m}_V^+\mathbf{c}^2 = const$ is a total actual energy of rotating with angular frequency $\omega$ sub-elementary particle with actual mass of its torus $\mathbf{m}_V^+$.

In conservative system the doubled potential energy of unpaired sub-elementary fermion of elementary fermion with actual and complementary mass of torus and antitorus: $\mathbf{m}_V^+$ and $\mathbf{m}_V^-$:

$$\mathbf{2V} = \mathbf{2}(\mathbf{E}_{tot} - \mathbf{T}_k) = \mathbf{m}_V^+(2\mathbf{c}^2-\mathbf{v}^2) = (\mathbf{m}_V^+ + \mathbf{m}_V^-)\mathbf{c}^2 \qquad 13.20a$$

In the case of harmonic oscillation or standing wave, when $\mathbf{E}_{tot} = \mathbf{V} + \mathbf{T}_k = 2\mathbf{V}$ and $\mathbf{V} = \mathbf{T}_k$, the characteristic time of rotating with angular frequency ($\omega = \mathbf{v}/\mathbf{r}$) particle is dependent only on the ratio of its absolute velocity to the light one $(\mathbf{v}/\mathbf{c})^2$.

*13.1 The application of new time concept for explanation of Fermat principle*

The Fermat principle states that light waves of a given frequency traverse the path between two points which takes the least time. Its modern form is "A light ray, going between two points, must follow optical path length which is stationary with respect to variations of the path."

In this formulation, the paths may be maxima, minima, or saddle points.

The most obvious example of this is the passage of light through a homogeneous medium in which the speed of light doesn't change with position. In this case shortest time is equivalent to the shortest distance between the points, which is a straight line. The examples are existing that time of light passage, including reflected beam, can be minimum or maximum like for light beams from source in the center of ellipsoid with mirror internal surface. There can be a number of trajectories of light beams with the same time of passion. For example, it is true for different beams from one focal point to another passing throw the lens on different distance from lens center. The most important condition for realization of Fermat principle is $\mathbf{t} = \mathbf{const}$. This principle explains the *law of reflection*, as the equality of angles of incidence and angle of reflection: $\boldsymbol{\theta}_I = \boldsymbol{\theta}_R$ and Snell's law of refraction: $\mathbf{sin}\,\boldsymbol{\theta}_I = \mathbf{n}\,\mathbf{sin}\,\boldsymbol{\theta}_R$.

However, it is not yet clear why the Fermat principle is working. Let us analyze the application of Fermat principle to light refraction, using our formula for time (13.6). In accordance to Fermat principle the variation of action time for photons at: $\mathbf{E}_{tot} = \mathbf{V} + \mathbf{T}_k = \hbar\boldsymbol{\omega}_{ph} = const$ (condition of conservative system) should be zero: $\Delta\mathbf{t} = \mathbf{0}$.

The ratio of velocity of light in vacuum/bivacuum to its velocity ($\mathbf{v} \lesssim \mathbf{c}$) in gas, liquid or transparent solid determines the refraction index of corresponding medium: $(\mathbf{v}/\mathbf{c})^2 = 1/\mathbf{n}$. Taking this into account, the variation of (13.6) in [W] and [C] phase of photon can be presented as:

$$\Delta\mathbf{t} = \Delta\left[-\frac{\vec{\mathbf{v}}}{\vec{\mathbf{a}}}\,\frac{1-(1/\mathbf{n})}{2-(1/\mathbf{n})}\right]_{W,C} = 0 \qquad 13.21$$

After differentiation (13.21), we get:

$$\frac{\Delta\mathbf{n}}{\mathbf{n}-1} - \frac{2\Delta\mathbf{n}}{2\mathbf{n}-1} = \frac{\Delta\mathbf{a}}{\mathbf{a}} - \frac{\Delta\mathbf{v}}{\mathbf{v}} \qquad 13.22$$

At the conditions, when velocity of light in medium is close to this velocity in empty space: $\mathbf{n} = (\mathbf{c}/\mathbf{v})^2 \gtrsim 1$ we have $\frac{\Delta\mathbf{n}}{\mathbf{n}-1} \gg \frac{2\Delta\mathbf{n}}{2\mathbf{n}-1}$ and (13.22) turns to:

$$\Delta\mathbf{n} \cong (\mathbf{n}-1)\left[\left(-\frac{\mathbf{v}_2-\mathbf{v}_1}{\mathbf{v}_1}\right) + \frac{\Delta\mathbf{a}}{\mathbf{a}}\right]_W \qquad 13.23$$

The relative change of acceleration $\Delta\mathbf{a}/\mathbf{a}$ describes the jump of light velocity on the interface between two different homogeneous medium.

It is easy to see from this formula, that if the light velocity in 2nd medium is lower, than in 1st and $(\mathbf{v}_2 - \mathbf{v}_1) < 0$, the refraction index will increase: $\Delta\mathbf{n} > \mathbf{0}$. *This is in total accordance with empirical data and explains why the Fermat principle is working in geometrical optics.*

Formula (13.23) describes the change of photon parameters it its Wave [W] phase.

The centripetal acceleration of photon in *Corpuscular [C] phase* can be expressed via tangential velocity and rotation radius of photon (Fig.8) as: $\mathbf{a}_{cp} = -\frac{\vec{\mathbf{v}}^2}{\vec{\mathbf{r}}} = -\boldsymbol{\omega}^2\mathbf{r}$ and

$$\frac{\Delta\mathbf{a}}{\mathbf{a}} = \left(\frac{2\Delta\boldsymbol{\omega}}{\boldsymbol{\omega}} + \frac{\Delta\mathbf{r}}{\mathbf{r}}\right)_C \qquad 13.24$$

The relative jump of tangential velocity of photon ($\vec{\mathbf{v}} = \boldsymbol{\omega}\vec{\mathbf{r}}$) on the interphase between two mediums is:

$$\frac{\Delta \mathbf{v}}{\mathbf{v}} = \left(\frac{\Delta \boldsymbol{\omega}}{\boldsymbol{\omega}} + \frac{\Delta \mathbf{r}}{\mathbf{r}}\right)_C \qquad 13.25$$

Consequently, the difference in relative increments for [C] phase of photon is:

$$\left(\frac{\Delta \mathbf{a}}{\mathbf{a}} - \frac{\Delta \mathbf{v}}{\mathbf{v}}\right)_C = \left[\frac{\Delta \boldsymbol{\omega}}{\boldsymbol{\omega}}\right]_C \qquad 13.26$$

Putting this expression to (13.23), we get the increment of refraction index for photon in Corpuscular phase via relative jump of its angular frequency:

$$\Delta \mathbf{n} \cong (\mathbf{n} - \mathbf{1})\left[\frac{\Delta \boldsymbol{\omega}}{\boldsymbol{\omega}}\right]_C \qquad 13.27$$

This angular frequency of photon rotation coincides with frequency of its [$\mathbf{C} \rightleftharpoons \mathbf{W}$] pulsation only in symmetric primordial Bivacuum. In the volume of liquids or solids the symmetry of Bivacuum dipoles and their dynamics are changed by elementary particles of medium. From 13.27 we get, that this should be accompanied by increasing of rotational frequency of photon in its [C] phase.

Our Unified theory, in contrast to relativistic one, considers the velocity of particles as the *absolute* parameter, relative to primordial symmetric Bivacuum matrix (see eq. 6.3). The light velocity (**c**) is also the absolute parameter, determined by properties of Bivacuum (ether) and independent on velocity of source of photons.

### *13.2 The quantitative evidence in proof of new theory of time and elementary particles fusion from Bivacuum dipoles*

Using eq. (13.6 and 13.9a), it is possible to calculate the centrifugal acceleration ($a_{cf}^{\phi}$) in fast rotating Cooper pairs of sub-elementary fermions $[\mathbf{F}_\downarrow^- \bowtie \mathbf{F}_\uparrow^+]_C$ in triplets $< [\mathbf{F}_\downarrow^- \bowtie \mathbf{F}_\uparrow^+]_C + (\mathbf{F}_\updownarrow^\pm)_W >$, when paired sub-elementary fermions are rotating in corpuscular [C] phase and unpaired $(\mathbf{F}_\updownarrow^\pm)_W >$ is in the wave [W] phase.

Let us consider the condition of the rest state of the electron, when its *external translational* velocity is equal to zero and internal tangential velocity of sub-elementary fermion and antifermion $[\mathbf{F}_\downarrow^- \bowtie \mathbf{F}_\uparrow^+]_C$ rotation in triplet around common axis follows Golden mean condition:

$$(\mathbf{v}/\mathbf{c})_\phi^2 = \phi = 0.618 \qquad 13.28$$

$$\mathbf{v}^\phi = \mathbf{c}\,(0.618)^{1/2} = 2.358 \times 10^7\ m/s$$

In accordance to our theory of these conditions stand for the rest mass ($\mathbf{m}_0$) and charge ($\mathbf{e}_0$) origination (see chapter 6). The life-time $\mathbf{t}_\mathbf{C}$ of Corpuscular phase of rotating $[\mathbf{F}_\downarrow^- \bowtie \mathbf{F}_\uparrow^+]_C$ of the electron is equal to semiperiod of [$\mathbf{C} \rightleftharpoons \mathbf{W}$] pulsation of pair and triplet itself, determined by Compton angular frequency $\boldsymbol{\omega}_0^e = \boldsymbol{\omega}_{\mathbf{C}\rightleftharpoons\mathbf{W}}^e$ :

$$\mathbf{t}_\mathbf{C}^e \equiv \frac{1}{2}\mathbf{T}_{\mathbf{C}\rightleftharpoons\mathbf{W}}^e = \frac{1}{2\boldsymbol{\nu}_{\mathbf{C}\rightleftharpoons\mathbf{W}}^e} = \frac{\pi}{\boldsymbol{\omega}_0^e} = 4.02 \times 10^{-21}\ s \qquad 13.29$$

$$where:\quad \boldsymbol{\omega}_0^e = \mathbf{m}_0^e \mathbf{c}^2/\hbar \qquad 13.29a$$

Putting 13.28 and 13.29 in 13.6, we get for internal centrifugal acceleration of each of paired electronic sub-elementary fermions in [C] phase at Golden mean condition:

$$\left[a_{cf}^{\phi} = (\mathbf{dv}/\mathbf{dt})^{\phi}\right]^{e} = \frac{\mathbf{v}^{\phi}}{\mathbf{t}_C^{e}}\frac{1-\phi}{2-\phi} = 1.62 \times 10^{28}\ m/s^2 \qquad 13.30$$

For comparisons, the free fall acceleration in gravitational field of the Earth is only: $g = 9.81$ m/s$^2$.

The corresponding centrifugal force is equal to product of acceleration (13.30) on the rest mass of rotating paired sub-elementary fermion:

$$\mathbf{F}_{cf}^{\phi} = \mathbf{m}_0 a^{\phi} = (9.1 \times 10^{-31})\,(0.162 \times 10^{29}) = 1.47 \times 10^{-2}\ kg\ m/s^2 \qquad 13.31$$

From conventional expression for centrifugal force in such a system and Golden mean conditions, we get:

$$\mathbf{F}_{cf}^{\phi} = \frac{2\mathbf{m}_0\,\phi\mathbf{c}^2}{\mathbf{L}_0} = \frac{2}{3.83 \times 10^{-13}} \times 9.1093897 \cdot 10^{-31} \times 5.56 \cdot 10^{14} = \qquad 13.32$$
$$= 0.264 \times 10^{-2}\ kg \cdot m/s^2$$

This value is about 5.5 times less, than obtained using our expression for time and acceleration (13.30).

The condition of the electrons stability is that this centrifugal force is compensated by the opposite centripetal force in rotating pairs $[\mathbf{F}_{\downarrow}^{-} \bowtie \mathbf{F}_{\uparrow}^{+}]_C^{\phi}$. This compensation can be provided by Coulomb and in much less extent by gravitational attraction between torus and antitorus of paired sub-elementary fermion in triplets $<[\mathbf{F}_{\downarrow}^{-} \bowtie \mathbf{F}_{\uparrow}^{+}]_C + (\mathbf{F}_{\updownarrow}^{\pm})_W>$:

$$\mathbf{F}_{Coul}^{\phi} = \frac{\mathbf{e}_+\mathbf{e}_-}{\boldsymbol{\varepsilon}^{\phi}(\mathbf{L}^{\phi})^2} = \frac{\mathbf{e}_0^2}{\boldsymbol{\varepsilon}_0^{\phi}\mathbf{L}_0^2} = 1.98 \times 10^{-2}\ kg\ m/s^2 \qquad 13.33$$

$$\mathbf{F}_G^{\phi} = \mathbf{G}\frac{\mathbf{m}_0^2}{\mathbf{L}_0^2} = 6.67259 \times 10^{-11}\frac{(9.1093897 \times 10^{-31})^2}{(3.83 \times 10^{-13})^2} = 3.76 \times 10^{-46}\ kg\ m/s^2 \qquad 13.33a$$

where: $\mathbf{e}_-$ and $\mathbf{e}_+$ are the charges of $\mathbf{F}_{\downarrow}^{-}$ and $\mathbf{F}_{\uparrow}^{+}$ at Golden mean (GM) conditions, equal to the rest charge $\mathbf{e}_0 = 1.602 \times 10^{-13}$ C of the electron and positron, correspondingly, in accordance to our model of elementary particles..

The radius of rotation of this pair is equal to Compton radius at GM conditions: $\mathbf{L}^{\phi} = \mathbf{L}_0 = \hbar/\mathbf{m}_0\mathbf{c} \simeq 3.83 \times 10^{-13}\ m$. Assuming, that permittivity of Bivacuum between charges in pair $[\mathbf{F}_{\downarrow}^{-} \bowtie \mathbf{F}_{\uparrow}^{+}]_C$ is close to that of vacuum: $\boldsymbol{\varepsilon}^{\phi} \simeq \boldsymbol{\varepsilon}_0 = 8.85 \times 10^{-12}$ F m$^{-1}$, we get for Coulomb attraction force $\mathbf{F}_{Coul}^{\phi} = 1.98 \times 10^{-2}\ kg\ m/s^2$.

The gravitational constant in (12.28a) $\mathbf{G} = 6.67259 \times 10^{-11}$ m$^3$ kg$^{-1}$ s$^{-2}$ and the rest mass of the electron squared: $\mathbf{m}_0^2 = \left(9.1093897 \times 10^{-31}\ kg\right)^2$. It is easy to see, that gravitational attraction is negligible small as respect to Coulomb one.

The calculated Coulomb force (13.33) is close to the opposite centrifugal force (13.31), providing stabilization of pairs $[\mathbf{F}_{\downarrow}^{-} \bowtie \mathbf{F}_{\uparrow}^{+}]_C^{\phi}$ in triplets of the electrons:

$$\frac{\mathbf{F}_{Coul}^{\phi}}{\mathbf{F}_{cf}^{\phi}} = \frac{1.98 \times 10^{-2}}{1.47 \times 10^{-2}} = 1.343 \qquad 13.34$$

A possible explanation of this small disbalance in Coulomb and centrifugal forces, can be a bigger permittivity of Bivacuum in the internal space of this pairs, as respect to empty Bivacuum/vacuum: $\boldsymbol{\varepsilon}^{\phi}/\boldsymbol{\varepsilon}_0 = 1.343$. The reason of bigger internal permittivity $\boldsymbol{\varepsilon}^{\phi} = 1/\boldsymbol{\mu}_0\mathbf{c}_{\phi}^2$ can be a bigger refraction index in space between two sub-elementary fermions in pairs

$[\mathbf{F}_{\downarrow}^{-} \bowtie \mathbf{F}_{\uparrow}^{+}]_{C}^{\phi}$.

Like in the case of proton, stabilization of electronic triplet in its [W] phase can be realized via electronic gluons, i.e. superposition of their Cumulative virtual clouds $[\mathbf{CVC}^{+} \bowtie \mathbf{CVC}^{-}]^{e}$ between paired sub-elementary fermions in [W] phase.

The close values of centrifugal and Coulomb interaction for the electrons and positrons, calculated on the base of parameters of paired sub-elementary fermions in their Corpuscular phase (angular frequency of $[\mathbf{C} \rightleftharpoons \mathbf{W}]$ pulsation and tangential velocity of their rotation), following from our model of elementary particles, is important fact, confirming our Unified theory of Bivacuum, the new model of stable elementary particles and time.

For much less stable doublets, like muon + antimuon, the centrifugal force at Golden mean conditions (13.31) exceeds many times the Coulomb attraction between its sub-elementary fermion and antifermion:

$$\mathbf{F}_{cf}^{\phi} = \frac{2\mathbf{m}_0\,\phi\mathbf{c}^2}{\mathbf{L}_0} = \frac{2}{\hbar}\mathbf{m}_0^2\,\phi\mathbf{c}^3 >> \frac{\mathbf{e}^2}{\boldsymbol{\varepsilon}_0\mathbf{L}_0^2} = \mathbf{F}_{Coul} \qquad 13.35$$

This inequality is a result of the same charges of muon and electron at the mass of former exceeding the mass of latter about 200 times. It is a reason of [*muon* + *antimuon*] pairs much less stability and life-time, than that of sub-elementary [*fermion* + *antifermion*] pairs of the electrons.

### *13.3 Shift of the period of elementary oscillations in gravitational field*

The decreasing of the wavelength of photons (EM waves) and corresponding decreasing of their period in a gravitational field, predicted by general relativity theory (GRT), is dependent on mass (**M**) and distance (**r**) from center of mass to photons location and detection as:

$$\frac{\boldsymbol{\lambda}_G}{\boldsymbol{\lambda}_0} = \frac{\mathbf{T}_G}{\mathbf{T}_0} = \sqrt{1 - \frac{2GM}{c^2 r}} \qquad 13.36$$

$$or: \quad \mathbf{T}_G \simeq \mathbf{T}_0\left(1 - \frac{GM}{c^2 r}\right) \quad \text{at} \quad \frac{2GM}{c^2 r} << 1 \qquad 13.36a$$

A heuristic Newtonian derivation gives similar result as (13.36a):

$$\frac{T_G}{T_0} = \frac{\nu_0}{\nu_G} = \frac{\lambda_G}{\lambda_0} = \frac{hc}{\lambda_0}\frac{\lambda_G}{hc} = \qquad 13.37$$

$$= \frac{E_0}{E_G} = \frac{m_G c^2 - \frac{GMm_G}{r}}{m_G c^2} = 1 - \frac{GM}{c^2 r} \qquad 13.37a$$

where: $T_G$, $\nu_G$ and $\lambda_G$ are the shifted by G - field period, frequency and wave length of elementary wave; $h$ is Planck's constant, $c$ is the speed of light, $E_0$ is the unperturbed energy, $E_G$ is the shifted energy; $m_G$ is the effective mass of photon in field.

In the absence of gravitational field, when $M = 0$ or $r = \infty$, the period of oscillation is maximum $\mathbf{T}_G \simeq \mathbf{T}_0$.

As far the Newtonian gravitational force can be expressed via gravitational acceleration $(a_G = G\frac{M}{r^2})$ as:

$$\mathbf{F}_G = G\frac{Mm}{r^2} = a_G m \qquad 13.38$$

$$where: \quad a_G = G\frac{M}{r^2} = g \qquad 13.38a$$

Near surface of the Earth this acceleration is equal to free fall acceleration: $a_G = g = 9.8$ m/s$^2$.

Using (13.37a), formula (13.36a) can be presented as:

$$\mathbf{T}_G \simeq \mathbf{T}_0\left(1 - \frac{a_G r}{c^2}\right) = \mathbf{T}_0\left(1 - \frac{GM}{c^2 r}\right) \qquad 13.39$$

In accordance to this formula, the period of oscillation ($\mathbf{T}_G$) of test system, like photon or electron $[\mathbf{C} \rightleftharpoons \mathbf{W}]$ pulsation period, should increase with increasing of separation ($r$) between the test system and center of gravitating body: $\mathbf{T}_G \to \mathbf{T}_0$

The same result we get from our (13.6) in nonrelativistic conditions: $(\mathbf{v}/\mathbf{c})^2 << 1$.

For the other hand, from (13.36a) it follows that increasing of ($r$) at permanent $M$ should increase the period of pulsation ($\mathbf{T}_G$) and *decrease its frequency* - red Doppler shift.

The experiment for confirmation of described above consequences of General relativity theory (GR) was set up by Pound and Rebka (1959) in the Harvard tower, using Mössbauer effect. The Harvard tower is just 22.6 m, so the fractional gravitational red shift between the frequency $\nu^{bottom}$ of $\gamma$ –quantum *emitted at the bottom* of tower and frequency $\nu^{top}$ *absorbed at the top* of tower predicted by GRT, similar to simple classical approach (13.37), is given by the formula:

$$\frac{\Delta E}{E} = \frac{\mathbf{v}^{bottom} - \mathbf{v}^{top}}{\mathbf{v}^{top}} = \frac{T^{top} - T^{bottom}}{T^{top}} = \frac{Gl}{c^2} = 2.45 \times 10^{-15} \qquad 13.40$$

where: **G** is the gravitational constant; $l = r_2 - r_1 = 22.6\ m$ is the tower height and $c$ is the speed of light.

Pound and Rebka used the 14.4 keV gamma ray from the iron-57 isotope that has a high enough resolution to detect such a small difference in energy and frequency: $\Delta E = h(\mathbf{v}^{bottom} - \mathbf{v}^{top})$. In other set of experiments the source of $\gamma$ –quantum was placed at the top of tower and detector at the bottom.

The predicted theoretically relative frequency shifts on the upward and downward paths where opposite by sign, but the same by absolute values. Their sum: $4.9 \times 10^{-15}$ appears to be very close to measured: $5.1 \times 10^{-15}$. Consequently, as it follows from our formula for period of elementary pulsations (13.6), it is smaller in locations, where gravitational or centrifugal accelerations are bigger.

The coincidence of quantitative experimental relative shifts values with theoretical ones, following from GTR and simple classical Newton's formalism (13.37a) is excellent.

However, it does not contain a strong evidence that GTR works better, than classical Newtonian approach.

### *13.4 The explanation of Hefele-Keating experiments*

The additional confirmation of validity of our formula for time (13.6) is its ability to explain well known experiments of Hefele-Keating (1971) for verification of special and general theories of relativity (SR and GR).

They flew four *cesium atomic clocks* around the Earth in jets, first eastbound, then westbound. These experiments proved that atomic clocks period is dependent on the direction, velocity and altitude of jet airplanes. The direction and velocity of the airplanes where factors of the SR and the altitude was a factor of GR.

Compared to the time kept by control atomic clock fixed on the ground (USA), the *eastbound* clocks on the jets where slower (period of oscillation bigger) and *westbound clocks* - faster (period of oscillation shorter).

The velocity of *eastbound* clocks are the sum of tangential velocity of jet and tangential velocity of atmosphere at the altitude of jet flight: $\mathbf{v}_{res}^{east} = \mathbf{v}_{jet}' + \mathbf{v}_{at}$. For the other hand, the resulting velocity of *westbound* clock is a difference of these velocities: $\mathbf{v}^{west} = \mathbf{v}_{jet}' - \mathbf{v}_{at}$. The correct position of reference clock (non rotating) should be at the axes of the earth rotation (i.e. poles) of the earth. The velocity of the earth orbiting around the Sun and Sun system velocity in the universe was not taken into account.

Webster Kehr (2002) in his book "The detection of Ether" point out, that in original version of special relativity (1905) each of jets flying with permanent velocity should be considered as the *rest* reference frames.

However, even in such approximate approach, where the *local reference frames* instead Universal reference frame (URF) was used, Hafele and Keating found out, that the time effects, *calculated* using relativity theory, *coincide well* with experimental ones.

We will show below, that these experiments can be explained also on the base of our theory of time and simple Newtonian formula for gravitation and free fall acceleration, as a part of Unified theory.

The free fall acceleration following from Newton formula (13.38a) is:

$$a_G = (d\mathbf{v}/\mathbf{dt})_G = G\frac{M}{r^2} = g \qquad 13.41$$

Formula (13.6) can be presented in form, interrelating characteristic time of object with gravitational free fall acceleration ($a_G = g$), velocity of object and the increments of these parameters at permanent velocity:

$$\frac{\mathbf{T}^{ext}}{4\pi} \overset{\mathbf{v}<<\mathbf{c}}{\simeq} \mathbf{t}^{ext} = \frac{\vec{\mathbf{v}}\, r^2}{\mathbf{GM}} \frac{1-(\mathbf{v}/\mathbf{c})^2}{2-(\mathbf{v}/\mathbf{c})^2} \qquad 13.42$$

$$or: \quad \left[\frac{1}{4\pi}\Delta\mathbf{T}^{ext}\right]_{\mathbf{v}=const} \overset{\mathbf{v}<<\mathbf{c}}{\simeq} \frac{1}{2\mathbf{GM}}(2\vec{\mathbf{v}}\, r\Delta r + r^2\Delta\vec{\mathbf{v}}) \qquad 13.42a$$

where: $T = 2\pi/\omega$ is the period of elementary oscillation in external reference frame (atomic clock in private case).

*Formula (13.42) interrelate our concept of time with gravitation, however, in different way, than general theory of relativity.*

At permanent tangential velocity of jets respectively to the Earth surface: $\mathbf{v} = const$, $\Delta\mathbf{v} = 0$ for nonrelativistic case: $\mathbf{v} << \mathbf{c}$ we get from (13.42a) the confirmation of (13.40), that the external period is increasing and frequency decreasing with distance from the earth center:

$$[\Delta T = -\Delta\mathbf{v}]^{ext}_{\mathbf{v=const}} \overset{\mathbf{v}<<\mathbf{c}}{\simeq} 4\pi\frac{\mathbf{v}\, r\Delta r}{\mathbf{GM}} = 4\pi\frac{\mathbf{v}}{\mathbf{g}}\frac{\Delta r}{r} \qquad 13.43$$

where: $\Delta r = r_2 - r_1$ in private case corresponds to $l$ in eq.(12.33).

For the other case of permanent distance to the Earth center and surface: $r = const$; $\Delta r = 0$ and (13.42a) turns to:

$$[\Delta T = -\Delta\mathbf{v}]_{r=\mathbf{const}} \overset{\mathbf{v}<<\mathbf{c}}{\simeq} 2\pi\frac{r^2\Delta\vec{\mathbf{v}}}{\mathbf{GM}} = 2\pi\frac{\Delta\vec{\mathbf{v}}}{\mathbf{g}} \qquad 13.44$$

where: $\mathbf{G} = 6.67259 \times 10^{-11}\, \mathrm{m^3\, kg^{-1}\, s^{-2}}$; $\mathbf{M} = 5.9742 \times 10^{24}$ kg is the earth mass;

$r = 6.378164 \times 10^6$ m is the equatorial radius of the Earth; $\mathbf{g} = \mathbf{GM}/r^2 = 9.8$ m/s$^2$ free fall acceleration.

From this formula we can see, that as far velocity of *eastbound* clocks are the sum of tangential velocity of jet and tangential velocity of atmosphere at the altitude of jet flight: $\mathbf{v}_{res}^{east} = \mathbf{v}_{jet}' + \mathbf{v}_{at}$, the period of atomic clock should increase - time is slowing down. For the *westbound* clock the decreasing of actual velocity of clock: $\mathbf{v}^{west} = \mathbf{v}_{jet}' - \mathbf{v}_{at}$ should decrease the period of atomic clock and they show 'faster' time. These consequences are in total accordance with experiment of Hafele-Keating (1971).

*13.5 Interrelation between period of the Earth rotation, its radius, tangential velocity, and free fall acceleration*

If we take the local reference frame, as a center of Earth, where the tangential velocity is zero ($\mathbf{v}_{tn} = 0$; $\Delta\mathbf{v}_{tn} = 0$), then the time and frequency increments should be also zero , as it follows from both formulas (13.43 and 13.44): $[\Delta T = -\Delta\mathbf{v}]_{\mathbf{v}=\mathbf{0};\, r=\mathbf{0}} = 0$

The tangential velocity of the point on the Earth surface rotation is:

$$\mathbf{v}_{Earth}^{tn} = 2\pi r/T_{Earth} = \frac{6.28 \times 6.378164 \times 10^6 \text{ m}}{24 \times 60 \times 60 \text{ s}} = \frac{4.0 \times 10^7}{0.864 \times 10^5} = 4.63 \times 10^2 \text{ m/s} \qquad 13.45$$

where: $T_{Earth} = 24\ h = 8.64 \times 10^4 s$ is the period of the Earth rotation.

We may assume, that the atmosphere of the Earth has the same tangential velocity, i.e. rotate with Earth.

The velocity of jet as respect to this rotating atmosphere is about $\mathbf{v}_{jet} = 700 km/h = 2 \times 10^2$ m/s.

Putting value (13.45) and others in (13.43) and assuming $(\Delta r/r) = 1$, we get for corresponding increment of period, corresponding to change of the radius of rotation from zero to the earth radius:

$$T_{Earth}^{cal} \sim [\Delta T]_{r=\mathbf{const}} = 4\pi \frac{\mathbf{v}}{\mathbf{g}} \frac{\Delta r}{r} = 12.56 \frac{4.63 \times 10^2}{9.8} = 5.93 \times 10^3 \text{s} \qquad 13.46$$

*This calculated value is about 15 times less*, than real period of the Earth rotation: $T_{Earth}/T_{Earth}^{cal} \simeq 15$. This discrepancy may be a result of following factors:

1) The opposite direction of rotation of the inner volumes of the earth, for example its nuclear, as respect to its surface core, keeping the resulting angular momentum equal to zero:

$$M_{ext}\mathbf{v}_{ext}\, \mathbf{r}_{ext} + M_{in}\mathbf{v}_{in}\, \mathbf{r}_{in} = 0 \qquad 13.47$$

where $M_{ext}$; $\mathbf{v}_{ext}$; and $\Delta\mathbf{r}_{ext}$ are the averaged mass, velocity and effective radius of corresponding regions of the earth, rotation in opposite direction.

This factor may strongly increase the effective tangential velocity of the earth surface ($\mathbf{v}$) as respect to axis of its rotation in (13.46).

2) nonlinear dependence of ($\mathbf{g}$) on the distance from center of the Earth in the internal region of planet, i.e. $\mathbf{g} = \mathbf{f}(\Delta r/r)$;

3) contribution to ($\mathbf{v}$) in (13.46) the Earth velocity motion on the orbit around Sun ($30 \times 10^3$ m/s) and Solar system in the Universe ($370 \times 10^3$ m/s);

4) slowing down the frequency of the Earth rotation with time (billions of years) due to different kind of energy dissipation, like interaction with moon, etc.

Formula (13.43) points to qualitatively similar time effects, as general relativity and our formula (13.44) to the same effects, as special relativity when $\mathbf{v} \ll \mathbf{c}$.

Consequently, our Unified theory, including new approach to time problem and accepting simple Newtonian formula for gravitational force, can explain all most important experiments, which where used for confirmation of special and general relativity.

The time in our approach is a characteristic parameter of any closed system (classical and quantum) dynamics, involving not only velocity but also acceleration. In contrast to time definition, following from special relativity (13.10), the time in our Unified theory is infinitive and independent on velocity in any inertial system of particles, when $(\mathbf{dv/dt}) = \mathbf{0}$.

However, at any nonzero acceleration (dv/dt) = const > 0 the time is dependent on velocity of these objects in more complex manner, than it follows from special relativity. In fact, there are no physical systems in our expanding with acceleration Universe, formed by rotating galactics and stabilized by gravitational field, which can be considered, as perfectly inertial, i.e. where the acceleration is absent totally. This means, that conventional relativistic formula for time (13.10) is not applicable for real physical systems.

## 14. Theory of Virtual Replica of material objects and its multiplication in space and time

The theory of Virtual Replica (VR) of macroscopic objects in Bivacuum and primary VR multiplication in space and time $\mathbf{VRM(r,t)}$ is proposed.

The primary Virtual Replica of the object can be subdivided on two kinds: the surface $\mathbf{VR}_{sur}$ and the volume $\mathbf{VR}_{vol}$. Their superposition contains the total information about any material object:

$$\mathbf{VR} = \mathbf{VR}_{sur} \bowtie \mathbf{VR}_{vol} \qquad 14.1$$

The surface $\mathbf{VR}_{sur}$, like the regular optical hologram, reflects a three-dimensional (3D) shape of the object. It represents the interference pattern of modulated by the surface particles of the object Bivacuum virtual pressure waves $\mathbf{VPW}_m^{\pm}$ (the surface object waves) with basic reference virtual pressure waves of Bivacuum: $\mathbf{VPW}_{q=1}^{\pm}$.

The volume $\mathbf{VR_{vol}}$ reflects the internal properties: spatial and dynamic inhomogeneous in the volume of macroscopic object. The volume $\mathbf{VR_{vol}}$ is a result of 3D interference of modulated by the particles in the object's volume $\mathbf{VPW}_m^{\pm}$, named the volume object waves with all pervading reference virtual waves of Bivacuum $\mathbf{VPW}_{q=1}^{\pm}$. The regular optical holograms do not contain information about the internal properties of the object, like the volume Virtual Replica.

The interference of primary Virtual Replica, located in the volume of the object, with Bivacuum reference waves - provides **VR** multiplication in space $\mathbf{VRM(r)}$. To keep the energy, charge and spin conservation, it was assumed, that the mass/energy symmetry shifts of Bivacuum dipoles, involved in $\mathbf{VR}_{sur}$ and $\mathbf{VR}_{vol}$ formation and their multiplication (iteration), should compensate each other. This condition is satisfied by assumption, that the nodes and antinodes of virtual standing waves of multiple Virtual Replicas are formed by the Cooper pairs of Bivacuum fermions and antifermions:

$$\mathbf{VRM(r)} = \sum \mathbf{N}[\mathbf{BVF}^{\uparrow} \bowtie \mathbf{BVF}^{\downarrow}] \qquad 14.2$$

where **N** tends to infinity.

The absolute value of paired symmetry shifts can change from very small to very big, reflecting the properties of the object. However, this do not violate the energy and charge conservation of Bivacuum, since the mass-energy and charge symmetry shifts are equal but opposite in Bivacuum fermions and antifermions, forming virtual Cooper pairs.

The Virtual Replica spatial Multiplication can be named the Holoiteration (in Greece 'holo' means the 'whole' or 'total'). If the properties of the object vary in time, its primary

VR and its spatial multiplication also will be dependent on time: $\mathbf{VRM(r,t)}$.

If the object is changing with time, its primary $\mathbf{VR(t)}$ is also time-dependent. In this case Virtual Replica Multiplication: $\mathbf{VRM(r,t)}$ can be formally described as a result of linear superposition of primary VR of the object with corresponding amplitude of probability $\mathbf{c}_n$, like the orthogonal eigen values of wave function of elementary particle:

$$\mathbf{VRM(r,t)} = \sum c_n[\mathbf{VR}_n(\mathbf{r,t})] \qquad 14.3$$

The condition of orthogonality means that only one of the infinitive numbers of virtual replicas can be displayed at the certain time, depending on the conditions of **VR** detection.

The stability of the most probable secondary Virtual Replica as a hierarchical system of virtual standing waves could be responsible for *so-called phantom or ghost effect* of the object after its destruction or replacing. For individual elementary particles the notion of secondary virtual replica, as a result of primary VR multiplication, coincides with notion of secondary 'anchor sites', important for particle jump-way propagation in space, described above and in http://arxiv.org/abs/physics/0207027.

The Virtual Replica Multiplication $\mathbf{VRM(r,t)}$ is a process, filling all the volume around the object with secondary Virtual Replicas $\mathbf{VR(t)}$.

The multiplication of $\mathbf{VR(t)}$ is a spatially isotropic process in Bivacuum, like excitation of spherical waves, propagating with permanent velocity in all directions from the primary VR tending to conditions of virtual standing waves formation.

Each selected region of this Holoiteration interference pattern of $\mathbf{VRM(r,t)}$ contains information about the external - shape/surface and the internal - volume properties of macroscopic object changing with time. The possibility of feedback action of Virtual Replica, generated by the object, on physical properties of this object or similar kinds of objects can be tested experimentally.

### *14.1 Superposition of the Internal and Surface Virtual Replicas of the object, as the "Ether Body"*

The superposition of the time-dependent surface $\mathbf{VR}^{sur}_{n=1}(t)$ and volume $\mathbf{VR}^{vol}_{m=1}(t)$ primary virtual replicas of macroscopic object, like human being, we define, as the "Ether Body", known from the Eastern philosophy.

The overall shape of primary surface $\mathbf{VR}^{sur}_{n=1}(t)$ contain the information about shape of the object, like do the optical hologram.

For the other hand, the primary volume $\mathbf{VR}^{vol}_{m=1}(t)$ should reflect also the structural and dynamic properties of the object's internal components.

The time-dependent *ether body* can be presented as a result of superposition of microscopic virtual replicas of the atoms, forming the surface and the volume of macroscopic object:

$$\textbf{Ether Body}(\mathbf{t}) \equiv \sum \mathbf{VR}^{res}(\mathbf{t}) = \sum^{\mathbf{N_{sur}}} \mathbf{VR}^{sur}_n(\mathbf{t}) + \sum^{\mathbf{M}_{vol}} \mathbf{VR}^{vol}_m(\mathbf{t}) \qquad 14.4$$

Stability of hierarchic system of virtual standing waves, forming Ether Body of macroscopic object, like a hierarchical system of *curls* in superfluid $^4\mathbf{He}$, could be responsible for so-called *"phantom effect"* of this object.

### *14.2 The "Astral" and "Mental" bodies definitions, based on Unified theory*

It is assumed in our approach, that:

**a**) the Astral Body: $\mathbf{VRM}^{VPW}_{ast}(\mathbf{r,t})$ is produced by the iteration of time-dependent Ether body $\sum \mathbf{VR(t)}$ in space as a result of its interference with the reference virtual pressure waves ($\mathbf{VPW}^+ \bowtie \mathbf{VPW}^-$);

**b**) the Mental Body: $\mathbf{VRM}_{men}^{VirSW}(\mathbf{r},\mathbf{t})$ is produced by the the iteration of the Ether body in space as a result of its interference with the reference virtual spin waves ($\mathbf{VirSW}^+ \bowtie \mathbf{VirSW}^-$).

The dielectric permittivity ($\varepsilon_0$) and permeability ($\mu_0$) of Bivacuum in the volume of the Astral bodies (astral phantoms) may differ from their averaged values in Bivacuum because of small charge symmetry shift $\Delta e = |e_+ - e_-| > 0$ in separated and virtual Cooper pairs of Bivacuum fermions $[\mathbf{BVF}^{\uparrow} \bowtie \mathbf{BVF}^{\downarrow}]^{sur,vol}$, forming the Astral body: $\mathbf{VRM}_{ast}^{VPW}(\mathbf{r},\mathbf{t})$.

Consequently, the probability of atoms and molecules excitation and ionization (dependent on Coulomb interaction between electrons and nuclear), as a result of their thermal collisions with excessive kinetic energy, may be higher in the volumes of the Astral bodies, than outside of them.

This may explain a shining of some $\mathbf{VRM}_{ast}^{VPW}(\mathbf{r},\mathbf{t})$ in IR and visible range, representing phantoms (ghosts) and their photos and spectrograms.

The sensitivity of Kirlian effect or Gas Discharge Visualization (GDV) to the certain processes in human body can be explained by specific properties of the Ether and Astral bodies, changing the probability of the air molecules excitation/ionization in the process of gas discharge visualization (GDV).

The informational properties of the mental body: $\mathbf{VRM}_{ast}^{VirSW}(\mathbf{r},\mathbf{t})$ is a result of nonlocal properties of Virtual Spin Waves ($\mathbf{VirSW}^{\pm}$), acting in symmetric Bivacuum instantly without light velocity limitation.

The Mental Body of human being is dependent on the neuronal activity of the brain and nerve system. It follows from our theory of elementary act of consciousness, that the librational dynamics of water clusters in microtubules in state of mesoscopic and macroscopic Bose condensation (mBC) can contribute to the Ether, Astral and Mental bodies properties.

The Hierarchical superposition of huge number of Astral and Mental Bodies of all human population on the Earth can be responsible for Global Informational Field origination, named Noosphere.

The Astral and Mental bodies are interrelated with the Ether body. This provide a possibility of the exchange interaction and feedback reaction between all three virtual bodies of macroscopic object: Ether, Astral and Mental.

### *14.2 Mind - Matter Interaction and Virtual Replicas Imprinting*

The perturbation of the Ether body of the object - Receptor by the Astral or Mental body of the other object - Sender can be imprinted in properties of physical properties of Receptor as stable structural perturbations.

The stability of such kind of informational 'taping' is determined by specific properties of material of Receptor. These imprinting properties are dependent on the dimensions, stability and interaction of coherent molecular clusters of this material in state of mesoscopic Bose condensation (mBC) with VR. The mBC are revealed and evaluated in our Hierarchic theory of condensed matter (Kaivarainen, 1995, 2000, 2008). The VR is composed from dynamic system of Bivacuum dipoles with small symmetry shift, providing small uncompensated charge. Consequently, these all matter pervading tiny dipoles with Compton radius of muon or tauon - may partly screen Coulomb interaction between charged elementary particles, like electrons, protons and other ions of matter.

The ice, water and aqueous systems are very good for imprinting of virtual information and energy via influence on strength of hydrogen bonds: [$O^-$ - $H^+$....$O^-$ - H], depending on electric polarization of water molecules: H-O-H.

The introduced *VRM(r,t)* and *Virtual Guides* (see next section) can be responsible for turning the mesoscopic Bose condensation to non-continuous macroscopic BC and macroscopic entanglement.

The perturbations of ’sensitive’ to imprinting rigid materials, like ice or crystals containing larger domains of mBC, stabilized by H-bonds, ion-dipole and ion-ion interaction, should have much longer life-time or ’memory’ than liquids. This consequence of our theory can be verified experimentally.

For example, the ***Ghost phenomena*** can be explained by reproducing of such imprinted in walls, cells and floor information, realized via distant virtual replica multiplication ($VRM^{dis}$).

The reproduction of VR from imprinted in condensed matter $VRM^{dis}$ is a process, similar to treatment of regular hologram by the reference waves.

In the case of ’Ghost’ the reference waves are presented by the basic $\mathbf{VPW}^{\pm}_{q=1}$ and $\mathbf{VirSW}^{\pm 1/2}_{q=1}$ propagating in Bivacuum. This reproduction - replay of imprinted VR in sensitive materials can be modulated by superposition of Virtual replicas of cosmic objects, like Earth, Moon and Sun.

The *Mental body* formation in living organisms and humans in accordance to our approach (Kaivarainen, 2006, 2008) is related to equilibrium shift of [assembly ⇌ disassembly] of coherent water clusters in microtubules (MT) of the neurons (librational effectons), accompanied the elementary acts (cycles) of consciousness in *processes* of meditation, intention and braining.

The corresponding coherent alternations of kinetic energy and momentum of water molecules in MT can be transmitted from Sender to remote Receiver via nonlocal virtual spin-momentum-energy guides $\mathbf{VirG}_{SME}$ (**S** <=> **R**), described in next chapter.

In complex process of Psi phenomena, the important stages, starting by Sender (psychic) are:

**1**) *’target searching’* by the [*mental body*] of psychic, then formation of $\mathbf{VirG}_{SME}(\mathbf{S} \Leftrightarrow \mathbf{R})^i$;

**2**) *activation of psychic’s [astral body]* by its [*ether body*].

The latter can be interrelated with specific processes of physical body of psychic, like dynamics of water in microtubules of neurons ensembles, realizing elementary acts of perception and consciousness, in accordance to our model (Kaivarainen, 2000; 2005).

The possible mechanism of entanglement between remote microscopic and macroscopic objects, connected by the bundles of Virtual Guides of spin, momentum and energy will be described below.

### *14.3 Virtual Guides in Bivacuum, providing remote macroscopic entanglement and nonlocal interaction*

The Virtual Guides of spin, momentum and energy, connecting pairs of remote fermions of opposite spins defined as Sender (**S**) and Receiver (**R**), represent virtual filaments, formed by Cooper pairs of Bivacuum fermions and antifermions $\mathbf{N}\left(\mathbf{BVF}^{\uparrow} \bowtie \mathbf{BVF}^{\downarrow}\right)$ or pairs of Bivacuum bosons with opposite polarization $\mathbf{N}[\mathbf{BVB}^{-} \bowtie \mathbf{BVB}^{+}]$ interacting with each other side-by-side (⇔=⇔):

$$\frac{<[\mathbf{F}_{\downarrow}^{+} \bowtie \mathbf{F}_{\uparrow}^{-}]_C + (\mathbf{F}_{\downarrow}^{-})_W>}{<[\mathbf{F}_{\downarrow}^{+} \bowtie \mathbf{F}_{\uparrow}^{-}]_W + (\mathbf{F}_{\downarrow}^{-})_C>}[\mathbf{S}] \rightarrowtail \left[\begin{array}{c} \mathbf{VPW}^{-} \bowtie \mathbf{VPW}^{+} \\ \Longleftarrow\Leftrightarrow\Leftrightarrow=\Leftrightarrow\Leftrightarrow=\Leftrightarrow\Leftrightarrow=\Leftrightarrow\Leftrightarrow=\Leftrightarrow\Leftrightarrow=\Leftrightarrow\Leftrightarrow\Longrightarrow \\ \mathbf{N}\left(\mathbf{BVF}^{\uparrow} \bowtie \mathbf{BVF}^{\downarrow}\right) \ or\ \mathbf{N}[\mathbf{BVB}^{-} \bowtie \mathbf{BVB}^{+}] \end{array}\right] \rightarrowtail \frac{<(\mathbf{F}_{\uparrow}^{-})_C + [\mathbf{F}_{\downarrow}^{-} \bowtie \mathbf{F}_{\uparrow}^{+}]_W>}{<(\mathbf{F}_{\uparrow}^{-})_W + [\mathbf{F}_{\downarrow}^{-} \bowtie \mathbf{F}_{\uparrow}^{+}]_C>}[\mathbf{R}]$$

The radiuses and gaps between tori and antitori of virtual Cooper pairs $(\mathbf{BVF}^{\uparrow} \bowtie \mathbf{BVF}^{\downarrow})$, which assembly virtual guides **VirG** are self-adapted to Compton radius of elementary particles, which they are connecting.

The longitudinal momentum of pairs $[\mathbf{BVF}^{\uparrow} \bowtie \mathbf{BVF}^{\downarrow}]^{s}_{S=0}$ along the main axe of virtual filaments is close to zero ($\mathbf{p} \to \mathbf{0}$) and corresponding de Broglie wave length tends to infinity: $\boldsymbol{\lambda}_B = \mathbf{h}/\mathbf{p} \to \infty$, providing conditions for quasi - one - dimensional virtual Bose condensate. The notion of time in such conditions, when external velocity is permanent and equal to zero: $\mathbf{v} = \mathbf{0} = \mathbf{const}$ is absent or uncertain, as it follows from our time theory in chapter 13:

$$\mathbf{t} = \left[ -\frac{\vec{\mathbf{v}}}{\vec{\mathbf{a}}} \frac{1-(\mathbf{v}/\mathbf{c})^2}{2-(\mathbf{v}/\mathbf{c})^2} \right]_{x,y,z} = \frac{0}{0}$$

Two remote coherent triplets - elementary particles, like: [electron - electron], [proton - proton], [neutron-neutron] or [photon-photon] of Sender and Receiver with similar frequency of $[\mathbf{C} \rightleftharpoons \mathbf{W}]_{e,p}$ pulsation and opposite phase (spins) can be connected by Virtual guides: $\mathbf{VirG}_{SME}(\mathbf{S} <\!\!=\!\!\Leftrightarrow\!\!=\!\!> \mathbf{R})$ of spin (S), momentum (M) and energy (E).

The spin - information (qubits), momentum and kinetic energy instant exchange via such Virtual Guide between [S] and [R] is possible, as a consequence of coherent pulsation of the gap between torus and antitorus of Bivacuum dipoles forming $\mathbf{VirG}^{i}_{SME}(\mathbf{S} <\!\!=\!\!> \mathbf{R})$ with the same frequency as that of $[\mathbf{C} \rightleftharpoons \mathbf{W}]_{e,p}$ pulsation of the entangled elementary particles.

The Resonance Nonlocal Interaction Between
Distant Electrons of Opposite Spins and Phase of [C ⇌ W] Pulsation
via Virtual Spin - Momentum - Energy Guide $[\mathrm{VirG_{SME}}]$

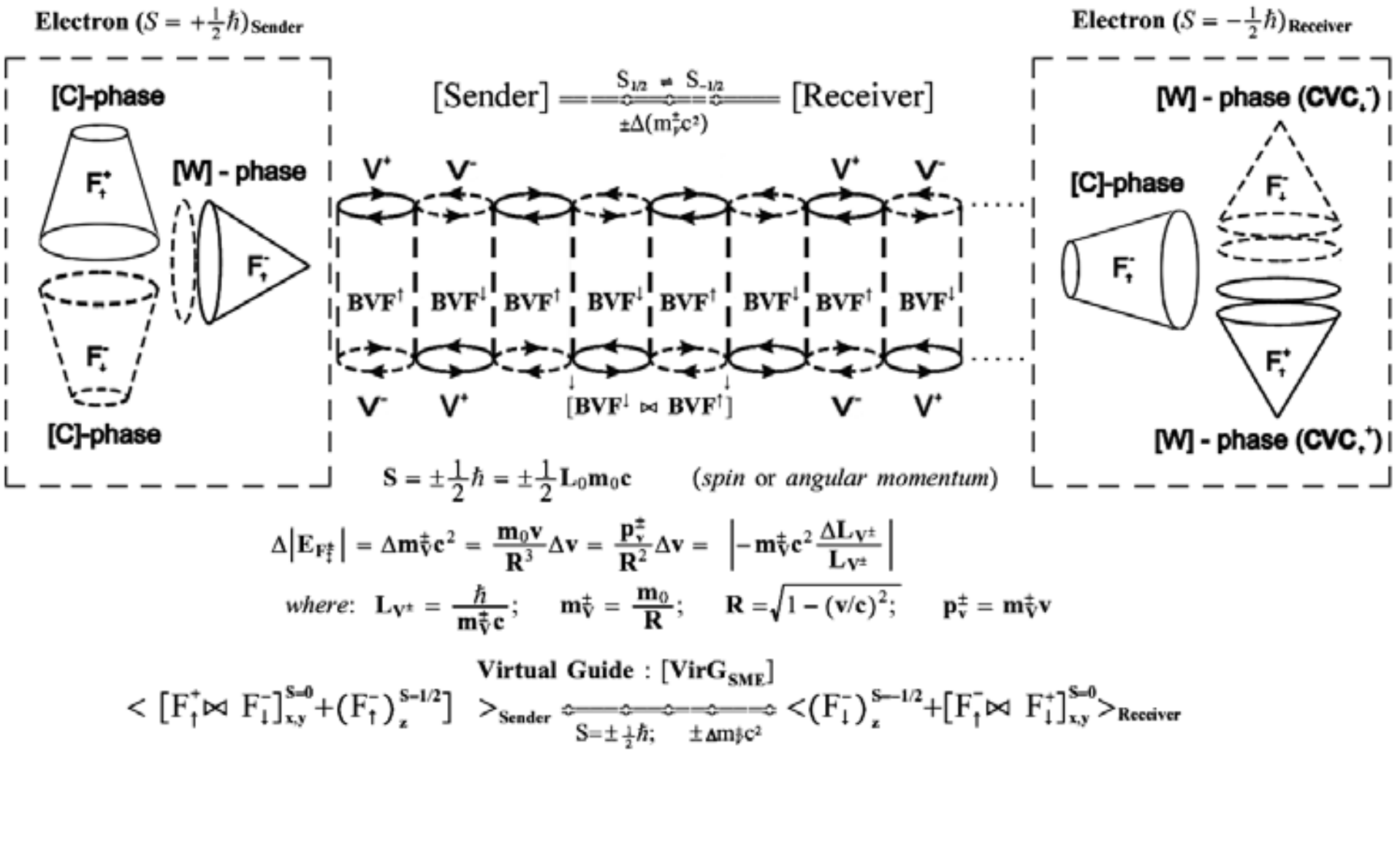

**Fig. 9**. The mechanism of nonlocal Bivacuum mediated exchange interaction (entanglement) between two distant unpaired sub-elementary fermions of 'tuned' elementary triplets (particles) of the opposite spins $< [\mathbf{F}_\uparrow^+ \bowtie \mathbf{F}_\downarrow^-] + \mathbf{F}_\uparrow^- >^{i}_{\mathbf{Sender}}$ and $< [\mathbf{F}_\downarrow^+ \bowtie \mathbf{F}_\uparrow^-] + \mathbf{F}_\downarrow^- >^{i}_{\mathbf{Receiver}}$, with close frequency of $[\mathbf{C} \rightleftharpoons \mathbf{W}]$ pulsation and momentum (i.e. de Broglie wave length $\lambda_\mathbf{B} = \mathbf{h}/\mathbf{m}_V^{+}\mathbf{v}$) of particles. The tunnelling of momentum and energy increments: $\boldsymbol{\Delta}|\mathbf{m}_\mathbf{V}^{\pm}\mathbf{c}^2| \sim \boldsymbol{\Delta}|\mathbf{VirP}^{+}| \pm \boldsymbol{\Delta}|\mathbf{VirP}^{-}|$ from Sender to Receiver and vice-verse via Virtual spin-momentum-energy Guide $\mathbf{VirG_{SME}}$ is accompanied by instantaneous pulsation of the gap between tori and antitori ($\Delta\mathbf{d}$), interrelated with simultaneous radius pulsation: $\Delta\mathbf{d}/\boldsymbol{\Delta}\mathbf{L}_\mathbf{V}^{\pm} = \boldsymbol{\pi}$ of Bivacuum dipoles, assembling $\mathbf{VirG}^{i}_{\mathbf{SME}}$.

When the unpaired sub-elementary of fermion - Sender: $\mathbf{F}_\uparrow^- >^{i}_{\mathbf{Sender}}$ is in the waves

[W] - phase and that Receiver: $\mathbf{F}_{\downarrow}^{-} >^{i}_{\mathbf{Receiver}}$ in [C] phase, the *momentum and energy* is transmitted from [S] to [R]. In the opposite phase, when the Receiver is in the [W] phase the momentum and energy is 'teleportated' from [R] to [S].

The jump-way decreasing of the gap between tori ($NV^+$) and antitori ($NV^-$) of Bivacuum dipoles, composing **VirG**, in the process of teleportation, is stimulated by [W] phase and virtual pressure jump: $\Delta|\mathbf{m}_{\mathbf{V}}^{\pm}\mathbf{c}^2| \sim \Delta|\mathbf{VirP}^+| \pm \Delta|\mathbf{VirP}^-|$ of Sender or Receiver, alternatively.

The nonlocal *spin state* exchange between [S] and [R] can be the consequence of the violation of symmetry of positive and negative energy distribution in the process of $\mathbf{VirG}^{i}_{\mathbf{SME}}$ dipoles pulsation: $\Delta\mathbf{L}_{\mathbf{V}}^{+} \neq \Delta\mathbf{L}_{\mathbf{V}}^{-}$, when $[\mathbf{C} \rightleftharpoons \mathbf{W}]$ pulsation of Sender and Receiver triplets are in-phase, i.e. in the same spin state.

### *14.4 The role of tuning force ($\mathbf{F}_{\mathbf{VPW}^{\pm}}$) of virtual pressure waves $\mathbf{VPW}_q^{\pm}$ of Bivacuum in entanglement*

The tuning between two similar elementary particles: 'sender (S)' and 'receiver (R)' via $\mathbf{VirG}_{SME}(\mathbf{S} \Leftarrow\Rightarrow \mathbf{R})^i$ may be qualitatively described, using well known model of *damped harmonic oscillators,* interacting with all-pervading virtual pressure waves ($\mathbf{VPW}_{q=1}^{\pm}$) of Bivacuum with fundamental frequency $\boldsymbol{\omega}_0 = \mathbf{m}_0\mathbf{c}^2/\hbar$. The criteria of tuning - synchronization of [S] and [R] is the equality of the amplitude probability of resonant energy exchange of Sender and Receiver with virtual pressure waves ($\mathbf{VPW}_{q=1}^{\pm}$): $\mathbf{A}^{S}_{C\rightleftharpoons W} = \mathbf{A}^{R}_{C\rightleftharpoons W}$, resulting from minimization of frequency difference $(\boldsymbol{\omega}_S - \boldsymbol{\omega}_0) \rightarrow 0$ and $(\boldsymbol{\omega}_R - \boldsymbol{\omega}_0) \rightarrow 0$:

$$\mathbf{A}^{S}_{C\rightleftharpoons W} \sim \left[ \frac{1}{(\mathbf{m}_V^+)_S} \frac{\mathbf{F}_{\mathbf{VPW}^{\pm}}}{(\boldsymbol{\omega}_S^2 - \boldsymbol{\omega}_0^2) + \mathrm{Im}\,\boldsymbol{\gamma}\boldsymbol{\omega}_S} \right] \qquad 14.5$$

$$[\mathbf{A}^{R}_{C\rightleftharpoons W}]_{x,y,z} \sim \left[ \frac{1}{(\mathbf{m}_V^+)_R} \frac{\mathbf{F}_{\mathbf{VPW}^{\pm}}}{(\boldsymbol{\omega}_R^2 - \boldsymbol{\omega}_0^2) + \mathrm{Im}\,\boldsymbol{\gamma}\boldsymbol{\omega}_R} \right] \qquad 14.5a$$

where the frequencies of $\mathbf{C} \rightleftharpoons \mathbf{W}$ pulsation of particles of Sender ($\boldsymbol{\omega}_S$) and Receiver ($\boldsymbol{\omega}_R$) are:

$$\boldsymbol{\omega}_R = \boldsymbol{\omega}_{\mathbf{C}\rightleftharpoons\mathbf{W}} = \boldsymbol{\omega}_0^{in} + (\boldsymbol{\omega}_B^{ext})_R \qquad 14.6$$

$$\boldsymbol{\omega}_S = \boldsymbol{\omega}_{\mathbf{C}\rightleftharpoons\mathbf{W}} = \boldsymbol{\omega}_0^{in} + (\boldsymbol{\omega}_B^{ext})_S \qquad 14.6a$$

$\boldsymbol{\gamma}$ is a damping coefficient due to *decoherence effects*, generated by local fluctuations of Bivacuum deteriorating the phase/spin transmission via $\mathbf{VirG}_{SME}$; $(\mathbf{m}_V^+)_{S,R}$ are the actual mass of (S) and (R); $[\mathbf{F}_{\mathbf{VPW}}]$ is a *tuning force of virtual pressure waves* $\mathbf{VPW}^{\pm}$ *of Bivacuum with tuning energy* $\mathbf{E}_{VPW} = \mathbf{q}\mathbf{m}_0\mathbf{c}^2$ *and wave length* $\mathbf{L}_{VPW} = \hbar/\mathbf{m}_0\mathbf{c}$

$$\mathbf{F}_{\mathbf{VPW}_q^{\pm}} = \frac{\mathbf{E}_{VPW_q}}{\mathbf{L}_{VPW_q}} = \frac{\mathbf{q}}{\hbar}\mathbf{m}_0^2\mathbf{c}^3 \qquad 14.7$$

The most probable Tuning force has a minimum, corresponding to $\mathbf{q} = \mathbf{j} - \mathbf{k} = \mathbf{1}$.

The influence of *virtual pressure force* ($\mathbf{F}_{\mathbf{VPW}_q}$) stimulates the synchronization of [S] and [R] pulsations, i.e. $\omega_R \rightarrow \omega_S \rightarrow \omega_0$. This fundamental frequency $\boldsymbol{\omega}_0 = \mathbf{m}_0\mathbf{c}^2/\hbar$ is the same in any space volume, including those of Sender and Receiver.

The $\mathbf{VirG}_{SME}$ represent quasi **1D** macroscopic virtual Bose condensate with a configuration of single microtubules, formed by Bivacuum bosons ($\mathbf{BVB}^{\pm}$) or with configuration of double microtubules, composed from Cooper pairs as described in previous section.

The effectiveness of entanglement between number of similar elementary particles of Sender and Receiver - free or in composition of atoms and molecules via highly anisotropic nonlocal virtual guide bundles

$$\left[\mathbf{N}(\mathbf{t},\mathbf{r}) \times \sum^{\mathbf{n}} \mathbf{VirG}_{SME}\,(\mathbf{S} <=> \mathbf{R})\right]^{i}_{x,y,z} \qquad 14.7a$$

is dependent on synchronization of $[\mathbf{C} \rightleftharpoons \mathbf{W}]$ pulsation frequency of these particles.

In this expression $(\mathbf{n})$ is a number of pairs of similar tuned elementary particles (protons, neutrons and electrons) in atoms/molecules of **S** and **R**; $\mathbf{N}(\mathbf{t},\mathbf{r})$ is a number of coherent atoms/molecules in the coherent molecular clusters - mesoscopic BC (Kaivarainen, 2001; 2004).

The 'tuning' of particles phase and frequency pulsation occur under the forced resonance exchange interaction between virtual pressure waves $\mathbf{VPW}_q^+$; $\mathbf{VPW}_q^-$ and pulsing particles.

The mechanism proposed may explain the experimentally confirmed nonlocal interaction between coherent elementary particles (Aspect and Gragier, 1983), atoms and their remote coherent clusters.

Our theory predicts that the same mechanism, involving nonlocal bundles $[\mathbf{N}(\mathbf{t},\mathbf{r}) \times \sum \mathbf{VirG}_{SME}\,(\mathbf{S} <=> \mathbf{R})]^{i}_{x,y,z}$, may provide the entanglement between macroscopic systems, including biological ones.

*14.5 The vortical filaments in superfluids, as the analogs of virtual guides of Bivacuum*

When the rotation velocity of a cylindrical vessel containing **He II** becomes high enough, then the emergency of so-called vortex filaments becomes thermodynamically favorable. The filament is formed by the superfluid component of **He II** in such a way that their momentum of movement decreases the total energy of **He II** in a rotating vessel. The shape of filaments in this case is like a straight rod and their *thickness* is of the order of atom's dimensions, increasing with lowering the temperature at $T < T_\lambda$.

Vortex filaments are continuous. They may be closed or limited within the boundaries of vessel.

The hydrodynamics of normal and superfluid components of He II in container of radius $(\mathbf{r})$, rotating with angular frequency $\Omega$ are characterized by two velocities, correspondingly

$$\mathbf{v}_n = \Omega\,\mathbf{r} \qquad 14.8$$

$$\mathbf{v}_{sf} = \frac{\hbar}{\mathbf{m}}\nabla\phi = N\frac{\hbar}{\mathbf{m\,r}} \qquad 14.8a$$

where $\nabla\phi \sim k_{sf} = 1/\mathbf{L}_{sf}$ is a phase of Bose condensate wave function: $\boldsymbol{\Psi} = \boldsymbol{\rho}_s^{1/2} \times \mathbf{e}^{i\phi}$ ($\boldsymbol{\rho}_s$ is a density of superfluid component); $N$ is a number of rectilinear vortex lines.

The motion of superfluid component is potential, as far its velocity $(\mathbf{v}_{sf})$ is determined by eq. 14.8a and:

$$rot\,\mathbf{v}_{sf} = 0 \qquad 14.8b$$

The values of velocity of circulation of filaments are determined (Landau, 1941) as follows:

$$\oint \mathbf{v}_{sf} dl = 2\pi r\, \mathbf{v}_{sf} = 2\pi\kappa = \frac{\hbar}{m}\Delta\Phi \qquad 14.9$$

where: $\mathbf{\Delta\Phi} = \mathbf{n}\ \mathbf{2\pi}$ is a phase change as a result of circulation, $n = 1,2,3\ldots$ is the integer number.

and

$$\mathbf{v}_{sf} = \kappa / r \qquad 14.9a$$

Increasing the radius of circulation (r) leads to decreased circulation velocity ($\mathbf{v}_{sf}$).

Comparing (14.9a) and (14.9) gives:

$$\kappa = n\frac{\hbar}{m} \qquad 14.10$$

It has been shown that only vortical structures with $n = 1$ are thermodynamically stable.

Taking this into account, we have from (14.9a) and (14.10):

$$r = n\frac{\hbar}{\mathbf{m}\mathbf{v}_{sf}} \qquad 14.11$$

An increase in the angle frequency of rotation of the cylinder containing **HeII** results in the increased density distribution of vortex filaments on the cross-section of the cylinder.

As a result of interaction between the filament and the normal component of **HeII**, the filaments move in the rotating cylinder with normal liquid.

The flow of **He II** through the capillaries also can be accompanied by appearance of vortex filaments.

In ring-shaped vessels the circulation of closed vortex filaments is stable. Stability is related to the quantum pattern of circulation change (eqs. 14.9 and 14.10).

Let us consider now the phenomena of superfluidity in **He II** in the framework of our hierarchic concept (Kaivarainen, 2001).

### *14.6 Two stages of Virtual Guides of spin, momentum and energy ($VirG^i_{\mathbf{S,M,E}}$) formation between remote elementary particles*

The entanglement between two remote *similar* elementary particles (electrons, protons, neutrons, photons) of opposite spins, named [Sender (S)] and [Receiver (R)], is revealed in many experiments, starting from Aspect and Grangier (1983). The formal explanation of entanglement between remote particles is their unification/description by the same wave function. For example, this takes a place at superconductivity and superfluidity, i.e. in conditions of macroscopic Bose condensation. However, the mediator and concrete mechanism of entanglement is still obscure.

*In accordance to our theory, the entanglement between similar elementary particles can be realized via Virtual Guides of spin, momentum and energy ($VirG^i_{\mathbf{S,M,E}}$).*

We propose here two stages of Virtual Guides formation:

1. The tuning of the frequency and phase of $[C \rightleftharpoons W]$ pulsation of remote elementary particles, like photons electrons, protons, neutrons in isolated state or in composition of atoms and molecules. The 'tuning' occur under the action of the basic all-pervading Bivacuum virtual pressure waves: $\mathrm{VPW}^+_{\mathbf{q=1}}$ and $\mathrm{VPW}^-_{\mathbf{q=1}}$;

2. This stage stimulate self-assembly of Cooper pairs of Bivacuum fermions ($BVF^- \bowtie BVF^+$) to the $VirG_{SME}$ ($S \Leftrightarrow R$) filaments with properties of quasi one-dimensional virtual Bose condensate.

### *14.7 The mechanism of momentum and energy transmission between*

*similar elementary particles of Sender and Receiver via* $VirG_{\mathbf{SME}}(S <=> R)^i$

The increments or decrements of momentum $\pm\Delta\mathbf{p} = \mathbf{\Delta}(\mathbf{m}_V^+\mathbf{v})_{tr,lb}$ and kinetic $(\pm\Delta\mathbf{T}_k)_{tr,lb}$ energy transmission from [S] to [R] of *coherent elementary particles* is determined by the translational and librational velocity variation ($\mathbf{\Delta v}$) of nucleons of Sender. This means, that energy/momentum transition from [S] to [R] is possible, if they are in nonequilibrium state.

The variation of kinetic energy of atomic nuclei under external force application, induces nonequilibrium in a system ($\mathbf{S} + \mathbf{R}$) and decoherence of $[\mathbf{C} \rightleftharpoons \mathbf{W}]$ pulsation of protons and neutrons of [S] and [R]. The nonlocal energy transmission from [S] to [R] is possible, if the decoherence is not big enough for disassembly of the virtual microtubules and their bundles. The bundles of atomic Virtual Guides are named the *Entanglement channels:*

$$\mathbf{Entanglement\ channels} = \left[\mathbf{N}(\mathbf{t},\mathbf{r}) \times \sum^{\mathbf{n}} \mathbf{VirG}_{SME}(\mathbf{S} <=> \mathbf{R})\right]^i_{x,y,z} \qquad 14.12$$

where: $\mathbf{N}(\mathbf{t},\mathbf{r})$ is a number of virtual guides in the bundle, equal to number of coherent atoms/molecules in state of mesoscopic Bose condensation (mBC) in volume of remote Sender and Receiver; $\mathbf{n}$ - is a number of coherent elementary pairs of similar fermions (e, p, n) of opposite spins in atom/molecules, composing distant clusters in state of mesoscopic Bose condensation (mBC).

The instantaneous energy flux via each of $(\mathbf{VirG}_{SME})^i$ is mediated by pulsation of energy and radii of tori ($\mathbf{V}^+$) and antitori ($\mathbf{V}^-$) of Bivacuum bosons: $\mathbf{BVB}^+ = [\mathbf{V}^+\updownarrow \mathbf{V}^-]$ in single Virtual Guides or Bivacuum fermions of Cooper pairs $[\mathbf{BVF}^\uparrow \bowtie \mathbf{BVF}^\downarrow]^i$ in the case of double Virtual Guides.

The corresponding energy increments of the actual torus and complementary antitorus of $\mathbf{BVB}^\pm$, forming $(\mathbf{VirG}_{SME})^i$, are directly related to increments of Sender particle external velocity ($\Delta\mathbf{v}$):

$$\Delta\mathbf{E}_{V^+} = +\Delta\mathbf{m}_V^+c^2 = \left(+\frac{\mathbf{p}^+}{\mathbf{R}^2}(\Delta\mathbf{v})_{\mathbf{F}_\uparrow^+}^{[\mathbf{F}_\uparrow^+ \bowtie \mathbf{F}_\downarrow^-]} = \mathbf{m}_V^+\mathbf{c}^2\frac{\Delta\mathbf{L}_{V^+}}{\mathbf{L}_{V^+}} = \mathbf{m}_V^+\mathbf{c}^2\frac{\Delta\mathbf{d}_{V^+\Leftrightarrow V^-}}{\mathbf{d}_{V^+\Leftrightarrow V^-}}\right) \quad \text{actual} \qquad 14.$$

$$\Delta\mathbf{E}_{V^-} = -\Delta\mathbf{m}_V^-c^2 = \left(-\frac{\mathbf{p}^-}{\mathbf{R}^2}(\Delta\mathbf{v})_{\mathbf{F}_\uparrow^+}^{[\mathbf{F}_\uparrow^+ \bowtie \mathbf{F}_\downarrow^-]} = -\mathbf{m}_V^-\mathbf{c}^2\frac{\Delta\mathbf{L}_{V^-}}{\mathbf{L}_{V^-}} = -\mathbf{m}_V^-\mathbf{c}^2\frac{\Delta\mathbf{d}_{V^+\Leftrightarrow V^-}}{\mathbf{d}_{V^+\Leftrightarrow V^-}}\right) \quad \text{complementary} \qquad 14.$$

where: $\mathbf{p}^+ = \mathbf{m}_V^+\mathbf{v}$; $\mathbf{p}^- = \mathbf{m}_V^-\mathbf{v}$ are the actual and complementary momenta; $\mathbf{L}_{V^+} = \hbar/\mathbf{m}_V^+\mathbf{c}$ and $\mathbf{L}_{V^-} = \hbar/\mathbf{m}_V^-\mathbf{c}$ are the radii of torus and antitorus of $\mathbf{BVB}^\pm = [\mathbf{V}^+ \Updownarrow \mathbf{V}^-]$, forming $\mathbf{VirG}_{SME}(\mathbf{S} <=> \mathbf{R})^i$.

The nonlocal energy exchange between [S] and [R] is accompanied by the *instant pulsation of radii* of tori ($V^+$) and antitori ($V^-$) of Cooper pairs of Bivacuum fermions and $BVB^\pm$, accompanied by corresponding pulsation $|\Delta\mathbf{L}_{V^\pm}/\mathbf{L}_{V^\pm}|$ of the whole virtual microtubule $\mathbf{VirG}_{SME}$ (Fig.9).

*The nonequilibrium state* of similar elementary particles of [S] and [R] with similar rest mass ($\mathbf{m}_0^S\mathbf{c}^2 = \mathbf{m}_0^R\mathbf{c}^2$), connected by $\mathbf{VirG}_{S,M,E}$, means the difference in their external energies and frequency of the empirical de Broglie waves and that of $[\mathbf{C} \rightleftharpoons \mathbf{W}]$ pulsation. The consequence of this difference are beats between states of [S] and [R], equal to frequency of $\mathbf{VirG}_{SME}$ radius pulsation. Using eqs. 14.13 and 14.13a, we get:

$$\Delta\mathbf{v}_{\mathbf{VirG}}^{S,R} = \mathbf{v}_{\mathbf{C}\rightleftharpoons\mathbf{W}}^S - \mathbf{v}_{\mathbf{C}\rightleftharpoons\mathbf{W}}^R == \frac{1}{h}\left[\frac{h^2}{(\mathbf{m}_V^+\boldsymbol{\lambda}_B^2)^S} - \frac{h^2}{(\mathbf{m}_V^+\boldsymbol{\lambda}_B^2)^R}\right] \qquad 14.14$$

The beats between the total frequencies of [S] and [R] states (electrons, protons or neutrons), connected by $\mathbf{VirG}_{S,M,E}$ and different excitation states $(j-k)$ of $[\mathbf{BVF}^{\uparrow} \bowtie \mathbf{BVF}^{\downarrow}]_{j-k}$ are accompanied by *emission* $\rightleftharpoons$ *absorption* of positive and negative virtual pressure waves: $\mathbf{VPW}^{+}$ and $\mathbf{VPW}^{-}$, generating positive and negative virtual pressure: $\mathbf{VirP}^{+}$ and $\mathbf{VirP}^{-}$.

The difference between actual energies of elementary particles of Sender and Receiver can be expressed via these virtual pressures, using eq.14.13 and 14.14, as:

$$\mathbf{E}^{S}_{tot} - \mathbf{E}^{R}_{tot} = h\Delta\mathbf{v}^{S,R}_{\mathbf{VirG}} = (\mathbf{m}^{+}_{V} - \mathbf{m}^{+}_{V})^{S,R}\mathbf{c}^{2} + (\mathbf{m}^{-}_{V} - \mathbf{m}^{-}_{V})^{S,R}\mathbf{c}^{2} \sim \qquad 14.15a$$

$$\sim (\mathbf{VirP}^{\pm} + \mathbf{VirP}^{\pm})^{S,R} \qquad 14.15.b$$

If the temperature or kinetic energy of [S] is higher, than that of [R]: $\mathbf{T}_{S} > \mathbf{T}_{R}$, then $\Delta\mathbf{v}^{S,R}_{\mathbf{VirG}} > 0$ and the *direction* of momentum and energy flux, mediated by positive and negative virtual pressure of subquantum particles and antiparticles: $\Delta\mathbf{VirP}^{+}$ and $\Delta\mathbf{VirP}^{-}$, is from [S] *to* [R].

The length of $\mathbf{VirG}_{SME}$(**S** <=> **R**), connecting tuned elementary particles, also can vary in the process of [S] *and* [R] interaction because of immediate self-assembly of Bivacuum dipoles into virtual guides.

### *14.8 The mechanism of spin exchange between tuned particles of Sender and Receiver via VirG*SME

Most effectively the proposed mechanism of spin (information), momentum and energy exchange can work between Sender and Receiver, containing coherent molecular clusters with dimensions of 3D standing de Broglie waves of molecules in state of mesoscopic Bose condensate (mBC) (Kaivarainen, 2001, 2008).

*The nonlocal spin/qubit exchange* between [S] and [R] via single or double $\mathbf{VirG}^{i}_{SME}(\mathbf{S} <=> \mathbf{R})^{i}$ does not need the radius pulsation, but only the instantaneous polarization change of Bivacuum bosons $(\mathbf{BVB}^{+} \rightleftharpoons \mathbf{BVB}^{-})^{i}$ of single virtual guides or instant spin state exchange of two Bivacuum fermions, forming virtual Cooper pairs via intermediate stage $[\mathbf{BVB}^{+} \bowtie \mathbf{BVB}^{-}]^{i}$ in virtual guide (Fig.9):

$$[\mathbf{BVF}^{\uparrow} \bowtie \mathbf{BVF}^{\downarrow}]^{i} \rightleftharpoons [\mathbf{BVB}^{+} \bowtie \mathbf{BVB}^{-}]^{i} \rightleftharpoons [\mathbf{BVF}^{\downarrow} \bowtie \mathbf{BVF}^{\uparrow}]^{i} \qquad 14.16$$

The instantaneous spin state/information exchange frequency is determined by frequency of spin change of fermion of Sender, accompanied by counterphase spin state change of fermion of Receiver.

The described above mechanisms of nonlocal/instant transmission of spin/information, momentum and energy between coherent clusters of elementary particles and atoms of Sender and Receiver, connected by Virtual Guides, may describe a lot of unconventional experimental results, like Kozyrev, Tiller ones and lot of Psi phenomena.

In virtual microtubules $\mathbf{VirG}^{i}_{SME}(\mathbf{S} <=> \mathbf{R})^{i}$ the time and its 'pace' are uncertain: $\mathbf{t} = \mathbf{0}/\mathbf{0}$, if the external translational or tangential velocities (**v**) and accelerations (**dv**/**dt**) of Bivacuum dipoles, composing them, are zero.

### *14.9 Activation of Bivacuum mediated interaction (BMI) between Sender and Receiver*

Theories of the surface and volume Virtual Replica ($\mathbf{VR}^{sur,vol}$) of material objects in Bivacuum (ether body) and primary **VR** multiplication **VRM**(**r**,**t**) (astral and mental bodies) in combination with theory of Virtual Guides ($\mathbf{VirG}_{SME}$) are the background for explanation of different kind of paranormal phenomena.

The primary $\mathbf{VR}^{sur,vol}$ represents a result of interference of basic Bivacuum virtual waves (reference waves) with similar by nature $\mathbf{VPW}^{\pm}_{m}$ and $\mathbf{VirSW}^{\pm 1/2}_{m}$, modulated by $[\mathbf{C} \rightleftharpoons \mathbf{W}]$ pulsation of elementary particles and translational and librational de Broglie waves of molecules of macroscopic object (object waves).

Such interference of the *reference waves* with the *surface* de Broglie waves gives the surface $VR^{sur}$. The interference of reference virtual waves with the volume de Broglie waves gives the volume $VR^{sur}$.

The infinitive multiplication of primary $\mathbf{VR}^{sur}$ and $\mathbf{VR}^{vol}$ in space and time: **VRM(r,t)** in form of 3D packets of virtual standing waves, represents the *self- iteration* of primary $\mathbf{VR}^{sur,vol}$, stimulated by its interference with all pervading *reference waves* - Bivacuum Virtual Pressure Waves ($\mathbf{VPW}^{\pm}_{q=1}$) and Virtual Spin Waves ($\mathbf{VirSW}^{\pm 1/2}_{q=1}$).

It is possible to name the **VRM**, as **Holoiteration** by analogy with regular optical hologram.

Depending on the type modulation (section 13.2) the primary **VR** and its multiplication **VRM(r,t)** can be subdivided on the:

a) frequency modulated;

b) amplitude modulated;

c) phase modulated;

d) polarization modulated.

Only their superposition contains all the information about positions and dynamics of atoms/molecules, composing object's volume and surface.

*14.10 Stages of the Entanglement Channels formation between macroscopic objects*

A possible mechanism of nonlocal exchange of the spin (qubits), momentum and energy between macroscopic Sender [S] and Receiver [R] involves following *three stages:*

*Stage 1.* Superposition of nonlocal fractions of Virtual Replicas Multiplication **VRM(r,t)**, formed by standing Virtual spin waves of remote [Sender] and [Receiver]:

$$\mathbf{VRM}_S^{\mathbf{VirSW}}(\mathbf{r},\mathbf{t}) \bowtie \mathbf{VRM}_R^{\mathbf{VirSW}}(\mathbf{r},\mathbf{t}) \qquad 14.17$$

In other terms, this stage represents the superposition of the mental bodies of Sender and Receiver.

*Stage 2.* Superposition of fraction of Virtual Replicas Multiplication of [S] and [R], formed by standing virtual pressure waves, modulated by [S] and [R]:

$$\mathbf{VRM}(\mathbf{r},\mathbf{t})_S^{dis} \bowtie \mathbf{VRM}(\mathbf{r},\mathbf{t})_R^{dis} = \sum \{[\mathbf{VPW}_m^+ \bowtie \mathbf{VPW}_m^-]_S^i <\!=\!\Leftrightarrow\!=\!> [\mathbf{VPW}_m^+ \bowtie \mathbf{VPW}_m^-]_R^i\} \qquad 14.18$$

In other words, this stage represents the superposition of the astral bodies of Sender and Receiver.

The described above three stages of [S] and [R] Bivacuum mediated interaction (**BMI**) involves formation of Entanglement channel as a second stage. For activation of this channel, the whole system: ([**S**] + [**R**]) should be in nonequilibrium state.

*Stage 3.* Formation of the Entanglement channels, as a number of bundles of nonlocal Virtual guides $\mathbf{VirG}^i_{SME}$ (**S** <=> **R**) of spin, momentum and energy, connecting nucleons and electrons of coherent atoms of [S] and [R] in state of correlated mesoscopic Bose condensates:

$$\left[\textbf{Entanglement channel} = \mathbf{N}(\mathbf{t},\mathbf{r}) \times \sum^{\mathbf{n}} \mathbf{VirG}_{SME}\,(\mathbf{S} <\!=\!> \mathbf{R})\right]^i_{x,y,z} \qquad 14.18a$$

where: **n** is number of correlated pairs of similar elementary particles of Sender and Receiver with opposite spins; $\mathbf{N(t,r)}$ is a number of coherent atoms/molecules in the coherent molecular clusters - mesoscopic BC (Kaivarainen, 2001; 2004).

$\mathbf{VirG}_{SME}$ (**S** <=> **R**) is a quasi one-dimensional virtual Bose condensate, formed by virtual Cooper pairs of $[\mathbf{BVF}^{\uparrow} \bowtie \mathbf{BVF}^{\downarrow}]^i$ or Bivacuum bosons $(\mathbf{BVB}^{+})^i$ *and* $(\mathbf{BVB}^{-})^i$.

We put forward a conjecture, that even the exchange of macroscopic number of coherent atoms between very remote regions of the Universe is possible via the Entanglement channels.

If this consequence of Unified theory will be confirmed, we get a method for the instant remote transportation of macroscopic objects (teleportation) in state of Bose condensation.

*For special case, if Sender [S] is a psychic, the double highly ordered conducting membranes of the coherent nerve cells (like in axons) may provide the cumulative Casimir effect.*

This coherent effect may increase the fraction of water in the internal nonpolar regions of the membranes in state of *mesoscopic Bose condensation* (mBC), contributing Virtual Replica of [S] and [R].

The quantum neurodynamics processes in Sender (Healer) may be accompanied by radiation of electromagnetic waves or magnetic impulses, propagating in Bivacuum via virtual guides: $\mathbf{VirG}_{SME}$(**S** <=> **R**). Such kind of radiation from different regions of body of Healer has been revealed experimentally.

The important role in Bivacuum mediated Mind-Matter and Mind-Mind interaction, plays the coherent fraction of water in microtubules of neurons in state of mesoscopic molecular Bose condensate (mBC) (Kaivarainen: http://arxiv.org/abs/physics/0102086).

This fraction of **mBC** is a variable parameter, dependent on structural state of microtubules and number of simultaneous elementary acts of consciousness (Kaivarainen, 2008 and http://arxiv.org/abs/physics/0003045). It can be modulated not only by excitation of nerve cells, but also by specific interaction with virtual replica of one or more chromosomes ($VR^{DNA}$) of the same or other cells.

*The change of frequency of selected kind of thermal fluctuations, like cavitational ones, in the volume of receiver [R], including cytoplasm water of nerve cells, is accompanied by reversible disassembly of microtubules and actin' filaments, i.e.* [**gel** $\rightleftharpoons$ **sol**] transitions. *These reactions, responsible for elementary act of consciousness,* are dependent on the changes of corresponding activation barriers.

### *14.11 The interrelation between macroscopic entanglement and acts of consciousness*

*The mechanisms of macroscopic entanglement, proposed in our work, is* responsible for change of intermolecular Van der Waals interaction in the volume of [R] and probability of selected thermal fluctuations (i.e. cavitational fluctuations), induced by [S]. In this case, realization of certain series of elementary acts of consciousness of [S] will induce similar series in nerve system of [R]. This means informational exchange between $VR^R$ *and* $VR^S$ of two psychics via Virtual Guides: $\mathbf{VirG}^i_{SME}$(**S** <=> **R**), and their bundles, forming the Entanglement channels (14.18a).

The *specific character* of telepathic signal transmission from [S] to [R] may be provided by modulation of $\mathbf{VRM}^S_{MT}$ of microtubules by $\mathbf{VRM}^S_{DNA}$ of DNA of Sender's chromosomes in neuron ensembles, responsible for subconsciousness, imagination and consciousness. The resonance - the most effective remote informational/energy exchange between two psychics is dependent on corresponding 'tuning' of their nerve systems. As a background of this tuning can be the described Bivacuum mediated interaction (BMI) between the crucial neurons components of [S] to [R]:

$$\sum\left[\mathbf{2\ centrioles + chromosomes}\right]_{\mathbf{S}} \overset{\mathbf{BMI}}{<===>} \sum\left[\mathbf{2\ centrioles + chromosomes}\right]_{\mathbf{R}} \qquad 14.19$$

In accordance to our theory of elementary cycle of consciousness and three stages of entanglement channel formation, described above, the modulation of dynamics of [assembly $\rightleftharpoons$ disassembly] of microtubules by influence on probability of cavitational fluctuations in the nerve cells and corresponding [*gel* $\rightleftharpoons$ *sol*] transitions by directed mental activity of [Sender] can provide telepathic contact and remote viewing between [Sender] and [Receiver].

The mechanism of remote healing could be the same, but the local targets in the body of patient [R] are not necessarily the MTs and chromosomes of the nerve cells, but centrioles + chromosomes of the ill organs (heart, liver, etc.).

The telekinesis, as example of mind-matter interaction, should be accompanied by significant nonequilibrium process in the nerve system of Sender, related to increasing of kinetic energy of coherent molecules in neurons of Sender, like cumulative momentum of water clusters, coherently melting in microtubules of centrioles and inducing their disassembly. Corresponding momentum and kinetic energy are transmitted to 'receiver - target' via multiple correlated bundles of $\mathbf{VirG}_{SME}$ in superimposed $\mathbf{VRM}_{S,R}$ (Psi-channels).

The specific magnetic potential exchange between [S] and [R] via *Virtual tunnel* can be generated by the nerve impulse regular propagation along the axons and depolarization of nerve cells membranes (i.e. electric current) in the 'tuned' ensemble of neuron cells of psychic - [Sender], accompanied by magnetic flux. These processes are accompanied by $\mathbf{BVF}^{\uparrow} \rightleftharpoons \mathbf{BVB}^{\pm} \rightleftharpoons \mathbf{BVF}^{\downarrow}$ equilibrium shift to the right or left, representing magnetic field excitation.

The evidence are existing, that *Virtual tunnel* between [S] and [R] works better, if the frequencies of geomagnetic Schumann waves - around 8 Hz (close to brain waves frequency) are the same in location of [S] and [R]. However, the main coherence factor in accordance to our theory, are all-pervading Bivacuum virtual pressure waves ($\mathbf{VPW}^{\pm}_{q=1}$), with basic Compton frequency $[\omega_0 = \mathbf{m}_0\mathbf{c}^2/\hbar]^i$, equal to carrying frequency of [Corpuscle $\rightleftharpoons$ Wave] pulsations of the electrons, protons, neurons, composing real matter and providing entanglement. The macroscopic Bivacuum flicker fluctuation, activated by non-regular changes/jumps in properties of complex Hierarchical Virtual replica of Solar system and even galactic, related to *sideral time*, also may influence on quality of Entanglement Channels between Sender and Receiver.

**TABLE**

**The Entanglement Channel formation between Sender [S] and Receiver [R]:**

$$\left[\mathbf{N}(\mathbf{t},\mathbf{r}) \times \sum^{\mathbf{n}} \mathbf{VirG}_{SME}\,(\mathbf{S} <=> \mathbf{R})\right]^{i}_{x,y,z}.$$

**The role of paired and unpaired sub-elementary particles of the electron's [Corpuscle $\rightleftharpoons$ Wave] pulsation and its rotation:**

$$\langle[\mathbf{F}^{+}_{\uparrow} \bowtie \mathbf{F}^{-}_{\downarrow}]_W + (\mathbf{F}^{-}_{\uparrow})_C\rangle \rightleftharpoons \langle[\mathbf{F}^{+}_{\uparrow} \bowtie \mathbf{F}^{-}_{\downarrow}]_C + (\mathbf{F}^{-}_{\uparrow})_W\rangle$$

**in Bivacuum - mediated interaction between sender [S] and receiver [R]**

**Pair of sub-elementary particle and antiparticle pulsation and rotation**:

$$[\mathbf{F}_{\uparrow}^{+} \bowtie \mathbf{F}_{\downarrow}^{-}]_W \rightleftharpoons \langle[\mathbf{F}_{\uparrow}^{+} \bowtie \mathbf{F}_{\downarrow}^{-}]_C$$

1. Virtual Pressure Waves: $\left[\mathbf{VPW}^{+} \bowtie \mathbf{VPW}^{-}\right]$
2. Total Virtual Pressure energy increment, equal to that of unpaired ($\Delta\mathbf{E}_{\mathbf{F}_{\uparrow}^{+}}$):

$$\Delta\mathbf{VirP}^{\pm}_{\mathbf{F}_{\uparrow}^{+}\bowtie\mathbf{F}_{\downarrow}^{-}} = \frac{1}{2}\left|\mathbf{VirP}^{+}_{\mathbf{F}_{\uparrow}^{+}} \bowtie \mathbf{VirP}^{-}_{\mathbf{F}_{\downarrow}^{-}}\right|^{[\mathbf{F}_{\uparrow}^{+}\bowtie\mathbf{F}_{\downarrow}^{-}]}$$

3. Virtual Replica of the Object ($\mathbf{VR} = \mathbf{VR}^{in} + \mathbf{VR}^{sur}$)
4. Virtual Replicas of [S] and [R] and their Multiplication:

$$\mathbf{VRM}_S(\mathbf{r,t}) = \sum \mathbf{VR}_S <\rightleftharpoons> \sum \mathbf{VR}_R = \mathbf{VRM}_R(\mathbf{r,t})$$

**Unpaired sub-elementary fermion pulsation and rotation**:

$$(\mathbf{F}_{\uparrow\downarrow}^{\pm})_C \overset{\mathbf{BvSO}}{\rightleftharpoons} (\mathbf{F}_{\uparrow\downarrow}^{\pm})_W \overset{\mathbf{CVC}^{\circlearrowright\circlearrowleft}}{<==>} \mathbf{VirSW}^{\circlearrowright\circlearrowleft}$$

1. Virtual Spin waves: $\mathbf{VirSW}^{\circlearrowright\circlearrowleft}$

$$\mathbf{I}_S \equiv \mathbf{I}_{\mathbf{VirSW}^{\pm 1/2}} \sim \mathbf{K}_{BVF^{\uparrow}\rightleftharpoons BVF^{\downarrow}}(\mathbf{t}) = (\mathbf{K}_{BVF^{\uparrow}\rightleftharpoons BVF^{\downarrow}})_0\,[\sin(\omega_0^i t) + \gamma\omega_B^{lb}\sin(\omega_B^{lb} t)]$$

2. Electromagnetic potential:

$$\mathbf{E}_{EM} = \boldsymbol{\alpha}|\mathbf{m}_V^{+} - \mathbf{m}_V^{-}|\mathbf{c}^2 \sim \sim \frac{1}{2}\alpha\left|\mathbf{VirP}^{+}_{\mathbf{F}_{\uparrow}^{+}} - \mathbf{VirP}^{-}_{\mathbf{F}_{\downarrow}^{-}}\right|^{[\mathbf{F}_{\uparrow}^{+}\bowtie\mathbf{F}_{\downarrow}^{-}]}$$

3. Gravitational potential:

$$\mathbf{E}_G = \boldsymbol{\beta}[\mathbf{m}_V^{+} - |\mathbf{m}_V^{-}|]\mathbf{c}^2 \sim \sim \frac{1}{2}\boldsymbol{\beta}\left|\mathbf{VirP}^{+}_{\mathbf{F}_{\uparrow}^{+}} - \mathbf{VirP}^{-}_{\mathbf{F}_{\downarrow}^{-}}\right|^{[\mathbf{F}_{\uparrow}^{+}\bowtie\mathbf{F}_{\downarrow}^{-}]}$$

4. Virtual Spin Waves (**VirSW**):
5. The Entanglement Channel formation

$$\left[\mathbf{N}(\mathbf{t},\mathbf{r}) \times \sum^{\mathbf{n}} \mathbf{VirG}_{SME}\,(\mathbf{S} <=> \mathbf{R})\right]^{i}_{x,y,z}$$

between the remote [S] and [R]:

$$\mathbf{VirSW}_S^{S=+1/2} \underset{\mathbf{BVF}^{\uparrow}\bowtie\mathbf{BVF}^{\downarrow}}{\overset{\mathbf{BVB}^{\pm}}{<=\rightleftharpoons=>}} \mathbf{VirSW}_R^{S=-1/2}$$

Pauli attraction (Cooper pairs formation) or repulsion between $\mathbf{BVF}^{\updownarrow}$ of the opposite or similar spins is provided by paired and unpaired sub-elementary fermions

---

*One of the result of the Entanglement Channel formation, as a superposition of $VRM_{S,R}$ and bundles of $VirG^{ext}_{SME}$, is a change of permittivity $\varepsilon_0$ and permeability $\mu_0$ of Bivacuum $[\varepsilon_0 = n_0^2 = 1/(\mu_0 c^2)]$. In turn, $(\pm\Delta\varepsilon_0)$ influence Van-der-Waals interactions in condensed matter, changing the probability of defects origination in solids and cavitational fluctuations in liquids. The bidirectional change of pH of water via Virtual tunnel can be a consequence of $\pm\Delta VP^{\pm}$ and $\pm\Delta\varepsilon_0$ influence on cavitational fluctuations, accompanied by shift of dynamic equilibrium: $H_2O \rightleftharpoons HO^- + H^+$ and assembly $\rightleftharpoons$ disassembly of actin filaments in the nerve cells and $\left[gel \rightleftharpoons sol\right]$ transitions as a part of act of consciousness.*

The coherency between signals of [S] and [R] can be provided by forced resonance of basic Virtual Pressure Waves ($VPW^{\pm}$) with elementary particles of [S] and [R].

The modulation of Bivacuum mediated interaction by the cosmic macroscopic fluctuations, Shumann resonance and geophysical magnetic flicker noise also increase the probability of macroscopic entanglement.

The [*dissociation* $\rightleftharpoons$ *association*] equilibrium oscillation of coherent water clusters in state of molecular Bose condensate (mBC) in the microtubules of the nerve cells is a crucial factor for realization of quantum Psi phenomena. This follows from our model of elementary act of consciousness.

*14.12 The examples of Bivacuum Mediated Interaction (BMI) between macroscopic objects, realized via Entanglement channels*

In accordance to our approach, the remote interaction between macroscopic Sender [S] and Receiver [R] can be realized, as a result of *Bivacuum mediated interaction (BMI)*, like superposition of distant and nonlocal components of their Virtual Replicas Multiplication ($\mathbf{VRM}_S = \Leftrightarrow = \mathbf{VRM}_R$), described in previous sections.

Nonequilibrium processes in [Sender], accompanied by acceleration of particles, like evaporation, heating, cooling, melting, boiling etc. may stimulate the *nonelastic effects* in the volume of [Receiver] and increments of modulated virtual pressure and spin waves ($\Delta\mathbf{VPW}^{\pm}_{\mathbf{m}}$ and $\Delta\mathbf{VirSW}^{\pm 1/2}_{\mathbf{m}}$), accompanied $[\mathbf{C} \rightleftharpoons \mathbf{W}]$ pulsation of triplets $[\mathbf{F}^{+}_{\uparrow} \bowtie \mathbf{F}^{-}_{\downarrow}] + \mathbf{F}^{\pm}_{\updownarrow} >^{i}$ , formed by sub-elementary fermions of different generation, representing electrons, protons and neutrons.

The following unconventional kinds of effects of non-electromagnetic and non-gravitational nature can be anticipated in the remote interaction between **macroscopic** nonequilibrium [Sender] and sensitive detector [Receiver] via multiple Virtual spin and energy guides $\mathbf{VirG}_{SME}$ (Fig.9), if our theory of nonlocal spin, momentum and energy exchange between [S] and [R], described above is correct:

I. Weak repulsion and attraction between 'tuned' [S] and [R] and rotational momentum in [R] induced by [S], as a result of transmission of momentum/kinetic energy and angular momentum (spin) between elementary particles of [S] and [R]. The probability of such 'tuned' interaction between [S] and [R] is dependent on dimensions of coherent clusters of atoms and molecules of condensed matter in state of mesoscopic Bose condensation (**mBC**) (Kaivarainen, 1995; 2001; 2003; 2004). The number of atoms in such clusters $\mathbf{N}(\mathbf{t},\mathbf{r})$ is related to number of $\mathbf{VirG}_{SME}$ in the Entanglement channels connecting tuned **mBC** in [S] and [R]:

$$\mathbf{Entanglement\ \ channel} = \left[ \mathbf{N}(\mathbf{t},\mathbf{r}) \times \sum^{\mathbf{n}} \mathbf{VirG}_{SME}\,(\mathbf{S} <=> \mathbf{R}) \right]^{i}_{x,y,z}$$

. The number of coherent pairs of atoms in Sender and Receiver $\mathbf{N}(\mathbf{t},\mathbf{r})$ may be regulated by temperature, ultrasound, etc.

*The kinetic energy distant transmission from atoms of [S] to atoms of [R] may be accompanied by the temperature variation and local pressure/sound effects in [R];*

II. Increasing the probability of thermal fluctuations in the volume of [R] due to decreasing of Van der Waals interactions, because of charges screening effects, induced by overlapping of distant virtual replicas of [S] and [R] and increasing of dielectric permittivity of Bivacuum.

In water the variation of probability of cavitational fluctuations should by accompanied by the in-phase variation of pH and electric conductivity due to shifting the equilibrium: $H_2O \rightleftharpoons H^+ + HO^-$ to the right or left;

III. Small changing of mass of Receptor in conditions, changing the probability of the inelastic recoil effects in the volume of [R] under influence of [S];

IV. Registration of metastable virtual particles, as a result of Bivacuum symmetry perturbations.

The first kind (I) of new class of listed above interactions between coherent fermions of [S] and [R] is a result of huge number of correlated Entanglement channels formation between coherent regions of Sender and Receiver.

*These Entanglement channels can be responsible for:*

a) the momentum and kinetic energy exchange between [Sender] and [Receiver] providing attraction or repulsion between **S and R**;

b) regulation of Pauli attraction (spin-spin exchange) effects between fermions of [S] and [R];

c) transmission of macroscopic rotational momentum from the clusters of coherent atoms of [S] to [R], representing primary librational effectons (mesoscopic Bose condensate).

*The second kind (II) of phenomena*: influence of [S] on probability of thermal fluctuations in [R], - is a consequence of the additional symmetry shift in Bivacuum fermions ($\mathbf{BVF}^{\updownarrow}$), induced by superposition of distant and nonlocal multiplicated Virtual Replicas of [S] and [R]: $\mathbf{VRM}^{S} \bowtie \mathbf{VRM}^{R}$, which is accompanied by increasing of Bivacuum fermions ($\mathbf{BVF}^{\updownarrow} = [\mathbf{V}^{+} \Updownarrow \mathbf{V}^{-}]$) virtual charge: $\Delta\mathbf{e} = (\mathbf{e}_{V^+} - \mathbf{e}_{V^-}) << \mathbf{e}_0$ in the volume of [R]. Corresponding increasing of Bivacuum permittivity ($\boldsymbol{\varepsilon}_0$) and decreasing magnetic permeability ($\boldsymbol{\mu}_0$) : $\boldsymbol{\varepsilon}_0 = 1/(\boldsymbol{\mu}_0\mathbf{c}^2)$ is responsible for the charges screening effects in volume of [R], induced by [S]. This weakens the electromagnetic Van der Waals interaction between molecules of [R] and increases the probability of defects origination and cavitational fluctuations in solid or liquid phase of Receiver.

*The third kind of phenomena (III)*: reversible decreasing of mass of rigid [R] can be a result of reversible lost of energy of Corpuscular phase of particles, as a consequence of inelastic recoil effects, following the in-phase $[\mathbf{C} \rightarrow \mathbf{W}]$ transition of $\mathbf{N}_{coh}$ coherent nucleons in the volume of [R].

The probability of recoil effects can be enhanced by heating the rigid object or by striking it by another hard object. This effect can be registered directly - by the object mass decreasing. In conditions, close to equilibrium, the Matter - Bivacuum energy exchange relaxation time, following the process of coherent $[\mathbf{C} \rightleftharpoons \mathbf{W}]$ pulsation of macroscopic fraction of atoms is very short and corresponding mass defect effect is undetectable. *Such collective recoil effect of coherent particles* could be big in superconducting or superfluid systems of macroscopic Bose condensation or in crystals, with big domains of atoms in state of Bose condensation.

Another possible explanation of the Receptor mass change is a shifting of [Corpuscle $\rightleftharpoons$ Wave] dynamic equilibrium to the massless Wave phase under the influence of Sender.

*The fourth kind of the listed above phenomena* - increasing the probability of virtual particles and antiparticles origination can be a result of asymmetric Bivacuum dipoles acceleration in conditions of forced resonance with high-frequency Bivacuum virtual waves.

The listed above nontrivial consequences of Unified theory (I - IV) are consistent with unusual data, obtained by groups of Kozyrev (1984; 1991) and Korotaev (1999; 2000).

It is important to note, that these experiments are incompatible with current paradigm. It is timed out and should be replaced by the new one.

### *14.13 The effects of the Virtual Replica of asymmetric constructions, like pyramids, on the matter-target properties*

Our theory of Virtual Replica material objects and its interaction with sensitive targets is confirmed by number of experiments with asymmetric inorganic constructions, like pyramids, cones, etc.

In works of Adamenko, Levchook (1994), Narimanov (2001) and Miakin (2002) such 'pyramid effects' has been demonstrated on examples of the following test-systems, placed

inside the pyramids:
- the cultures of microbes (dynamic behavior),
- water (pH, $O_2$ concentration),
- polymers solution enable to thixotropic structures formation (optical density),
- benzene acid (UV absorption).

*The Virtual Replicas of the pyramids or cones* should be much more asymmetric, than VR generated by cube or sphere. Consequently, the effects of corresponding virtual replicas of hollow of filled structure on very sensitive test systems, like water or biosystems, can be different also.

This consequence of our theory was confirmed experimentally by Narimanov (2001). He keeps a flask with water under the *pyramid with dimensions of tens of centimeters* during few days. This interaction makes pH of water lower, than in control flask, placed under *cube* in the same room and temperature.

The ice, formed from the *'pyramid - treated water*' melts about 10% faster, than the control ice. These results point to increasing the concentration of the defects in the pyramid - treated ice structure.

The perturbation of the vacuum screening properties by the different shapes of cavities is confirmed also by different Lamb shifts in atomic spectra of samples in such cavities.

It is known, that the Lamb shift is determined by screening of the electrons and nuclear charges by the charged virtual particles and antiparticles - asymmetric Bivacuum dipoles in our terminology, induced by Virtual Replica of asymmetric objects.

## 15. Unification of Actual and Virtual Self. The Conjectures of the Soul, Global Consciousness & Superconsciousness

### *15.1 Elementary act of Consciousness or "Cycle of Mind", Involving Distant and Nonlocal Interaction*

Each macroscopic process can be subdivided on the quantum and classical stages.

*Particularly, the quantum stages of Cycle of Mind involves:*

1) the stimulation of dynamic correlation between water clusters in the same and remote microtubules (MTs) in state of mesoscopic Bose condensation (mBC) by phonons (acoustic waves) and by librational IR photons (electromagnetic-EM waves) distant exchange;

2) the transition from distant EM interaction between remote MTs to nonlocal quantum interaction, induced by IR photons exchange between clusters, the clusters Virtual Replicas multiplication (VRM) and virtual guides (VirG) formation between elementary particles of remote coherent water molecules. This process represents transition from mesoscopic Bose condensation to macroscopic nonuniform Bose condensation, oscillating between coherent and decoherent states (Kaivarainen, 2006, http://arxiv.org/abs/physics/0207027);

3) the collapsing of corresponding macroscopic wave function, as a result of the optical bistability of the entangled water clusters and their disassembly to noncoherent molecules due to librational photons pumping, shifting clusters to less stable state;

4) turning the clusterphilic interaction between water clusters in the open state of cavities between alpha and beta tubulins to hydrophobic one and the in-phase shift of these cavities to the closed state due to clusters disassembly (Kaivarainen, http://arxiv.org/abs/physics/0102086).

*The classical stages of our Hierarchic model of elementary act of "Cycle of Mind" are following:*

a) the nerve cells membrane depolarization;

b) the gel $\rightarrow$ sol transition induced by disconnection of microtubules (MTs) with

membranes and disassembly of the actin filaments;

c) the shape /volume pulsation of dendrites of big number of coherently interacting nerve cells, accompanied by jump-way reorganization of synaptic contacts on the dendrites surface;

d) the back sol → gel transition in corresponding cells ensembles, stabilizing (memorizing) of new state by formation of new system of microtubules and MTs associated proteins in the dendrites (MAPs).

The 'Period of Cycle of Mind' is determined by the sum of the life-time of quantum phase - macroscopic Bose condensation of the entangled water clusters in a big number of microtubules, providing coherence and macroscopic entanglement and the life-time of collapsed mesoscopic Bose condensation, induced by decoherence factors. The life-time of entangled coherent phase can be many times shorter than classical phase. Consequently, the $\left[ coherence \rightleftharpoons decoherence \right]$ dynamic equilibrium in macroscopic system of neurons is strongly shifted to the right. However, even extremely short time of macroscopic entanglement between microtubules is enough for nonlocal remote interaction. This is true not only for systems under consideration, but for any kind of oscillating macroscopic entanglement, possible even at physiological temperatures.

Our approach to elementary act of consciousness has some common features with well-known Penrose - Hameroff model, interrelated act of consciousness with the wave function of microtubules collapsing. So we start from description of this Orchestrated objective reduction (Orch OR) model.

*Two triggering mechanisms of elementary act of consciousness*

In some cases, the excitation/depolarization of nerve cells by external factors (sound, vision, smell, tactical feeling) are the triggering - *primary* events of consciousness and [gel→sol] transitions in nerve cells are *secondary* events.

However, the opposite 2nd mechanism, when [gel → sol] transitions are the *primary* events, for example, as a result of thinking/meditation and nerve cells depolarization are *secondary* events, is possible also.

The 1st mechanism includes following stages of elementary act of consciousness:

a) simultaneous depolarization of big enough number of neurons, forming ensemble, accompanied by opening the potential-dependent channels and increasing the concentration of $Ca^{2+}$ in cytoplasm of neurons body;

b) collective disassembly of actin filaments, accompanied by [gel → sol] transition of big group of depolarized neurons stimulated by $Ca^{2+}-$ activated proteins like gelsolin and villin. Before depolarization the concentration of $Ca^{2+}$ outside of cell is about $10^{-3}M$ and inside about $10^{-7}M$. Such strong gradient provide fast increasing of these ions concentration in cell till $10^{-5}M$ after depolarization.

c) strong decreasing of cytoplasm viscosity and disjoining of the (+) ends of MTs from membranes, making possible the spatial fluctuations of MTs orientations inducing decoherence and switching off the entanglement between mBC;

d) volume/shape pulsation of neuron's body and dendrites, inducing reorganization of ionic channels activity and synaptic contacts in the excited neuron ensembles. These volume/shape pulsations occur due to reversible decrease of the intra-cell water activity and corresponding swallow of cell as a result of increasing of passive osmotic diffusion of water from the external space into the cell.

In opposite, 2nd mechanism of elementary act of consciousness, the depolarization of nerve membranes and axonal firing is a *secondary* event and *gel → sol* transition a *primary* one, stimulated in turn by simultaneous disassembly of big number of correlated water clusters to relatively independent molecules.

The electrical recording of human brain activity demonstrate a coherent (40 to 70 Hz) firing among widely distributed and distant brain neurons (Singer, 1993). Such synchronization in a big population of groups of cells points to possibility of not only the regular axon-mediated interaction, but also to fields-mediated interaction and quantum entanglement between water clusters in MTs of remote neurons bodies.

*The comparison of Hierarchic model of consciousness with Quantum brain dynamics model*

Our approach to Quantum Mind problem has some common features with model of Quantum Brain Dynamics (QBD ), proposed by L.Riccardi and H.Umezawa in 1967 and developed by C.I.Stuart, Y.Takahashi, H.Umezava (1978, 1979), M.Jibu and K.Yasue (1992, 1995).

In addition to traditional electrical and chemical factors in the nerve tissue function, this group introduced two new types of *quantum* excitations (ingredients), responsible for the overall control of electrical and chemical signal transfer: *corticons and exchange bosons* (dipolar phonons).

The *corticons* has a definite spatial localization and can be described by Pauli spin matrices. The *exchange bosons*, like phonons are delocalized and follow Bose-Einstein statistics.

By absorbing and emitting bosons coherently, *corticons* manifest global collective dynamics…, providing systematized brain functioning" (Jibu and Yasue,1993).

In other paper (1992) these authors gave more concrete definitions: "*Corticons* are nothing but quanta of the molecular vibrational field of biomolecules (quanta of electric polarization, confined in protein filaments). *Exchange bosons* are nothing but quanta of the vibrational field of water molecules… ".

It is easy to find analogy between spatially localized "corticons" and our primary effectons as well as between "exchange bosons" and our primary - electromagnetic and secondary - acoustic deformons.

However, Hierarchic theory considers 24 collective excitations in any condensed matter (including water and biosystems) and can analyze each of them quantitatively.

Jibu, Yasue, Hagan and others (1994) discussed a possible role of quantum optical coherence in microtubules for brain function. They considered MTs as a *wave guides* for coherent superradiation. They also supposed that coherent photons, penetrating in MTs, lead to *"self-induced transparency*". Both of these phenomena are well known in fiber and quantum optics.

We also use these phenomena for explanation of transition from mesoscopic entanglement of water clusters in MTs to macroscopic one, as a result of IR photons exchange between coherent clusters (mBC). However, we have to note, that our mechanism of transition of mBC to macroscopic BC in 'tuned' MTs do not need self-induced transparency.

It follows also from our approach that the mechanism of macroscopic, but noncontineous Bose condensation of water clusters do not need the hypothesis of Frölich that the proteins (tubulins of MTs) are pumped into coherent macroscopic quantum states by biochemical energy.

We also do not use the idea of Jibu et al. that the MTs works like the photons wave - guides without possibility of side radiation throw the walls of MTs. The latter in our approach increases the probability of macroscopic entanglement between remote MTs and cells of the organism.

*The Properties of Microtubules and*

*Internal Water*

Let us consider the properties of microtubules (MT) as one of the most important component of cytoskeleton, responsible for spatial organization and dynamic behavior of the cells.

The microtubules (MTs) are the nanostructures of cells, interconnecting the quantum and classical stages of the Cycle of Mind.

The [assembly ⇔ disassembly] equilibrium of microtubules composed of $\alpha$ and $\beta$ tubulins is strongly dependent on internal and external water activity ($a$), concentration of $Ca^{2+}$ and on the electric field gradient change due to MTs piezoelectric properties.

The $\alpha$ and $\beta$ tubulins are globular proteins with equal molecular mass ($MM = 55.000$), usually forming $\alpha\beta$ dimers with linear dimension 8 nm. Polymerization of microtubules can be stimulated by NaCl, $Mg^{2+}$ and GTP (1:1 tubulin monomer) (Alberts *et al.*, 1983). The presence of heavy water (deuterium oxide) also stimulates polymerization of MTs.

In contrast to that the presence of ions of $Ca^{2+}$ even in micromolar concentrations, action of colhicine and lowering the temperature till $4^0$C induce disassembly of MT.

Due to multigenic composition, $\alpha$ and $\beta$ tubulins have a number of isoforms. For example, two-dimensional gel electrophoresis revealed 17 varieties of $\beta$ tubulin in mammalian brain (Lee *et al.*, 1986). Tubulin structure may also be altered by enzymatic modification: addition or removal of amino acids, glycosylation, etc.

*Microtubules* are hollow cylinders, filled with water. Their internal diameter about $d_{\text{in}}$ =140Å and external diameter $d_{\text{ext}} = 280Å$ (Figure 2). These data, including the dimensions of $\alpha\beta$ dimers were obtained from x-ray crystallography (Amos and Klug, 1974). However we must keep in mind that under the conditions of crystallization the multiglobular proteins and their assemblies tends to more compact structure than in solutions due to lower water activity.

This means that in natural conditions the above dimensions could be a bit bigger.

The length of microtubules (MT) can vary in the interval:

$$l_t = (1 - 20) \times 10^5 Å \qquad 19$$

The spacing between the tubulin monomers in MT is about 40 Å and that between $\alpha\beta$ dimers: 80 Å are the same in longitudinal and transversal directions of MT.

Microtubules sometimes can be as long as axons of nerve cells, *i.e.* tenth of centimeters long. Microtubules (MT) in axons are usually parallel and are arranged in bundles. Microtubules associated proteins (MAP) form a ”bridges”, linking MT and are responsible for their interaction and cooperative system formation. Brain contains a big amount of microtubules. *Their most regular length of MTs is about* $10^5 Å$, *i.e. close to librational photon wave length.*

The viscosity of ordered water in such narrow microtubules seems to be too high for transport of ions or metabolites at normal conditions.

All 24 types of quasi-particles, introduced in Hierarchic theory of condensed matter (Kaivarainen, 2008), also can be pertinent for ordered water in the microtubules (MT). However, the dynamic equilibrium between populations of different quasi-particles of water in MT must be shifted towards primary librational effectons, comparing to bulk water because of interfacial effects (Kaivarainen, 2000, 2008). The dimensions of internal primary librational effectons have to be bigger than in bulk water as a consequence of stabilization of MT walls the mobility of water molecules, decreasing their momentum and increasing their most probable de Broglie wave length.

The interrelation must exist between properties of internal water in MT and structure and dynamics of their walls, depending on $[\alpha - \beta]$ tubulins interaction. Especially important can be a quantum transitions like convertons $[tr \Leftrightarrow lb]$. The convertons in are accompanied by [dissociation/association] of primary librational effectons, i.e. flickering of coherent water clusters, followed by the change of angle between $\alpha$ and $\beta$ subunits in tubulin dimers.

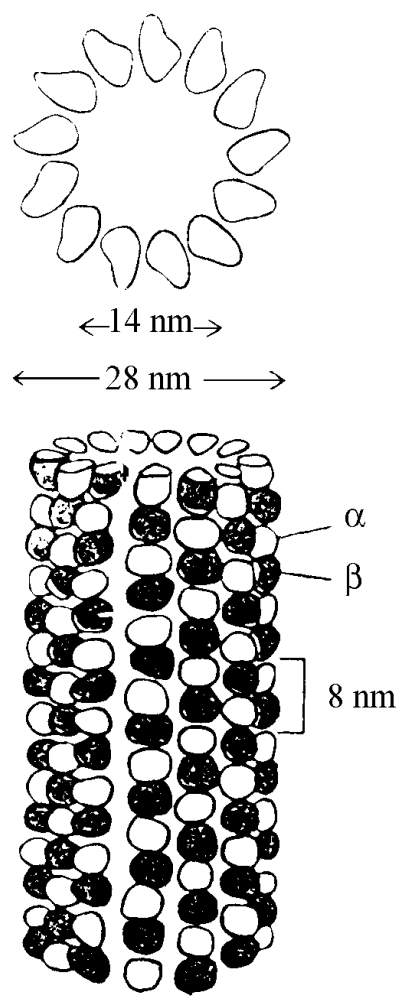

**Figure 10**. Construction of microtubule from $\alpha$ and $\beta$ tubulins, globular proteins with molecular mass 55 kD, existing in form of dimers $(\alpha\beta)$. Each $\alpha\beta$ dimer is a dipole with negative charges, shifted towards $\alpha$ subunit (De Brabander, 1982). Consequently, microtubules, as an oriented elongated structure of dipoles system, have the piezoelectric properties (Athestaedt, 1974; Mascarennas, 1974).

The equilibrium of "closed" (A) and "open"(B) states of nonpolar cavities between $\alpha$ and $\beta$ tubulins in $(\alpha\beta)$ dimers can be shifted to the (B) one under the change of external electric field in a course of membrane depolarization. It is a consequence of piezoelectric properties of MTs and stimulate the formation of coherent water clusters in the open cavities of $(\alpha\beta)$ dimers. The open cavities serve as a centers of water cluster formation and molecular Bose condensation.

The parallel orientation of MT in different cells, optimal for maximum [MT-MT] resonance exchange interaction could be achieved due to twisting of centrioles, changing spatial orientation of MT. However, it looks that the normal orientation of MT as respect to each other corresponds to the most stable condition, *i.e.* minimum of potential energy of interaction (see Albreht-Buehner, 1990).

It is important to stress here that the orientation of two centrioles as a source of MT bundles in each cell are always normal to each other, providing conditions for IR photons (librational and translational) 3-D standing waves formation.

The linear dimensions of the primary librational effectons edge $(l_{ef}^{lb})$ in pure water at physiological temperature $(36^0C)$ is about 11 Å and in the ice at $0^0C$ it is equal to 45 Å.

We assume that in the rigid internal core of MT, the linear dimension (edge length) of librational effecton, approximated by cube is between 11Å and 45 Å *i.e.* about $l_{ef}^{lb} \sim$ 23Å. It will be shown below, that this assumption fits the spatial and symmetry properties of MT very well.

The most probable group velocity of water molecules forming primary *lb* effectons is:

$$\mathbf{v}_{gr}^{lb} \sim h/(m_{H_2O} \times l_{ef}^{lb}) \qquad 20.$$

The librational mobility of internal water molecules in MT, which determines $(\mathbf{v}_{gr}^{lb})$ should be about 2 times less than in bulk water at $37^0C$, if we assume for water in microtubules: $l_{ef}^{lb} \sim 23\text{Å}$.

Results of our computer simulations for pure bulk water shows, that the distance between centers of primary [lb] effectons, approximated by cube exceed their linear dimension to about 3.5 times (Fig 11b). For our case it means that the average distance between the effectons centers is about:

$$d = l_{ef}^{lb} \times 3.5 = 23 \times 3.5 \sim 80\text{Å}$$

This result of our theory points to the equidistant (80 Å) localization of the primary *lb* effectons in clefts between $\alpha$ and $\beta$ tubulins of each $(\alpha\beta)$ dimer in the internal core of MTs.

In the case, if the dimensions of librational effectons in MTs are quite the same as in bulk water, i.e. 11 Å, the separation between them should be:
$d = l_{ef}^{lb} \times 3.5 = 11 \times 3.5 \sim 40$ Å.

This result points that the coherent water clusters can naturally exist not only between $\alpha$ and $\beta$ subunits of each pair, but also between pairs of $(\alpha\beta)$ dimers.

In the both cases the regular distribution of the internal flickering water clusters in MT may serve as an important factor for realization of the signal propagation along the MT (conformational wave), accompanied by correlated process of closing and opening the clefts between neighboring $\alpha$ and $\beta$ tubulins pairs.

This large-scale protein dynamics is regulated by dissociation $\rightleftharpoons$ association of water clusters in the clefts between $(\alpha\beta)$ dimers of MT (Fig.10) due to $[lb/tr]$ convertons excitation and librational photons and phonons exchange between primary and secondary effectons, correspondingly.

The dynamic equilibrium between *tr* and *lb* types of the intra MT water effectons must to be very sensitive to $\alpha - \beta$ tubulins interactions, dependent on nerve cells excitation and their membranes polarization.

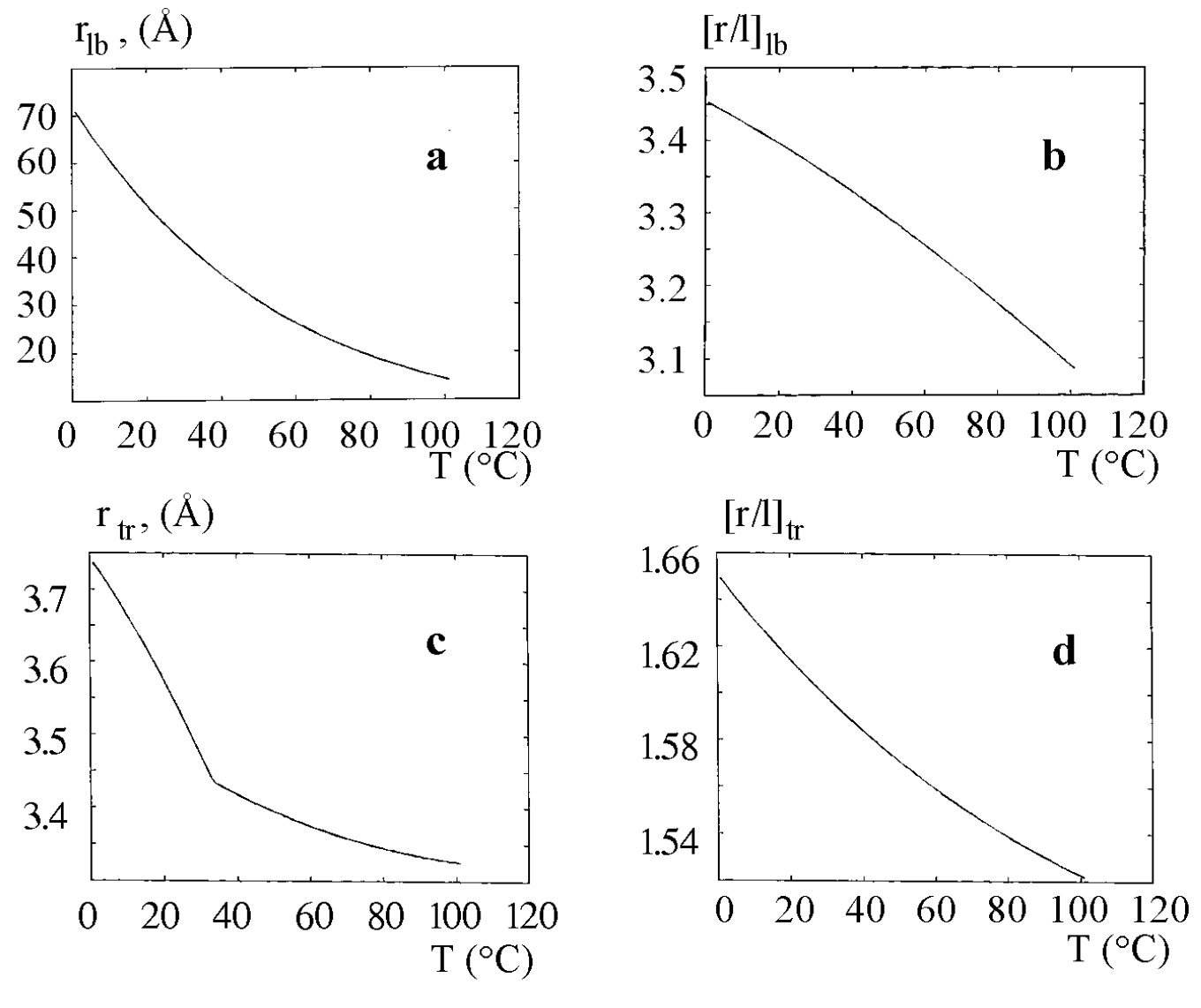

**Figure 11**. Theoretical temperature dependencies of:

(a) - the space between centers of primary [lb] effectons (calculated in accordance to eq.4.62 of http://arxiv.org/abs/physics/0102086);

(b) - the ratio of space between primary [lb] effectons to their length (calculated, using eq.4.63 of http://arxiv.org/abs/physics/0102086);

(c) - the space between centers of primary [tr] effectons (in accordance to eq.4.62);

(d) - the ratio of space between primary [tr] effectons to their length (eq.4.63).

**Two statements of our Hierarchic model of consciousness are important**:

1. The ability of intra-MT primary water effectons (tr and lb) for superradiation of six coherent IR librational photons from each of the effectons side, approximated by parallelepiped:

two identical - "longitudinal" IR photons, propagating along the core of microtubule, forming the longitudinal standing waves inside it and two pairs of identical - "transverse" IR photons, responsible for remote exchange interaction between microtubules. *In accordance to superradiation mechanism the intensity of longitudinal radiation of MTs is much bigger than that of transverse one;*

2. The parameters of the water clusters radiation (frequency and intensity of librational photons) are regulated by the interaction of the internal water with open and closed states of cavities between $\alpha$ and $\beta$ tubulins of MTs.

In normal cells, microtubules grow from pair of centriole in cell center to cell's periphery. In the center of cells of plants the centrioles are absent. Two centrioles in cells of animals are always oriented at the right angle with respect to each other. The centrioles represent a construction of 9 triplets of microtubules, i.e. two centriole contain $(2 \times 27 = 54)$ microtubules. The centriole length is about 3000 Å and its diameter is $1000\,\text{Å}$.

These dimensions mean that all 27 microtubules of each centrioles can be correlated in the volume ($\mathbf{v}_d$) of one translational or librational electromagnetic deformon:

$$[\mathbf{v}_d = \frac{9}{4\pi}\lambda_p^3]_{tr,lb}$$

$$\text{where: } (\lambda_p)_{lb} \sim 10^5\,\text{Å} \quad \text{and} \quad (\lambda_p)_{tr} \sim 3.5 \times 10^5\,\text{Å}$$

*The transition from electromagnetic interaction between remote water clusters (mBC) to macroscopic Bose condensation and nonlocal interaction*

The role of Virtual Replica of clusters in state of mBC spatial multiplication VRM(r) (see chapter 14) is to create the virtual connections between remote actual clusters. The subsequent formation of Virtual Channels between pairs of coherent elementary particles (electrons, protons and neutrons) of opposite spins, - turns the mesoscopic BC to macroscopic one. This is a result of unification of remote clusters wave function to integer linear superposition of its eigenvalues, described by macroscopic wave function.

Part of these eigenvalues of the integer wave function describe virtual replica of the cluster and other eigenvalues - clusters themselves. Consequently, the corresponding nonuniform macroscopic Bose condensate is partly virtual and partly actual.

The transition from distant EM interaction between remote MTs to nonlocal quantum interaction is a result of entanglement between clusters - mBC, stimulated by librational IR photons exchange and Virtual Replicas multiplication (VRM) of the clusters. This transition is accompanied by nonuniform macroscopic virtual Bose condensation (VirBC). *The nonuniform VirBC become possible only at certain spatial orientation of coherent water clusters as mesoscopic Bose condensate (mBC) in 'tuned' microtubules.*

The mechanism of this transition is based on our theories of Virtual Replica multiplication in space - **VRM(r)**, virtual Bose condensation (**VirBC**) and nonlocal virtual guides (**VirG**$_{S,M,E}$) of spin, momentum and energy (chapter 14).

The Virtual Guides have a shape of virtual microtubules with properties of

quasi-1-dimensional virtual Bose condensate (Fig.9). The VirG are constructed from 'side-to-side' polymerized Bivacuum bosons $BVB^{\pm}$ or Cooper pairs of Bivacuum fermions $[\mathbf{BVF}^{\uparrow} \bowtie \mathbf{BVF}^{\downarrow}]$. The bundles of $\mathbf{VirG}_{S,M,E}$, connecting coherent atoms of Sender (S) and Receiver (S), named the *Entanglement Channels*, are responsible for macroscopic entanglement, providing nonlocal interaction, telepathy, remote healing and telekinesis. The poltergeist can be considered as a private case of telekinesis, realized via Entanglement Channels, connecting coherent elementary particles of psychic and the object.

The quantum stage of Cycle Mind involves the collapsing of unified wave function. It represents the reversible dissipative process.

The Hameroff - Penrose model considers only the coherent conformational transition of cavities between big number of pairs of tubulins between open to closed states in remote entangled MTs as the act of wave function collapsing.

In our model this transition is only the trigger, stimulating quantum transition of big number of entangled water clusters in state of noncontineous macroscopic Bose condensation to selected state of mesoscopic Bose condensation (mBC).

*Consequently, the Cycles of Mind can be considered as a reversible transitions of certain part of brain between coherence and decoherence, involving quantum and classical stages.* The non-uniform coherent state of macroscopic Bose Condensation of water in microtubules system differs from continuous macroscopic Bose Condensation, pertinent for superfluidity and superconductivity.

The *bistability* is water clusters polarization change as a result of $a \rightleftharpoons b$ equilibrium shift between their optic (a) and acoustic (b) states to the right. This shift is a consequence of librational IR photons pumping due to exchange between MTs by librational photons.

The related phenomena: *self-induced transparency* is a consequence of saturation of b-state by primary librational effectons.

The pike regime - the light emission pulsation, after $b$ –state saturation represents super-radiation of big number of entangled water clusters in the process of their correlated $\sum(\mathbf{b} \rightarrow \mathbf{a})$ transitions.

Such cycles of quantum mechanical and nonlinear classical events can be considered as elementary acts of memory and consciousness realization. One act can be as long as 500 ms, *i.e.* half of second, like proposed in Hameroff-Penrose model.

The elementary act of consciousness include a stage of coherent electric firing in brain (Singer, 1993) of distant neurons groups with period of about 1/40 sec.

The characteristic frequency of pike regime (Bates, 1978; Andreev et al.,1988) can be correlated with frequency of [gel-sol] transitions of neuronal groups in the head brain.

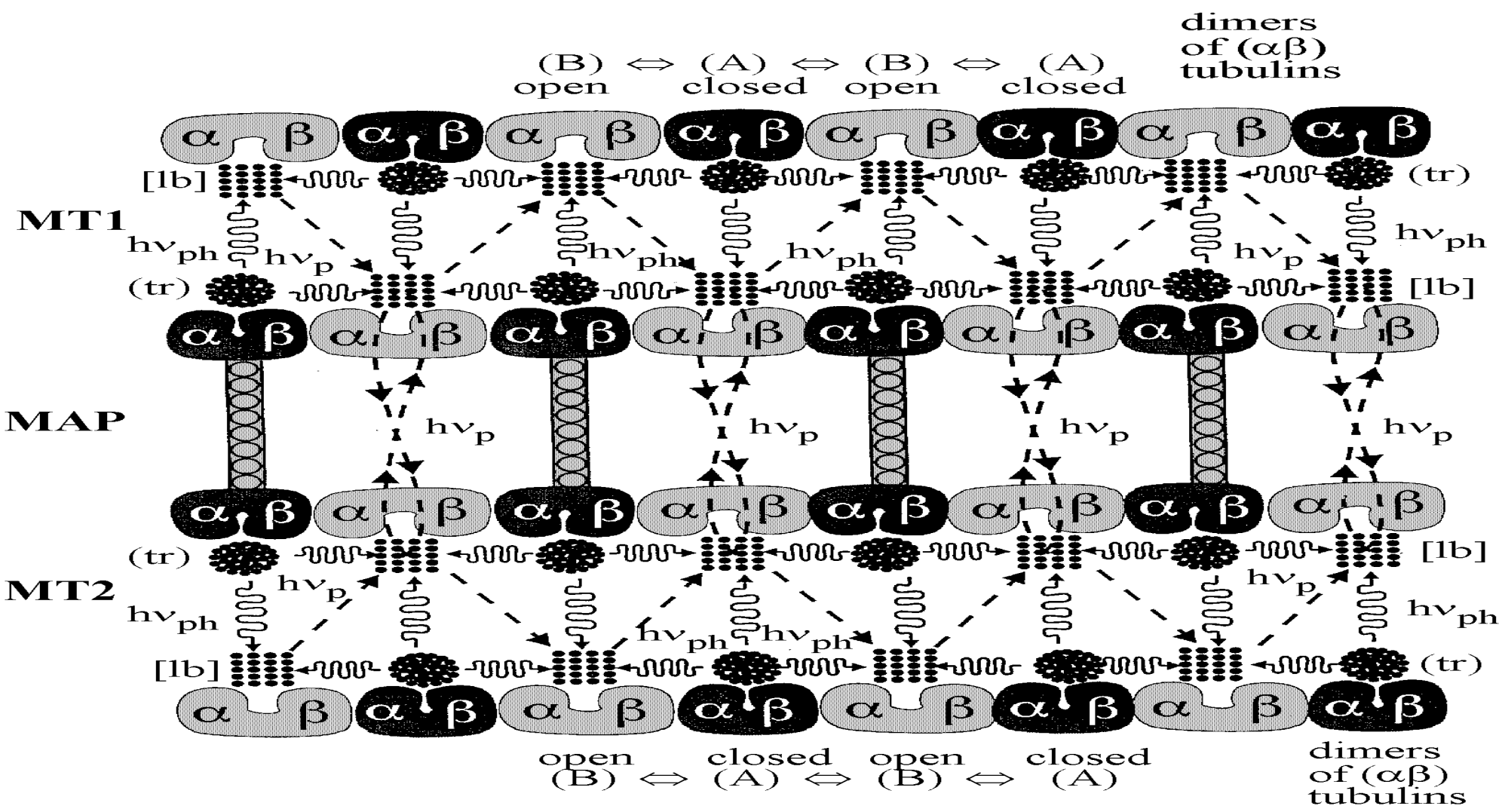

**Figure 12**. The correlation between local, conformational and distant - electromagnetic interactions between pairs of tubulins and microtubules (MT1 and MT2), connected by MAP by mean of librational IR photons exchange.
The dynamics of $\left[ increasing \rightleftharpoons decreasing \right]$ of the entangled water mass in state of macroscopic BC in the process of elementary act of consciousness is a result of correlated shift of dynamic equilibrium between open and closed cavities between alpha and beta tubulins. As a result of these cavities transition from the open to closed state the primary *librational (lb) effectons* (coherent water clusters in state of mesoscopic Bose condensation - mBC) disassembly to small primary *translational (tr) effectons* (independent water molecules), induced by transition of the open states of cavities to the closed one. The nonuniform macroscopic entanglement between the remote water clusters in state of mBC is stimulated by coherent IR photons exchange and vibro-gravitational interaction between these clusters.

MAP– microtubules associated proteins stabilize the overall structure of MTs. They prevent the disassembly of MTs in bundles of axons and *cilia* in a course of their coherent bending. In neuron's body the concentration of MAP and their role in stabilization of MTs is much lower than in cilia (Kaivarainen, 1995, 2003).

The distant electromagnetic and vibro-gravitational interactions between different MT are the consequence of IR photons and coherent vibro-gravitational waves exchange. The corresponding two types of waves are excited as a result of correlated ($a \Leftrightarrow b$) transitions of water primary librational effectons, localized in the open B- states of ($\alpha\beta$) clefts. Frequency of ($a \Leftrightarrow b$) transitions and corresponding superradiated IR photons is about $2 \times 10^{13}\ s^{-1}$. It is much higher, than frequency of transitions of clefts of $\alpha\beta$ tubulin dimers between open and closed states.

Max Tegmark (2000) made evaluation of decoherence time of neurons and microtubules for analyzing the correctness of Hameroff-Penrose idea of wave function collapsing as a trigger of neurons ensembles axonal firing.

The following three sources of decoherence for the ions in the act of transition of neuron between 'resting' and 'firing' are most effective:

1. Collisions with other ions
2. Collisions with water molecules
3. Coulomb interactions with more distant ions.

The coherence time of such process, calculated from this simple model appears to very short: about $10^{-20}$ s.

The electrical excitations in tubulins of microtubules, which Penrose and others have suggested may be relevant to consciousness also where analyzed. Tegmark considered a simple model of two separated but superimposed (entangled) positions of kink, travelling along the MT with speed higher than 1 m/s , as it supposed in Hameroff-Penrose (H-P) model. The life-time of such quantum state was evaluated as a result of long-range electromagnetic interaction of nearest ion with kink.

His conclusion is that the role of quantum effects and wave function collapsing in Hameroff-Penrose model is negligible because of very short time of coherence: $10^{-13}$ s for microtubules.

Hagan, Hameroff and Tuszynski (2002) responded to this criticism, using the same formalism and kink model. Using corrected parameters they get the increasing of life-time of coherent superposition in H-P model for many orders, up to $10^{-4}$ s. This fits the model much better.

Anyway the approach, used by Tegmark for evaluation of the time of coherence and entanglement is not applicable to our model.

It follows from our approach that even very short life-time of oscillating semivirtual macroscopic Bose condensation can be effective for realization of entanglement and nonlocal interaction in the brain.

*The additional experimental verification of the Hierarchic Model of Consciousness "in vitro"*

It is possible to suggest some experimental ways of verification of Elementary Act of Consciousness using model systems. The important point of Elementary Act of Consciousness is stabilization of highly ordered water clusters (primary librational effectons) in the hollow core of microtubules. One can predict that in this case the IR librational bands of water in the oscillatory spectra of model system, containing sufficiently high concentration of MTs, must differ from IR spectra of bulk water as follows:

- the shape of librational band of water in the former case must contain 2 components: the first one, big and broad, like in bulk water and the second one small and sharp, due to increasing coherent fraction of librational effectons. The second peak should disappeared after disassembly of MTs with specific reagents;
- the velocity of sound in the system of microtubules must be bigger, than that in disassembled system of MTs and bulk pure water due to bigger fraction of ordered ice-like water;
- all the above mentioned parameters must be dependent on the applied electric potential, due to piezoelectric properties of MT;
- the irradiation of MTs system in vitro by ultrasonic or electromagnetic fields with frequency of super-deformons excitation of the internal water of MTs at physiological temperatures $(25 - 40^0 C)$ :

$$\nu_s = (2 - 4) \times 10^4 \ Hz$$

have to lead to increasing the probability of disassembly of MTs, induced by cavitational fluctuations. The corresponding effect of decreasing turbidity of MT-containing system could be registered by light scattering method.

Another consequence of super-deformons stimulation by external fields could be the increasing of intensity of radiation in visible and UV region due to emission of corresponding ”biophotons” as a result of recombination reaction of water molecules:

$$HO^- + H^+ \overset{h\nu}{\rightleftharpoons} H_2O$$

Cavitational fluctuations of water, representing in accordance to our theory super-deformons excitations, are responsible for dissociation of water molecules, *i.e.*

elevation of protons and hydroxyls concentration. These processes are directly related to sonoluminiscence phenomena.

The coherent transitions of $(\alpha\beta)$ dimers, composing MTs, between "closed" (A) and "open" (B) conformers with frequency $(\nu_{mc} \sim 10^7\ s^{-1})$ are determined by frequency of water macroconvertons (flickering clusters) excitation, localized in cavity between $\alpha$ and $\beta$ tubulins. If the charges of (A) and (B) conformers differ from each other, then the coherent $(A \rightleftharpoons B)$ transitions generate the vibro-gravitational and electromagnetic field with the same radio-frequency. The latter component of biofield could be detected by corresponding radio waves receiver.

*15.2 The entropy-driven information processing*

It follows from our Cycle of Mind, that changes of system of electromagnetic, acoustic and vibro-gravitational 3D standing waves, stimulating interconversions between mesoscopic and macroscopic Bose condensation of the internal water of MTs change the informational content of this water.

This process induces redistribution of probabilities of different water excitations in huge number of microtubules. It means corresponding change of informational entropy $< I >$, related to microtubules in accordance with known relations (Kaivarainen 2006; 2008; http://arXiv.org/abs/physics/0003107):

$$< I > \; = \sum_i P_i \lg(1/P_i) = -\sum_i P_i \lg(P_i) \qquad 15.1$$

where: $P_i$ is a probability of the $(i)$ state (excitation) with energy $(E_i)$, defined as:

$$P_i = \frac{\exp(-\frac{E_i}{kT})}{\sum_i \exp(-\frac{E_i}{kT})} \qquad 15.2$$

For the total system the well known relation between entropy (S) and information (I) is:

$$S(e.u.) \; = k\, lnW = (k\, ln2)I = 2.3 \times 10^{-24} I \;\; (bit) \qquad 15.3$$

where statistical weight of macrosystem:

$$W = \frac{N!}{N_1!N_2!\ldots N_i!} \qquad 15.4$$

the total number of internal water molecules in macrosystem of interacting MT is: $N = N_1 + N_2 + \ldots + N_i$;

$[i]$ is number of non degenerated states of 24 quasiparticles of intra MT water.

The information amplitude of condensed matter [http://arXiv.org/abs/physics/0003107] we introduce as a product of the number of molecules $(N_i)$ in each of $[i]$ collective excitations:

$$N_m^i = \mathbf{V}_i/\mathbf{v}_{H_2O} = (1/n_i)/(V_0/N_0) \qquad 15.5$$

and informational entropy (15.1):

$$< \mathbf{IA} > \; = \; < I > N_m^i = -\frac{N_0}{V_0} \sum_i \frac{P_i}{n_i} \lg_2(P_i) \qquad 15.6$$

where $N_0$ and $V_0$ are the Avogadro number and molar volume of water; $\mathbf{V}_i = 1/n_i$ is a volume of excitation of ($i$)-type; $n_i$ is concentration of corresponding excitations.

The distant IR photons exchange between water in MT and oscillation between mesoscopic and nonuniform macroscopic Bose condensation, accompanied by the change of probability $P_i$ of different excitations can be considered as an informational exchange between nerve cells. It is accompanied by change of fractions of water excitations in system of interacting MTs, providing conditions for macroscopic entanglement between coherent water clusters.

### *15.3 Actual and Virtual Self.*
### *Definitions of Soul, Global Consciousness & Superconsciousness*

The individual actual consciousness can be considered as a result of dynamic exchange interference between the *Actual Self and Virtual Self.*

*The Actual Self* can be defined as our complex *[brain + nerve system + body]* system.

*The Virtual Self* is a multiplicated Virtual Replicas (VR) of the Real Self, representing superposition of the ether virtual replica ($\mathbf{VR}_{eth}$) or *Ether body*, the astral virtual replica ($\mathbf{VR}_{ast}$) or *Astral body* and mental one ($\mathbf{VR}_{men}$) or *Mental body*, described earlier:

$$\mathbf{Virtual\ Self} = \mathbf{VR}_{eth} + \mathbf{VR}_{ast} + \mathbf{VR}_{men} \qquad 15.7$$

On molecular level the Actual Self can be characterized by the interrelated sub-systems of DNA and microtubules of each cell of the whole organism, especially neurons:

$$\sum(\mathbf{DNA} + \mathbf{MT}s) \qquad 15.8$$

The interaction of the Real Self (physical body) with Virtual Self can be modulated by Virtual Replica (VR) of the Universe:

$$\mathbf{Virtual\ Replica\ of\ the\ Universe\ (VRU)} = \sum^{i=\infty} \mathbf{VR}_i\ \mathbf{of\ all\ objects\ of\ the\ Universe} \qquad 15.9$$

**VRU is** a superposition of virtual replicas of all material objects, including living organisms, planets, stars, galactics with their black holes, as a powerful source of nonlocal interaction. We suppose, that **VRU** works as a quantum supercomputer or *Superconsciousness* or *Universe Self* (God). Schematically interaction between Real Self, Virtual Self and the Universe Self can be presented as:

$$\frac{\sum(\mathbf{DNA} + \mathbf{MT}s)}{\mathbf{Real\ Self}} \rightleftharpoons \frac{[\mathbf{VR}_{eth} + \mathbf{VR}_{ast} + \mathbf{VR}_{men}]}{\mathbf{Virtual\ Self}} \rightleftharpoons \frac{\mathbf{VRU} = \sum^{i=\infty} \mathbf{VR}_i}{\mathbf{Universe\ Self\ (GOD)}} \qquad 15.10$$

The notion of '*Soul or Spirit'* we introduce as a result of interference of *Virtual Self* with virtual replica of the Universe (**VRU**) or *Universe Self*:

$$\mathbf{Soul} = \frac{[\mathbf{VR}_{eth} + \mathbf{VR}_{ast} + \mathbf{VR}_{men}]}{\mathbf{Virtual\ Self}} \rightleftharpoons \frac{\mathbf{VRU} = \sum^{i=\infty} \mathbf{VR}_i}{\mathbf{Universe\ Self\ (GOD)}} \qquad 15.11$$

It follows from this definition of Soul, that it can be independent on Real Self and exist even after death. Its living time can be as long as one of the Virtual Self, representing complex superposition of standing virtual pressure waves ($\mathbf{VPW}^{\pm}$), modulated by Real Self.

The *Virtual Self*, modulated by *Superconsciousness and global Virtual Replica of the*

*Earth*, acts on elementary particles of

$$\sum[\mathbf{microtubules\ of\ centrioles} + \mathbf{internal\ water\ \ clusters}\ (\mathbf{mBC})] \qquad 15.12$$

of the nerve cells of Real Self via paired sub-elementary particles of triplets of the electrons, protons and neutrons of coherent water clusters in MTs in the process of their $\mathbf{C} \rightleftharpoons \mathbf{W}$ pulsation:

$$\langle[\mathbf{F}_{\uparrow}^{+} \bowtie \mathbf{F}_{\downarrow}^{-}]_{W} + (\mathbf{F}_{\uparrow}^{-})_{C}\rangle \rightleftharpoons \langle[\mathbf{F}_{\uparrow}^{+} \bowtie \mathbf{F}_{\downarrow}^{-}]_{C} + (\mathbf{F}_{\uparrow}^{-})_{W}\rangle_{e,p,n} \qquad 15.13$$

This back action of modulated Virtual Self may increase or decrease the probability of superdeformons (cavitational fluctuations) in the internal water of microtubules. It shifts the equilibrium of *assembly* $\rightleftharpoons$ *disassembly* of microtubules in 'tuned' ensemble of neurons to the right or to the left, correspondingly.

This shift is accompanied by the in-phase *gel→ sol* transitions in cytoplasm of nerve cells bodies and their shrinking or swelling, providing Cycle of Mind described above.

The mechanism proposed realize the interaction between Virtual and Actual Worlds.

### *15.4 The Global Virtual Replica, Noosphere and Global Consciousness*

The Global Virtual Replica (GVR) of the Earth is a part of Superconsciousness. The GVR is a result of superposition of all VR of the oceans, mountains, rocks and the internal structure of the planet. The GVR includes all kind of biosystems: human beings, animals, fishes, plants, microorganisms, etc. is a part of Superconsciousness. The sum of *Virtual Selves* of humans creates the *Noosphere,* responsible for *Global Consciousness* as a part of Global Virtual Replica.

The existence of Global Consciousness or Noosphere is confirmed by convincing statistical data, obtained by Global Consciousness Project, headed by Roger Nelson in Princeton (USA). In this very interesting and important project, the global net of Random Event Generators (REG) is used.

Different kind of REG rely on quantum tunneling to produce an unpredictable, broad-spectrum *white noise* in the form of low-amplitude voltage fluctuations in resistors. The Princeton Engineering Anomalies Research (PEAR) laboratory use portable REG, which was designed by John Bradish of the PEAR team. This device uses "thermal noise" in resistors which is a quantum level phenomenon that produces broad-spectrum voltage fluctuation.

On numerous examples it was confirmed, that the strong emotions of big population of people, for example, at the Earthquakes, tsunami, hurricanes or dramatic events, like September 11 and death of princess Diana - influence on the output of REG, mostly nearby the region of event, but not necessarily, if such event is broadcasting widely by numerous TV stations at same time.

*There is no explanation for these amazing results in the framework of existing paradigm. However, we may try to do it in the framework of our approach, using introduced notion of the Global virtual replica (GVR) of the Earth and Noosphere, as a part of it.*

In accordance to our theory of Virtual Replica of any material object - from elementary particle to the neuron, it represents a result of interference of modulated by de Broglie waves of the object basic virtual pressure waves $(\mathbf{VPW}_0^{\pm})_m$ with non modulated $\mathbf{VPW}_0^{\pm}$. Like the regular optical hologram, this Virtual Replica (quantum hologram) is a systems of standing waves. However, these special virtual standing waves are the result of correlated transitions of Bivacuum dipoles with nonzero symmetry shift between one of the excited and ground states (see section 'Virtual pressure waves'). The reasons of minor symmetry shift of Bivacuum dipoles of Global VR, different in remote location on the earth surface, can be at least following:

a) induced by different gravitational and magnetic potentials of the Earth in these locations, for example before and in the process of eruptions or earthquakes;

b) induced by specific excitations of neurons and microtubules + internal water clusters in states of strong emotional stress, correlated in big population of humans. This population have to experience simultaneously the different kind of emotional event, like in New York at September 11 or at final game of World cup in football, translated in real time over lot of countries;

c) the relative position of Sun, Moon, other planets and Earth also influence GVR;

d) under the influence of asymmetric structures like pyramids or cones.

In accordance to our theory of mass and charge origination without Higgs bosons, even small symmetry shift in Bivacuum dipoles is accompanied by nonzero mass and charge origination. These all-penetrating dipoles can screen the Coulomb interaction between the electrons and ions in lattice of materials, including resistors as a sensitive part of REG. In fact, this means increasing the permittivity of Bivacuum.

As a result, the random "thermal noise" in resistors, resulting voltage fluctuation in range: 1000 - 30.000 Hz becomes less random. It can be a consequence of creation of preferable direction of quantum tunneling under the influence of Bivacuum dipoles with certain degree and selected direction of symmetry shift. In such a way the deviations of normal statistical distribution registered by net of REGs, distributed over the planet, can be explained.

To verify this conjecture of correlation between REG outcome and asymmetry of VR, interacting with its resistor, this author propose to compare REG data inside and outside the pyramids (asymmetric geometry and VR) and balls (symmetric geometry and VR) of different dimensions, materials and walls dimensions. If our conjecture is correct a small difference of these data should be revealed.

### *15.5. Possible mechanism of remote IDEAS exchange via Bivacuum Mediated Interaction*

The experimental data are existing, that the conscious evaluation of some event by one brain tends to reduce the element of surprise for other brains, observing the same event. These most interesting data can be explained in framework of our approach, as two stage process, involving hierarchy of virtual replicas.

*The 1st stage* is represented by following scheme of modification of Star system virtual replica (SVR), including and depending on the Earth virtual replica (**EVR**) and Sun system Superconsciousness (SSC): [$\mathbf{SVR} = \mathbf{f(EVR)} \rightarrow \mathbf{SVR}^*$ **and** $\mathbf{SSC} \rightarrow \mathbf{SSC}^*$], induced by change of mental and internal-ether physical VR of human brain and nerve system, as a result of 'leaning - getting the same new information' by one or few human beings (*j*) via external astral and mental secondary **VRM**:

$$\sum_j [\mathbf{VRM}_{mental} + \mathbf{VR}_{ether}] + \mathbf{Info} \rightarrow \sum_j [\mathbf{VRM}^*_{mental} + \mathbf{VR}^*_{ether}] \rightarrow \qquad 15.14$$

$$\rightarrow \sum_j \mathbf{VRM}^*_{astral} \rightarrow [\mathbf{EVR}^* \bowtie \mathbf{SVR}^*] \rightarrow \mathbf{SSC}^*$$

*The 2nd stage* is related with action of modified Solar system Superconsciousness ($\mathbf{SSC}^*$) and the Earth virtual replica (EVR) on the individuals human consciousness via their external astral ($VRM_{astral}$), mental ($VRM_{mental}$) and corresponding internal - ether ($VR_{ether}$) virtual replicas of another ($k$) persons, accompanied by transmission of Information:

$$[\mathbf{SSC}^* \bowtie \mathbf{EVR}^*] \overset{Info}{\rightarrow} \sum_k [\mathbf{VRM}_{astral} + \mathbf{Info}] \rightarrow \sum_k \mathbf{VRM}^*_{astral} \rightarrow \sum_k [\mathbf{VRM}^*_{mental} + \mathbf{VR}^*_{ether}] \qquad 15.15$$

The mental and ether virtual replicas: $\mathbf{VR}_{mental}$ **and** $\mathbf{VR}_{ether}$ are the external and internal reflections of quantum neurodynamics process in brain and the nerve system, correspondingly. Each elementary act of consciousness, as a transition between discreet neuron's ensemble states, should be accompanied by corresponding transition between mental and ether VR, following by emission or absorption of number ($N$) of quantum of information (*qubit* – **qB**) quantum states:

$$[\mathbf{VRM}_{mental} + \mathbf{VR}_{ether}]_j - [\mathbf{VRM}_{mental} + \mathbf{VR}_{ether}]_k = \mathbf{N\,qB} \qquad 15.16$$

We may postulate, that $N_{\min}qB$ is a minimum information ($N_{\min}qB = \min Info$), which can irreversibly change the mental and astral virtual replica:
$\left[(\mathbf{VRM}_{mental} \bowtie \mathbf{VRM}_{astral}) + N_{\min}qB\right] \rightarrow VMR^*_{astral}$ and perturbation of the Earth virtual replica ($\mathbf{EVR}^*$), then Solar virtual replica ($\mathbf{SVR}^*$) and finally variation of Superconsciousness ($\mathbf{SC}^*$):

$$[\mathbf{EVR} \bowtie \mathbf{SVR} \bowtie \mathbf{SC}] \overset{N_{\min}qB}{\rightarrow} [\mathbf{EVR} \bowtie \mathbf{SVR}^* \bowtie \mathbf{SC}^*] \qquad 15.17$$

If such a model is right, then we share with mankind all well formulated by individuals IDEAS. The IDEA, as a coherent and stable superposition of $N_{\min}qB$ :

$$\mathbf{IDEA} = \sum_k c_k N_{\min} qB_k$$

may influence on the Earth and Solar virtual replicas (EVR and SVR), then Superconsciousness and then, via feedback reaction it affect the individual consciousness of all the rest of human beings.

Our description of the Earth Virtual Replicas of the Universe, Solar system and Earth, their interrelation with duality of matter and corresponding levels of Superconsciousness, have some common features with One Mind Model of Mark Germine (2002) and Henry Stapp (1982) hypothesis of interrelation between collapsing of superimposed mental states and actual world.

### *15.6 Evolution of Superconsciousness*

A complex Hierarchical system of Virtual Replicas Multiplication: $\sum \mathbf{VRM}(\mathbf{r},\mathbf{t})$ of the planets of Star systems, black holes, the galactics and their clusters may be considered as the Hierarchical quantum Supercomputer or Superconsciousness, able to simulate all probable situations of virtual future and past.

This becomes possible in conditions of time uncertainty: $t = 0/0$, when the translational velocity of Bivacuum bosons or Cooper pairs of Bivacuum fermions, composing Virtual Replicas is zero ($\mathbf{v} = \mathbf{0}$) and the accelerations of these dipoles in the volume of $\sum \mathbf{VRM}(\mathbf{r},\mathbf{t})$ are also zero ($\mathbf{dv/dt} = \mathbf{0}$). This follows from our formula for time, including not only the velocity, like in special relativity, but also acceleration:

$$\mathbf{t} = -\frac{\vec{\mathbf{v}}}{\mathbf{d}\vec{\mathbf{v}}/\mathbf{dt}} \frac{1-(\mathbf{v}/\mathbf{c})^2}{2-(\mathbf{v}/\mathbf{c})^2} \qquad 15.18$$

Our theory admits a possibility of feedback reaction between the iterated $\mathbf{VRM}(\mathbf{r},\mathbf{t})$ and primary one **VR** and then between primary VR and the object itself. If this mechanism works, then most probable events anticipation by sensitive physical detectors and human beings (psychics) is possible.

This may explain the reproducible results of the unconsciousness response (by changes of human skin conductance) of future events (presponse), obtained by Dick Bierman and Dean Radin (2002).

However, in these experiments the possibility of influence of intention of participant on random events generator (REG), choosing next photo (calm or emotional) also should be taken into account. Corresponding weak influence of humans intention on REG was demonstrated in long term studies of Danne and Jahn (2003).

*In contrast to virtual time, the reversibility of real time looks impossible,* as far it needs the reversibility of all dynamic process in Universe due to interrelations between closed systems of different levels of hierarchy. It is evident that such 'play back' of the Universe history needs the immense amount of energy.

*All three described Virtual Replicas: Ether, Astral and Mental bodies are interrelated with each other and physical body.*

The experimental evidences are existing, that between properties of the *Ether* bodies and corresponding physical bodies of living organisms or inorganic matter, the correlation takes a place. It is confirmed by the Kirlian effect, reflecting the ionization/excitation threshold of the air molecules in the volume of Ether and Astral bodies.

The primary virtual replica multiplications: distant ($\mathbf{VRM}^{dis}$) and nonlocal ($\mathbf{VRM}^{nl}$) (see section 8) of human's brain and nerve system provide the *astral and mental memories*, interrelated with *ether memory*.

In turn, the combination of the individual *astral and mental memories* can be incorporated into complex Virtual Replica Multiplication (**VRM**) of the whole Earth/Planet: $\mathbf{VRM}_{plan}$ .

It means that, the individual human's $\mathbf{VRM}(\mathbf{r},\mathbf{t})$, related to elementary act of memorization and consciousness, can be incorporated in holographic structure of planetary $\mathbf{VRM}_{plan}$.

Such a mechanism stands for a imprinting of the individual mental activity to planetary $\mathrm{VRM}_{planet}$, i.e. the NOOSPHERE formation.

*A kind of filters and thresholds*, preventing 'downloading' to **Earth VRM** (Noosphere) the information in form of standing $\mathbf{VPW}^{\pm}$, of already existing or 'destructive' info, should exist. The principle of selection of "valuable" new information/perturbation can be based, for example, on criteria of Golden mean or Hidden harmony of primary VRM construction, as a background principle of [Bivacuum +Matter] dynamic and spatial self-organization.

The notion of "Collective Informational Bank" in such approach means formation of **Virtual Replica of the Universe** (**VRU**) from huge number of planets ($\mathbf{VRM}_{plan}$) and stars ($\mathbf{VRM}_{star}$) of the Universe.

The ability of processing of information may be a consequence of ability of the Universe Virtual Replica: to self-organization, evolution, devolution and formation of metastable states.

The feedback reaction between different components of Hierarchical **UVR**, can be considered, as a Superconsciousness (**SC**) of the Universe".

*Taking into account a huge scale of the Universe, the effective interaction between*

*remote 'nerve centers' of the universe can be realized only using the mechanism of nonlocality.*

*The Superconsciousness (SC) have a properties of Quantum Supercomputer*, able to processing of information, memorizing and organizing it by certain principles. This can be considered as the evolution of Superconsciousness.

**TABLE**

**The Stages of Self-Organization and Evolution of the Universe, Mediated by Virtual Pressure Waves ($\mathbf{VPW}_q^{\pm}$)**

---

$$\text{(I)}\ \ \mathbf{BIVACUUM} \overset{\mathbf{Bivacuum\ Symmetry\ Breach}}{<\!=\!=\!=\!=\!=\!=\!=\!=\!=\!=\!=\!=\!>} [\mathbf{MATTER + FIELDS} + \textstyle\sum \mathbf{VRM\ OF\ MATTER}] \rightleftharpoons$$

$$\text{(II)}$$

$$\mathbf{BIOSYSTEMS} \overset{\mathbf{Consciousness}}{<\!=\!=\!=\!=\!=\!=\!=\!>} [\mathbf{COMPLEX\ ORGANISMS\ (CO)} + \textstyle\sum \mathbf{VRM(r,t)\ of\ CO}] \rightleftharpoons$$

$$\text{(III)}\ \ [\textstyle\sum \mathbf{VRM\ OF\ MATTER} + \sum \mathbf{VRM\ of\ CO}] \overset{\mathbf{self\!-\!organization\ of\ Bivacuum+Matter}}{<\!=\!=\!=\!=\!=\!=\!=\!=\!=\!=\!=\!=\!=\!=\!=\!=\!=\!=\!>} \ (\mathbf{IV})$$

$$\text{(IV)}\ \ \mathbf{UNIVERSE\ VR\ (VRU)} \overset{\mathbf{self\!-\!organization\ of\ UVR\ to\ Quant\ Supercomp}}{<\!=\!=\!=\!=\!=\!=\!=\!=\!=\!=\!=\!=\!=\!=\!=\!=\!=\!=\!=\!=\!=\!>} \mathbf{SUPERCONSCIOUSNESS\ (SC)}$$

---

where: **VRM(r,t)** is virtual replicas multiplication in space and time; **CO** means complex organisms; **UVR** means the Universe Virtual Replica, self-organizing to Superconsciousness (SC) with properties of Quantum Supercomputer.

The important contribution to realization of *Superconsciousness, as a self-organizing quantum supercomputer,* is related to superposition of nonlocal and temporal components of hierarchical systems of Virtual Replicas Multiplications **VRM(r, t)** and Virtual Guides ($\mathbf{VirG}_{SME}$) formation, providing different kind of Bivacuum mediated interaction (**BMI**).

The complex system of bundles of virtual microtubules, assembled from Bivacuum dipoles, like the axons in humans body, exchanging the information, momentum and energy between remote complex objects of the Universe, containing coherent regions, i.e. neutron stars, black holes in centers of galactics, etc. is a crucial part of the Universe Quantum Supercomputer or Superconsciousness.

The possibility of feedback reaction between *Superconsciusness* and *Planet Noosphere*, following from our concept, means that at proper conditions of Universe Virtual Replica (**UVR**) high instability (bifurcation point), even small perturbation of planet $\mathbf{VRM}_{planet}$ may influence Superconsciousness.

The uncertainty of time is valid also in part of Universe Virtual Replica (**UVR**), presented by bundles of filaments of Virtual Guides: $\mathbf{N}(\mathbf{VirG}_{SME})$, formed by polymerized symmetric $\sum \mathbf{BVB}^{\pm}$ or Cooper pairs of Bivacuum fermions $\sum[\mathbf{BVF}^{\uparrow} \bowtie \mathbf{BVF}^{\downarrow}]$, making it possible simulation of past and future events by Quantum supercomputer, as a part of Superconsciousness.

*Variation of Bivacuum symmetry shift by physical fields (electrostatic, magnetic or gravitational) may influence the value and even sign of pace of time of virtual replicas, i.e. their future or past.*

The time-dependent superposition of individual virtual replicas of inorganic matter and living organisms (if any) of each star (solar) system forms the Star system Virtual Replica (**SVR**) and a Star system consciousness (**SSC**). In turn, superposition of all stars systems virtual replica **SVR** and **SSC**, as a function (**F**) of **SVR**, of the Galactic can be responsible for formation of Galactic Consciousness (**GC**):

$$\mathbf{Galactical\,Consciousness\,(t)} = \mathbf{F(GVR(t)]} = \sum \mathbf{SSC(t)} \qquad 15.19$$

$$\mathbf{where}:\ \mathbf{SSC(t)} = \mathbf{F(SVR)} = \mathbf{F}\left\{\sum\ [\mathbf{VRM}_M(t) \Leftrightarrow \mathbf{VRM_L}(t)]\right\}$$

where: $\mathbf{VRM}_M(t)$ and $\mathbf{VRM_L}(t)$ are the selected/individual multiplicated virtual replicas of material objects and living organisms, in the case of their existence in given star system.

Due to virtuality of superposition of $[\mathbf{VRM}_{M,L} = \mathbf{VRM}_M(\mathbf{t}) \Leftrightarrow \mathbf{VRM_L}(\mathbf{t})]$ and $\mathbf{SVR(t)}$, they do not follow special theory of relativity, by definition, and consequently the causality principle.

As a result of ability of *star virtual replicas (SVR)* to evolution/self-organization in Bivacuum, as in active medium in each selected current moment of time $t = t_C$, we have an infinitive number of discreet projections of current *SVR* (most stable in this moment) and states of *star systems consciousness (SSC)* on the Future time ($t_F$) :

$$\mathbf{SSC}_{Future} = \mathbf{F}\left\{\sum^{\infty} SVR_{Future}(t_F)\right\} = \mathbf{F}\left\{\sum^{\infty}\sum^{\infty} VR_{Future}(t_F)\right\} \qquad 15.20$$

In similar way we have the infinitive number of metastable 'memorized' multiplicated Virtual replicas $VRM_P(t)$ of the past time ($t_P$):

$$\mathbf{SSC}_{Past} = \mathbf{F}\left\{\sum^{\infty} SVR_{Past}(t_P)\right\} = \mathbf{F}\left\{\sum^{\infty}\sum^{\infty} VRM_{Past}(t_P)\right\} \qquad 15.20$$

Each selected $\mathbf{SSC}(\mathbf{t}_F, \mathbf{t}_P)$ of the future and past time (evolution and devolution of virtual hologram is dependent on the virtual future and past time quantization with current time ($t_C$), as a reference point:

$$t_F = t_C + \sum_{q=1}^{q=\infty} \Delta t_q \qquad 15.21$$

$$t_P = t_C - \sum_{q=1}^{q=\infty} \Delta t_q \qquad 15.21a$$

where $\Delta t_q$ is a quantum of time (minimum time interval) separating stable states of standing virtual waves, forming instant Star system virtual replicas.

The quanta of information is emitted or absorbed in the process of quantum jump between $SVR_{q+1}$ and $SVR_q$ by analogy with emission/absorption of quanta of energy (photons), as a result of quantum transitions between different electronic states of atoms and molecules.

*In any kind of virtual systems, like VR and VRM(r,t), the simulation of past and future events is possible by Quantum supercomputer or Superconsciousness (see Table 2), because by definition, the relativistic mechanics do not work for them and the causality principle is absent.*

Shifting of Bivacuum dipoles symmetry by physical fields (electric or gravitational), accompanied by their acceleration/deceleration and acquiring the external kinetic energy, may influence the value and even sign of pace of time of virtual replicas, i.e. their future or past.

The time-dependent superposition of individual virtual replicas multiplication of inorganic objects $\mathbf{VRM_{Ob}}(t)$ and living organisms $\mathbf{VRM_L}(t)$ of each star (solar) system forms the Star system Virtual Replica (**SVR**). In turn, superposition of all stars systems virtual replica SVR of Galactic can be responsible for formation of Galactic Virtual Replica.

We put forward a conjecture, that as a result of ability of complex virtual replica, like human's one or Galactic virtual replica **GVR(t)** with active medium properties to evolution/self-organization and informational processing, this process can be considered as virtual supercomputer or Galactic Virtual Consciousness (**GVC**):

$$\mathbf{GVC\,(t)} = \mathbf{F(GVR(t)]} \;= \sum \mathbf{SSC(t)} \qquad 15.22$$

$$where: \;\; \mathbf{SVR} = \mathbf{F}\left\{\sum \,[\mathbf{VRM_{Ob}}(t) \Leftrightarrow \mathbf{VRM_L}(t)]\right\}$$

where: $\mathbf{VRM_{Ob}}(t)$ and $\mathbf{VRM_L}(t)$ are selected virtual replicas multiplication of inorganic objects and living organisms, correspondingly, not only in space, but also in time.

It follows from our approach, that Superconsciousness of the Universe is a huge hierarchical virtual supercomputer, resulting from nonlocal Bivacuum mediated interaction (BMI) between Virtual Consciousness of individual planets (like Earth Noosphere), stars and Galactic virtual consciousness, based on their ability to self-organization.

Our description of Virtual Replicas of the Earth, Solar system and Universe and their feedback action on real systems - have some common features with Henry Stapp (1982) hypothesis of interrelation between collapsing of superimposed mental states and actual world.

In each selected current moment of time $\mathbf{t} = \mathbf{t}_C$, as a result of *Virtual consciousness (VC)* activity, a big number of discreet time - *evolution* versions $\mathbf{VR}^i_F(\mathbf{t}_F)$ of current $\mathbf{VR}^0$ with different probability ($\mathbf{P}_i$) of realization appears. The most probable future system of virtual replica multiplication $\mathbf{VR}_{Future}$ can be calculated as:

$$\mathbf{VR}_{Future} = \frac{\sum P_i\;[VRM^i_F(t_F)]}{\sum P_i} \qquad 15.23$$

In similar way we have the most probable virtual replica of the past infinitive number of possible metastable Virtual replicas $VRM^i_P(t_P)$ of past, as a result of time - *involution* of current $\mathrm{VRM}^0$:

$$\mathbf{VC}_{Past} = \frac{\sum P^i_P\;[VRM^i_P(t_P)]}{\sum P^i_P} \qquad 15.24$$

*Clairvoyance or anticipation* is a result of ability of gifted psychic [Sender] to 'search', using specific 'key images' at first stage - $\mathbf{VRM}^0$ at current time and then select from this future or past set of secondary virtual replicas the most probable one. This complex process includes very 'tuned' interaction of the astral and mental bodies (distant VRM) and ether (local VR) bodies of [Sender] with $\sum^{\infty} SVR_F(t)$. Similar mechanism works, in accordance to our approach, in extra-perception by Sender - psychic of the past of some individual $VRM_P(t)$, as a selected component of $\sum^{\infty} SVR_P(t)$.

The proposed in our work mechanism of interaction of human consciousness of Receptor and related virtual replica with virtual replicas of Sender consciousness, corresponding to certain notions and images, means *telepathy*. Similar mechanism can be responsible for possibility of sharing of well formulated by individuals new *IDEAS* with certain population of mankind, sensitive enough for perception of specific **VRM**, corresponding to certain mental state of this individuals, induced by such ideas or notions.

*The 'phantom' effects* where revealed in a system of interacting 'charged' by intention vessel of water and few other distant vessels with aqueous solutions, surrounding the 'charged' vessel (Tiller, Dibble and Kohane, 2001 and *www.tiller.org*). After replacing the 'charged' vessel far out of system, the *'memory' of its presence* remains for a long time. The presence and orientation of large quartz crystal strongly affected the amplitude of 'phantom' effect. In experiment, described, screening of the target [R] from electromagnetic fields by Faraday's cage did not influence on the distant interaction between [S] and [R] and the phantom effect.

Consequently, there are a lot of experimental evidence already, confirming the existence of new fundamental remote Bivacuum Mediated Interaction (BMI) between Sender and Receiver, following from our approach.

### 16. The examples of Bivacuum mediated remote mental action of Psychic on different physical targets and their explanation

A big number of examples demonstrate how Bivacuum mediated interaction (BMI), generated by gifted psychic - 'Sender' [S], can interfere with real physical fields and targets [R] of non-biological (like water) and biological nature:

*1. Speeding up and slowing down the rate of americium* $^{241}$*Am nuclear decay.*

This points to quantum nature of Paranormal phenomena the influence of Entanglement Channels between the nuclear of neurons of psychic and quarks and gluons of the target in the process of quarks (sub-elementary fermions) [$\mathbf{C} \rightleftharpoons \mathbf{W}$] pulsation.

*2. Rotation of the plane of polarization of laser beam by 7-30*$^0$.

This effect points to perturbation of Bivacuum optical isotropic properties due to opposite change of dynamic equilibrium between Bivacuum fermions and antifermions density in different volumes of space:

$$\mathbf{BVF}^{\uparrow} \rightleftharpoons \mathbf{BVB}^{\pm} \rightleftharpoons \mathbf{BVF}^{\downarrow} \tag{16.1}$$

under the psychic influence, accompanied by magnetic field excitation in Bivacuum, like in Faraday optical effect;

*3. Induction of a pulse magnetic field (100 nT and up to 27x106 nT) by psychic, accompanied by rotation of a compasses needle.*

The explanation of this effect is similar to that, described in item 2.

*4. Deviation of the electrical resistance of thermoresistor.*

This can be explained by Bivacuum mediated interaction (**BMI**) influence on resonance exchange interaction of the electrons with Bivacuum, as a result of deviation of frequency of $\mathbf{VPW}_q^{\pm}$, induced by psychic, from frequency of [$\mathbf{C} \rightleftharpoons \mathbf{W}$] pulsation of the electrons, increasing or decreasing their kinetic energy and velocity.

*5. Induction of a periodic electrical signal from a piezoelectric sensor.*

It can be the result of effect, like described above, combined with influence of BMI on Van der Waals interaction between atoms/ions of lattice, changing it rigidness and probability of thermal fluctuations, accompanied by defects origination in lattice, via resonance exchange of $\mathbf{VPW}_q^{\pm}$ and [$\mathbf{C} \rightleftharpoons \mathbf{W}$] pulsation of the electrons, protons and neutrons.

*6. Moving the plate of an encased precise analytical balance equivalent to 100 mg force.*

This can be a result of momentum and kinetic energy transmission from the nerve cells of psychic to selected coherent atoms of the balance via Virtual Guides of spin, momentum and energy (**VirG**$_{SME}$), forming the Entanglement Channels.

*7. Induction of a temporary peak in the Raman spectrum of tap water at 2200 cm$^{-1}$.*

This can be a consequence of mechanism described above, confirming the existence of **VirG**$_{SME}$, connecting atoms (protons) of water in microtubules of nerve cells of psychic and in water - target (Receiver) and the ability of the *Entanglement Channels* (**EntCh**):

$$\mathbf{EntCh} = \sum^{N} [\mathbf{VirG}_{SME}(\mathbf{S} <==> \mathbf{R})]_n \qquad 16.2$$

influence the internal and external properties of water molecules;

*8. Temporary changes in the microstructure of water as observed through scattering of laser beam ($\lambda = 632.8$ nm) at various angles.*

The same mechanism, as described above.

*9. Deviation of UV adsorption spectra of DNA - water solution in the area of 220- 280 nm.*

The same mechanism as above;

*10. Predetermined by operator deviation from randomness of various random number generators has been revealed in Princeton group.*

This can be a result of asymmetric influence of the Entanglement Channels, generated by the operator on elementary particles of electronics, responsible for randomization of results.

*11. Increasing of the concentration of dislocations (missing atoms in microcrystalline structure) in "metal bending" experiments with local increase of surface hardness.*

The same mechanism, as in the example 4 above, describing the BMI influence on piezoelectric materials.

**Marcel Vogel** claimed that *quartz crystals of certain shape could amplify the mental intent* action on water. He demonstrated that in water, circulating around an intent-charged crystal, the following changes are revealed:

- decreasing in surface tension, increasing of conductivity;
- a significant drop in the freezing point (as low as -30 degrees);
- bidirectional alterations in the pH up to 3 points;
- the appearance of two new bands in the IR and UV absorption spectrum, etc.

These important experiments point to ability of certain materials (crystal in this case) for taping the properties of multiplied Virtual Replica (i.e. secondary VR) of psychic. This process and the following processing/reading of taped properties, have a similarity with fixation and reproduction of regular holograms from photo-materials. *The role of 'reference waves' in reproducing of taped Virtual Replica in crystal play basic Virtual pressure waves ($VPW_q^{\pm}$) of Bivacuum.*

Dean and Brame found that the water, treated by healer, demonstrated changes with both IR spectrophotometry (indicating altered hydrogen bonding) and specific peaks in UV spectra of water, meaning change of the electronic state of water molecules. The half-life for these effects lasted from three days to three years.

It was demonstrated, that mental influence, as a carrier of information, cannot be significantly blocked by any physical screening and that the effect does not depend on the distance. This point to non-electromagnetic nature of mind-matter interaction. The nature of Bivacuum mediated interaction (BMI), provided by the Entanglement Channels satisfy this condition.

The remote influence of psychic on the permanent magnetic field sensor (ferroprobe magnetometer), screened from the external regular EM influence - was revealed by Ageev, Dulnev, Kolmakov, etc. (2003).

The detector of alternating magnetic field did not revealed any significant changes, induced by remote ’sender/psychic’. This in accordance with mechanism, described by equilibrium shift between Bivacuum fermions and antifermions with opposite spins, mentioned above to the left or right (eq. 16.1).

*The active Psi-channel seems to be non-isotropic and strictly directed in space.* The same group revealed that good psychic (sender) is able selectively change the electric potential of the electrodes in the aqueous solution of sodium chloride (100 ml of 0.9% NaCl solution) in one of two vessels, separated from each other for 2 meters only. The distance between psychic and vessels was about 500 kilometers.

Well registered phenomena of remote viewing (RV) and precognition by Hal Puthoff and Rassel Targ at Stanford Research Institute (1996) are confirmed in other laboratories. Our theory of BMI explains the RV phenomena as a result of superposition of Virtual replicas multiplication of Sender (psychic) and Receiver (target):

$$\mathbf{VRM}_S \bowtie \mathbf{VRM}_R \qquad 16.3$$

as a stage, necessary for *Entanglement Channels* formation, providing feedback reaction between target and nerve system of the psychic, including visual centers.

Numerous studies have demonstrated (Targ & Katra, 1998) that size of the target (down to 1 mm square) and distance between sender [S] and target (up to 10,000 miles) do not appear to significantly impair signal perception. The electromagnetic shielding by Faraday cage or sea water does not negatively impact remote viewing ability.

These data, pointing to anisotropic component of BMI, are in accordance with proposed mechanism of nonlocal interaction via the *Entanglement Channels* (see eq. 16.2), connecting coherent elementary particles of Sender and Receiver.

We can see, that the proposed in this work new resonant Bivacuum mediated interaction (BMI) between psychic and target can explain all the above described phenomena. The existing currently paradigm, failed to do this.

### *16.1 The Biological and Biochemical effects of Bivacuum Mediated Interaction between Sender and Receiver*

There are two classes of remote healing (RH):

1) the healer (Sender [**S**]) find the target (Receiver [**R**]) on the basis of a name, location, birth date, etc. (in remote viewing terms, this is ”coordinate”). This process involves superposition of multiplicated Virtual Replica of [**S**] and [**R**]: $\{\mathbf{VRM}_S \bowtie \mathbf{VRM}_R\}$, necessary for recognition and targeting, following by formation of *Entanglement Channels* $= \sum^N [\mathbf{VirG}_{SME}(\mathbf{S} \Longleftrightarrow \mathbf{R})]_n$, connecting coherent clusters of atoms of [*S*] and [*R*] in state of mesoscopic Bose condensation (Kaivarainen, 2008);

2) other way for remote healing is using the *adjunct* object, previously treated by the healer, such as: water, cloth, a crystal, etc., by the patient with or without the healer’s knowledge. In this case the imprinting of VR of healer in adjunct object should take a place. In this case the role of primary VR of the of the healer/sender is realized by secondary VR, imprinted in the adjunct object.

In a 1991 Chien & al. report following biochemical effects, when studying influence of a qigong master Psi-field on the culture of human fibroblasts: 1.8% increase in cell growth rate in 24 hrs; 10-15% increase in DNA synthesis and 3-5% increase in cell protein synthesis during 2 h period.

When the master radiate the ”inhibiting” biofield with corresponding intention, the cell growth decreased by 6%, while DNA and protein synthesis decreased by 20-23%, respectively 35-48%. Intent-modulated emission of *biophotons* from the hands of *qigong* practitioners is a well-known phenomenon that has often been reported in the scientific literature. Wallace reported measuring up to 100 time stronger emissions from the hands of gifted persons compared to controls.

A study by Nakamura & al. (2000) reports an increase in subject’s hand biophotons intensity associated with a drop in skin surface temperature during Qigong practice. The significantly higher (up to 105nT) magnetic signals during Qi emission from hands of qigong practitioners, as compared to the controls were revealed (Lin & Chen, 2001).

It was proposed by Oschman (2000) that under special conditions, resonant brainwaves may entrain the body’s neural system to deliver healing frequencies to diseased tissues, or become coupled to the Schumann resonance and thus transmit distant healing effects to the target.

This author consider Schumann resonance as a non-dissipative factor enhancing ’tuning’ of the macroscopic entanglement between Sender and Receiver.

However, just electromagnetic mechanism of paranormal phenomena fail to explain such effects, as influence of mental activity on internuclear and gravitational forces, described above. *Only multistage fundamental Bivacuum mediated interaction, introduced in this work, may fill the gap between paranormal and normal.*

## 17. The nature of electrostatic and magnetic interaction, based on Unified theory

### *17.1 Electromagnetic dipole radiation as a consequence of charge oscillation*

The [*emission* $\rightleftharpoons$ *absorption*] of photons in a course of elementary fermions - triplets $<[\mathbf{F}_{\uparrow}^{-} \bowtie \mathbf{F}_{\downarrow}^{+}]_{S=0} + (\mathbf{F}_{\uparrow}^{+})_{S=\pm 1/2}>^{e,\tau}$ vibrations can be described by known mechanism of the electric dipole radiation ($\boldsymbol{\varepsilon}_{EH}$), induced by charge acceleration ($a$), following from Maxwell equations (Berestetsky, et. al.,1989):

$$\boldsymbol{\varepsilon}_{EH} = \frac{2e^2}{3c^3}a^2 \qquad 17.1$$

The resulting frequency of $[C \rightleftharpoons W]$ pulsation of each of three sub-elementary fermions in triplets is a sum of internal frequency contribution ($\boldsymbol{\omega}_0^{in}$) and the external frequency ($\boldsymbol{\omega}_B$) of de Broglie wave from:

$$[\boldsymbol{\omega}_{C\rightleftharpoons W} = \boldsymbol{\omega}_0^{in} + \boldsymbol{\omega}_B]^i \qquad 17.2$$

The acceleration can be related only with external translational dynamics which determines the empirical de Broglie wave parameters of particles. Acceleration is a result of alternating change of the charge deviation from the position of equilibrium: $\Delta\boldsymbol{\lambda}_B(\mathbf{t}) = (\boldsymbol{\lambda}_B^t - \boldsymbol{\lambda}_0)\sin\boldsymbol{\omega}_B\mathbf{t}$ with de Broglie wave frequency of triplets: $\omega_B = \hbar/(m_V^+ L_B^2)$, where $L_B = \hbar/m_V^+\mathbf{v}$. It is accompanied by oscillation of the instant de Broglie wave length ($\boldsymbol{\lambda}_B^t$).

The acceleration of charge in the process of $\mathbf{C} \leftrightharpoons \mathbf{W}$ pulsation of the anchor $\mathbf{BVF}_{anc}^{\updownarrow}$ can be expressed as:

$$\mathbf{a} = \boldsymbol{\omega}_B^2 \Delta\boldsymbol{\lambda}_B(\mathbf{t}) \qquad 17.3$$

$$\mathbf{a} = \omega_B^2(\lambda_B^t - \lambda_0)\sin\,\omega_B \mathbf{t} \qquad 17.4$$

where: $\lambda_B^t = \mathbf{2\pi L}_B^t$ is the instant de Broglie wave length of the particle and $\lambda_0 = \mathbf{h}/\mathbf{m}_0\mathbf{c}$ is the Compton length of triplet of the electron, proton or neutron.

The intensity of dipole radiation of pulsing $\mathbf{BVF}_{anc}^{\updownarrow}$ from 17.2 and 17.4 is:

$$\varepsilon_{EM} = \frac{2}{3c^3}\omega_B^4\,(\mathbf{d}_\mathbf{E}^t)^2 \qquad 14.5$$

where the oscillating electric dipole moment is: $\mathbf{d}_\mathbf{E}^t = \mathbf{e}(\lambda_B^t - \lambda_0)$.

Consequently, in accordance with our model of duality, the EM dipole radiation is due to modulation of the frequency of $\mathbf{C} \rightleftharpoons \mathbf{W}$ pulsation of three sub-elementary fermions of the electron or proton by $[\mathbf{C} \rightleftharpoons \mathbf{W}]_{anc}$ frequency of anchor Bivacuum bosons $\mathbf{BVB}_{anc}^{\pm}$, related to thermal vibrations of elementary particles. These vibrations are are accompanied by creation of secondary *anchor sites* (**AS**), described in chapter 10. When the accelerations and final kinetic energy of elementary charges are big enough, one or number of **AS** transform to photons.

The electromagnetic field, is a result of correlated Corpuscle - Wave pulsation of group of photons with wavelength $\lambda_p = \mathbf{c}/\mathbf{v}_p$, equal to their amplitude in 'empty' space, when $\mathbf{c} = \mathbf{v}$ :

$$\mathbf{A}_p = \lambda_p(\mathbf{c}/\mathbf{v}) = \lambda_p$$

The superposition of clockwise or counterclockwise direction of photon's rotation as respect to direction of their propagation, determines their polarization and integer spin.

### 17.2 *The basis of energy conservation law*

The fundamental law of the actual energy conservation can be reformulated in terms of our Unified theory. The additivity of different forms of the actual energy - means the additivity of sub-elementary particles symmetry shift, responsible for local energy of elementary particle and collective symmetry shift of infinitive number of Bivacuum dipoles in particle environment, responsible for fields origination.

The condition of energy conservation in Bivacuum means, that the sum of local symmetry shifts of group of **N** similar sub-elementary and elementary particles: $\mathbf{N}|\mathbf{m}_V^- - \mathbf{m}_V^+|^{loc}$ should be equal to the sum of distant symmetry shifts of Bivacuum dipoles around these particles: $\sum|\mathbf{m}_V^- - \mathbf{m}_V^+|^{dist}$, compensating local symmetry shifts, i.e. opposite by sign:

$$\mathbf{N}|\mathbf{m}_V^+ - \mathbf{m}_V^-|_{rot}^{loc}\mathbf{c}^2 + \sum|\mathbf{m}_V^+ - \mathbf{m}_V^-|_{tr}^{dist}\mathbf{c}^2 = 0 \qquad 17.6$$

or in another shape, representing extending formulas (8.1 - 8.1b):

$$\mathbf{N} \times \left[\mathbf{m}_0\mathbf{L}_0^2\omega_0^2 + \frac{\mathbf{h}^2}{\mathbf{m}_V^+\lambda_B^2} + \mathbf{E}_{El} + \mathbf{E}_H + \mathbf{E}_G\right]^{loc} + \frac{\vec{r}}{\mathbf{r}}\sum^{\infty} |\mathbf{m}_V^- - \mathbf{m}_V^+|_{tr}^{dist}\mathbf{c}^2 = 0 \qquad 17.7$$

where: $\omega_0 = \mathbf{m}_0\mathbf{c}^2/\hbar$ is the angular frequency of rotating triplet - fermion, providing its rest mass and charge origination with Compton radius $\mathbf{L}_0 = \hbar/\mathbf{m}_0\mathbf{c}$; $\vec{r}$ is unitary radius-vector; **r** is distance from the fermion.

$\mathbf{E}_{El}$, $\mathbf{E}_H$ and $\mathbf{E}_G$ are contributions of electric, magnetic and gravitational energies in total energy of the charged fermion, correspondingly, which will be considered in next chapter;

In accordance to our theory of elementary particles (sections 5.3 and 6.1), symmetry shift in sub-elementary particles, forming elementary fermions is a result of their rotation with tangential velocity $(\mathbf{v}/\mathbf{c})^2 = \varphi = 0.618$ and translation, responsible for empirical de Broglie wave length $\boldsymbol{\lambda}_B$:

$$\boldsymbol{\lambda}_B = h/\mathbf{p} \;\; \textit{and}$$

$$\textit{amplitude} \;\; \mathbf{A}_B = \boldsymbol{\lambda}_B(\mathbf{c}/\mathbf{v}) = \frac{\mathbf{h}}{\mathbf{p}}\frac{\mathbf{c}}{\mathbf{v}} \qquad 17.7a$$

Symmetry shift in surrounding Bivacuum dipoles can be result of two factors:

a) rotation of virtual Cooper pairs of Bivacuum fermions and antifermions $[\mathbf{BVF}^{\uparrow} \bowtie \mathbf{BVF}^{\downarrow}]$ around common axis in direction, opposite to that of rotation of elementary particles - fermions, like free electrons, protons, neutrons, etc. for compensation of the resulting angular momentum;

b) translational vibrations of Bivacuum dipoles (unpaired or paired), as a result of recoil ⇌ antirecoil effects, induced in surrounding medium by emission ⇌ absorption of cumulative virtual clouds $\mathbf{CVC}^{\pm}$, accompanied Corpuscle ⇌ Wave pulsation of sub-elementary fermions of elementary particles and assembly⇌disassembly of virtual Cooper pairs of asymmetric Bivacuum fermions and antifermions, compensating each other.

Such matter - fields energy interconversions in the Universe, following from our Unified Theory, can be considered, as a background for the energy conservation law.

### *17.3 The electric field origination*

In the process of $[\mathbf{C} \leftrightharpoons \mathbf{W}]$ pulsation of sub-elementary particles in triplets $< [\mathbf{F}_{\uparrow}^{+} \bowtie \mathbf{F}_{\downarrow}^{-}] + \mathbf{F}_{\updownarrow}^{\pm} >^{e,p}$ , accompanied by *recoil ⇌ antirecoil* effects the reversibility of

$$[\textit{local}(\textit{internal}) \;\Leftrightarrow\; \textit{distant}(\textit{external})]$$

symmetry compensation effects is in accordance with energy conservation law, described in section 17.2.

The *local* symmetry effects are pertinent for corpuscular [C] phase of elementary particles. They are confined in the volume of unpaired sub-elementary fermions or antifermions and stabilized by the Coulomb, magnetic and gravitational attraction between opposite charges and mass of asymmetric torus and antitorus of sub - elementary fermions.

The $[\mathbf{C} \to \mathbf{W}]$ transitions of unpaired/uncompensated $\mathbf{F}_{\updownarrow}^{\pm} >^{e,p}$ of elementary particles are accompanied by the *diverging* effects - translational and rotational (angular), accompanied by distant elastic deformation of Bivacuum matrix due to recoil, shifting the corresponding symmetry (charge and energy equilibrium) of Bivacuum dipoles.

The reverse $[\mathbf{W} \to \mathbf{C}]$ transition represents the *converging* effect. It is accompanied by getting back the energy and momentum, *diverged* in previous phase (antirecoil) and restoration of the unpaired sub-elementary fermion *local/enfolded* asymmetric properties. The recoil ⇌ antirecoil effects of paired sub-elementary fermions $[\mathbf{F}_{\uparrow}^{+} \bowtie \mathbf{F}_{\downarrow}^{-}]$ of the triplets compensate each other.

The [*divergence* ⇌ *convergence*] of the energy, charge and spin equilibrium in surrounding medium of Bivacuum dipoles in form of spherical elastic waves are opposite for particles of opposite charges.

Corresponding *charge symmetry shift* between torus and antitorus of big number of Bivacuum dipoles located in space between charges is dependent on separation ($R$) of pulsing triplets, as $(\vec{r}/R)$. The direction of symmetry shift, induced by positive and negative charges is also opposite, i.e. positively charged sub-elementary fermion shift symmetry toward the negative antitori and negatively charged sub-elementary fermion - toward the positive tori.

This compensation effect is increasing with decreasing the separation between charges ($R \to 0$) and provide increasing force of opposite charges attraction. The corresponding self-assembly of Cooper pairs of Bivacuum fermions and antifermions of opposite symmetry shift into virtual microfilaments, minimizing potential energy of the system: [elementary fermions + Bivacuum] stands for *electrostatic field* and its *force lines* origination.

The Coulomb *repulsion* between similar charges is the consequence of increasing the resulting Bivacuum asymmetry of the same sign (positive or negative). The enhanced by presence of similar charges the density of energy of Bivacuum dipoles population in space between charges decreases with increasing the separation between charges ($R \to \infty$).

Both - the electric attraction and repulsion force are dependent on the distance as ($\mathbf{1/r^2}$).

*The electrostatic field tension*, produced by charged particles can be expressed via gradients of charge symmetry shift of Bivacuum dipoles of surrounding medium, in selected direction (x,y,z):

$$\mathbf{E}_E = -grad\,|e_+ - e_-|_{BVD} = -grad\,|\mathbf{m}_V^+ - \mathbf{m}_V^-|\mathbf{c}_{BVD}^2 \qquad 17.8$$

$$\alpha \mathbf{T}_k^{\mathbf{F}_\updownarrow^\pm >^{e,p}} = \alpha \frac{1}{2}|\mathbf{m}_V^+ - \mathbf{m}_V^-|\mathbf{c}^2 \qquad 17.8a$$

The validity of 17.8a and physical sense of $\alpha$ will be presented in the next sections.

Possible mechanisms of Coulomb interaction can be determined by positive and negative subquantum particles density oscillation, representing virtual pressure waves: $\mathbf{VPW}^+$ and $\mathbf{VPW}^-$, excited by charged triplets - elementary fermions. The effect of superposition of $\mathbf{VPW}^+$ and $\mathbf{VPW}^-$ on attraction or repulsion of charges also can be explained in terms of tending of system: $\left[\text{Particles + Bivacuum}\right]$ to minimum of asymmetry and energy density of Bivacuum dipoles in space between charges.

*17.4. The magnetic field origination*

The curled magnetic field around electric wire is a consequence of fast rotation of pairs of sub-elementary fermion and antifermion $[\mathbf{F}_\uparrow^+ \bowtie \mathbf{F}_\downarrow^-]$ of triplets $< [\mathbf{F}_\uparrow^+ \bowtie \mathbf{F}_\downarrow^-] + \mathbf{F}_\updownarrow^\pm >^{e,p}$ of opposite charges - clockwise or counterclockwise, reinforced by rotation of unpaired $\mathbf{F}_\updownarrow^\pm >$. The direction of rotation is dependent on direction of triplets propagation in electric wire, in accordance of *right hand screw rule.*

The rotation of charged fermions induce spin equilibrium shift between populations of Bivacuum fermions and antifermions $[\mathbf{BVF}^\uparrow \leftrightharpoons \mathbf{BVB}^\pm \rightleftharpoons \mathbf{BVF}^\downarrow]$ to the left or right in different space volumes. The sign of shift is dependent on direction of triplets propagation.

The shift of spin equilibrium of Bivacuum dipoles and their spatial separation is a consequence of partial disassembly of virtual Cooper pairs, induced by electric current:

$$\mathbf{n}_\updownarrow[\mathbf{BVF}^\uparrow \bowtie \mathbf{BVF}^\downarrow] \to \mathbf{n}_\uparrow \mathbf{BVF}^\uparrow + \mathbf{n}_\downarrow \mathbf{BVF}^\downarrow \qquad 17.9$$

In the absence of magnetic field the densities of Bivacuum fermions and antifermions in each micro volume of space are equal to each other $\mathbf{n}_\uparrow = \mathbf{n}_\downarrow$ and all of them compensate each other spins.

Let's assume, that the *increasing* of $\mathbf{BVF}^\uparrow$ density ($\mathbf{n}_\uparrow$) and corresponding *decreasing* of $\mathbf{BVF}^\downarrow$ density ($\mathbf{n}_\downarrow$) corresponds to the *North (N)* magnetic pole formation. The opposite to that, Bivacuum dipoles densities shifts stands for South (S) pole formation, i.e. when $\mathbf{n}_\downarrow$ is *increasing* and $\mathbf{n}_\uparrow$ *decreasing* in different volumes of space:

$$\textbf{N pole:}\quad \mathbf{n}_\uparrow > \mathbf{n}_\downarrow \qquad 17.10$$

$$\textbf{S pole:}\quad \mathbf{n}_\downarrow > \mathbf{n}_\uparrow$$

The *attraction* between opposite poles N and **S** reflects the tendency of $\mathbf{BVF}^\uparrow$ and $\mathbf{BVF}^\downarrow$ to form stable Cooper pairs, equalizing the symmetry shift between densities of Bivacuum dipoles of opposite spins:

$$\left[\text{ attraction: } (\mathbf{n}_\uparrow\mathbf{BVF}^\uparrow)^{\mathbf{N}} \bowtie (\mathbf{n}_\downarrow\mathbf{BVF}^\downarrow)^{\mathbf{S}}\right] \qquad 17.11$$

For the other hand, the *repulsion* between similar magnetic poles is a consequence of Pauli principle of spatial incompatibility of two fermions (real or virtual) of the same spins (see section ...............):

$$\text{repulsion: } (\mathbf{n}_\uparrow\mathbf{BVF}^\uparrow)^{\mathbf{N}} \Leftrightarrow (\mathbf{n}_\uparrow\mathbf{BVF}^\uparrow)^{\mathbf{N}} \qquad 17.12$$

$$\text{repulsion: } (\mathbf{n}_\downarrow\mathbf{BVF}^\downarrow)^{\mathbf{S}} \Leftrightarrow (\mathbf{n}_\downarrow\mathbf{BVF}^\downarrow)^{\mathbf{S}} \qquad 17.12a$$

The magnetic attraction and repulsion between unpaired Bivacuum fermion is increasing with their densities $\mathbf{n}_\uparrow = \mathbf{n}_\downarrow$. Like in the case of Coulomb interaction, the dependence of interaction force between magnetic poles on separation between them is $(\mathbf{1}/\mathbf{r}^2)$.

Consequently, just the dynamic equilibrium shift between density of unpaired Bivacuum fermions: $\mathbf{BVF}^\uparrow + \mathbf{BVF}^\downarrow$ because of dissemble of virtual Cooper pairs $\mathbf{BVF}^\uparrow \bowtie \mathbf{BVF}^\downarrow$ and polarization of their spatial separation. This process is dependent on direction of electric current and rotation of charged fermions in wire. It stands for the poles and intensity of curled magnetic field around the directed electric current.

The thermal motion of conducting electrons in metals or ions in plasma became more ordered in electric current, increasing correspondingly the magnetic cumulative effect. This is a result of increasing of probability and number of triplets, rotating in the same plane and direction. The bigger is velocity and kinetic energy of triplets, the faster is their rotation, the higher is the probability of spatial separation of Bivacuum fermions and antifermions and bigger magnetic field tension, excited by this rotation:

$$\mathbf{H} = \mathbf{grad}\ \frac{\mathbf{BVF}^\uparrow \Leftrightarrow \mathbf{BVF}^\downarrow}{\mathbf{BVF}^\uparrow \bowtie \mathbf{BVF}^\downarrow} = (\vec{r}/\mathbf{R})\mathbf{K}_{unpaired \rightleftharpoons paired} \qquad 17.13$$

where the equilibrium constant between independent Bivacuum dipoles and in composition of Cooper pairs is:

$$\mathbf{K}_{unpaired \rightleftharpoons paired} = \frac{\mathbf{BVF}^\uparrow \Leftrightarrow \mathbf{BVF}^\downarrow}{\mathbf{BVF}^\uparrow \bowtie \mathbf{BVF}^\downarrow} \qquad 17.13a$$

## 18. Unification of electric, magnetic and gravitational fields, as a consequence of Corpuscle - Wave duality of particles

The total energy of the fermions, equal to that of unpaired sub-elementary fermion:

$$\mathbf{E}_{tot} = |\mathbf{m}_V^+ - \mathbf{m}_V^-|\mathbf{c}^2 = \mathbf{m}_V^+\mathbf{v}_{res}^2 \qquad 18.1$$

where $\mathbf{v}_{res}$ is resulting velocity of rotation and translation of the triplets.

We present $\mathbf{E}_{tot}$ as a sum of all contributions, like in left part of 17.7, using extending

formulas (8.1 - 8.1b):

$$\mathbf{E}_{tot} = \mathbf{m}_0\mathbf{L}_0^2\boldsymbol{\omega}_0^2 + \frac{\mathbf{h}^2}{\mathbf{m}_V^+\boldsymbol{\lambda}_B^2} + \mathbf{E}_{El} + \mathbf{E}_H + \mathbf{E}_G^{loc} \qquad 18.2$$

where: $\boldsymbol{\omega}_0 = \mathbf{m}_0\mathbf{c}^2/\hbar$ is the angular frequency of rotating triplet - fermion, providing its rest mass and charge origination with Compton radius $\mathbf{L}_0 = \hbar/\mathbf{m}_0\mathbf{c}$; $\boldsymbol{\lambda}_B = h/\mathbf{m}_\mathbf{V}^+\mathbf{v}_{tr}$ is de Broglie wave length, determined by value of external translational momentum.

$\mathbf{E}_{El}$, $\mathbf{E}_H$ and $\mathbf{E}_G$ are contributions of electric, magnetic and gravitational energies in total energy of the charged fermion, correspondingly.

Reversible oscillation between local (*Loc*) and distant (*Dist*) contributions to the total energy are responsible for electric $\mathbf{E}_E$, magnetic $\mathbf{E}_H$ and gravitational $\mathbf{E}_G$ fields. The properties of paired sub-elementary fermion in triplets are responsible for these fields origination. The [*recoil* ⇌ *antirecoil*] effects stimulated by [**C** ⇌ **W**] pulsation of paired sub-elementary fermions of the triplets have the opposite energy and compensate each other.

The eq. 18.2 can be presented as:

$$\mathbf{E}_{tot} = \mathbf{m}_0\mathbf{L}_0^2\boldsymbol{\omega}_0^2 + \frac{\mathbf{h}^2}{\mathbf{m}_V^+\boldsymbol{\lambda}_B^2} + \frac{|\mathbf{e}_+\mathbf{e}_-|}{\mathbf{A}_B} + \mathbf{K}_\mathbf{H}\frac{|\boldsymbol{\mu}_+\boldsymbol{\mu}_-|}{\mathbf{A}_B} + \mathbf{G}\frac{|\mathbf{m}_\mathbf{V}^+\mathbf{m}_\mathbf{V}^-|}{\mathbf{A}_B} \qquad 18.3$$

where:

$$\mathbf{A}_B = \frac{\hbar}{(\mathbf{m}_\mathbf{V}^+ - \mathbf{m}_\mathbf{V}^-)\mathbf{c}} = \mathbf{L}_B(\mathbf{c}/\mathbf{v}) \qquad 18.4$$

is the amplitude of cumulative virtual cloud ($\mathbf{CVC}^\pm$) of de Broglie wave with length $\boldsymbol{\lambda}_B = 2\pi\mathbf{L}_B = h/\mathbf{m}_\mathbf{V}^+\mathbf{v}_{tr}$; $|\mathbf{e}_+\mathbf{e}_-| = \mathbf{e}_0^2$ is a rest charge squared.

The amplitude separates positive and negative charges of $\mathbf{CVC}^\pm$ as virtual dipole

In the process of [**C** ⇌ **W**] pulsation, fermion, like the electron or proton (Figure 8) all three fields are excited:

$$\mathbf{E}_{El} = \left[\frac{|\mathbf{e}_+\mathbf{e}_-|}{\mathbf{A}_T}\right]_W \underset{\substack{\mathbf{C\to W}\\ \textbf{Recoil}}}{\overset{\substack{\textbf{Antirecoil}\\ \mathbf{W\to C}}}{\rightleftarrows}} \left[\frac{\mathbf{e}^2}{\hbar\mathbf{c}}(\mathbf{m}_\mathbf{V}^+ - \mathbf{m}_\mathbf{V}^-)\mathbf{c}^2\right]_C \qquad 18.5$$

Taking into account obtained earlier relation (6.7) between mass and charge symmetry shifts

$$\mathbf{m}_V^+ - \mathbf{m}_V^- = \mathbf{m}_V^+\frac{\mathbf{e}_+^2 - \mathbf{e}_-^2}{\mathbf{e}_+^2} \qquad 18.6$$

the *Loc* ⇌ *Dist* oscillations in (18.5), responsible for electric field $\mathbf{E}_E$ excitation, can be presented as:

$$\mathbf{E}_E = \left[\frac{|\mathbf{e}_+\mathbf{e}_-|}{\mathbf{A}_T}\right]^{Loc} \underset{\substack{\mathbf{W\to C}\\ \textbf{Antirecoil}}}{\overset{\substack{\textbf{Recoil}\\ \mathbf{C\to W}}}{\rightleftarrows}} \left[\boldsymbol{\alpha}\times\mathbf{m}_V^+\mathbf{c}^2\frac{\mathbf{e}_+^2 - \mathbf{e}_-^2}{\mathbf{e}_+^2}\right]_{tr}^{Dist} \qquad 18.7$$

where well known dimensionless electromagnetic fine structure constant:

$$\alpha = \mathbf{e}^2/\hbar\mathbf{c} \qquad 18.8$$

The right part of (18.6) characterizes the electric dipole moment of triplet, equal to that of unpaired sub-elementary fermion $\left(\mathbf{F}_\updownarrow^\pm\right)$.

The local *internal* Coulomb interaction between opposite and asymmetric charges of torus and antitorus of unpaired sub-elementary fermions (antifermions) $\left(\mathbf{F}_\updownarrow^\pm\right)$ turns reversibly to the *external* electric field. This happens because of elastic $\left[\textit{diverging} \rightleftharpoons \textit{converging}\right]$ effects, induced by $\mathbf{C} \rightleftharpoons \mathbf{W}$ pulsation of $\left(\mathbf{F}_\updownarrow^\pm\right)_{S=\pm1/2}$ and emission ⇌.absorption of $\mathbf{CVC}^\pm$, exciting, in turn, positive and negative virtual pressure waves: $\mathbf{VPW}^+$ and $\mathbf{VPW}^-$.

*The oscillation of magnetic dipoles radiation* in the process of $[C \leftrightharpoons W]$ pulsations between local and distant modes is described by expression 18.9.

These oscillations do not accompanied by magnetic moments symmetry change, but only by the oscillation of the amplitude $\mathbf{A}_B$ of Cumulative Virtual Cloud ($\mathbf{CVC}^\pm$), separating $\boldsymbol{\mu}_+$ and $\boldsymbol{\mu}_-$ and rotational (torsion) energy of $\mathbf{CVC}^\pm$, $\left[\textit{emitted} \rightleftharpoons \textit{absorbed}\right]$ in the process of $[\mathbf{C} \leftrightharpoons \mathbf{W}]$ pulsation of sub-elementary particles in triplets (18.9a).

This contribution, responsible for magnetic field $\mathbf{E}_H$, can be described as:

$$\mathbf{E}_H = \left[\mathbf{K}_H^i \frac{|\boldsymbol{\mu}_+\boldsymbol{\mu}_-|}{\mathbf{A}_B}\right]_{[W]} \overset{\mathbf{C}\to\mathbf{W}}{\underset{\mathbf{W}\to\mathbf{C}}{\rightleftarrows}} \left[\mathbf{K}_H^i \frac{\boldsymbol{\mu}_0^2}{\hbar\mathbf{c}}(\mathbf{m}_\mathbf{V}^+ - \mathbf{m}_\mathbf{V}^-)\mathbf{c}^2\right]_{[C]} \qquad 18.9$$

$$or: \left[\mathbf{K}_H^i \frac{|\boldsymbol{\mu}_+\boldsymbol{\mu}_-|}{\mathbf{A}_B}\right]_{[C]}^{Loc} \overset{\mathbf{C}\to\mathbf{W}}{\underset{\mathbf{W}\to\mathbf{C}}{\rightleftarrows}} \left[\mathbf{K}_H^i \frac{\boldsymbol{\mu}_0^2}{\hbar\mathbf{c}}\, \mathbf{m}_\mathbf{V}^+\boldsymbol{\omega}_\mathbf{T}^2\mathbf{L}_T^2\right]_{[W]}^{Dis} \qquad 18.9a$$

where new notion of *magnetic fine structure constant*, introduced in our approach is:

$$\frac{\boldsymbol{\mu}_0^2}{\hbar\mathbf{c}} = \gamma$$

The magnetic conversion coefficient $\mathbf{K}_H$ we may find from the equality of the electrostatic and magnetic energy contributions, determined by *recoil* ⇌ *antirecoil* effects:

$$\mathbf{E}_E = \mathbf{T}_{rec} = \frac{1}{2}\frac{\mathbf{e}^2}{\hbar\mathbf{c}}(\mathbf{m}_\mathbf{V}^+ - \mathbf{m}_\mathbf{V}^-)\mathbf{c}^2 = \frac{1}{2}\alpha\, \mathbf{m}_\mathbf{V}^+\mathbf{v}_{res}^2 = \frac{1}{2}\mathbf{K}_H^i \frac{\boldsymbol{\mu}_0^2}{\hbar\mathbf{c}}(\mathbf{m}_\mathbf{V}^+ - \mathbf{m}_\mathbf{V}^-)\mathbf{c}^2 = \mathbf{E}_H \qquad 18.10$$

This formula is a consequence of *equal probability* of energy distribution between translational (electrostatic) and rotational (magnetic) independent degrees of freedom of an unpaired sub-elementary fermion in triplets and its cumulative virtual cloud ($\mathbf{CVC}^\pm$).

From the above condition it follows, that:

$$\mathbf{K}_H \frac{\boldsymbol{\mu}_0^2}{\hbar\mathbf{c}} = \mathbf{K}_H \frac{\hbar\mathbf{e}_0^2}{4\mathbf{m}_0^2\mathbf{c}^3} = \frac{\mathbf{e}_0^2}{\hbar\mathbf{c}} \qquad 18.11$$

where $\boldsymbol{\mu}_0^2 = |\boldsymbol{\mu}_+\boldsymbol{\mu}_-| = \left(\frac{1}{2}\mathbf{e}_0\frac{\hbar}{\mathbf{m}_0\mathbf{c}}\right)^2$ is the Bohr magneton.

The introduced *magnetic conversion coefficient* $\mathbf{K}_H$ with dimension of wave number squared, can be obtained from 18.11 as:

$$\mathbf{K}_H^{e,p} = \left(\frac{\mathbf{m}_0^{e,p}\mathbf{c}}{\hbar/2}\right)^2 = \left(\frac{2}{\mathbf{L}_0^{e,p}}\right)^2 \qquad 18.12$$

where $\mathbf{L}_0^{e,p} = \hbar/\mathbf{m}_0^{e,p}\mathbf{c}$ is the Compton radius of the electron or proton.

In Golden mean conditions, corresponding just to rest mass and charge origination due to tangential velocity of triplets rotation, when external translational motion of particles is absent, the formula 18.10 gives:

$$\mathbf{E}_H + \mathbf{E}_E = \alpha\,(\mathbf{m}_\mathbf{V}^+\mathbf{v}_\mathbf{res}^2)^\phi = \alpha\,\mathbf{m}_0\mathbf{c}^2 \qquad 18.13$$

*Origination of magnetic field* can be a result of dynamic equilibrium shift between Bivacuum fermions and Bivacuum antifermions density to the left or right as described in previous section.

The chaotic thermal velocity of the 'free' conductivity electrons in metals and ions at room temperature is very high even in the absence of current, and follows Maxwell-Boltzmann distribution:

$$\mathbf{v}_T = \sqrt{\frac{\mathbf{kT}}{\mathbf{m}_\mathbf{V}^+}} \sim 10^7\ cm/s \qquad 18.14$$

It proves, that not the velocity, but the ordering of the electrons translational and rotational dynamics in space, provided by current, is a main reason of the curled magnetic field excitation around the conductor.

### *18.1 Interpretation of the Maxwell displacement current, based on Bivacuum model*

The magnetic field origination in Bivacuum can be analyzed also on the base of more conventional approach.

Let us analyze the 1st Maxwell equation, interrelating the circulation of vector of magnetic field tension **H** along the contour **L** with the conduction current (**j**) and *displacement current* $\mathbf{j}_d = \frac{1}{4\pi}\frac{\partial \mathbf{E}_{BVF}}{\partial t}$ through the surface, limited by **L** :

$$\oint_\mathbf{L} \mathbf{H}dl = \frac{4\pi}{c}\int_\mathbf{S}\left(\mathbf{j} + \frac{1}{4\pi}\frac{\partial \mathbf{E}_{BVF}}{\partial t}\right)d\mathbf{s} \qquad 18.15$$

where (**s**) is the element of surface, limited with contour ($l$).

The existence of the displacement current: $\mathbf{j}_d = \frac{1}{4\pi}\frac{\partial \mathbf{E}}{\partial t}$ is in accordance with our approach is a result of oscillating virtual charge - dipoles ($\mathbf{BVF}^\updownarrow$ and $\mathbf{BVB}^\pm$) of Bivacuum.

In condition of *primordial* Bivacuum of the ideal virtual dipoles symmetry (i.e. in the absence of matter and fields) the charges of torus and antitorus totally compensate each other. However, even in primordial symmetric Bivacuum the oscillations of distance between opposite charges of torus and antitorus of Bivacuum dipoles, following by the energy gap oscillation (chapter 1), are responsible for *displacement current.* This alternating current generates corresponding *displacement magnetic field:*

$$H_d = \frac{4\pi}{c}\int_\mathbf{S}\frac{1}{4\pi}\frac{\partial \mathbf{E}_{BVF}}{\partial t}d\mathbf{s} \qquad 18.16$$

Bivacuum dipoles zero-point oscillations are the consequence of the in-phase transitions of $\mathbf{V}^+$ and $\mathbf{V}^-$ between the excited and ground states, compensating each other. These transitions are accompanied by spontaneous emission and absorption of positive and negative virtual clouds $\mathbf{VC}^+$ and $\mathbf{VC}^-$, exiting in turn, virtual pressure waves: $\mathbf{VPW}^+$ and $\mathbf{VPW}^-$. The stimulation of such transitions, for example, by pulsing electric field, should influence on interaction of Bivacuum with elementary particles.

It follows also from our interpretation of displacement current, that the displacement magnetic field can be enhanced by elementary particles, pulsing between Corpuscular and Wave phase, even in the absence of translational motion.

*18.2 New kind of "recoil current" in Bivacuum, increasing the displacement one. The velocity of zero-point oscillation and physical sense of electric charge*

This additional new current is a consequence of vibrations of $\mathbf{BVF}^{\updownarrow}$, induced by recoil-antirecoil effects, accompanied $[\mathbf{C} \leftrightharpoons \mathbf{W}]$ transitions of unpaired sub-elementary fermion of triplets $< [\mathbf{F}_{\uparrow}^{-} \bowtie \mathbf{F}_{\downarrow}^{+}]_{S=0} + (\mathbf{F}_{\uparrow}^{+})_{S=\pm 1/2} >^{e,p}$ - fermions.

The corresponding elastic symmetry oscillations of Bivacuum fermions $(\mathbf{BVF}^{\updownarrow}) \equiv [\mathbf{V}^{+} \Updownarrow \mathbf{V}^{-}]$ are followed by small charge-dipole symmetry oscillations with amplitude, determined by the most probable resulting translational - rotational recoil velocity $(\mathbf{v}_{rec})$.

At conditions $\mathbf{e}_{+} \simeq \mathbf{e}_{-} \simeq \mathbf{e}_{0}$ and $|\mathbf{e}_{+} - \mathbf{e}_{-}| << \mathbf{e}_{0}$, meaning small symmetry vibrations of $\mathbf{V}^{+}$ and $\mathbf{V}^{-}$, we have for the charge shift oscillation eq.(6.8):

$$\Delta\mathbf{e}_{\pm} = \mathbf{e}_{+} - \mathbf{e}_{-} = \frac{1}{2}\mathbf{e}_{0}\frac{\mathbf{v}_{rec}^{2}}{\mathbf{c}^{2}} \qquad 18.17$$

The most probable recoil kinetic energy $\mathbf{T}_{rec}$ and recoil velocity $\mathbf{v}_{rec}$, standing for electromagnetism (18.10), can be defined as:

$$2\mathbf{T}_{k} = \alpha(\mathbf{m}_{\mathbf{V}}^{+} - \mathbf{m}_{\mathbf{V}}^{-})\mathbf{c}^{2} = \alpha\, \mathbf{m}_{\mathbf{V}}^{+}\mathbf{v}_{\mathbf{res}}^{2} = \mathbf{m}_{\mathbf{V}}^{+}\mathbf{v}_{\mathbf{rec}}^{2} \qquad 18.18$$

$$\mathbf{v}_{rec}^{2} = \alpha\, \mathbf{v}_{\mathbf{res}}^{2} \qquad 18.19$$

The relation between resulting velocity $\mathbf{v}_{res}$, the internal tangential velocity $\mathbf{v}_{\mathbf{in}}^{\phi} = \phi\mathbf{c}^{2}$, responsible for rest mass and charge origination and the external translational velocity $\mathbf{v}_{\mathbf{ext}}$, standing for empirical de Broglie wave length, was introduced in chapter 8 as:

$$\mathbf{v}_{\mathbf{res}}^{2} = \phi\mathbf{c}^{2} + \mathbf{v}_{\mathbf{ext}}^{2} \qquad 18.20$$

Using interrelation between mass and charge symmetry shifts (6.7), formula (18.18) for recoil kinetic energy can be presented as:

$$\mathbf{T}_{rec} = \frac{1}{2}\alpha\, \mathbf{m}_{\mathbf{V}}^{+}\mathbf{v}_{\mathbf{res}}^{2} = \frac{1}{2}\alpha\, \mathbf{m}_{\mathbf{V}}^{+}\mathbf{c}^{2}\frac{\mathbf{e}_{+}^{2} - \mathbf{e}_{-}^{2}}{\mathbf{e}_{+}^{2}} \qquad 18.21$$

In presence of matter and fields, when primordial Bivacuum turns to secondary one, composed from Bivacuum dipoles of small asymmetry: $\mathbf{e}_{+} \simeq \mathbf{e}_{-} \simeq \mathbf{e}_{0}$, we may assume, that:

$$\mathbf{e}_{+}^{2} - \mathbf{e}_{-}^{2} = (\mathbf{e}_{+} + \mathbf{e}_{-})(\mathbf{e}_{+} - \mathbf{e}_{-}) \simeq 2\mathbf{e}_{0}(\mathbf{e}_{+} - \mathbf{e}_{-}) \qquad 18.22$$

and right part of (18.21) turns to formula, interrelating the recoil kinetic energy of asymmetric Bivacuum dipoles with their charge symmetry shift:

$$\mathbf{T}_{rec} = \frac{1}{2}\alpha\, \mathbf{m}_{\mathbf{V}}^{+}\mathbf{v}_{\mathbf{res}}^{2} = \alpha\, \mathbf{m}_{\mathbf{V}}^{+}\mathbf{c}^{2}\frac{\mathbf{e}_{+} - \mathbf{e}_{-}}{\mathbf{e}_{0}} \qquad 18.23$$

This new formula reflects the oscillation of Bivacuum dipoles charge symmetry shift near particle, generated by recoil kinetic energy oscillation, accompanied $\mathbf{C} \rightleftharpoons \mathbf{W}$ pulsation of elementary particle.

The minimum value of recoil velocity, corresponding to zero *external* translational velocity of triplets, like electrons, positrons and protons, can be evaluated from the internal velocity of sub-elementary fermions, determined by Golden mean conditions $(\mathbf{v}_{rec}/\mathbf{c})^2 = \phi = 0.61803398$.

At this conditions $(\mathbf{m}_V^+)^{\phi} = \mathbf{m}_0/\phi$ and left part of 18.23 can be presented as:

$$\mathbf{T}_{rec} = \frac{1}{2}\alpha\, \mathbf{m}_0\mathbf{c}^2 \qquad 18.23a$$

The minimum recoil velocity can be considered as the *velocity of zero-point oscillations* of elementary charge position, related to *recoil* $\rightleftharpoons$ *antirecoil* effect, accompanied $C \rightleftharpoons W$ pulsation of the unpaired sub-elementary fermion of the triplets (electrons or protons) in the absence of translational motion:

$$(\mathbf{v}_{rec}^2)^{\min} \equiv (\mathbf{v}_0^2)_{H,E}^{\min} = \boldsymbol{\alpha\phi}\, \mathbf{c}^2 \qquad 18.24$$

$$or: \; \frac{(\mathbf{v}_{rec}^2)^{\min}}{\mathbf{c}^2} = \boldsymbol{\alpha\phi} \qquad 18.24a$$

where: $\alpha = e^2/\hbar c = 0{,}0072973506$; $\boldsymbol{\alpha\phi} = (\mathbf{v}_{rec}^2)^{\min}/\mathbf{c}^2 = 4.51 \times 10^{-3}$ and

$$\mathbf{v}_{rec}^{\min} = 6.716 \times 10^{-2}\, \mathbf{c} \qquad 18.25$$

**The physical sense of the electric charge follows from 17.38**.

This formula can be presented as:

$$\alpha = \frac{\mathbf{e}^2}{\hbar c} = \frac{1}{\phi}\left(\frac{\mathbf{v}_{rec}^{\min}}{\mathbf{c}}\right)^2 \qquad 18.26$$

$$or: \quad \mathbf{e}^{\pm} = \mathbf{v}_{rec}^{\min} \times \left(\frac{\hbar}{\phi c}\right)^{1/2} \qquad 18.27$$

Formula 18.27 means, that zero-point electric charge $< [\mathbf{F}_\uparrow^- \bowtie \mathbf{F}_\downarrow^+]_{S=0} + (\mathbf{F}_\uparrow^+)_{S=\pm 1/2} >^{e,p}$, as a permanent source of energy, is a consequence of unpaired sub-elementary fermion $(\mathbf{F}_\uparrow^+)_{S=\pm 1/2}>^{e,p}$ reversible Corpuscle $\rightleftharpoons$ Wave pulsation with high frequency: $\nu_0^{e,p} = \mathbf{m}_0^{e,p}\mathbf{c}^2/h$.

These pulsation, resulting from *emission*$\rightleftharpoons$*absorption* of cumulative virtual clouds ($\mathbf{CVC}^{\pm}$) by $(\mathbf{F}_\uparrow^+)_{S=\pm 1/2}>^{e,p}$, are accompanied by *recoil*$\rightleftharpoons$*antirecoil* vibrations of the triplet and corresponding elastic vibrations of surrounding continuum of Bivacuum dipoles with zero-point velocity (18.24).

We suppose that $\mathbf{C} \rightleftharpoons \mathbf{W}$ pulsation of positive charges influence more on positively charges tori ($\mathbf{V}^+$) and pulsation of negative charges - on negatively charged antitori ($\mathbf{V}^-$) of Bivacuum dipoles. So, if we have separated positive and negative charges, their influence on symmetry of Bivacuum dipoles in space between them compensate each other. As a consequence, the *uncompensated* energy of virtual pressure waves $\mathbf{VPW}^+$ and $\mathbf{VPW}^-$ outside charges is higher, than between them. This imitates the attraction effect between pulsing charges.

On contrary, in system of two charges of the same sign, their $\mathbf{C} \rightleftharpoons \mathbf{W}$ pulsation shift symmetry of Bivacuum dipoles in the same direction - positive or negative. As a result, the *uncompensated* energy of $\mathbf{VPW}^+$ and $\mathbf{VPW}^-$ between charges is higher, than outside them, and we get the repulsion force between charges.

The alternating *recoil current* ($j_{rec}^{EH}$), additional to that of Maxwell *displacement current* ($j_d$,) existing in presence of charged particles even in the absence of conducting current ($\mathbf{j} = \mathbf{0}$) is equal to product of (18.17) and square root of (18.24). At Golden mean conditions $(\mathbf{v}/\mathbf{c})^2 = \phi$ this new *recoil current,* following from our approach, is:

$$(\mathbf{j}_{rec}^{\phi})^{EH} = (\Delta\mathbf{e}_{\pm})^{\phi}(\mathbf{v}_{rec})^{\min} = \frac{1}{2}\boldsymbol{\alpha}^{1/2}\boldsymbol{\phi}^{3/2}\,\mathbf{e}_0\mathbf{c} \qquad 18.28$$

Consequently, the Maxwell equation (18.15) can be modified, taking into account the electromagnetic (EH) recoil current, as

$$\oint_{\mathbf{L}} \mathbf{H}dl = \frac{4\pi}{c}\int_{\mathbf{S}}\left(\mathbf{j} + \frac{1}{4\pi}\frac{\partial\mathbf{E}}{\partial t} + \mathbf{j}_{rec}^{EH}\right)d\mathbf{s} = \mathbf{I}_{tot} \qquad 18.29$$

where: $\mathbf{I}_{tot}$ is the total current throw the surface (**S**).

We have to note, that $\mathbf{j}_{rec}^{EH}$ is nonzero not only in the vicinity of particles, but as well in any remote space regions of Bivacuum, perturbed by electric and magnetic potentials, decreasing, however, with distance from particle as (**1**/**r**). This consequence of our theory coincides with the extended electromagnetic theory of Bo Lehnert (2004, 2004a), also considering current in vacuum, additional to displacement one.

In accordance with the known Helmholtz theorem, each kind of vector field (**F**), tending to zero at infinity, can be presented, as a sum of the gradient of some scalar potential ($\phi$) and a rotor of vector potential (**A**):

$$\mathbf{F} = \mathbf{grad}\,\varphi + \mathbf{rot}\,\mathbf{A} \qquad 18.30$$

To explain the *ability of secondary (perturbed by fields) Bivacuum to keep the average (macroscopic) mass and charge equal to zero,* we have to postulate, that the mass and charge symmetry shifts oscillations of Bivacuum fermions and antifermions, forming virtual Cooper pairs:

$$(\mathbf{BVF}^{\uparrow})^{\pm}_{S=+1/2} \equiv [\mathbf{V}^{+} \upuparrows \mathbf{V}^{-}]^{\pm} \bowtie [\mathbf{V}^{+} \downdownarrows \mathbf{V}^{-}]^{\mp} \equiv (\mathbf{BVF}^{\downarrow})^{\mp}_{S=-1/2} \qquad 18.31$$

are opposite by sign, but equal by the absolute value. Consequently, the polarized secondary Bivacuum (i.e. perturbed by matter and field) can be considered, as a *plasma of the in-phase oscillating virtual dipoles (BVF)* of opposite resulting charge and mass/energy.

### *18.3 The Link Between Maxwell's Formalism and our model of photon*

The quantization rule for energy of each of two sub-elementary fermions, forming photons, can be expressed as:

$$\mathbf{E_p} = \mathbf{n}\mathbf{E}_{C\rightleftharpoons W} = \mathbf{n}\hbar\boldsymbol{\omega}_{C\rightleftharpoons W} = \boldsymbol{\alpha}\,\mathbf{n}\hbar[\boldsymbol{\omega}_V^{+} - \boldsymbol{\omega}_V^{-}] = \qquad 18.32$$

$$= \boldsymbol{\alpha}\,\mathbf{n}(\mathbf{m}_V^{+} - \mathbf{m}_V^{-})\mathbf{c}^2 = \frac{\alpha}{2}\hbar[rot\vec{\mathbf{V}}^{+} - rot\vec{\mathbf{V}}^{-}] \qquad 18.32a$$

where: $\mathbf{m}_V^{+}\mathbf{c}^2 = \hbar\boldsymbol{\omega}_V^{+}$ and $\mathbf{m}_V^{-}\mathbf{c}^2 = \hbar\boldsymbol{\omega}_V^{-}$ are the quantized energies of the actual and complementary torus $\vec{\mathbf{V}}^{+}$ and antitorus $\vec{\mathbf{V}}^{-}$ of sub-elementary fermion or antifermion of the photon in [C] phase;

The energy of photon is a result of quantum beats with frequency: $\boldsymbol{\omega}_{C\rightleftharpoons W} = \boldsymbol{\omega}_V^{+} - \boldsymbol{\omega}_V^{-}$ between the actual and complementary corpuscular states of two uncompensated sub-elementary fermions with additive spins in composition of photons (Fig.8).

It is assumed, that all of subquantum particles and antiparticles, forming actual and complementary vortical excitation of [C] phase of sub-elementary fermions, have the same angle frequency: $\boldsymbol{\omega}_V^{+}$ and $\boldsymbol{\omega}_V^{-}$, correspondingly.

We can express the divergency of Pointing vector: $\mathbf{P} = (c/4\pi)[\mathbf{EH}]$ via difference of contributions, related to actual and complementary tori, using known relation of vector

analysis:

$$div[\mathbf{EH}] = \frac{4\pi}{c} div\mathbf{P} = \mathbf{H}\,rot\mathbf{E} - \mathbf{E}\,rot\mathbf{H} \qquad 18.33$$

where **H** and **E** are the magnetic and electric energy contributions of subquantum particles, radiated and absorbed in a course of correlated $[C \rightleftharpoons W]$ pulsation of two uncompensated sub-elementary fermions of photon.

The analogy between (18.32) and (18.32a), illustrating the dynamics of [torus + antitorus] dipole, is evident, if we assume:

$$\hbar\omega_V^+ \sim \mathbf{H}\,rot\mathbf{E} \sim \frac{\alpha}{2}\hbar\, rot\vec{\mathbf{V}}^+ \qquad 18.34$$

$$\hbar\omega_V^- \sim \mathbf{E}\,rot\mathbf{H} \sim \frac{\alpha}{2}\hbar\, rot\vec{\mathbf{V}}^- \qquad 18.34a$$

Then, the divergence of Pointing vector will take a form:

$$\frac{4\pi}{c} div\mathbf{P} = \frac{\alpha}{2}\hbar\left[rot\vec{\mathbf{V}}^+ - rot\vec{\mathbf{V}}^-\right] \sim \boldsymbol{\alpha}[\mathbf{m}_V^+ - \mathbf{m}_V^-]c^2 \qquad 18.35$$

We can see from 18.34 and 18.34a, that the properties of both: magnetic and electric fields are implemented in each of our torus and antitorus of Bivacuum dipoles.

We may apply also the *Green theorems*, interrelating the volume and surface integrals, to our duality model. One of known Green theorems is:

$$\int_V (\Psi\nabla^2\Phi - \Phi\nabla^2\Psi)\, dV = \int_S dS \times (\Psi\nabla\Phi - \Phi\nabla\Psi)\, dV \qquad 18.36$$

If we define the scalar functions, as the instant energies of the actual and complementary states of [C] phase of sub-elementary particles as $\Phi = \mathbf{m}_V^+\mathbf{c}^2$ and $\Psi = \mathbf{m}_V^-\mathbf{c}^2$, then, taking into account that

$$\nabla^2\Phi = div\,grad\,\Phi = div\,grad(\mathbf{m}_V^+\mathbf{c}^2) \qquad 18.37$$

$$\nabla^2\Psi = div\,grad\,\Psi = div\,grad(\mathbf{m}_V^-\mathbf{c}^2) \qquad 18.37a$$

formula (18.36) can be presented in form:

$$\int_V [(\mathbf{m}_V^-\mathbf{c}^2)\nabla^2(\mathbf{m}_V^+\mathbf{c}^2) - (\mathbf{m}_V^+\mathbf{c}^2)\nabla^2(\mathbf{m}_V^-\mathbf{c}^2)]\, dV \qquad 18.38$$

$$= \int_S dS\ [(\mathbf{m}_V^-\mathbf{c}^2)\nabla(\mathbf{m}_V^+\mathbf{c}^2) - (\mathbf{m}_V^+\mathbf{c}^2)\nabla(\mathbf{m}_V^-\mathbf{c}^2)]\, dV \qquad 18.38a$$

The upper part (18.38) represents the energy of sub-elementary fermion in [C] phase and the lower part (18.38a) - the energy of cumulative virtual cloud (CVC), corresponding to [W] phase of the same particle.

### *18.4 New approach to quantum gravity and antigravity, dark matter and inertia*

The local *internal* gravitational interaction between the opposite mass poles of the mass-dipoles of unpaired sub-elementary fermions (antifermions) $(\mathbf{F}_\updownarrow^\pm)_{S=\pm 1/2}$ in **[C]** phase turns reversibly to the *external* interaction between mass poles of $\mathbf{CVC}^\pm$ in **[W]** phase (see

eqs. 18.2 and 18.3).

The corresponding dynamic equilibrium between the *diverging* and *converging* flows of gravitational energy, following [**C** ⇌ **W**] pulsation, can be described as:

$$\mathbf{E}_G = \left[ \mathbf{G} \frac{\left|\mathbf{m}_V^+ \mathbf{m}_V^-\right|^{e,p,n}}{\mathbf{L_A}} \right]_{[W]} \underset{\substack{\mathbf{W \to C} \\ \textbf{Antirecoil}}}{\overset{\substack{\textbf{Recoil} \\ \mathbf{C \to W}}}{\rightleftarrows}} \left[ \boldsymbol{\beta} \, |\mathbf{m}_\mathbf{V}^+ - \mathbf{m}_\mathbf{V}^-| \mathbf{c}^2 \right]_{[C]} \qquad 18.39$$

where: $\mathbf{L_A} = \hbar/(\mathbf{m}_\mathbf{V}^+ - \mathbf{m}_\mathbf{V}^-)\mathbf{c}$ is a characteristic radius of $\mathbf{CVC}^{\pm}$; $\frac{\mathbf{r}}{r}$ is ratio of unitary vector to distance from particle; $(\mathbf{m}_0^{e,p,n})^2 = \left|\mathbf{m}_V^+ \mathbf{m}_V^-\right|^{e,p,n}$ is a rest mass of the electron, proton or neutron squared.

The introduced dimensionless gravitational fine structure constant $\boldsymbol{\beta}^{e,p,n}$ (Kaivarainen, 2006; 2008) is a ratio of the rest mass of elementary particle squared (electron, proton, neutron, etc.) to the Planck mass $\mathbf{M}_{Pl}^2 = \hbar\mathbf{c}/\mathbf{G}$, squared:

$$\boldsymbol{\beta}^i = \left( \frac{\mathbf{m}_0^{e,p,n}}{\mathbf{M}_{Pl}} \right)^2 \qquad 18.40$$

For the electron $\beta^e = 1.739 \times 10^{-45}$ and $\sqrt{\beta^e} = \frac{\mathbf{m}_0^e}{\mathbf{M}_{Pl}} = 0.41 \times 10^{-22}$.

The effective recoil velocity of particle's squared $(\mathbf{v}_G^2)_{eff}$, accompanied **C** ⇌ **W** pulsations, characteristic for excitation of gravitational waves, can be introduced as

$$\mathbf{E}_G = \mathbf{m}_\mathbf{V}^+ (\mathbf{v}_G^2)_{eff} = \beta\,(\mathbf{m}_\mathbf{V}^+ - \mathbf{m}_\mathbf{V}^-)\mathbf{c}^2 \qquad 18.41$$

At the Golden mean conditions, when $(\mathbf{v}^2/\mathbf{c}^2) = \phi = 0.618$, $(\mathbf{m}_\mathbf{V}^+ - \mathbf{m}_\mathbf{V}^-)^\phi = \mathbf{m}_0$ and $(\mathbf{m}_\mathbf{V}^+)^\phi = \mathbf{m}_0/\phi$ we get from (18.41) the reduced value of characteristic gravitational velocity of zero-point oscillation, of the electron in state of rest:

$$(\mathbf{v}_G)_{eff}^\phi = \mathbf{c}\,\sqrt{\beta\phi} = \mathbf{c} \times 0.41 \times 10^{-22} \times 0.786 = 0.322 \times 10^{-22}\,\mathbf{c} \qquad 18.42$$

The energy of gravitational field (18.41) at Golden mean conditions for rest mass and charge origination, in the absence of external translational motion turns to

$$\mathbf{E}_G^0 = \left[ \frac{\mathbf{m}_0}{\phi} (\mathbf{v}_G^2)_{eff} \right]^\phi = \beta\,\mathbf{2m}_0\mathbf{c}^2 \qquad 18.43$$

This is the gravitational energy of Bivacuum dipole of torus and antitorus, pulsing in-phase between symmetric excited and ground symmetric states of opposite energy.

The excitation of spherical virtual pressure waves of positive and negative energy: $\mathbf{VPW}_q^+$ and $\mathbf{VPW}_q^-$ is a result of torus and antitorus beats, accompanied [**C** ⇋ **W**] counterphase pulsation of unpaired $\mathbf{F}_\updownarrow^\pm >^{e,p}$ and paired sub-elementary fermions $[\mathbf{F}_\uparrow^- \bowtie \mathbf{F}_\downarrow^+]_{S=0}$ of triplets.

It is important to note, that the energy of introduced gravitational field does not depend on charge of triplets, determined by unpaired sub-elementary fermion of triplets $< [\mathbf{F}_\uparrow^- \bowtie \mathbf{F}_\downarrow^+]_{S=0} + \left(\mathbf{F}_\updownarrow^\pm\right)_{S=\pm 1/2} >$, in contrast to electrostatic and magnetic field.

*It follows from our approach, that the gravitational energy is existing even in 'empty' secondary Bivacuum, when the influence of matter and fields on symmetry of Bivacuum dipoles is small, however* $\mathbf{m}_\mathbf{V}^+ \neq \mathbf{m}_\mathbf{V}^-$ and

$$\mathbf{E}_G = \beta\,(\mathbf{m}_\mathbf{V}^+ - \mathbf{m}_\mathbf{V}^-)\mathbf{c}^2 > 0 \qquad 18.44$$

This condition can be responsible for attraction effect of '*cold dark matter*' of the Universe.

The *cold dark matter* phenomena is a consequence of simultaneous transitions -beats between asymmetric and symmetric ground states of huge number of Bivacuum dipoles, slightly asymmetric in secondary Bivacuum, composing big virtual domains of nonlocality in state of Bose condensation (see section 3.1).

These Bivacuum domains can't be detected by any astronomical methods and behave, as extended in space gravitating mass with huge resulting energy of gravitation.

From the proposed mechanism of gravitation and similar values of $(\mathbf{m}_V^+)$ in the left and right parts of eq. (18.41), it follows *the equality of gravitational and inertial mass*.

The *inertia* itself can be defined, as a *resistance to additional symmetry shift* between the actual and complementary masses/energy of sub-elementary fermions ($\mathbf{m}_V^+$ and $\mathbf{m}_V^-$) of elementary particles and surrounding Bivacuum dipoles, accompanied particles acceleration.

The *generalized Le Chatelier's Principle* can be introduced, as a resistance of any system, containing sub-elementary fermions of elementary fermions (triplets) in state of equilibrium, to additional symmetry shift, accompanied particles acceleration, increasing its actual mass and decreasing complementary one.

The antigravitation or *dark energy* phenomena can be a consequence of energy conservation law in our formulation (section 17.2). This formulation means that shift of mass-energy symmetry towards the positive energy in huge amount of Bivacuum dipoles $(\mathbf{m}_V^+ - \mathbf{m}_V^-) > 0$ in one domains should be compensated by the opposite symmetry shift in other domains in state of BC: $(\mathbf{m}_V^+ - \mathbf{m}_V^-) < 0$. It follows from 18.44, that sign of gravitational energy changes from positive to negative and gravitation turns to antigravitation:

$$\mathbf{E}_G = -\mathbf{E}_G = \beta\,(\mathbf{m}_V^+ - \mathbf{m}_V^-)\mathbf{c}^2 < 0 \qquad 18.45$$

Such conversion of symmetry of Bivacuum dipoles can be induced even in "normal" secondary Bivacuum - for example, by combination of strong magnetic and electric fields.

### *18.5 The hydrodynamic mechanism of gravitational attraction and repulsion*

In accordance to our conjecture (Kaivarainen, 1995; 2000; 2006), the mechanism of gravitation and antigravitation is similar to Bjerknes attraction/repulsion between pulsing spheres in liquid medium of Bivacuum. The dependence of Bjerknes force on distance between centers of pulsing objects is quadratic: $\mathbf{F}_{Bj} \sim \mathbf{1/r}^2$:

$$\mathbf{F}_G = \mathbf{F}_{Bj} = \frac{1}{\mathbf{r}^2}\pi\rho_\mathbf{G}\mathbf{R}_1^2\mathbf{R}_2^2\mathbf{v}^2\cos\beta \qquad 18.46$$

where $\rho_G$ is density of liquid, i.e. virtual density of secondary Bivacuum, dependent on Bivacuum dipoles symmetry shift; $\mathbf{R}_1$ and $\mathbf{R}_2$ radiuses of gravitating spheres, composed from pulsing particles; $\mathbf{v}$ is velocity of spheres surface oscillation (i.e. velocity of virtual pressure waves $\mathbf{VPW}_q^\pm$, excited by $[\mathbf{C} \rightleftharpoons \mathbf{W}]$ pulsation of elementary particles, which can be assumed to be equal to light velocity: $\mathbf{v} = \mathbf{c}$; $\beta$ is a phase shift between pulsation of spheres or system of coherent elementary particles.

On the big enough distances the *attraction* may turn to *repulsion*, i.e. gravitation turns to antigravitation. The latter effect is dependent on the phase shift of coherent $[\mathbf{C} \rightleftharpoons \mathbf{W}]$ pulsation of interacting remote triplets ($\beta$) in separate macroscopic objects. This repulsion effect can explain revealed experimentally *acceleration of the Universe expansion*.

The corresponding *antigravitation* energy or *negative pressure energy (dark energy)*, is about 70% of the total Universe energy.

The possibility of artificial phase shift of [$\mathbf{C} \rightleftharpoons \mathbf{W}$] pulsation of coherent elementary particles of any object may (for example by magnetic field) may change its gravitational attraction to repulsion and vice versa.

The volume and radius of pulsing spheres ($R_1$ and $R_2$) in such approach is determined by sum of volume of hadrons, composing gravitating systems in solid, liquid, gas or plasma state. The gravitational attraction or repulsion is a result of increasing or decreasing of virtual pressure of subquantum particles between interacting systems as respect to its value outside them. This model can serve as a background for new *quantum gravity theory.*

The effective radiuses of gravitating objects $\mathbf{R}_1$ and $\mathbf{R}_2$ can be calculated from the effective volumes of the objects:

$$\mathbf{V}_{1,2} = \frac{4}{3}\pi\mathbf{R}_{1,2}^3 = \mathbf{N}_{1,2}\frac{4}{3}\pi\mathbf{L}_{p,n}^3 \qquad 18.47$$

where: $\mathbf{N}_{1,2} = \mathbf{M}_{1,2}/\mathbf{m}_{p,n}$ is the number of protons and neutrons in gravitating bodies with mass $\mathbf{M}_1$ and $\mathbf{M}_2$; $\mathbf{m}_{p,n}$ is the mass of proton and neutron; $\mathbf{L}_{p,n} = \hbar/\mathbf{m}_{p,n}\mathbf{c}$ is the Compton radius of proton and neutron.

From (18.47) we get for effective radiuses:

$$\mathbf{R}_{1,2} = \left(\frac{\mathbf{M}_{1,2}}{\mathbf{m}_{p,n}}\right)^{1/3}\mathbf{L}_{p,n} = \left(\frac{\mathbf{M}_{1,2}}{\mathbf{m}_{p,n}}\right)^{1/3}\frac{\hbar}{\mathbf{m}_{p,n}\mathbf{c}} \qquad 18.48$$

Putting this expression to (18.46) we get for gravitational force between two macroscopic objects, each of them formed by atoms with coherently pulsing protons and neutrons:

$$\mathbf{F}_G = \frac{1}{\mathbf{r}^2}\pi\rho_{Bv}\frac{(\mathbf{M}_1\mathbf{M}_2)^{2/3}}{\mathbf{m}_{p,n}^{4/3}}\left(\frac{\hbar}{\mathbf{m}_{p,n}}\right)^4\frac{1}{\mathbf{c}^2} \qquad 18.49$$

Equalizing this formula with Newton's one: $\mathbf{F}_G^N = \frac{1}{\mathbf{r}^2}\mathbf{G}(\mathbf{M}_1\mathbf{M}_2)$, we get the expression for gravitational constant:

$$\mathbf{G} = \pi\frac{\rho_G}{\sqrt[3]{\mathbf{M}_1\mathbf{M}_2}}\frac{\hbar^2/\mathbf{c}^2}{\sqrt[3]{\mathbf{m}_{p,n}^{16}}} \qquad 18.50$$

The condition of gravitational constant permanent value following from (18.50), is a permanent ratio of virtual density $\rho_G$ of asymmetric Bivacuum dipoles in gravitational field to cube root of product of mass of gravitating bodies $\sqrt[3]{\mathbf{M}_1\mathbf{M}_2}$:

$$\mathbf{G} = \mathbf{const}, \quad \textit{if} \quad \frac{\rho_G}{\sqrt[3]{\mathbf{M}_1\mathbf{M}_2}} = \mathbf{const} \qquad 18.51$$

## 19. The mechanisms, increasing the refraction index of Bivacuum

By definition, the *torus* is a figure, formed by rotation of a circle with radius $\mathbf{L}_{\mathbf{V}^\pm}^i$, around the axis, shifted from the center of the circle at the distance $\pm\Delta\mathbf{L}_{E,H,G}$. The electric, magnetic ($E$, $H$) and gravitational ($G$) vibrations of positions $(\pm\Delta\mathbf{L}_{E,H,G})_{V^\pm}$ of the big number of recoiled Bivacuum fermions $\mathbf{BVF}_{rec}$, induced by the elastic *recoil⇋antirecoil* deformations of Bivacuum matrix, are accompanied by vibrations of square and volume of their torus ($\mathbf{V}^+$) and antitorus ($\mathbf{V}^-$).

The electric, magnetic and gravitational increments of square $(\Delta\mathbf{S}_{\mathbf{V}^{\pm}}^{E,G})$ and volume $(\Delta\mathbf{V}_{\mathbf{V}^{\pm}}^{E,G})$ of tori and antitori of $(\mathbf{BVF}_{rec}^{\updownarrow})^i$, as a consequence of their center vibrations, can be presented as:

$$\Delta\mathbf{S}_{\mathbf{V}^{\pm}}^{E,H,G} = 4\pi^2|\Delta\mathbf{L}_{EH,G}|_{V^{\pm}}^{EH,G} \times \mathbf{L}_{\mathbf{V}^{\pm}} \qquad 19.1$$

$$\Delta\mathbf{V}_{\mathbf{V}^{\pm}}^{E,H,G} = 4\pi^2|\Delta\mathbf{L}_{EH,G}|_{V^{\pm}}^{EH,G} \times \mathbf{L}_{\mathbf{V}^{\pm}}^2 \qquad 19.2$$

At conditions of zero-point oscillations, corresponding to Golden Mean (GM), when the ratio $(\mathbf{v}_0/\mathbf{c})^2 = \boldsymbol{\phi}$ and external tangential or translational velocity of Bivacuum dipoles ($\mathbf{v}$) is zero, the maximum shifts of center of secondary Bivacuum dipoles *in vicinity of pulsing elementary particles* due to electromagnetic and gravitational recoil-antirecoil (zero-point) vibrations are, correspondingly:

$$(\Delta\mathbf{L}_{\mathbf{E,H}}^{\mathbf{i}})_{V^{\pm}}^{\phi} = \left(\boldsymbol{\tau}_{C\rightleftharpoons W}^{\phi}\ \mathbf{v}_{EH}^{\phi}\right)^i = \frac{\hbar}{\mathbf{m}_0^i\mathbf{c}}(\boldsymbol{\alpha\phi})^{1/2} = 0,067\,(\mathbf{L}_{\mathbf{V}^{\pm}}^i) \qquad 19.3$$

$$(\Delta\mathbf{L}_{\mathbf{G}}^{\mathbf{i}})_{V^{\pm}}^{\phi} = \left(\boldsymbol{\tau}_{C\rightleftharpoons W}^{\phi}\ \mathbf{v}_{G}^{\phi}\right)^i = \frac{\hbar}{\mathbf{m}_0^i\mathbf{c}}\sqrt{\beta\phi} = 3,22 \times 10^{-23}\,(\mathbf{L}_{\mathbf{V}^{\pm}}^i) \qquad 19.4$$

where: the period of recoil $\leftrightharpoons$ antirecoil oscillation, induced by $\mathbf{C} \rightleftharpoons \mathbf{W}$ pulsation of elementary particle with rest mass $\mathbf{m}_0^i$ is $\left[\boldsymbol{\tau}_{C\rightleftharpoons W}^{\phi} = \mathbf{1}/\boldsymbol{\omega}_{C\rightleftharpoons W}^{\phi} = \hbar/\mathbf{m}_0^i\mathbf{c}^2\right]^i$; the recoil$\leftrightharpoons$antirecoil most probable velocity of zero-point oscillations, as a background of electric and magnetic fields (14.38) is: $\mathbf{v}_{E,H}^{\phi} = \mathbf{c}(\boldsymbol{\alpha\phi})^{1/2} = 0.201330447 \times 10^8\,\mathrm{m\,s^{-1}}$ where $(\boldsymbol{\alpha\phi})^{1/2} = 0,067$;

the corresponding zero-point velocity of the electron (18.42), which determines gravitational field is:

$$(\mathbf{v}_G)_{eff}^{\phi} = \mathbf{c}\sqrt{\beta_e\phi} = 0.322 \times 10^{-22}\,\mathbf{c}$$

where $\boldsymbol{\beta}_e^{1/2} = 0.48 \times 10^{-22}$.

The dielectric permittivity of Bivacuum and corresponding refraction index, by analogy with our theory of refraction index of matter (Kaivarainen, 2008), can be presented as a ratio of free volume of Bivacuum fermions and bosons in symmetric *primordial* Bivacuum ($\mathbf{V}_{\mathbf{pr}}$) to their volume in *secondary* Bivacuum (i.e. Bivacuum in presence of matter and fields): $\mathbf{V}_{\mathbf{sec}} = \mathbf{V}_{BVF} - (\mathbf{r}/r)\Delta\mathbf{V}_{\mathbf{BVF}_{rec}}^{E,G}$, perturbed by matter and fields. The secondary Bivacuum is optically more dense, than primary, if we assume that the volume, occupied by Bivacuum fermions, as a pair of tori and antitori, is excluded for photons.

The electric, magnetic and gravitational potentials and the related excluded volumes of perturbed Bivacuum fermions/antifermions decline with distance ($r$) as:

$$(\vec{\mathbf{r}}\,/r)\Delta\mathbf{V}_{\mathbf{BVF}_{rec}}^{E,H} \quad \text{and} \quad (\vec{\mathbf{r}}\,/r)\Delta\mathbf{V}_{\mathbf{BVF}_{rec}}^{G}$$

where: ($r$) is a distance from the charged and/or gravitating particle and $\vec{\mathbf{r}}$ is the unitary radius vector. Taking all this into account, we get for permittivity of secondary Bivacuum (see chapter 8 in book by Kaivarainen, 2008):

$$\boldsymbol{\varepsilon} = \mathbf{n}^2 = \left(\frac{\mathbf{c}}{\mathbf{v}_{EH,G}}\right)^2 = \frac{N\,\mathbf{V}_{\mathbf{pr}}}{N\,\mathbf{V}_{\mathbf{sec}}} = \frac{\mathbf{V}_{BVF}}{\mathbf{V}_{BVF} - (\mathbf{r}/r)\Delta\mathbf{V}_{\mathbf{BVF}rec}^{EH,G}} = \frac{1}{(1-\mathbf{r}/r)\Delta\mathbf{V}_{\mathbf{BVF}rec}^{EH,G}/\mathbf{V}_{BVF}} \qquad 19.5$$

$$\mathbf{n}^2 = \frac{\mathbf{1}}{\mathbf{1} - (\mathbf{r}/r)\, 3\pi|\mathbf{\Delta L}|_{V^\pm}^{EH,G} \times \mathbf{L}_{\mathbf{V}^\pm}} \tag{19.6}$$

where: the velocity of light propagation in asymmetric secondary Bivacuum of higher virtual density, than in primordial one, is notated as: $\mathbf{v}_{E,H,G} = \mathbf{c}_{E,H,G}$; the volume of primordial Bivacuum fermion is $\mathbf{V}_{BVF} = (4/3)\pi\mathbf{L}^3_{V^\pm}$ and its increment in secondary Bivacuum: $\mathbf{\Delta V}^{E,G}_{\mathbf{BVF}_{rec}} = \mathbf{\Delta V}^{E,G}_{\mathbf{V}^\pm}$ (19.2);

($\mathbf{r}/r$) is a ratio of unitary radius-vector to distance between the source of $[\mathbf{C} \leftrightharpoons \mathbf{W}]$ pulsations (elementary particle) and point, perturbed by the electrostatic, magnetic and gravitational potential $\mathbf{BVF}^{E,H,G}_{rec}$.

Putting (19.3) into formula (19.6) we get for refraction index of Bivacuum and relativistic factor ($\mathbf{R}_{E,H}$) in presence of charged elementary particle (electron, positron or proton, antiproton) the following expression:

$$\left[\varepsilon = \mathbf{n}^2 = \left(\frac{\mathbf{c}}{\mathbf{c}_{E,H}}\right)^2\right]_{E,H} = \frac{1}{1 - (\mathbf{r}/r)\, 3\pi(\alpha\phi)^{1/2}} \lesssim 2.71 \tag{19.7}$$

where: $1 \lesssim \mathbf{n}^2 \lesssim 2,71$ is tending to 1 at $r \to \infty$.

The relativistic factor, determined by contribution of electric and magnetic fields we introduce as:

$$\mathbf{R}_{EH} = \sqrt{1 - \frac{(\mathbf{c}_{E,H})^2}{\mathbf{c}^2}} = \sqrt{(\mathbf{r}/r)\, 0,631} \lesssim (\mathbf{r}/r)^{1/2}\, 0.794 \tag{19.8}$$

$0 \lesssim \mathbf{R}_{E,H} \lesssim 0,794$ is tending to zero at $r \to \infty$.

In similar way, using (19.4) and (19.6), for the refraction index of Bivacuum and the corresponding relativistic factor ($\mathbf{R}_G$) of gravitational vibrations of Bivacuum fermions ($\mathbf{BVF}^\updownarrow$) in the vicinity of pulsing elementary particles at zero-point conditions, we get:

$$\left[\boldsymbol{\varepsilon} = \mathbf{n}^2 = \left(\frac{\mathbf{c}_G}{\mathbf{c}}\right)^2\right]_G = \frac{1}{1 - (\mathbf{r}/r)3\pi(\beta^e\phi)^{1/2}} \succcurlyeq 1 \tag{19.9}$$

where

$$\sqrt{\beta_e\phi} = 0.322 \times 10^{-22}\, \mathbf{c}$$

The gravitational relativistic factor:

$$\mathbf{R}_G = \sqrt{1 - \left(\frac{\mathbf{c}_G}{\mathbf{c}}\right)^2} = \sqrt{(\mathbf{r}/r)\, 0,322 \times 10^{-22}} \lesssim (\mathbf{r}/r)^{1/2}\, 0,567 \times 10^{-11} \tag{19.10}$$

Like in previous case, the Bivacuum refraction index, increased by gravitational potential, is tending to its minimum value: $\mathbf{n}^2 \to 1$ at the increasing distance from the source: $r \to \infty$.

The charge - induced refraction index $\mathbf{n}_{E,H}$ in contrast to the mass - induced one, is independent of lepton generations of Bivacuum dipoles $(e, \mu, \tau)$.

The formulas (19.7) and (19.9) for Bivacuum dielectric permittivity and refraction index in presence of charged elementary particles point that bending probability of photons trajectory in presence of charged particles is much higher, than in presence of neutral particles with similar mass.

We have to stress, that the *light velocity* in conditions: $[\mathbf{n}^2_{E,H,G} = \mathbf{c}/\mathbf{v}_{E,H,G} = \mathbf{c}/\mathbf{c}_{E,H,G}] > 1$ is not longer a scalar, but a vector, determined by the gradient of Bivacuum fermion symmetry shift:

$$grad\,\mathbf{c} \sim grad\,\Delta|\mathbf{m}_V^+ - \mathbf{m}_V^-|_{E,H,G}\,\mathbf{c}^2 = grad\,\Delta(\mathbf{m}_V^+\mathbf{v}^2) \qquad 19.11$$

and corresponding gradient of torus and antitorus mass symmetry shift, characterized by the expression: $\mathbf{1} - \mathbf{m}_V^-/\mathbf{m}_V^+ = (\mathbf{c}_{E,H,G}/\mathbf{v})^2$:

$$grad[\mathbf{1} - \mathbf{m}_V^-/\mathbf{m}_V^+] = grad\left(\frac{\mathbf{c}_{E,H,G}}{\mathbf{c}}\right)^2 = grad\frac{1}{\mathbf{n}^2} \qquad 19.12$$

So, our approach explains bending of the light beams, under influence of strong gravitational potential in another way, than general theory of relativity.

Similar to our idea of polarizable vacuum and its permittivity variations has been developed by Dicke (1957), Fock (1964) and Puthoff (2001).

For spherically symmetric star or planet it was shown, using Dicke model (Dicke, 1957), that the dielectric constant **K** of polarizable vacuum is given by exponential form:

$$\mathbf{K} = \mathbf{exp(2GM/rc^2)} \qquad 19.13$$

where **G** is gravitational constant, **M** is mass, and **r** is the distance from the mass center.

The approximation of the formula above, given by Puthoff (2001):

$$\mathbf{K} \approx \mathbf{1} + \frac{2GM}{rc^2} + \frac{1}{2}\left(\frac{2GM}{rc^2}\right)^2 \qquad 19.14$$

Our approach propose physical mechanism of Bivacuum optical density increasing near charged and gravitating particles, inducing light beams bending.

The increasing of the excluded for photons volume of tori and antitori due to their rotations and vibrations, enhance the refraction index of Bivacuum and decrease the light velocity near gravitating and charged objects.

## 20. Application of angular momentum conservation law for evaluation of curvatures of electric and magnetic potentials

From the formulas of recoil energy responsible for electric and magnetic fields (18.18 and 18.23a) of the triplet $< [\mathbf{F}_\uparrow^- \bowtie \mathbf{F}_\downarrow^+]_{S=0} + \left(\mathbf{F}_\updownarrow^\pm\right)_{S=\pm 1/2} >^{e,\tau}$, we can find out the relations between the internal and external angular recoil momentums of $\mathbf{CVC}^\pm$.

Total longitudinal recoil momentum $\mathbf{M}_{tot}$ for electric and magnetic contributions is a sum of the internal $\mathbf{M}_0^{in}$ (zero-point) and external $\mathbf{M}_\lambda^{ext}$ angular momentums:

$$\mathbf{M}_{tot}(E,H) = \mathbf{M}_0^{in}(E,H) + \mathbf{M}_\lambda^{ext}(E,H) \qquad 20.1$$

From the law of angular momentum conservation it follows, that the angular momentums of Cumulative virtual cloud ($\mathbf{CVC}^\pm$) - right part of (20.2) and the recoil (*rec*) momentum - left part of 20.2 $[\alpha\mathbf{m}_0\mathbf{c}\mathbf{L}_\mathbf{E}^0]_{rec}$, accompanied $[\mathbf{C} \rightleftharpoons \mathbf{W}]$ pulsation of sub-elementary fermions in triplets should be equal.

For internal electric and magnetic recoil momentums, related to the rest mass and charge of elementary fermions origination, we get from (18.23a):

$$\mathbf{M}_0^{in}(E,H) = [\alpha\mathbf{m}_0\mathbf{c}\mathbf{L}_\mathbf{E}^0]_{rec} = [\mathbf{m}_0\mathbf{c}\mathbf{L}_0 - \alpha\mathbf{m}_0\mathbf{c}\mathbf{L}_0]_{\mathbf{CVC}} \qquad 20.2$$

$$or: \quad \alpha\mathbf{L}_\mathbf{E}^0 = \mathbf{L}_0 - \alpha\mathbf{L}_0 \qquad 20.2a$$

$$or: \quad \alpha\mathbf{L}_\mathbf{E}^0 = \mathbf{L}_0(1 - \alpha) \qquad 20.2b$$

where the Compton radius of elementary particle (electron, for example) is $\mathbf{L_0} = \hbar/\mathbf{m}_0\mathbf{c}$ and momentum is $\mathbf{m}_0\mathbf{c}$.

For external contribution to recoil angular momentum, responsible for electric and magnetic fields we get from (18.18):

$$\mathbf{M}_{\lambda}^{ext} = [\alpha \mathbf{m}_V^{+}\mathbf{v}\mathbf{L}_{\mathbf{E}}^{ext}]_{rec} = [\mathbf{m}_V^{+}\mathbf{v}\mathbf{L}_B - \alpha \mathbf{m}_V^{+}\mathbf{v}\mathbf{L}_B]_{\mathbf{CVC}} \qquad 20.3$$

$$or: \quad \alpha \mathbf{L}_{\mathbf{E}}^{ext} = \mathbf{L}_B - \alpha \mathbf{L}_B \qquad 20.3a$$

where the external momentum of particle $\mathbf{p}^{ext} = \mathbf{m}_V^{+}\mathbf{v} = \mathbf{h}/\boldsymbol{\lambda}_B = \frac{\hbar}{\mathbf{L}_B}$ is directly related to de Broglie wave length and external translational velocity of particle $\boldsymbol{\lambda}_B = 2\pi \mathbf{L}_B = \mathbf{h}/\mathbf{m}_V^{+}\mathbf{v}),$

The total momentum is a sum of the internal and external angular momentums:

$$\mathbf{M}_{tot} = \alpha(\mathbf{m}_0\mathbf{c}\mathbf{L}_{\mathbf{E}}^{0} + \mathbf{m}_V^{+}\mathbf{v}\mathbf{L}_{\mathbf{E}}^{ext})_{rec} = \qquad 20.4$$

$$= \mathbf{m}_0\mathbf{c}\mathbf{L_0} + \mathbf{m}_V^{+}\mathbf{v}\mathbf{L}_B - \alpha(\mathbf{m}_0\mathbf{c}\mathbf{L_0} + \mathbf{m}_V^{+}\mathbf{v}\mathbf{L}_B)_{\mathbf{CVC}}$$

From 20.2b and 20.4 we can see, that the space curvature, related to electric and magnetic interaction at zero-point condition ($\mathbf{v} = \mathbf{0}$), induced by $[C \rightleftharpoons W]$ pulsation of unpaired sub-elementary fermion in triplets (electrons or protons) is:

$$\mathbf{L}_{\mathbf{E}}^{0} = \frac{\mathbf{L}_0}{\alpha}(1-\alpha) = \mathbf{L}_0\left(\frac{1}{\alpha} - 1\right) = a_B - \mathbf{L}_0 = 136,036\mathbf{L}_0 \qquad 20.5$$

$$\alpha = 0.0072973506 \cong 1/137$$

So, we find out, that the space curvature, characteristic for electromagnetic potential of the electron at Golden Mean - internal rotational zero-point conditions ($\mathbf{L}_{\mathbf{E}}^{0}$) is very close to the radius of the *1st Bohr orbit* ($a_B$) in hydrogen atom:

$$a_B = \frac{1}{\alpha}\mathbf{L}_0 = 137,036\mathbf{L}_0 = 0.5291 \times 10^{-10}\, m \qquad 20.6$$

*Let us consider now the curvature of electric +magnetic potentials, determined by the external dynamics* of the charged particle and its de Broglie wave length from:

$$\mathbf{L}_{\mathbf{E}}^{ext} = \frac{\mathbf{L}_B}{\alpha}(1-\alpha) \cong \mathbf{L}_B\left(\frac{1}{\alpha} - 1\right) = 136.036\mathbf{L}_B = 136.036\frac{\lambda_B}{2\pi} \qquad 20.7$$

In most common nonrelativistic conditions the de Broglie wave length of elementary particle is much bigger than it its Compton length ($\mathbf{L}_B = \frac{\lambda_B}{2\pi} = \frac{1}{2\pi}\frac{\mathbf{h}}{\mathbf{mv}} >> \mathbf{L}_0 = \frac{\hbar}{\mathbf{m}_0\mathbf{c}}$) and, consequently, the effective external radius of Coulomb potential action is much bigger, than the minimum internal one: $\mathbf{L}_{\mathbf{E}}^{ext} >> \mathbf{L}_{\mathbf{E}}^{0}$.

## 21. Pauli principle: How it works ?

Let us consider the reasons why the Pauli principle "works" for *fermions* and do not work for *bosons*.

In accordance to our model of elementary particles, the numbers of sub-elementary fermions and sub-elementary antifermions, forming *bosons*, like photons (Fig.8), are equal.

Each of sub-elementary fermion and sub-elementary antifermion in symmetric pairs $[\mathbf{F}_{\updownarrow}^{+} + \mathbf{F}_{\updownarrow}^{-}]$ of *bosons* can pulsate between their [C] and [W] states in-phase ($S = \pm 1\hbar$) or counterphase, providing integer or zero spin ($S = 0$), correspondingly. In both cases the energies of positive and negative subquantum particles of $\mathbf{CVC}^{+}$ and $\mathbf{CVC}^{-}$ , emitted $\rightleftharpoons$ absorbed by $[\mathbf{F}_{\updownarrow}^{+} + \mathbf{F}_{\updownarrow}^{-}]$ excite positive and negative virtual pressure waves: $\mathbf{VPW}^{+}$ and $\mathbf{VPW}^{-}$ and compensate each other.

For the other hand, the oscillating density of sub-quantum particles and antiparticles in $\mathbf{CVC}^+$ and $\mathbf{CVC}^-$, emitted⇌absorbed by unpaired sub-elementary fermion or antifermion of elementary fermions $< [\mathbf{F}_\uparrow^+ \bowtie \mathbf{F}_\downarrow^-] + \mathbf{F}_\updownarrow^\pm >^i$ with the same spin and phase of $[\mathbf{C} \rightleftharpoons \mathbf{W}]$ pulsation are not equal to each other and do not compensated each other.

*In the framework of our model, Pauli repulsion effect between fermions with the same spin states and energy, meaning the same phase and frequency of* $[C \rightleftharpoons W]$ *pulsation, is similar to the effect of excluded volume.*

This effect is provided by spatial incompatibility of two cumulative virtual clouds: $\mathbf{CVC}_1^\pm$ and $\mathbf{CVC}_2^\pm$ from sub-elementary particles or antiparticles of triplets, emitted in the same moment of time in the same volume.

The effect of excluded volume works, if the distance between $\mathbf{CVC}_1^\pm$ and $\mathbf{CVC}_2^\pm$ is equal or less, than doubled de Broglie wave length of triplets: $2\boldsymbol{\lambda}_B = \mathbf{2}\boldsymbol{\lambda}_{CVC} = \mathbf{2h}/\mathbf{m}_V^+\mathbf{v}^{ext}$.

**Let us analyze this situation in more detail.**

The average *external* translational energy ($\overline{\mathbf{E}}_{E,H}^{\mathbf{C}\rightleftharpoons\mathbf{W}}$) of fermions $[\mathbf{F}_\uparrow^+ \bowtie \mathbf{F}_\downarrow^-] + \mathbf{F}_\updownarrow^\pm >^i$ is:

$$\overline{\mathbf{E}}_{E,H}^{\mathbf{C}\rightleftharpoons\mathbf{W}} = \mathbf{E}_{E,H}^{\mathbf{C}\rightleftharpoons\mathbf{W}} \pm [(\mathbf{E}_{E,H})_{[C]} \rightleftharpoons (\mathbf{E}_{E,H})_{[W]}]_{tr} \qquad 21.1$$

It involves opposite by sign oscillation of Coulomb *potential* interaction between opposite electric poles of $\mathbf{CVC}^\pm$ in [W] phase $\left[\frac{|\mathbf{e}_+\mathbf{e}_-|}{\mathbf{L}_T}\right]_W$ separated by the amplitude of $\mathbf{CVC}^\pm$ ($\mathbf{A}_{\mathbf{CVC}^\pm}$) of $\mathbf{F}_\updownarrow^\pm >^i$, transforming to *kinetic* recoil perturbation of Bivacuum matrix $\alpha[(\mathbf{m}_V^+ - \mathbf{m}_V^-)\mathbf{c}^2]_C$ in [C] phase in the process of $[\mathbf{C} \rightleftharpoons \mathbf{W}]$ pulsation, representing electric and magnetic field:

$$\mathbf{E}_{E,H} = \left[\frac{|\mathbf{e}_+\mathbf{e}_-|}{\mathbf{A}_{\mathbf{CVC}^\pm}}\right]_W \overset{[\mathbf{C}\rightleftharpoons\mathbf{W}]}{\rightleftharpoons} \alpha[(\mathbf{m}_V^+ - \mathbf{m}_V^-)\mathbf{c}^2]_C = \alpha[\mathbf{m}_V^+\mathbf{v}^2]^{Dis} = (\mathbf{E}_{E,H})_{[W]}^{Dist} \qquad 21.2$$

where the amplitude of de Broglie wave of the charged fermion is:

$$\mathbf{A}_{\mathbf{CVC}^\pm} = \frac{\hbar}{(\mathbf{m}_V^+ - \mathbf{m}_V^-)\mathbf{c}} = \frac{\hbar}{\mathbf{m}_V^+\mathbf{v}}\frac{\mathbf{c}}{\mathbf{v}} = \mathbf{L}_B\frac{\mathbf{c}}{\mathbf{v}} \qquad 21.2a$$

The electromagnetic energy of unpaired $\mathbf{F}_\updownarrow^\pm >^i$ is equal to longitudinal recoil energy:

$$(\mathbf{E}_{E,H}) = \alpha[(\mathbf{m}_V^+ - \mathbf{m}_V^-)\mathbf{c}^2]_C = \alpha[\mathbf{m}_V^+\mathbf{v}^2]_{rec} \sim \boldsymbol{\varepsilon}_C \ (\textit{electric field energy}) \qquad 21.3$$

The total energy of $\mathbf{CVC}^\pm$, equal to energy of empirical de Broglie wave, which is responsible for Pauli repulsion ($\boldsymbol{\varepsilon}_P$), is determined by its external translational momentum: $\frac{\mathbf{h}^2}{\mathbf{m}_V^+\boldsymbol{\lambda}_B^2} = \mathbf{m}_V^+\mathbf{v}^2$ minus recoil energy:

$$\mathbf{E}_B = \frac{\mathbf{h}^2}{\mathbf{m}_V^+\boldsymbol{\lambda}_B^2} - \alpha[\mathbf{m}_V^+\mathbf{v}^2] = \mathbf{m}_V^+\mathbf{v}^2(1-\alpha) \sim \boldsymbol{\varepsilon}_P \qquad 21.4$$

The Coulomb repulsion ($\boldsymbol{\varepsilon}_C$) between two similar elementary charge or attraction between opposite ones, is determined by electric field energy (21.3). For the other hand, the Pauli repulsion ($\boldsymbol{\varepsilon}_P$) between these charges, as a fermions, pulsing in the same phase and frequency on the distance, *equal or less than doubled de Broglie wave length*: $r \leq 2\lambda_B = h/\mathbf{m}_V^+\mathbf{v}$ is dependent on real energy of $\mathbf{CVC}^\pm$ (21.4).

The ratio between Pauli and Coulomb repulsion energies between two similar fermions on the distances about or less, than de Broglie wave length of these charges ($\boldsymbol{\lambda}_B$) is:

$$\frac{\boldsymbol{\varepsilon}_P}{\boldsymbol{\varepsilon}_C} = \frac{1-\alpha}{\alpha} = \frac{1}{\alpha} - 1 \simeq 136 \qquad 21.5$$

We can see, that it is close to reverse value of electromagnetic fine structure constant: $\mathbf{1/\alpha} \simeq 137$.

This means, that on these distances, comparable with linear dimensions of $\mathbf{CVC}^{\pm}$ usually much bigger than Compton length of charges: $\boldsymbol{\lambda}_B >> (\mathbf{L}_0 = \hbar/\mathbf{m}_0\mathbf{c})$, the Pauli non-electromagnetic repulsion is more than hundred times bigger, than Coulomb interaction.

Pauli repulsion is absent if $[\mathbf{C} \rightleftharpoons \mathbf{W}]$ pulsations of two unpaired sub-elementary fermions of two electron $< [\mathbf{F}_{\uparrow}^{+} \bowtie \mathbf{F}_{\downarrow}^{-}] + \mathbf{F}_{\uparrow}^{-} >$ have the same energy and frequency:

$$\nu = \frac{(\mathbf{m}_V^{+} - \mathbf{m}_V^{-})^{res}\mathbf{c}^2}{\mathbf{h}} \qquad 21.6$$

but opposite phase of pulsation.

The mechanism, proposed, explains the absence of the Pauli repulsion in systems of bosons and Cooper pairs, making possible their Bose condensation.

*21.1 Spatial compatibility of sub-elementary fermions of the same charge and opposite spins in triplets*

We postulate in our model, that $[C \Leftrightarrow W]$ pulsation of paired sub-elementary fermion and antifermion $[F_{\uparrow}^{+} \bowtie F_{\downarrow}^{-}]$ of opposite spins and charges in composition of the electron $< \left[(F_{\uparrow}^{+})^{[2]} \bowtie (F_{\downarrow}^{-})^{[1]}\right] + \mathbf{F}_{\uparrow}^{-} >$ or proton are counterphase with pulsation of unpaired $\mathbf{F}_{\uparrow}^{\pm} >$ (see Fig. 5).

It follows from above explanation of Pauli principle, that in the case of counterphase $[C \Leftrightarrow W]$ pulsations of one of paired $(F_{\downarrow}^{\pm})^{[1]}$ and unpaired $\mathbf{F}_{\uparrow}^{\pm} >$ with the same frequency (energy) and *opposite* spins, with similar charges, localized in the *same* energy realm, they are *spatially compatible*, as far their corpuscular [C] and wave [W] phase are realized alternatively in different semi-periods. Consequently, the Pauli repulsion between sub-elementary fermions of opposite spins in composition of elementary particles is absent.

*21.2 The reversibility of fermion properties after double turn* ($2\times360^0$) *in magnetic field, as a consequence of paired fermions spin state additivity*

It is known fact, that the total rotating cycle for spin of the electrons or positrons is not $360^0$, but $720^0$, i.e. *double turn* by external magnetic field of special configuration, is necessary to return elementary fermions to starting state (Davies, 1985). The correctness of any new model of elementary particles should be testified by its ability to explain this nontrivial fact.

We may propose *three* possible explanations, using our model of the electrons, positrons, protons and antiprotons, as a triplets of sub-elementary fermions/antifermions.

Let us analyze them on example of the electron:

$$< [F_{\uparrow}^{+} \bowtie F_{\downarrow}^{-}] + \mathbf{F}_{\uparrow}^{-} >^e \qquad 21.7$$

**1**. We may assume, that the direction of external magnetic field rotation acts *only on unpaired* sub-elementary fermion, as asymmetric pair [torus $(\mathbf{V}^{-})$ + antitorus $(\mathbf{V}^{+})$]:

$$\mathbf{F}_{\uparrow}^{-} = (\mathbf{V}^{-} \upuparrows \mathbf{V}^{+})_{as} \qquad 21.8$$

if the resulting magnetic moment of pair $[F_{\uparrow}^{+} \bowtie F_{\downarrow}^{-}]$ is zero and the pair do not interact with external magnetic field at all. In such conditions the 1st $360^0$ turn of external $\mathbf{H}$ field change the direction of rotation of one of two tori rotation to the opposite one: $(\mathbf{V}^{-})\uparrow \;\rightarrow (\mathbf{V}^{-})\downarrow$, transforming sub-elementary fermion to sub-elementary Bivacuum

boson ($\mathbf{B}^-$):

$$[\mathbf{F}_\uparrow^- \equiv (\mathbf{V}^- \upuparrows \mathbf{V}^+)] \stackrel{360^0}{\longrightarrow} [\mathbf{B}^- \equiv (\mathbf{V}^- \downarrow\uparrow \mathbf{V}^+)] \qquad 21.9$$

One more $360^0$ turn of the external magnetic field converts this sub-elementary boson and the triplet (21.9) to starting condition. The total cycle for unpaired $\mathbf{F}_\uparrow^- >$ of triplet can be presented as:

$$\text{(I)} \quad [\mathbf{F}_\uparrow^- > \ \equiv (\mathbf{V}^- \upuparrows \mathbf{V}^+)] \stackrel{360^0}{\longrightarrow} [\mathbf{B}^- \equiv (\mathbf{V}^- \downarrow\uparrow \mathbf{V}^+)] \stackrel{360^0}{\longrightarrow} [\mathbf{F}_\uparrow^- > \ \equiv (\mathbf{V}^- \upuparrows \mathbf{V}^+)] \qquad 21.10$$

**2**. *The second possible explanation* of double $720^0$ turn may be a consequence of following two stages, involving origination of pair of sub-elementary bosons ($B^\pm \bowtie B^\pm$) from *pair* of sub-elementary fermions, as intermediate stage and two full turns ($2 \times 360^0$) of unpaired sub-elementary fermion:

$$\text{(II)} \quad < [F_\uparrow^+ \bowtie F_\downarrow^-] + \mathbf{F}_\uparrow^- > \stackrel{360^0}{\rightarrow} < [B^\pm \bowtie B^\mp] + \mathbf{F}_\uparrow^- > \stackrel{360^0}{\rightarrow} < [F_\uparrow^+ \bowtie F_\downarrow^-] + \mathbf{F}_\uparrow^- > \qquad 21.11$$

Both of these mechanisms are not very probable, because they involve the action of external magnetic field on single or paired sub-elementary bosons with zero spin and, consequently, zero magnetic moment.

**3**. *The most probable third mechanism* avoids such strong assumption. The external rotating **H** field interact in two stage manner ($2 \times 360^0$) only with sub-elementary fermions/antifermions, changing their spins. However this mechanism demands that the angle of spin rotation of sub-elementary particle and antiparticles of neutral pairs $[F_\uparrow^+ \bowtie F_\downarrow^-]$ are the additive parameters. It means that turn of resulting spin of *pair* on $360^0$ includes reorientation spins of each $F_\uparrow^+$ and $F_\downarrow^-$ only on $180^0$. Consequently, the full spin turn of pair $[F_\uparrow^+ \bowtie F_\downarrow^-]$ resembles that of Mobius transformation.

The spin of unpaired sub-elementary fermion $\mathbf{F}_\uparrow^- >$, in contrast to paired ones, makes a *full turn* each $360^0$, i.e. twice in $720^0$ cycle:

$$< [(F_\uparrow^+)_x \bowtie (F_\downarrow^-)_y] + (\mathbf{F}_\uparrow^-)_z > \stackrel{360^0}{\longrightarrow} < [(F_\downarrow^+)_x \stackrel{180^0+180^0}{\bowtie} (F_\uparrow^-)_y] + (\mathbf{F}_\uparrow^-)_z > \rightarrow \qquad 21.12$$

$$\stackrel{360^0}{\longrightarrow} < [(F_\uparrow^+)_x \bowtie (F_\downarrow^-)_y] + (\mathbf{F}_\uparrow^-)_z >$$

The difference between the intermediate - 2nd stage and the original one in (9.8) is in opposite spin states of paired sub-elementary particle and antiparticle:

$$[(F_\uparrow^+)_x \bowtie (F_\downarrow^-)_y] \stackrel{360^0}{\longrightarrow} [(F_\downarrow^+)_x \stackrel{180^0+180^0}{\bowtie} (F_\uparrow^-)_y] \qquad 21.13$$

Because of Pauli repulsion (see previous section) between two sub-elementary fermions of the same spin state $(F_\uparrow^-)_y$ and $(\mathbf{F}_\uparrow^-)_z >$, in intermediate state of (21.12), the corresponding triplet configuration has deformed - stretched configuration, different from original and final ones.

In the latter - equilibrium configurations of triplet, the $[\mathbf{C} \rightleftharpoons \mathbf{W}]$ pulsation of unpaired sub-elementary fermion $(\mathbf{F}_\uparrow^-)_z >$ and paired $(F_\downarrow^-)_y$ is counterphase and spatially compatible due to the absence of Pauli repulsion.

One more known "strange" experimental feature of the electron can be explained by its triplet structure:

- the existence in triplets of paired, in-phase pulsating sub-elementary fermion and antifermion (21.13) with opposite charges, representing double electric dipoles (i.e. double charge) $[(F_\uparrow^+)_x \bowtie (F_\downarrow^-)_y]$, can be responsible for *two times stronger magnetic field*, generated by electron, as compared with those, generated by rotating sphere with single charge $|e^-|$.

*21.3. Bosons as a coherent system of sub-elementary and elementary fermions and antifermions forming Cooper pairs*

The spatial image of sub-elementary boson is a superposition of *strongly correlated* sub-elementary fermions with opposite charges and spin states with properties of Cooper pairs. In general case the elementary bosons are composed from the *integer* number of such pairs.

Bosons have zero or integer spin in the $\hbar$ units, in contrast to the half integer spins of fermions. In general case, bosons with $S = 0, \pm 1$ include: photons, gluons, mesons and boson resonances, phonons, pairs of elementary fermions with opposite spins, atoms and molecules.

We subdivide bosons into elementary and complex bosons:

1. *Elementary bosons* (like photons), composed from equal number of *sub-elementary* fermions and antifermions, moving with light velocity in contrast to complex bosons, like atoms;

2. *Complex bosons*, represent a coherent system of *elementary* fermions pairs: electrons and nucleons with resulting integer spin, like in neutral atoms and molecules.

Formation of stable *complex* bosons from elementary fermions with different actual masses: $(\mathbf{m}_V^+)_1 \neq (\mathbf{m}_V^+)_2$ is possible due to their electromagnetic attraction, like in *proton + electron* pairs in atoms. It may occur, if the length of their waves B are the same and equal to distance between them. These conditions may be achieved by difference in their external recoil velocities (see section 18.2), adjusting their momentums to the same value:

$$\mathbf{L}_1 = \hbar/(\mathbf{m}_V^+\mathbf{v})_1 = \mathbf{L}_2 = \hbar/(\mathbf{m}_V^+\mathbf{v})_2 \ldots = \mathbf{L}_n = \hbar/(\mathbf{m}_V^+\mathbf{v})_n \qquad 21.14$$

$$at: \quad \mathbf{v}_1/\mathbf{v}_n = (\mathbf{m}_V^+)_n/(\mathbf{m}_V^+)_1$$

The mentioned above conditions are the base for assembly of complex bosons, unified in the volume of 3D standing waves of fermions of the opposite or same spins.

The *hydrogen atom*, composing from two fermions: electron and proton is a simplest example of complex bosons. The heavier atoms also follow the same principle of self-organization.

The elementary boson, such as photon, represents dynamic superposition of sub-elementary fermions and antifermions, interacting via head-to-tail principle (Fig.8). Such composition determines the resulting external charge of photon, equal to zero and the value of photon's spin: $J = +1, 0, -1$.

Stability of all types of *elementary* particles: bosons and fermions (electrons, positrons etc.) is a result of superposition/exchange of cumulative virtual clouds $[\mathbf{CVC}^+ \bowtie \mathbf{CVC}^-]$ with gluon properties, emitted and absorbed in the process of in-phase $[C \rightleftharpoons W]$ pulsations of paired sub-elementary particles and sub-elementary antiparticles $[\mathbf{F}_\uparrow^+ \bowtie \mathbf{F}_\downarrow^-]$, like presented on Fig. 6.

## 22. The Mystery of Sri Yantra Diagram

In accordance to ancient archetypal ideas, geometry and numbers describe the fundamental energies in course of their dance - dynamics, transitions. For more than ten millenniums it was believed that the famous Tantric diagram-Sri Yantra contains in hidden form the basic functions active in the Universe (Fig. 13).

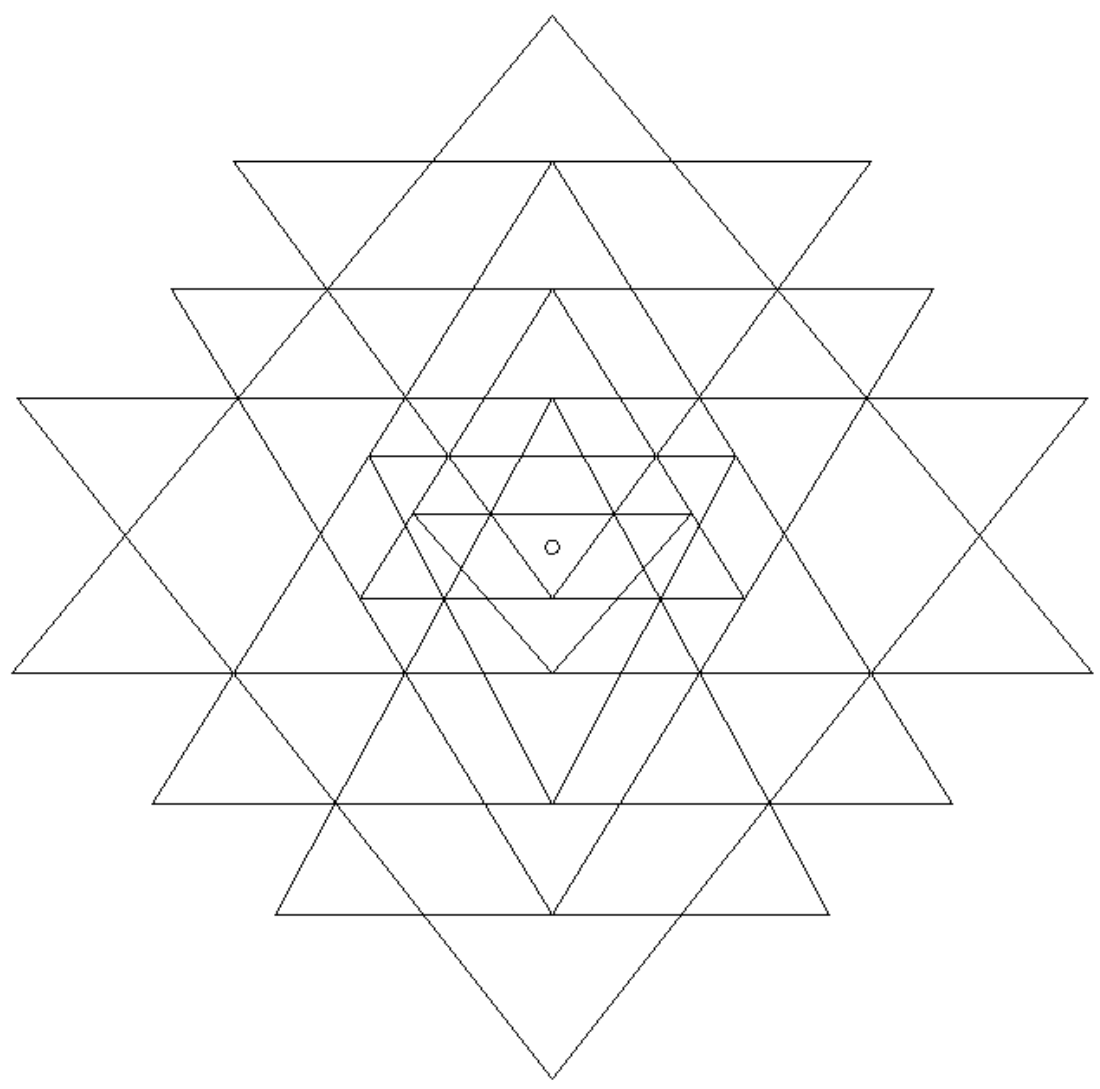

**Fig**. **13**. The Sri Yantra diagram is composed from nine triangles. Four of them are pointed up and five down.
In another way this diagram can be considered as superposition of:
a) the set of pairs of cones of opposite apex, corresponding to torus and antitorus of asymmetric Bivacuum fermions in [C] phase in different excitation states and
b) the set of diamonds, corresponding to [W] phase of corresponding excitation states of Bivacuum fermions (dashed lines).
Author is grateful to Patrick Flanagan for submitting of Sri Yantra diagram with precise coordinates of most important points, making possible its quantitative analysis.

Triangle is a symbol of a three-fold Nature. The Christian trinity, the symbol of God may be represented by triangle. The symbol of trinity is coherent to our idea of *triplets* of sub-elementary particles and antiparticles, as elementary particles. In Buddhism-Hindu triangle with *apex up* is a symbol of God-male and that with *apex down* is a symbol of God-female.

For millenniums it was believed, that Sri Yantra diagram represents geometric language, containing encrypted information about the principles of matter formation.

Let us analyze this diagram, using notions of our theory of elementary particles origination from Bivacuum dipoles and the mechanism of corpuscle - wave duality.

First of all, the ratio 5:4 between positive and negative triangles may reflect the primordial asymmetry of torus and antitorus of Bivacuum dipoles, as a condition of matter domination over antimatter.

We may see also, that Sri Yantra diagram contains the information about duality of sub-elementary fermions, forming elementary particles, i.e. their discrete corpuscular [C] and wave [W] phases. The diagram at Fig.13 can be considered as a superposition of:

*a) the [C] phase* is presented by set of pairs of cones of opposite apex, corresponding to asymmetric torus and antitorus of asymmetric Bivacuum fermions in different excitation states (see Fig. 11a, where the diameters of bases of pairs of cones correspond to diameters of torus and antitorus of Bivacuum fermions) and

*b) the [W] phase* is described by set of diamonds, corresponding to of Bivacuum fermions in different excitation states.

In accordance to our theory of sub-elementary fermion/antifermion origination, the former set (a) describes their [C] phase with different diameters of opposite cones bases, characterizing symmetry shift between torus ($V^+$) and antitorus ($V^-$), correspondingly. The asymmetry of torus and antitorus is increasing with Bivacuum fermion excitation state, accompanied by *decreasing* of spatial separation between them. From formula (1.4) for this

separation:

$$[\mathbf{d}_{\mathbf{V}^+ \Updownarrow \mathbf{V}^-}]^i_n = \frac{h}{\mathbf{m}^i_0 \mathbf{c}(1 + 2\mathbf{n})} \qquad 22.1$$

we can see, that the distance between torus and antitorus decreases with quantum number ($\mathbf{n}$) increasing, indeed.

It was astounding to find out, that at maximum excitation and maximum asymmetry of Bivacuum dipole, corresponding to minimum diamond dimension (Fig.14b), the ratio of *down* radius of cone/torus base to that of *upper* antitorus is $\mathbf{L}_+/\mathbf{L}_- = \mathbf{0.6}$, i.e. practically coincide with Golden mean ($\phi = \mathbf{0.618}$).

Just this critical ratio of torus and antitorus radiuses is a condition of the rest mass and charge origination, turning Bivacuum dipoles to muons and tauons, following by their fusion to triplets of electrons, protons, neutrons.

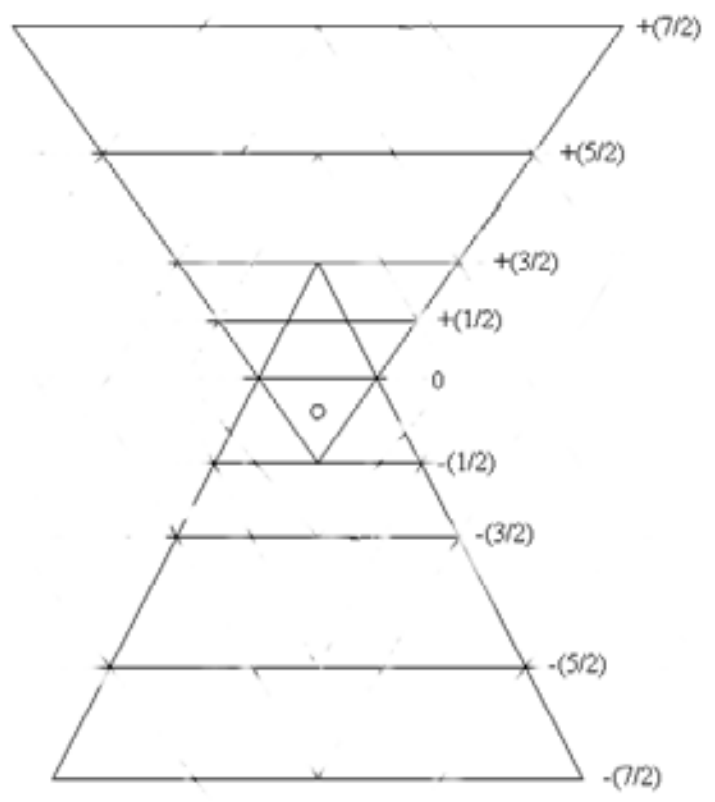

b
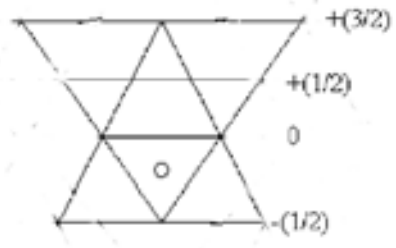

**Figure 14 a**. Part of Sri Yantra diagram, representing set of pairs of cones of opposite apex, corresponding to torus and antitorus of asymmetric Bivacuum fermions in [C] phase in different excitation states. The diameters of bases of pairs of cones corresponds to diameters of torus and antitorus of Bivacuum fermions.

**Figure 14b**. Superposition of [C] and [W] phase of asymmetric Bivacuum fermion, corresponding to critical state of excitation and asymmetry, determined by Golden mean condition. This state is characterized by origination of the rest mass and charge, turning Bivacuum fermion to sub-elementary fermion. The next stage of matter origination from Bivacuum is fusion of triplets of elementary fermions from sub-elementary fermions.

The diamonds of increasing as respect to Fig.15b dimensions, incorporated in Sri Yantra diagram (Fig.14), reflects [W] phase of Bivacuum dipoles of different excitation states in form of Cumulative Virtual Clouds [**CVC**], emitted and absorbed in the process of quantum beats between asymmetric states of torus and antitorus.

The probability of coincidental correlation of quantitative and qualitative features of Sri Yantra diagram properties with key features of our theory of elementary particles is very low. It is a mystery, indeed, that only 10 millenniums after famous Sri Yantra diagram became known in mankind history, we start to understand its encrypted information about fundamentals of quantum mechanics.

## 23 The examples of Bivacuum mediated interaction (BMI) between macroscopic objects

In accordance to our approach, the remote interaction between macroscopic Sender [S] and Receiver [R] can be realized, as a result of *Bivacuum mediated interaction (BMI)*, like superposition of distant and nonlocal components of their Virtual Replicas Multiplication ($\mathbf{VRM}_S = \Leftrightarrow = \mathbf{VRM}_R$), described in previous sections.

Nonequilibrium processes in [Sender], accompanied by acceleration of particles, like evaporation, heating, cooling, melting, boiling etc. may stimulate the *nonelastic effects* in the volume of [Receiver] and increments of modulated virtual pressure and spin waves ($\Delta\mathbf{VPW}^{\pm}_{\mathbf{m}}$ and $\Delta\mathbf{VirSW}^{\pm 1/2}_{\mathbf{m}}$), accompanied [$\mathbf{C} \rightleftharpoons \mathbf{W}$] pulsation of triplets $[\mathbf{F}^{+}_{\uparrow} \bowtie \mathbf{F}^{-}_{\downarrow}] + \mathbf{F}^{\pm}_{\updownarrow} >^{i}$ , formed by sub-elementary fermions of different generation, representing electrons, protons and neutrons.

The following unconventional kinds of effects of non electromagnetic and non-gravitational nature can be anticipated in the remote interaction between **macroscopic** nonequilibrium [Sender] and sensitive detector [Receiver] via multiple Virtual spin and energy guides $\mathbf{VirG}_{SME}$ (Fig.4), if our theory of nonlocal spin, momentum and energy exchange between [S] and [R], described above is correct:

**I**. Weak *repulsion and attraction* between 'tuned' [S] and [R] and rotational momentum in [R] induced by [S], as a result of transmission of momentum/kinetic energy and angular momentum (spin) between elementary particles of [S] and [R]. The probability of such 'tuned' interaction between [S] and [R] is dependent on dimensions of coherent clusters of atoms and molecules of condensed matter in state of mesoscopic Bose condensation (**mBC**) (Kaivarainen, 1995; 2001; 2003; 2004). The number of atoms in such clusters $\mathbf{N}(\mathbf{t},\mathbf{r})$ is related to number of $\mathbf{VirG}_{SME}$ in the bundles $\left[\mathbf{N}(\mathbf{t},\mathbf{r}) \times \sum^{\mathbf{n}} \mathbf{VirG}_{SME}(\mathbf{S} <=> \mathbf{R})\right]^{i}_{x,y,z}$ , connecting tuned **mBC** in [S] and [R]. The $\mathbf{N}(\mathbf{t},\mathbf{r})$ may be regulated by temperature, ultrasound, etc. The kinetic energy distant transmission from atoms of [S] to atoms of [R] may be accompanied by the temperature and local pressure/sound effects in [R];

**II**. Increasing the probability of *thermal fluctuations* in the volume of [R] due to decreasing of Van der Waals interactions, because of charges screening effects, induced by overlapping of distant virtual replicas of [S] and [R] and increasing of dielectric permittivity of Bivacuum. In water the variation of probability of cavitational fluctuations should by accompanied by the in-phase variation of pH and electric conductivity due to shifting the equilibrium: $H_2O \rightleftharpoons H^+ + HO^-$ to the right or left;

**III**. Small *changing of mass* of [R] in conditions, changing the probability of the inelastic recoil effects in the volume of [R] under influence of [S];

**IV**. Registration of metastable *virtual particles*, as a result of Bivacuum symmetry perturbations.

*The first kind* (I) of new class of interactions between coherent fermions of [S] and [R] is a result of huge number (bundles) of correlated virtual spin-momentum-energy guides $\mathbf{VirG}_{SME} \equiv [\mathbf{VirSW}_S^{\circlearrowright} <= \rightleftharpoons => \mathbf{VirSW}_R^{\circlearrowleft}]$ formation by standing spin waves ($\mathbf{VirSW}_{S,R}$).

These guides can be responsible for:

a) virtual signals (phase/spin), momentum and kinetic energy instant transmission between [S] and [R], meaning the nonlocal information and energy exchange;

b) the regulation of Pauli repulsion effects between fermions of [S] and [R] with parallel spins;

c) the transmission of macroscopic rotational momentum from [S] of [R]. This process provided by the entanglement channels $\left[\mathbf{N}(\mathbf{t},\mathbf{r}) \times \sum^{\mathbf{n}} \mathbf{VirG}_{SME}\,(\mathbf{S} <=> \mathbf{R})\right]^i_{x,y,z}$, is dependent on the difference between the *external* angular momentums of elementary fermions of [S] and [R].

*The second kind* (II) of phenomena: influence of [S] on probability of thermal fluctuations in [R], - is a consequence of the additional symmetry shift in Bivacuum fermions ($\mathbf{BVF}^{\updownarrow}$), induced by superposition of distant and nonlocal multiplicated Virtual Replicas of [S] and [R]: $\mathbf{VRM}^S \bowtie \mathbf{VRM}^R$, which is accompanied by increasing of Bivacuum fermions ($\mathbf{BVF}^{\updownarrow} = [\mathbf{V}^+ \Updownarrow \mathbf{V}^-]$) virtual charge: $\Delta\mathbf{e} = (\mathbf{e}_{V^+} - \mathbf{e}_{V^-}) << \mathbf{e}_0$ in the volume of [R]. Corresponding increasing of Bivacuum permittivity ($\varepsilon_0$) and decreasing magnetic permeability ($\mu_0$) : $\varepsilon_0 = 1/(\mu_0 \mathbf{c}^2)$ is responsible for the charges screening effects in volume of [R], induced by [S]. This weakens the electromagnetic Van der Waals interaction between molecules of [R] and increases the probability of defects origination and cavitational fluctuations in solid or liquid phase of Receiver.

*The third kind of phenomena (III)*: reversible decreasing of mass of rigid [R] can be a result of reversible lost of energy of Corpuscular phase of particles, as a consequence of inelastic recoil effects, following the in-phase $[\mathbf{C} \rightarrow \mathbf{W}]$ transition of $\mathbf{N}_{coh}$ coherent nucleons in the volume of [R].

The probability of recoil effects can be enhanced by heating the rigid object or by striking it by another hard object. This effect can be registered directly - by the object mass decreasing. In conditions, close to equilibrium, the Matter - Bivacuum energy exchange relaxation time, following the process of coherent $[\mathbf{C} \rightleftharpoons \mathbf{W}]$ pulsation of macroscopic fraction of atoms is very short and corresponding mass defect effect is undetectable. *Such collective recoil effect of coherent particles* could be big in superconducting or superfluid systems of macroscopic Bose condensation or in crystals, with big domains of atoms in state of Bose condensation.

*The fourth kind of the above listed phenomena* - increasing the probability of virtual particles and antiparticles origination in asymmetric Bivacuum in condition of forced resonance with exciting Bivacuum virtual waves will be discussed in section 16.2.

It will demonstrated also in chapter 17, that the listed above nontrivial consequences of Unified theory (I - IV) are consistent with unusual data, obtained by groups of Kozyrev (1984; 1991) and Korotaev (1999; 2000). It is important to note, that these experiments are incompatible with current paradigm. It means that it is timed out and should be replaced by the new one.

### *23.1 The idea of nonlocal signals transmitter and detector construction and testing*

Physical devices with function of [Sender] and [Receiver] for verification of nonlocal mechanism of communication via Virtual Guides of spin/information, momentum and energy, following from our Unified theory, were suggested (Kaivarainen, 2004a; 2004b). They represent two or more identical and 'tuned' to each other superconducting multi-coil systems.

The pair: [S] and [R] can be presented by two identical systems, composed from the same number of superconducting or superfluid coils of different radius - from few to tens of centimeters. enclosed in each other.

The ”tuning” of Virtual Replicas of [S] and [R] constructions in state of macroscopic Bose condensation (superconducting or superfluid) can be realized by keeping them nearby with parallel orientation of two set of rings during few hours for equalizing of their physical parameters, i.e. currents. After such tuning, they can be removed from each other, keeping their superconducting or superfluid state on at the same temperature, pressure and other conditions. The separation can be increased from hundreds of meters to hundreds of kilometers and tested for signals transmission in each equipped for such experiments laboratory.

The experiments for registration of nonlocal interactions could be performed, as follows. At the precisely fixed time moment, the superfluid or superconducting properties of one of rings of Sender [S], should be switched off by heating, ultrasound or magnetic field action. At the same moment of time the superconducting parameters of all rings of Receiver [R] should be registered. If the biggest changes will occur in the ring of [R]-system with the same radius, as that in [S]-system and faster, than light velocity, it will be a confirmation of possibility of nonlocal Bivacuum mediated information and momentum exchange - macroscopic entanglement, following from our theory and based on resonant principles. *The corresponding remote signals exchange via proposed in our work Virtual Guides ($VirG_{SME}$), should not be shielded by any screen.*

There are a number of laboratories over the World, capable to perform the proposed experimental project. In the case of success, such Nonlocal Signals Detector/Transmitter with variable parameters would be the invaluable tool for extraterrestrial civilizations search in projects, like SETI and for distant cosmos exploration (NASA). On the Earth, the Internet, radio and TV - nets also will get a strong challenge.

### *23.2 GeoNet of Detectors of Water Properties, as a Supersensor of Terrestrial and Extraterrestrial Coherent Signals*

Water is a sensitive detector for any kind of fields and virtual replicas of macroscopic objects. The Sun, Moon and, perhaps, the black hole in center of our galactic are the strongest sources of coherent oscillations of modulated virtual pressure waves of positive and negative energy ($\mathbf{VPW}^+$ and $\mathbf{VPW}^-$), interacting with protons and electrons of water molecules.

The induced by Bivacuum mediated interaction *coherent* changes of water physical properties on the remote points of the Earth surface, registered by CAMP devices, can be analyzed by the global - GeoNet system via the Internet.

The corresponding coherent variations of physical properties of standard aqueous solutions in screened by Faraday cages from EM fields vessels at constant temperature and pressure could be monitored by water properties detectors. Such aqueous samples detectors of certain physical parameters variations, like sound velocity density, refraction index or light scattering will be distributed over the surface of the Earth, forming a nodes of GeoNet.

This author propose to use such GeoNet on the Earth surface, like giant Supersensor for terrestrial and extraterrestrial coherent signals registration. For this end a hundreds of standard water-filled cells with detectors over the planet surface should be under permanent centralized monitoring, using satellites and the Internet.

The Fourier analysis of the terrestrial and extraterrestrial signals, accompanied by registered water perturbations, makes it possible to select only coherent patterns of dynamic changes of water properties in big number of water-filled cells over the Earth. These patterns will be analyzed for getting the detailed information about the amplitude and frequency of coherent signals.

Sensitivity of proposed global sensor system can be combined with existing currently net of random events generators (REG), used in Global Consciousness project (Roger Nelson, 2000 -2008), reinforcing the volume of outcome information.

The localization and forecast of the Earthquakes are a minimum results of such global project realization. This forecasting compensate quickly all related to project of GeoNet expenses.

The valuable knowledge about the influence of gravitational dynamics of Sun, Moon and planets of Solar system on virtual pressure waves ($VPW^+$ and $VPW^-$) of Bivacuum and geophysical process on the Earth could be obtained via proposed GeoNet of aqueous system detectors.

## 24. Experimental data, confirming Unified theory (UT)

### *24.1 Radiation of accelerating charges*

It follows from our theory, that the charged particles, *nonuniformly* accelerating in cyclotron, synchrotron or in undulators, could be a source of photons. It is a result of excitation of secondary *anchor sites* of elementary particles (section 7.5) turning their virtual photons properties to real ones.

The private case of undulator is a free electron laser (FEL): (http://en.wikipedia.org/wiki/Free_electron_laser. It generates tunable, coherent, high power radiation, currently ranging in wavelength from millimeters to the visible. In FEL a beam of electrons is accelerated to relativistic speeds. The beam passes through a periodic, transverse magnetic field. This field is produced by arranging magnets with alternating poles along the beam path. It forces the electrons in the beam to assume a sinusoidal path. The acceleration of the electrons along this path results in the release of a photon.

The *secondary anchor sites* (see section 7.5) of the electron in alternating magnetic field can be treated as a virtual photon (eq.7.46). The absorption of the electron's cumulative virtual cloud ($CVC^{\pm}$) by these *exited* virtual photons creates an actual photon. Adjusting either the speed/energy of the electrons or magnetic field strength tunes their de Broglie wavelength and frequency with their secondary *anchor sites asymmetry,* generating photons over a wide range of frequency. Similar mechanism may be responsible for EM emission in terahertz range by ceramic superconducting films, excited by femtosecond optical pulses (Tonouchi, et. al., 1997).

The energy of electromagnetic radiation $[\hbar\boldsymbol{\omega}_{ph}]$ is dependent on the doubled kinetic energy increment:

$$\Delta(2\mathbf{T}_k) \;=\Delta(\mathbf{m}_V^+\mathbf{v}^2) = \Delta\left(\frac{\mathbf{h}^2}{\mathbf{m}_V^+\boldsymbol{\lambda}_B^2}\right) \qquad 24.1$$

of alternately accelerated charged particles with undulator angular frequency ($\boldsymbol{\omega}_u = 2\pi\mathbf{v}_u$) and related *inelastic* recoil-antirecoil effects. These *local⇌nonlocal* effects with energy ($\pm\alpha\mathbf{m}_V^+\mathbf{v}^2$), accompanied [$\mathbf{C} \rightleftharpoons \mathbf{W}$] pulsation of particles, are responsible for activation of secondary *anchor sites* in Bivacuum matrix:

$$[\hbar\boldsymbol{\omega}_{ph}] \sim \Delta 2\mathbf{T}_k(t) = \frac{h^2}{\mathbf{m}_V^+\boldsymbol{\lambda}_B^2}(\sin\boldsymbol{\omega}_u\mathbf{t}) \;\pm\; \alpha\mathbf{m}_V^+\boldsymbol{\omega}_B^2\mathbf{L}^2(\sin\boldsymbol{\omega}_{C\rightleftharpoons W}\mathbf{t}) \qquad 24.2$$

*where*: $\boldsymbol{\lambda}_B$ and $\boldsymbol{\omega}_B$ are the electron's de Broglie wavelength and frequency;

$\alpha \mathbf{m}_V^+ \boldsymbol{\omega}_B^2 \mathbf{L}^2$ is the energy of the secondary anchor sites, determined by the energy of recoil effect.

We can see, that the alternation of kinetic energy of charged particle can be accompanied by electromagnetic radiation. This effect occur, if the alternation of kinetic energy: $\boldsymbol{\Delta}\mathbf{2T}_k(t)$ and corresponding *inelastic* recoil energy: $\boldsymbol{\Delta}[\alpha \mathbf{m}_V^+ \mathbf{v}^2](t) = \boldsymbol{\Delta}\alpha \mathbf{m}_V^+ \boldsymbol{\omega}_B^2 \mathbf{L}^2(t)$ exceeds the energetic threshold, necessary for photon origination: $\langle [\mathbf{F}_\uparrow^+ \bowtie \mathbf{F}_\downarrow^-]_W + (\mathbf{F}_\uparrow^-)_C \rangle_{p,e} \rightleftharpoons \langle [\mathbf{F}_\uparrow^+ \bowtie \mathbf{F}_\downarrow^-]_C + (\mathbf{F}_\uparrow^-)_W \rangle_{p,e}$,

The uniform acceleration, in contrast to alternative one, do not provide the fulfilment of condition of overcoming of corresponding threshold activation and the EM radiation is absent. Consequently, the real photon radiation by charged particles and other dissipation inelastic process in Bivacuum matrix, are possible only in the conditions of nonuniform particles acceleration.

Some similarity is existing between the mechanisms of *inelastic phonons* excitation in solids, detected by $\gamma$ –resonance spectroscopy, and photons excitation in Bivacuum by alternatively accelerated particle.

One more consequence of Unified Theory, coinciding with experiment, is that synchrotron and undulator radiation should be strongly asymmetric and coincide with direction of charged particle propagation in space.

Most of energy, emitted by relativistic particles is located in direction, close to their beam instant velocity ($\mathbf{v} = \mathbf{v}_{ext} \to \mathbf{c}$) in narrow angles range, determined by semi-empirical expression (Ginsburg, 1987):

$$\boldsymbol{\Delta\theta} \simeq [\mathbf{1} - (\mathbf{v}/\mathbf{c})^2]^{1/2} = \frac{\mathbf{m}_0 \mathbf{c}^2}{\mathbf{E}} \to 0 \qquad 24.3$$

where: $\mathbf{E} = \mathbf{mc}^2 = \mathbf{m}_V^+ \mathbf{c}^2$ is a total relativistic energy of the charged particle.

Our theory leads to same result. Formulas (4.2 and 4.2a) for relativistic condition ($\mathbf{v} \to \mathbf{c}$), can be easily transformed to:

$$[\mathbf{1} - (\mathbf{v}/\mathbf{c})^2]^{1/2} = \left| \frac{-\mathbf{m}_V^-}{\mathbf{m}_V^+} \right|^{1/2} = \frac{\mathbf{m}_0 \mathbf{c}^2}{\mathbf{m}_V^+ \mathbf{c}^2} = \frac{\mathbf{L}^+}{\mathbf{L}_0} \simeq \boldsymbol{\Delta\theta} \overset{\mathbf{v}\to\mathbf{c}}{\to} 0 \qquad 24.4$$

where, the radius of the actual torus, taking into account (4.3), is:

$$\mathbf{L}_V^+ = (\hbar/\mathbf{m}_V^+ \mathbf{v}_{gr}^{in}) \to 0 \quad at\ \mathbf{v} \to \mathbf{c} \qquad 24.5$$

$$\text{as far: } \mathbf{m}_V^+ = \mathbf{m} = \frac{\mathbf{m}_0}{\sqrt{1 - (\mathbf{v}/\mathbf{c})^2}} \to \infty \quad at \quad \mathbf{v} \to \mathbf{c}$$

and the Compton radius of sub-elementary particle is $L_0 = (\hbar/\mathbf{m}_0 \mathbf{c}) = const$

Their ratio determines the angle range of radiation of accelerating particle. As far, in accordance to our approach, the actual energy of particle is determined by the inertial mass: $\mathbf{E} = \mathbf{m}_V^+ \mathbf{c}^2 = \mathbf{mc}^2$, we can see that eq. 24.3 coincides with eq.24.4.

In the angle, defined by 24.4, the probability of excited *secondary anchor sites* is much higher than outside of corresponding cone of action.

*24.2 Artificial generation of unstable groups of virtual particles and antiparticles*

Let us consider the possible results of correlated symmetry shift in groups of virtual Cooper pairs $[\mathbf{BVF}_+^\uparrow \bowtie \mathbf{BVF}_-^\downarrow]_{S=0}^{as} \to [\mathbf{F}^\uparrow \bowtie \mathbf{F}^\downarrow]^{Vir}$ of Bivacuum fermions ($\mathbf{BVF}^\uparrow$) and antifermions ($\mathbf{BVF}^\downarrow$) with opposite spins, acquiring the opposite uncompensated mass: $\boldsymbol{\Delta}\mathbf{m}_\pm = \left(|m_V^+| - |m_V^-|\right)$ and charge: $\boldsymbol{\Delta}\mathbf{e}_\pm = (|e_+| - |e_-|)$ spontaneously or, most probable, in the local gravitational (G), electric (E), magnetic (H) and massless spin (S) fields. These

virtual groups can be considered as a *secondary anchor sites,* activated by the electrons and protons of the nearest material objects and their assembly and disassembly.

*The first stage* of virtual groups formation can be considered, as polymerization of virtual *Cooper pairs* of asymmetric Bivacuum fermions and antifermions to Virtual microtubules (chapter 14):

$$\mathbf{VirMT} = \mathbf{P}(\mathbf{r},\mathbf{t}) \times [\mathbf{F}_+^{\uparrow} \bowtie \mathbf{F}_-^{\downarrow}]^{Vir} \qquad 24.6$$

In primordial Bivacuum the symmetric Bivacuum dipoles of opposite polarization $\mathbf{P}(\mathbf{r},\mathbf{t}) \times [\mathbf{BVF}_+^{\uparrow} \bowtie\ \mathbf{BVF}_-^{\downarrow}]_{S=0}^{s}$, may rotate as respect to each other in opposite direction, keeping their resulting orientation in space permanent with their external tangential or translational velocity equal to zero ($\mathbf{v} = \mathbf{0}$). However, even small symmetry shift between properties of torus ($\mathbf{V}^+$) and antitorus ($\mathbf{V}^-$), caused by the external fields should be accompanied by external circulation with velocity ($\mathbf{v}$) around common axis ($\mathbf{v} > \mathbf{0}$). It follows from (3.11) that:

$$\mathbf{v}^2 = \mathbf{c}^2\left(1 - \frac{|-\mathbf{m}_V^-|}{\mathbf{m}_V^+}\right) > 0, \quad if \quad \mathbf{m}_V^+ > |-\mathbf{m}_V^-| \qquad 24.7$$

$$\mathbf{VirMT} = \mathbf{P}(\mathbf{r},\mathbf{t}) \times [\mathbf{BVF}_+^{\uparrow} \bowtie \mathbf{BVF}_-^{\downarrow}]_{S=0}^{s} \equiv \mathbf{P}(\mathbf{r},\mathbf{t}) \times [(\mathbf{V}^+ \upuparrows \mathbf{V}^-) \ \bowtie\ (\mathbf{V}^+ \downdownarrows \mathbf{V}^-)]_{S=0}^{s} \qquad 24.8$$

$$\overset{\mathbf{Fields}}{<======>} 2[\mathbf{F}_+^{\uparrow} \bowtie \mathbf{F}_-^{\downarrow}]^{Vir} <==> 3[\mathbf{F}_+^{\uparrow} \bowtie\ \mathbf{F}_-^{\downarrow}]^{Vir} <==> \mathbf{P}(\mathbf{r},\mathbf{t}) \times [\mathbf{F}_+^{\uparrow} \bowtie\ \mathbf{F}_-^{\downarrow}]^{Vir} \qquad 24.8a$$

where: $\mathbf{P}(\mathbf{r},\mathbf{t})$ is a number of Cooper pairs of Bivacuum dipoles in $\mathbf{VirMT}$, depending on their length ($\mathbf{r}$) and time ($\mathbf{t}$).

The formation of **VirMT** in symmetric primordial Bivacuum is self-organization process without consuming the external fields energy. However, the presence of fields, turning primordial Bivacuum to secondary one, induce the symmetry shift in pairs $[\mathbf{BVF}^{\uparrow} \bowtie \mathbf{BVF}^{\downarrow}]_{S=0}^{as}$ and *rotation* of **VirMT**, formed by them around central main axes between $\mathbf{BVF}^{\uparrow}$ and $\mathbf{BVF}^{\downarrow}$. The energy of relative rotation of asymmetric pairs around common axis in VirG is dependent on the energy of external field, inducing asymmetry.

*The second stage* - is a result of disassembly of the big coherent clusters (16.10a) to smaller ones, accompanied by violation of equilibrium between densities of virtual particles $\mathbf{n}_+(\mathbf{BVF}^{\uparrow})^{as} \equiv \mathbf{n}_+[\mathbf{F}_{\uparrow}^{+}]^{Vir}$ and antiparticles $\mathbf{n}_-(\mathbf{BVF}_-^{\downarrow})^{as} \equiv \mathbf{n}_-[\mathbf{F}_{\downarrow}^{-}]^{Vir}$, ($\mathbf{n}_+ \neq \mathbf{n}_-$) acquiring, consequently, the uncompensated charge and mass:

$$[\mathbf{n}_+\mathbf{F}_{\uparrow}^{+} \ \bowtie\ \mathbf{n}_-\mathbf{F}_{\downarrow}^{-}]^{Vir} \overset{grad(\mathbf{G},\mathbf{E},\mathbf{H})}{<======>} \qquad 24.9$$

$$\mathbf{n}_- < \mathbf{F}_{\updownarrow}^{-} >_i^{Vir} \ \lesseqgtr \mathbf{n}_+ < \mathbf{F}_{\updownarrow}^{+} >_i^{Vir} \qquad 24.9a$$

where: $i = e, \mu, \tau$ are three electron' generations and the total density of virtual sub-elementary fermions and antifermions is:

$$\mathbf{n} = \mathbf{n}_- + \mathbf{n}_+ \qquad 24.10$$

$$\mathbf{n}_- \neq \mathbf{n}_+ \qquad 24.10a$$

In strong electrostatic fields, like between condenser plates, the virtual Cooper like pairs from Bivacuum fermions of similar symmetry shift, i.e. similar charge, but with opposite direction of rotation (spin) may originate. The formation of corresponding *charged* clusters and $\mathbf{VirG}_{\mathbf{SME}}^{\pm}$ becomes possible in cases, when energy of spin-spin exchange between them exceeds the energy of Coulomb repulsion between Bivacuum fermions of opposite spins:

$$\mathbf{n}_-[\mathbf{BVF}_-^{\uparrow}\bowtie\mathbf{BVF}_-^{\downarrow}]_{S=0}^{as} \sim \mathbf{VirG}_{SME}^{-} \qquad 24.11$$

$$or: \quad \mathbf{n}_+[\mathbf{BVF}_+^{\uparrow}\bowtie\mathbf{BVF}_+^{\downarrow}]_{S=0}^{as} \sim \mathbf{VirG}_{SME}^{+} \qquad 24.11a$$

The shift of equilibrium between densities of asymmetric Bivacuum fermions and antifermions of opposite charges and mass-energy in strong anisotropic electric and gravitational fields is accompanied by generation of non zero difference of positive and negative virtual pressure of Bivacuum:

$$[\pm\Delta\mathbf{VirP}^{\pm} = \mathbf{n}_+(\mathbf{m}_V^+ - \mathbf{m}_V^-)\mathbf{c}^2 - \mathbf{n}_-(\mathbf{m}_V^- - \mathbf{m}_V^+)\mathbf{c}^2\,]^i \qquad 24.12$$

$$or: \ [\pm\Delta\mathbf{VirP}^{\pm} = \mathbf{n}_+(\mathbf{m}_V^+\mathbf{v}^2) - \mathbf{n}_-(\mathbf{m}_V^-\mathbf{v}^2)\,]^i \qquad 24.12a$$

The metastable virtual fermions may fuse to stable real fermions - triplets and photons, if the value of $(\mathbf{BVF}^{\updownarrow})^{as}$ symmetry shifts will increase to that, corresponding to Golden mean condition under the influence of high frequency $\mathbf{VPW}_{q=2,3}^{\pm}$ (see section 12.2).

The dissociation of metastable neutral Virtual Guides or Bivacuum fermions clusters, like secondary *anchor sites* of elementary particles to charged virtual fragments with fermion properties is energetically much easier, than that of stable photons, and may occur even in weak fields gradients.

Synchronization of $[C \rightleftharpoons W]$ pulsation of such virtual unstable fermions, as a condition of entanglement between them, provides their collective behavior even after big $\mathbf{VirG}_{SME} = \mathbf{P}(\mathbf{r},\mathbf{t})$ dissociation to coherent groups ($\mathbf{n}_-\mathbf{e}^-$ and $\mathbf{n}_+\mathbf{e}^+$, where $\mathbf{n}_\pm \succeq 10$) and their spatial separation.

The results, confirming our scenario of coherent groups of metastable charged particles origination from asymmetric **VirMT**, has been obtained in works of Keith Fredericks (2002) and Sue Benford (2001). Fredericks analyzed the trucks on Kodak photo-emulsions, placed in vicinity of human hands during 5-30 minutes. The plastic isolator was used between the fingers and the photographic emulsion. *The tracks in emulsions point to existing of correlation in twisting of trajectories of big group of charged particles (about 20) in a weak magnetic field.* The in-phase character of set of the irregular trajectories may reflect the influence of geomagnetic flicker noise on groups of correlated charged particles.

In these experiments the Bivacuum symmetry shift, necessary for dissociation of virtual Bivacuum dipoles clusters on charged virtual fermions, can be induced by the electric, magnetic fields and nonlocal spin/torsion field. These fields can be excited by 'flickering' water clusters in microtubules of the nerve cells bodies and axons of living organisms in the process of nerve excitation (Kaivarainen, 2002; 2003; 2004).

The corresponding [dissociation ⇌ association] of coherent water cluster in state of mesoscopic molecular Bose condensate (**mBC**) is accompanied by oscillation of the $H_20$ dipoles angular momentum vibration with the same frequency about $10^7\ s^{-1}$. If the flickering of water clusters in MTs of the same cell or between 'tuned' group of cells occurs in-phase, then the cumulative effect of modulated $\mathbf{VirSW}_{\mathbf{m}}^{\pm 1/2}$ and EM field generation by human's finger near photoemulsion can be strong enough for stimulation of dissociation of virtual vortices (16.11a) to virtual electrons and positrons, producing the observed tracks in photoemulsion or photofilm.

In work of Benford (2001) the special device - *spin field generator* was demonstrated to produce a tracks on the dental film, placed on a distance of 2 cm from generator and exposed to its action for 7 min. The spin field generator represents rotating hollow cylinder or ring made of ferrite-magnetic material with the axis of rotation coinciding with the cylinder's main symmetry axis. Four permanent (wedge-like) magnets are inserted into the cylinder. It rotates with velocity several thousand revolutions per minute.

The effect of this generator is decreasing with distance and becomes undetectable by the dental films after the distance from the top of cylinder bigger than 8 cm. The dots and tracks on dental X-ray films were reproduced over 200 trials. They are close to the regular charged particle tracks on surface emulsions. However, the more exact identification of particles failed. The uncommon features of these tracks may be a result of unusual properties of short-living virtual electrons, positrons, protons and antiprotons and their coherent clusters.

### *24.3 Michelson-Morley experiment, as a possible evidence of the Virtual Replica of the Earth*

The experiments, performed in 1887 by Michelson-Morley and similar later experiments of higher precision, has been based on checking the difference of light velocity in the direction of Earth orbiting around the Sun and in the direction normal or opposite to this one. In the case of *fixed ether* with certain medium properties, *independent of the Earth motion*, one may anticipate that the difference in these two light velocities should exist. The absence of any difference was interpreted by Einstein, as the absence of the ether. This conclusion was used in his Special Relativity (SR) theory for postulating of permanency of light velocity, *but different time* in different inertial systems. The time of inertial system in SR is dependent on system velocity as respect to the light velocity. The *principle of relativity* of SR states that, regardless of an observer's position or velocity in the universe, all physical laws will appear constant. From this principle, it follows that an observer cannot determine either his absolute velocity or direction of travel in space. This principle includes statement of the *absence of the absolute velocity.*

In accordance to our approach to time problem, the time is a characteristic parameter of conservative system, equal to infinity in the absence of acceleration at any permanent kinetic energy of particles, forming such systems. *So, in contrast to special relativity, the time in our theory is infinitive and independent on velocity in any inertial system.* For the other hand at any nonzero acceleration, for example, centripetal in the case of orbital rotation of particles/objects the time is dependent on tangential velocity of these objects. There are no physical systems in Nature, which can be considered, as perfectly inertial, i.e. where any acceleration is absent. However, the situations are possible where the opposite accelerations and forces compensate each other and the resulting one is zero.

For example, this takes a place in free-fall or satellite systems, when *centripetal, i.e. gravitational:* $a_{cp} = GM/r^2$ and centrifugal ($a_{cf}$) accelerations compensate each other:

$$a_{res} = a_{cp} + a_{cf} = 0 \qquad 24.13$$

It is so called *equivalence principle*, used in General Relativity (GR) theory. The kinetic energy of such mechanical system/object can be permanent, however the *potential energy* and force of stretching ($\mathbf{F}_{str}$) of object increases proportional to sum:

$$(|a_{cp}| + |a_{cf}|) \sim 2GM/r^2 \qquad 24.14$$

and elastic deformation of the object. At certain big enough stretching energy, equal to *stress-energy*, the object can be destroyed and the kinetic energy of such system will increase also.

The statement of General Relativity, that condition $(24.13)$, true for geodesic motion, is a condition of *inertial motion* of object, as defined by the 1st Newton law, is wrong. The Newton law of inertia is strictly applicable for ideal conditions, where any kind of forces, acting on material point/object's external or internal dynamics (kinetic or potential energy) are absent.

In General Relativity (GR), geodesics are the idealized world lines of a particle *free*

*from all external force*. In GR the gravity is not a force but a curved space-time geometry where the source of curvature is the stress-energy tensor. This means, that gravitational force do not act on particle itself, but on space curvature, changing correspondingly the trajectory of particle. This principle of GR looks very artificial and nonrealistic. In all known real examples of geodesic motion, the object/particle is not free *from all external force*, but is a result of opposite forces compensation of each other.

The conjecture of virtual replica (VR) allows another interpretation of Michelson-Morley experiments. The VR of the Earth or any other material object is a result of interference of basic $\mathrm{VPW}_0^{\pm}$ with positive and negative Bivacuum virtual pressure waves ($\mathrm{VPW}_m^{+}$ and $\mathrm{VPW}_m^{-}$), modulated by the object's particles de Broglie waves. The VR has properties of quantum hologram.

The Ether component of VR may have at least as big diameter, as the Earth atmosphere and it moves in space together with planet. It is obvious, that in such 'virtual shell' of the Earth the light velocity could be the same in any directions.

This author propose the experiment, which may confirm the existence of both: the VR and the Aether/Bivacuum, as a superfluid medium with certain mechanical properties, like compressibility providing the $\mathbf{VPW}^{\pm}$ existing. For this end we assume that the properties of VR on distance of about few hundred kilometers from the planet surface differs from that on the surface.

If we perform one series of the Michelson-Morley like experiments on the satellite, rotating with the same angular frequency and velocity as the Earth, i.e. fixed as respect to the Earth surface and another series of experiments on the surface, the *existence of difference* in results will confirm our Virtual Replica theory and the Bivacuum model with Ether properties.

The absence of difference in light velocity in opposite direction as respect to Earth trajectory in Michelson-Morley experiments can be explained in two different ways:

1. As a result of equality of light velocity in any directions, independently on direction of Earth translational propagation in space (confirmation of the Einstein relativity principle and the absence of the Ether);

2. As a result of correlation between magnetic and gravitational fields of the object, like Earth, and Bivacuum dipoles symmetry shift in surrounding object Bivacuum (ether virtual replica of the Earth), increasing the refraction index of Bivacuum and decreasing little bit light velocity isotropically and independently on the direction of the Earth motion. This explanation is compatible with the ether drug concept.

Consequently, the absence of difference in light velocity in Michelson-Morley like experiments, in is not a strong evidence against Ether with Bivacuum properties.